\documentclass{aa}  

\usepackage{graphicx}
\usepackage{txfonts}
\usepackage{lscape}
\usepackage{subfig}
\usepackage{amssymb,natbib,supertabular}
\usepackage{defs}
\usepackage{appendix}
\usepackage{color}
\usepackage{longtable}
\usepackage{supertabular}
\usepackage[breaklinks, colorlinks, citecolor=blue]{hyperref}
\usepackage{hyperref}
\begin{document}

   \title{Molecular gas properties of \planck-selected protocluster candidates at $z{\simeq}$1.3--3}

   \subtitle{}

\author{M.~Polletta\inst{\ref{inst1}}
\and H.~Dole\inst{\ref{inst2}} 
\and C.~Martinache\inst{\ref{inst2}} 
\and M.~D.~Lehnert\inst{\ref{inst3}}
\and B.~L.~Frye\inst{\ref{inst4}}
\and R.~Kneissl\inst{\ref{inst5},\ref{inst6}}
}
  \offprints{M. Polletta\\ \email{maria.polletta@inaf.it}}
\institute{INAF - Istituto di Astrofisica Spaziale e Fisica cosmica (IASF) Milano, via A. Corti 12, 20133 Milan, Italy\label{inst1}
\and Universit\'{e} Paris-Saclay, Institut d'Astrophysique Spatiale, CNRS, b\^at 121, 91400 Orsay, France\label{inst2}
\and Universit\'e Lyon 1, ENS de Lyon, CNRS UMR5574, Centre de Recherche Astrophysique de Lyon, F-69230 Saint-Genis-Laval, France\label{inst3}
\and Department of Astronomy/Steward Observatory, 933 North Cherry Avenue, University of Arizona, Tucson, AZ 85721, USA\label{inst4}
\and European Southern Observatory, ESO Vitacura, Alonso de Cordova 3107, Vitacura, Casilla, 19001 Santiago, Chile\label{inst5}
\and Atacama Large Millimeter/submillimeter Array, ALMA Santiago Central Offices, Alonso de Cordova 3107, Vitacura, Casilla, 763-0355 Santiago, Chile\label{inst6}
}

   \date{Submitted 19 September 2021}

  \abstract{We report on IRAM-30m/EMIR observations of 38 \herschel\ sources
chosen as the brightest \textit{red} sub-millimeter (sub-mm) sources in 18
\planck-selected fields drawn from the \planck\ high-$z$ (PHz) protocluster
candidates sample.  These fields host overdensities of red \herschel\
sources, with high star formation rates ($\sim$10,000\,\msun\,yr$^{-1}$) as
derived from \planck\ measurements.  The goals of these observations are to
measure spectroscopic redshifts of the \herschel\ sources in the PHz fields,
investigate the origin of their bright sub-mm emission, and find evidence of
their association with high-$z$ protoclusters.  We detect 40 CO lines on a
total of 24 bright ($S_{\rm 350\mu m}{>}$40\,mJy) \herschel\ sources in 14
of the 18 PHz fields.  The measured average redshift is ${<}z_{\rm
CO}{>}$\,=\,2.25$\pm$0.09, spanning a range from 1.32 to 2.75.  We measure
redshifts for multiple \herschel\ sources in close projected proximity in
eight PHz fields.  In half of those fields we detect from two to three
objects at similar redshifts, supporting the idea that the PHz fields
contain high-$z$ protoclusters.  The detection of sources at different
redshifts in the same field demonstrates that foreground and background
sources also contribute to the total sub-mm emission.  We compare the
properties of the molecular gas, and of the star formation activity of our
sources with samples of normal star-forming galaxies (SFGs), sub-mm galaxies
(SMGs), and CO-detected cluster and protocluster galaxies drawn from the
literature at similar redshifts.  We find that the PHz-IRAM sources are
mainly normal SFGs, with only $\sim$20\% undergoing a starburst phase.  The
PHz-IRAM sources are characterized by star formation rates
(${<}$SFR${>}$\,=\,1043$\pm$157\,\msun\,yr$^{-1}$) and gas masses
($<$M$_\mathrm{gas}{>}$\,=\,(4.0$\pm$0.7)$\times$10$^{11}$\,\msun) that are,
on average, eight and five times higher than those typical of normal SFGs at
similar redshifts.  Their dust temperatures
($<$T$_\mathrm{dust}{>}$\,=\,29.2$\pm$0.9\,K), and depletion timescales
(${<}\tau_\mathrm{dep}{>}{=}$0.47$\pm$0.07\,Gyr) are instead consistent with
those of normal SFGs.  The analysis of the CO spectral line energy
distribution, available for ten PHz-IRAM sources, peaks at a low quantum
rotation number ($J_\mathrm{up}$\,=\,3) in most of the cases, implying low
gas excitation. These properties imply that a significant fraction
of the PHz-IRAM sources contains extended, and cold molecular gas reservoirs
at low excitation, and that their star-formation is driven by secular
processes.  Multiplicity and moderate gravitational lensing might also play
a role in producing the observed properties.  Some of these properties are
also observed in other CO-detected infrared-luminous protocluster galaxies
at $z{\simeq}$1.3--3.  We find that the highest star-forming protoclusters
drawn from the largest volume simulations available to date, have similar
SFRs as the PHz protoclusters, but separate out into a larger number of
star-forming galaxies.  Millimeter and CO observations at higher spatial
resolution than those presented here would be necessary to further elucidate
the properties of our PHz-IRAM sources, and determine which mechanisms drive
star-formation in infrared-luminous protocluster galaxies.  }


   \maketitle
%

\section{Introduction}

The study of galaxy protoclusters probes the early formation of massive
structures and galaxy clusters, as well as the growth and evolution of the
massive quiescent galaxies that dominate the cluster population at
$z{<}$1.5~\citep[see ][for a review]{overzier16}.  
Protocluster are expected to be found as overdensities of sub-millimeter
(sub-mm) bright dusty star-forming galaxies~\citep[DSFGs, also known as
sub-mm galaxies, or SMGs][]{casey15} at $z\gtrsim$2 covering wide
areas~\citep[tens of Megaparsec; ][]{muldrew15}, and with total star
formation rates (SFRs) of thousands of \msun\,yr$^{-1}$.

Several observational works have reported the discovery of DSFG-rich
structures with large SFRs for single members and for the whole
structure~\citep{ivison13,chapman15,casey15,wang16,kato16,coogan18,lee17,oteo18,lacaille19,gomez19,
tadaki19,rotermund21,calvi21}.  Models struggle to reproduce the large
amount of on-going star formation measured in these
protoclusters~\citep{granato15,lim21}, and to explain the large number of
highly active star-forming members~\citep{casey15,hung16,casey16}.  The two
main proposed scenarios to explain their large SFRs are burst events, likely
driven by mergers, or secular processes, like cold gas
accretion~\citep{narayanan15,casey16}.  Cold gas accretion is expected to
fuel galaxies hosted by low mass halos (M$_{\rm
halo}{\lesssim}$10$^{12}$\,\msun) or through cold flows~\citep{keres05} in
massive haloes ($>$10$^{12}$\,\msun), but only above a certain redshift
(i.e., z$\gtrsim$2).  In massive halos at z$\lesssim$2 the gas is
expected to be shock-heated to the virial temperature of the halo and the
cooling time is too long to fuel star formation~\citep{dekel06}.  These
different cooling time regimes should leave an imprint in the cold gas
properties of protocluster galaxies.  If cold gas is replenished, the
activity could be sustained for longer timescales ($\sim$1\,Gyr), thus
increasing the probability of observing multiple active SFGs
simultaneously~\citep{narayanan15}.  The merger scenario would instead be
favored by measuring short gas depletion timescales ($\tau_{\rm dep}$), and
large star formation efficiencies (SFEs) in SFGs located in high-$z$
protoclusters.

The study of DSFGs in protoclusters offer also the opportunity to
investigate the mechanisms that halt their star formation.  The
fraction of quenched galaxies in high-$z$ ($z\geq$1.8) clusters is already
very high, reaching 100\% at large masses (M$_{\rm
star}{>}$10$^{11}$\,\msun)~\citep[see ][]{strazzullo18,newman14}.  This
large fraction suggests that such a transformation occurs in protocluster at
$z{\gtrsim}$2.  Indeed, there is observational evidence that protocluster galaxies
experience earlier quenching than field
galaxies~\citep{hatch11,zirm12,casey16}.  The dependence of quenching
efficiency with stellar mass, and with the environment in high-$z$
protoclusters is, however, still unconstrained.

Molecular gas studies of protocluster galaxies are crucial to establish
their assembly, growth, and evolution.  Measurements of the content and
distribution of cold gas in protocluster galaxies carry information on
processes of cold gas inflow, outflow, and consumption, whether for example
the cold gas is continuously replenished or quickly consumed, as well as on
the processes that trigger, regulate, and halt star formation.
As more and more protoclusters are being discovered, thanks to facilities
like ALMA and IRAM, the number of protocluster galaxies with cold gas mass
estimates has
burgeoned~\citep{aravena12,wagg12,casasola13,ivison13,tadaki14,umehata15,stach17,noble17,dannerbauer17,lee17,miller18,coogan18,tadaki19,kneissl19}. 
Some studies find that gas masses and fractions in protocluster members are
consistent with field scaling relations, implying a total gas density higher
inside the protocluster than in the field because of the large number
density~\citep{lee17}.
Other molecular gas studies of DSFG in protoclusters at $z\sim$2.4--2.5
reveal, instead, large SFEs powered by major
mergers~\citep{ivison13,tadaki14}.  In the protocluster around 4C\,23.56 at
$z\,{=}\,$2.49, \citet{lee19} find some evidence of broader CO lines in the
protocluster members with respect to field galaxies, possibly due to ongoing
mergers.  A study carried out on 16 CO-detected members, all H$\alpha$
emitters, in three protoclusters, reports a mass-dependent environment
effect on the gas fraction, with protocluster members with
M$_\mathrm{star}<$10$^{11}$\,M$_{\odot}$ having larger gas fractions than
field galaxies~\citep{tadaki19}.  By contrast, two independent studies of
CO-detected protocluster galaxies find lower gas fractions compared to field
galaxies in the members situated in the protocluster center~\citep{wang18},
and, more tentatively, in those with large stellar
masses~\citep[$\gtrsim$10$^{11}$\,\msun;][]{kneissl19}.  These results
suggest that the environment effects on the gas properties might be mass
dependent~\citep[e.g., ][]{tadaki19,kneissl19}.  A study of the molecular
gas in the XCS\,J2215.9$-$1738 cluster at $z$\,=\,1.46 find that galaxy
members are relatively gas-rich when they first enter the cluster, and they
become gas-poor as they approach the cluster centre~\citep{hayashi17}.  This implies that the
cooler dust and gas components within cluster galaxies may be influenced by
their environment, reducing the gas reservoir available for their subsequent
star formation.  In summary, the environment seems to play a role in halting
gas accretion (i.e.  through starvation), and/or reducing and removing gas
content~\citep[i.e.  through ram-pressure stripping;
][]{hayashi17,wang18,foltz18}.  To understand how a dense environment
affects star-formation (triggering, fueling, and quenching mechanisms) it is
crucial to study the member cold gas properties and compare them with well
matched field samples~\citep[see e.g., ][]{tadaki19}.

In this work, we present a study of the molecular gas properties derived
from observations with the IRAM/30-m telescope of 38 protocluster member
candidates.  These were selected as bright \textit{red}\footnote{A SPIRE source is defined
\textit{red} if detected in all three SPIRE bands and has flux
density ratios
S$_\mathrm{500}$/S$_\mathrm{350}{>}$0.6 and S$_\mathrm{350}$/S$_\mathrm{250}{>}$0.7~\citep[see ][]{planck15}.} \herschel/SPIRE sources
situated in 18 \planck-selected fields drawn from the \planck\ high-$z$
sample~\citep[PHz; ][]{planck15,planck16}.  The main goals of this work are
to measure their redshifts, and the properties of the molecular gas, and to
investigate the origin of their large SFRs.

The paper is structured as follows.  The PHz sample, and the IRAM-30m
observed subsample are described in Sect.~\ref{sec:phz}, and
Sect.~\ref{sec:phz_iram}.  In Sect.~\ref{sec:spire_density}, we quantify the
sub-mm galaxy overdensity in each field.  The IRAM observations and
strategy, the data reduction and the line measurements are described in
Sect.~\ref{sec:emir_observations}.  The analysis of the molecular gas
properties, dust temperatures, CO and IR luminosities, SFRs, molecular gas
masses, and $\tau_{\rm dep}$ are presented and compared with field galaxies
and other protocluster galaxies from the literature in
Sect.~\ref{sec:analysis}.  In Sect.~\ref{sec:discussion}, we discuss our
findings and interpretation.  Conclusions are given in
Sect.~\ref{sec:conclusion}.

Throughout this work we adopt a \citet{chabrier03} initial mass function
(IMF), and we denote the stellar mass with $\mathcal{M}$.  We assume a flat
$\Lambda$ cold dark matter (CDM) model, with cosmological parameters from
the Planck 2018 release~\citep[i.e., $\Omega_{\Lambda}$\,=\,0.685;
$\Omega_{\mathrm{m}}$\,=\,0.315; H$_{\mathrm{0}}$\,=\,67.4\,\kms\,Mpc;
][]{planck_cosmo18}.

\section{The \planck\ high-$z$ sources}\label{sec:phz}

\planck\footnote{\planck\ (http://www.esa.int/Planck) is a project of the
European Space Agency (ESA) with instruments provided by two scientific
consortia funded by ESA member states (in particular the lead countries
France and Italy), with contributions from NASA (USA), and telescope
reflectors provided by a collaboration between ESA and a scientific
consortium led and funded by Denmark.} all-sky observations have provided a
sample of protocluster candidates, called \planck\ high-$z$ sources, or
PHz\footnote{The PHz catalog is available at
https://archives.esac.esa.int/ doi/html/data/astronomy/planck/Catalogue\_PHZ.html.}.
These were selected as bright sub-mm sources with red sub-mm colors implying
$z$\,=\,2--4 and total SFRs of the order of several thousands of
M$_{\odot}$\,yr$^{-1}$~\citep{planck16}.  The PHz selection requires a
$>$5$\sigma$ detection in the so called {\tt red-excess} (RX) 550\um\
map\footnote{The RX map is obtained after subtracting from the map at
550\um\ the image obtained by linearly interpolating the signal in the
350\um, and 850\um\ maps.}, and a $>$3$\sigma$ detection in the {\tt
cleaned}\footnote{The \planck\ maps cleaning procedure consists in removing
emission from the cosmic microwave background and foreground Galactic
emission.} maps at 350, 550, and 850$\mu$m.  The final PHz position is
obtained through a double Gaussian fit in the {\tt cleaned} 550\um\ map. 
The PHz catalog contains 2151 sources~\citep[for more details on the
selection procedure and the catalog see ][]{planck16}.

\herschel\footnote{\herschel\ is
an ESA space observatory with science instruments provided by European-led
Principal Investigator consortia and with important participation from
NASA.} SPIRE~\citep{pilbratt10,griffin10} observations of 228 of these {\it
Planck} sources have shown that they contain concentrations ($\sim$10
sources on average within the \planck\ beam, i.e.,
FWHM\,=\,4.6\,arcmin at 350\um) of \textit{red} (S$_{\rm 350 \mu
m}$/S$_{\rm 250 \mu m}{>}$0.7, and S$_{\rm 500 \mu m}$/S$_{\rm 350 \mu
m}{>}$0.6) sub-mm galaxies consistent with overdensities of galaxies at
$z$\,$\simeq$\,2--3, and total SFRs of
3000--15,000\,\msun\,yr$^{-1}$~\citep{planck15,clements14}.  Follow-up
observations with {\it Spitzer}/IRAC of 82 PHz have shown that they are
among the regions in the known Universe with the largest concentrations
($\sim$15--25 sources arcmin$^{-2}$) of IRAC red ($S_{\rm 4.5\mu m}$/$S_{\rm
3.6\mu m}{>}$1) galaxies~\citep{martinache18}.  All of these observations
strongly suggest that the PHz are highly star-forming over-densities at
$z$\,=\,2--3.

To assess whether they contain galaxy overdensities that will collapse to
become galaxy clusters, and to understand the origin of their bright and red
sub-mm fluxes, it is necessary to resolve the sub-mm emission, identify the
single sub-mm galaxies and determine their redshifts and physical
properties~\citep[see e.g.][]{flores16,kneissl19,koyama21,polletta21}.  To
this end, we carried out IRAM 30-m observations of a sub-set of PHz fields
with the goals of determining their redshifts and investigating the origin
of the sub-mm emission.

\subsection{The PHz-IRAM targets}\label{sec:phz_iram}

\begin{figure} 
\centering
\includegraphics[width=9cm]{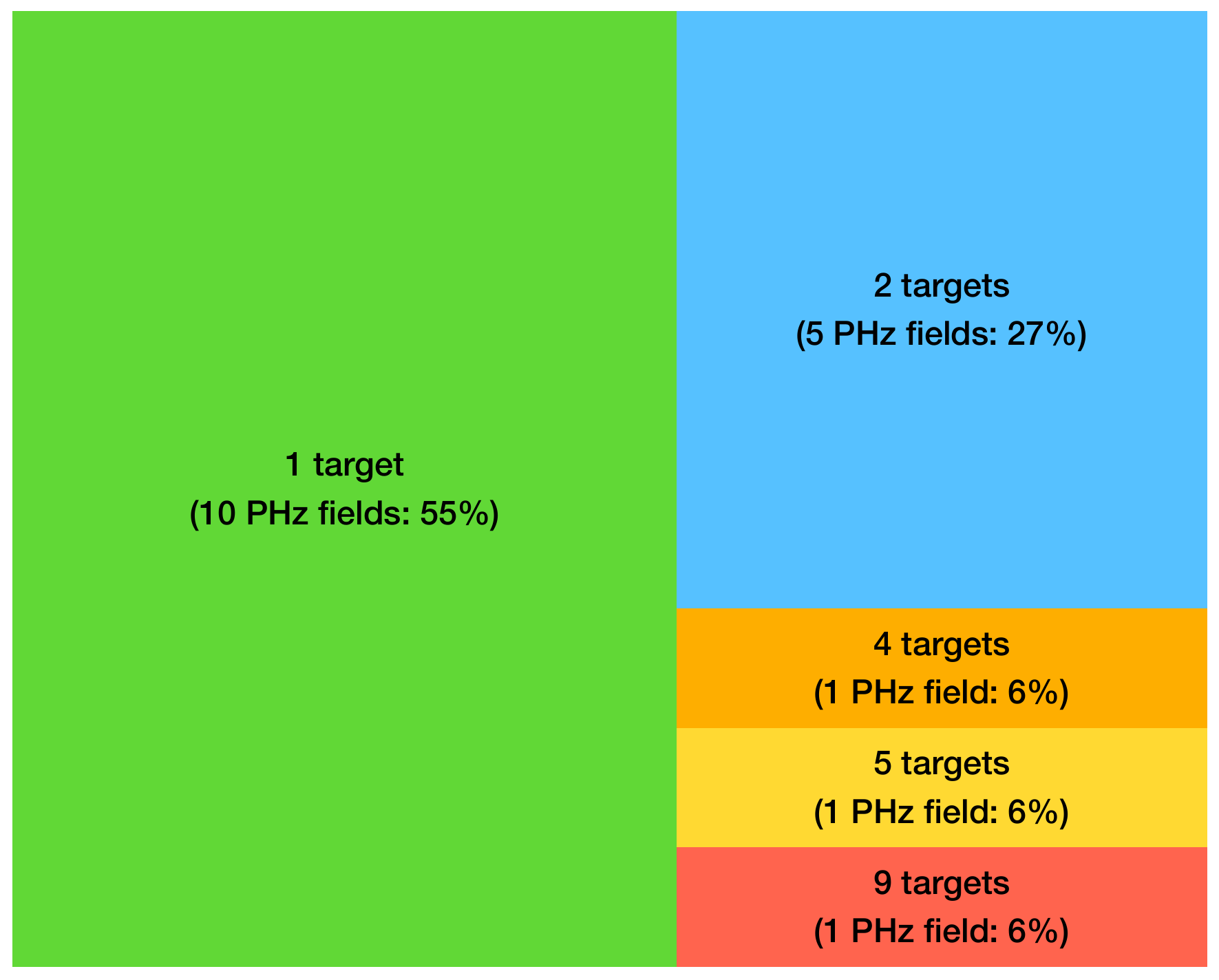}
\caption{{\small Scheme representing the number of fields with one (green),
two (light blue), four (orange), five (yellow), or nine (red) targets observed with
IRAM-30m.  The area of each rectangle is proportional to the number of
fields.}}
\label{fig:iram_sample}
\end{figure}

The targets selected for the IRAM-30m observations were drawn from a subset
of \planck\ sources in the northern hemisphere previously observed with both
\herschel\ and \spitzer.  The selected targets are 38 \textit{red} SPIRE
sources distributed over 18 PHz fields.  Multiple targets were selected in
the same field in eight cases (five fields with two targets, and one with 4,
5, and nine targets), and a single target was observed in ten fields (see
Fig.~\ref{fig:iram_sample}).  Four of the selected targets were also
detected by SCUBA-2 at 850\,$\mu$m~\citep{mackenzie17}.  The list of
targets, along with the measured SPIRE and SCUBA-2 flux densities are listed
in Table~\ref{tab:targets}.  The selected sample includes ten fields,
identified by a PHz ID$>$120000, that are not in the published PHz
catalog~\citep{planck16}, but were detected as bright and red sources in a
preliminary version of the PHz sample obtained from \planck\ maps convolved
with a 8\,arcmin spatial resolution, and a more relaxed masking criterion. 
Three of these additional sources (G112, G143, and G052) are also present in
the second Planck Catalogue of Compact
Sources~\citep[PCCS2;][]{planck_pccs2}.  \spitzer/IRAC data are available in
all of the fields, and ground-based multiband optical and near-infrared data
in a sub-set.  These data will be used to study the multiwavelength emission
of the SPIRE counterparts and will be presented in a forthcoming paper.

The selection of the PHz fields for the IRAM observations is the result of
ten different programs carried out during the past six years (2016-2021). 
The different number of targets per field is due to a change in the
observing strategies throughout the project.  Initially, we aimed at
measuring one redshift per field.  To this end, we selected fields with a
bright ($S_{\rm 350\mu m}{>}$75\,mJy) \textit{red} SPIRE source,
and with a significant overdensity of SPIRE sources~\citep{planck16}.  In
those fields, the observing strategy was to observe the brightest
\textit{red} SPIRE source using a wide frequency coverage to search for a CO
line.  Subsequently, we either looked for an additional CO line in the
detected targets, and/or observed additional SPIRE sources in the same field
where we had a redshift.  The goal of these additional observations was to
detect other sources at the same redshift.  For this purpose, we decreased
the density flux limit at 350$\mu$m from 75\,mJy to 40\,mJy, and gave
priority to fields with sufficiently bright secondary targets, and for which
ancillary multiwavelength data were available. We observed three additional
fields for specific reasons.  First, we observed G073 because previous ALMA
observations had serendipitously detected a CO line in this
field~\citep{kneissl19}.  In G073, we targeted a CO-detected source
(G073\,03) with the goal of confirming the redshift by detecting another CO
transition, and we also targeted another SPIRE source (G073\,15) with the
goal of identifying a CO line at the same redshift.  Second, we observed
G088 because of its significant overdensity of \textit{red} SPIRE sources
and the availability of HerMES/FLS data~\citep{oliver12}.  Last, we observed
G237 because it contains a spectroscopically confirmed
protocluster~\citep{polletta21, koyama21}.  We targeted a faint SPIRE source
(G237\,9741) because associated with a spectroscopic member, and the
brightest \textit{red} SPIRE source in the field (G237\,962).  All
these additional targets are \textit{red}, and four out of five also satisfy
the flux limit criterion ($S_{\rm 350\mu m}{>}$40\,mJy). In one
field (G112), a mistake was made in importing the coordinates during the
observations, and the pointing ended up being 17\arcsec\ from the selected
brightest SPIRE source, and closer (at 12\arcsec) to another SPIRE source
(G112\,06) with $S_{\rm 350\mu m}{=}$68\,mJy, but not \textit{red}.
Thus, all the observed targets have $S_{\rm 350\mu m}{>}$40\,mJy,
with the exception of G237\,9741, and 17 have $S_{\rm 350\mu m}{>}$75\,mJy.
In the following, we will refer to all the selected targets as the PHz-IRAM
sample.

\begin{figure} 
\centering
\includegraphics[width=\linewidth]{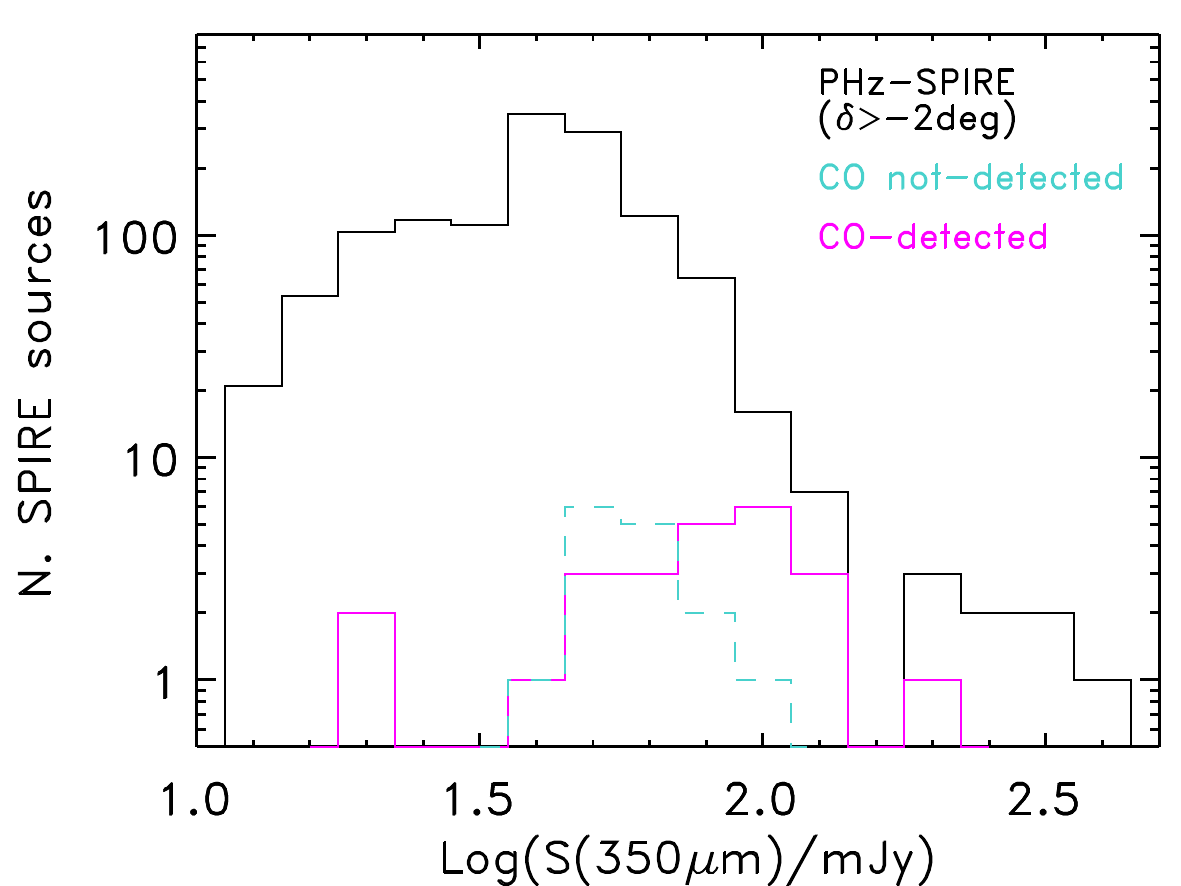}
\caption{{\small SPIRE-350$\mu$m flux density distribution of
the targets observed with IRAM in this study (magenta line: CO-detected, and
turquoise line: CO not-detected), and of all 
SPIRE sources (black line) situated in the PHz fields in the region of the sky visible
from the IRAM-30\,m telescope site ($\delta{\gtrsim}-$2\deg).}}
\label{fig:f350_histo}
\end{figure}

\begin{figure*} 
\centering
\includegraphics[width=15cm]{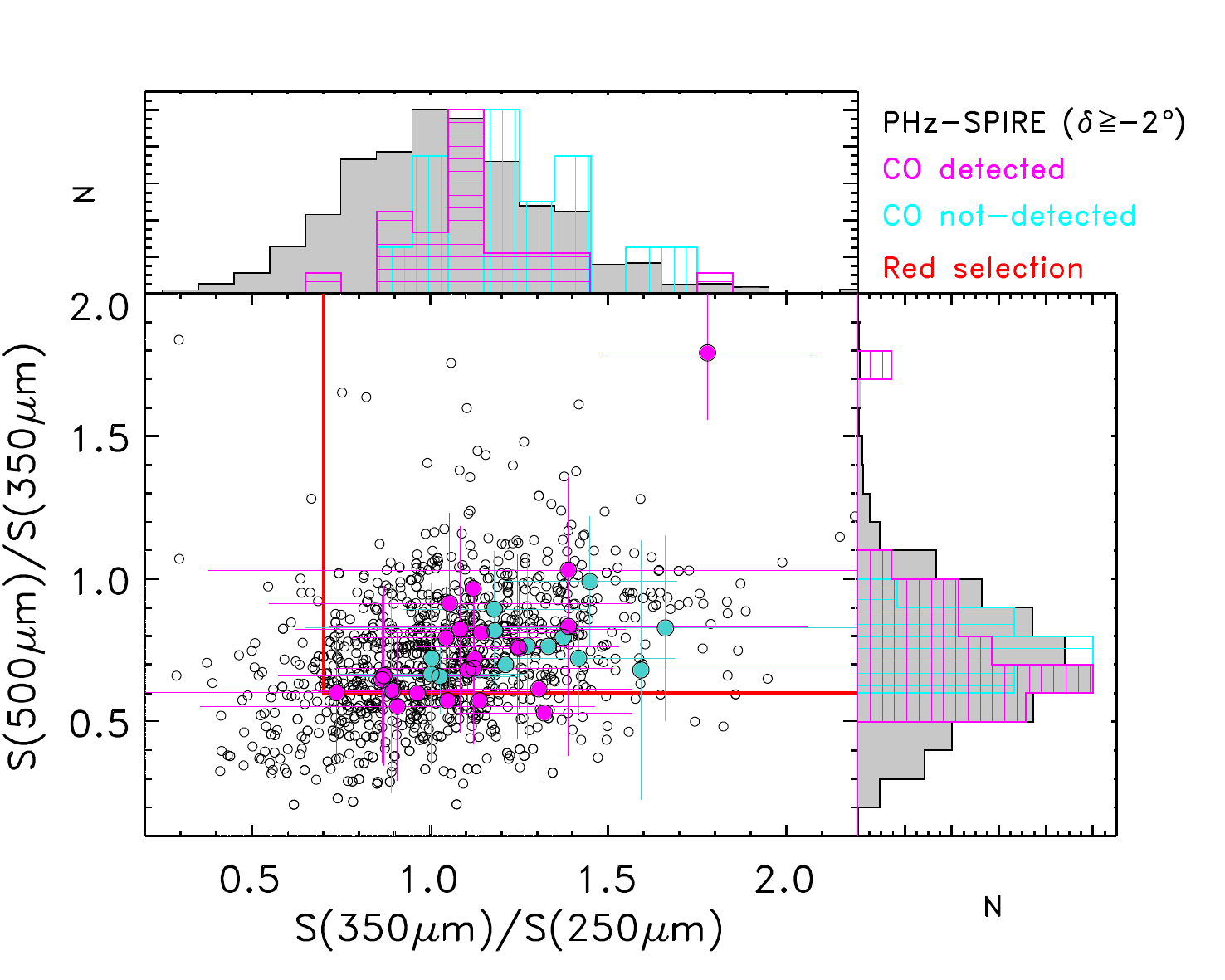}
\caption{{\small SPIRE colors of the targets selected for IRAM observations
(magenta full circles: CO-detected, and turquoise full circles: IRAM observed, but
not CO-detected), and of all SPIRE sources (black circles) situated in the PHz
fields in the region of the sky visible
from the IRAM-30\,m telescope site ($\delta{\geq}-$2\deg). The red lines
enclose the region of \textit{red} SPIRE colors. The \textit{top} and
\textit{right} panels show the sub-mm color distributions normalized at the peak
(grey: all PHz-SPIRE sources at $\delta{\geq}-$2}; magenta: CO detected, cyan: CO not-detected).}
\label{fig:phz_colors}
\end{figure*}

In Fig.~\ref{fig:f350_histo}, we compare the SPIRE flux density at
350$\mu$m of the PHz-IRAM sample (magenta and turquoise lines) with all SPIRE
sources in the PHz fields in the region of the sky visible from the
IRAM-30\,m telescope site ($\delta{\gtrsim}-$2\deg) and with \herschel\
coverage (1268 sources over 100 PHz fields).  The sub-mm colors of the
selected targets (magenta and turquoise full circles, and histograms) are
compared with those of the parent sample (black circles, and grey filled
histograms) in Fig.~\ref{fig:phz_colors}.  The selected targets are among
the brightest \textit{red} SPIRE sources located in the PHz fields, and
quite similar in terms of colors to the parent sample, with the only
difference that they lack the bluest sources.  In a handful of
cases, S$_\mathrm{500}$/S$_\mathrm{350}$ is slightly lower than 0.6, the
nominal threshold to classify a SPIRE source as red.  This is the case
of G112\,06, selected because of a mistake in importing the coordinates, of three
sources (G073\,03, G124\,01, G124\,02) where the flux density at 500$\mu$m had
not been correctly deblended at the time the sources were selected, and of
one case (G131\,15) where the target was a secondary target. In all these
sources the flux ratio is just below the threshold
(0.55${\leq}S_\mathrm{500}$/S$_\mathrm{350}{<}$0.6) as illustrated in
Fig.~\ref{fig:phz_colors}.

\subsection{SPIRE source density}\label{sec:spire_density}

We characterize the SPIRE source overdensity of each targeted field
following~\citet{clements14}, and~\citet{polletta21}.  Briefly, we build a
density map considering all \textit{red} SPIRE sources over a
20\arcmin$\times$20\arcmin\ region with ${>}1\sigma$ detection in all 3
SPIRE bands, and a ${>}3\sigma$ detection in at least one band. For each
selected SPIRE source, we compute a flux-weighted local density given by the distance
distribution to the nearest five neighbors as:
\begin{equation}
\delta_i  = \frac{W_i}{\pi\,d_{i,5}^2} \sum_{j=1}^{5}
\exp\left[-0.5\left(\frac{d_{i,j}}{d_{i,5}}\right)^2\right]
\label{clements_eq}
\end{equation}
where $d_{ij}$ is the distance to the $j^{th}$ source, $d_{i,5}$ is the
distance from the $i^{th}$ source to the 5$^{th}$ nearest neighbor, and
$W_i$ is a weight given by the ratio between the $i^{th}$ source
flux density at 350$\mu$m and the sum of the 350$\mu$m flux densities from
all sources in the field~\citep[adapted from ][]{clements14}.  From the
$\delta_i$ value of each SPIRE source, we create a map after convolution
with a Gaussian kernel with a FWHM of 3\,\arcmin.  From the map, we then
compute a mean background density ($\rho_\mathrm{bck}$) as the 3$\sigma$
clipped mean and the rms ($\rho_\mathrm{rms}$).  

A map of density contrast ($\delta_\mathrm{PHz}$)\footnote{The density
contrast is defined as $\delta_{\rm PHz}$\,=\,($\rho_{\rm max}-\rho_{\rm
bck}$)/$\rho_{\rm bck}$.}, and of overdensity significance\footnote{The
overdensity significance is defined as ($\rho_{\rm max}-\rho_{\rm
bck}$)/$\rho_\mathrm{rms}$.} are then obtained.  In the overdensity
significance map, we identify the region with adjacent pixels with a value
greater than the maximum significance expressed by the nearest
rounded-down integer number of $\sigma$, typically 5 , and compute the mean
density contrast in such a region.  The maximum overdensity significance,
the relative mean density contrast and region size for each PHz-IRAM field
are reported in Table~\ref{tab:her_density}.

\begin{table}[!ht]
\centering
\caption{\textit{Red} SPIRE overdensity\label{tab:her_density}}
\setlength{\tabcolsep}{4.pt}
\begin{tabular}{l ccc rccr} 
\hline \hline
  \planck\ &$\rho_{max}$ & $\rho_\mathrm{bck}$  & $\rho_\mathrm{rms}$ & ${<}\delta_\mathrm{PHz}{>}$ & $\sigma_{max}$ &\multicolumn{2}{c}{Area\tablefootmark{a}}   \\
   ID      &  \multicolumn{3}{c}{(N/arcmin$^2$)}                      & ${>}\sigma_{max}$           &                & ${>}\sigma_{max}$ & $>$3 \\
 \hline
   G176  &   0.30  &  0.003 &  0.05 &   86  &  5 &   2.88 &  18.6  \\
   G223  &   0.20  &  0.003 &  0.04 &   62  &  5 &   1.40 &  17.9  \\
   G173  &   0.18  &  0.002 &  0.03 &   69  &  5 &   2.88 &  17.1  \\
   G162  &   0.29  &  0.006 &  0.05 &   44  &  5 &   0.60 &   7.5  \\
   G006  &   0.40  &  0.005 &  0.06 &   73  &  6 &   0.40 &  14.4  \\
   G237  &   0.26  &  0.009 &  0.05 &   28  &  5 &   1.48 &  16.0  \\
   G191  &   0.17  &  0.005 &  0.04 &   33  &  4 &   0.92 &  14.2  \\
   G088  &   0.70  &  0.006 &  0.09 &  116  &  7 &   1.96 &  15.9  \\
   G059  &   1.47  &  0.002 &  0.18 &  734  &  8 &   0.32 &  15.0  \\ 
   G073  &   0.11  &  0.002 &  0.02 &   56  &  6 &   0.44 &   8.9  \\ 
   G124  &   0.56  &  0.003 &  0.08 &  160  &  6 &   1.60 &  13.6  \\ 
   G072  &   0.29  &  0.003 &  0.05 &   82  &  5 &   1.16 &   7.2  \\ 
   G112  &   0.25  &  0.004 &  0.04 &   62  &  5 &   1.52 &  10.0  \\ 
   G143  &   0.41  &  0.003 &  0.06 &  148  &  6 &   1.88 &  14.6  \\ 
   G131  &   0.21  &  0.005 &  0.05 &   40  &  4 &   1.72 &  10.1  \\ 
   G052  &   0.15  &  0.004 &  0.02 &   37  &  6 &   0.44 &  10.5  \\ 
   G068  &   0.13  &  0.002 &  0.02 &   49  &  5 &   0.88 &   6.8  \\ 
   G063  &   0.06  &  0.002 &  0.01 &   28  &  5 &   0.16 &  13.1  \\ 
\hline                
\end{tabular}                                                                                                              
\tablefoot{             
\tablefoottext{a}{\small Area in arcmin$^2$ of the regions where the
overdensity significance is greater than $\sigma_{max}$, and 3.}
}
\end{table}

Most of the selected fields exhibit significant ($>$5$\sigma$) overdensities of
\textit{red} SPIRE sources.  These overdensities are overlaid on the
\herschel/SPIRE multiband images in Fig.~\ref{fig:her_maps}.  We also show
the \planck\ {\tt red-excess} contours, and the position of all SPIRE
sources in the field.  The SPIRE sources observed with IRAM, shown as large
stars, are typically situated on the overdensity peak.  Note that,
typically, the \planck\ {\tt red-excess} emission well matches the SPIRE
overdensity, but in some cases the \planck\ {\tt red-excess} contours are
missing (G052, and G073) because those regions were masked in the 5\arcmin\
resolution \planck\ maps utilized to extract the PHz catalog.  In the
remaining eight PHz fields that are not included in the official PHz source
list~\citep{planck16}, a significant signal in the \planck\ {\tt red-excess}
map is still visible in six cases (i.e., in G124, but split in two less
significant blobs, in G072, G112, G143, G131, and in G063), and no {\tt
red-excess} signal is present in the remaining two (i.e., G059, and
G068).  We refer to Appendix A in~\citet{planck16} for a more detailed
comparison between the preliminary PHz extraction and the official catalog.

\begin{figure*}[h!]
\centering
  \subfloat[G176]{\includegraphics[height=5cm]{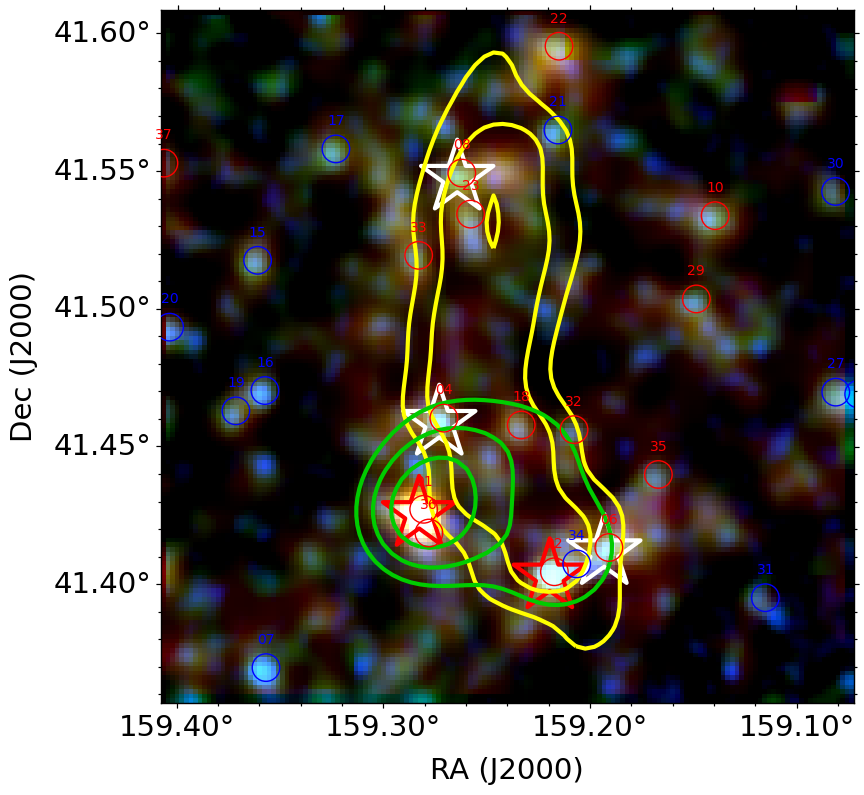}}
\;\subfloat[G223]{\includegraphics[height=5cm]{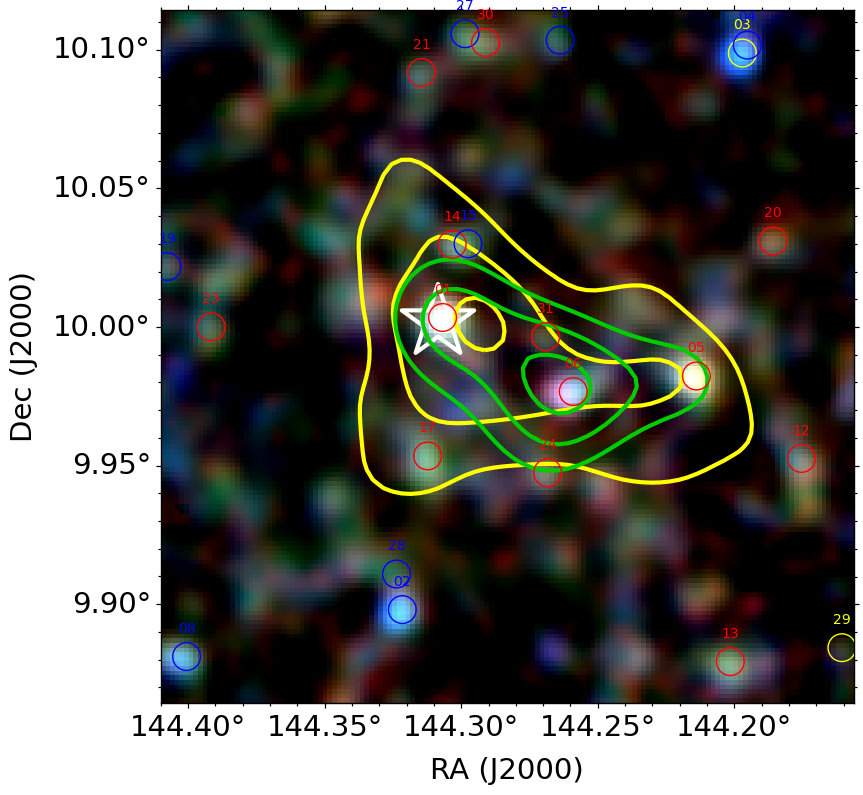}}
\;\subfloat[G173]{\includegraphics[height=5cm]{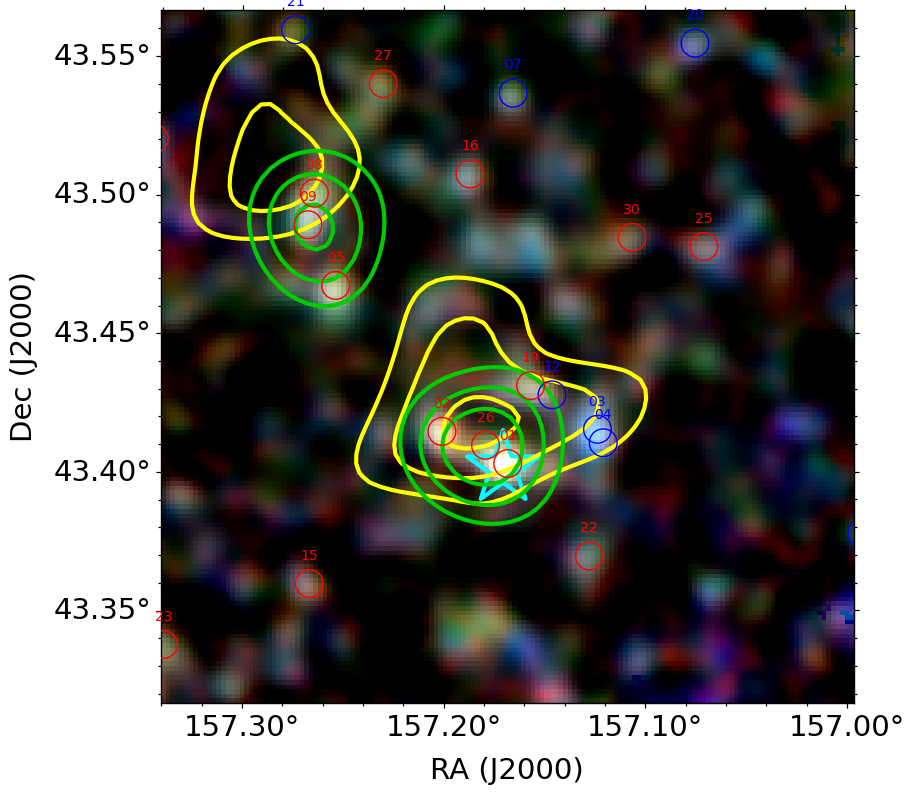}}\\
  \subfloat[G162]{\includegraphics[height=5cm]{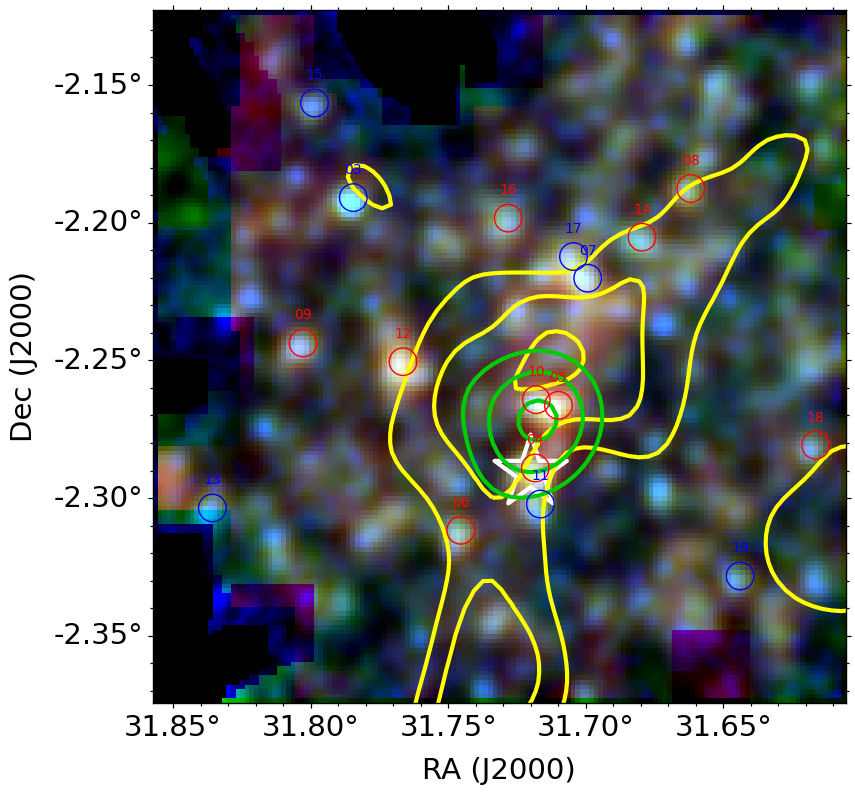}}
\;\subfloat[G006]{\includegraphics[height=5cm]{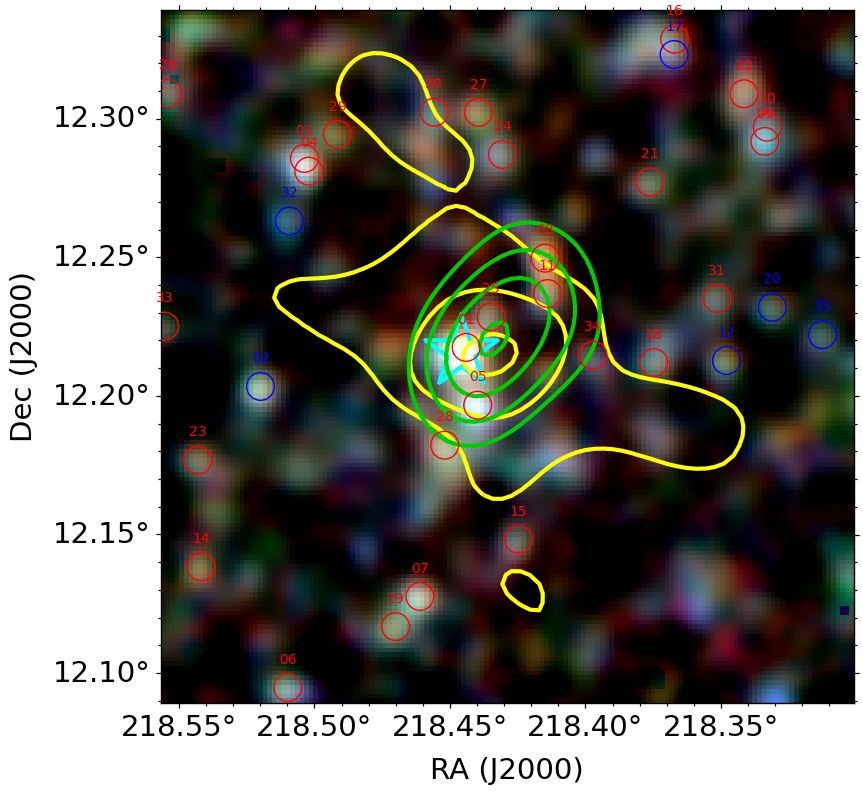}}
\;\subfloat[G237]{\includegraphics[height=5cm]{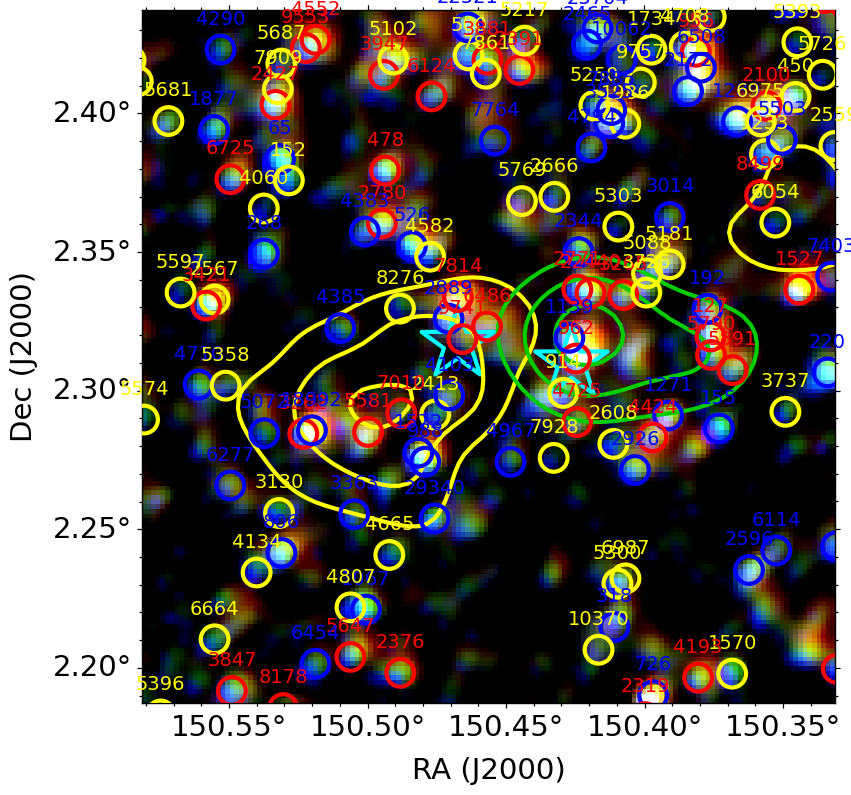}}\\
  \subfloat[G191]{\includegraphics[height=5cm]{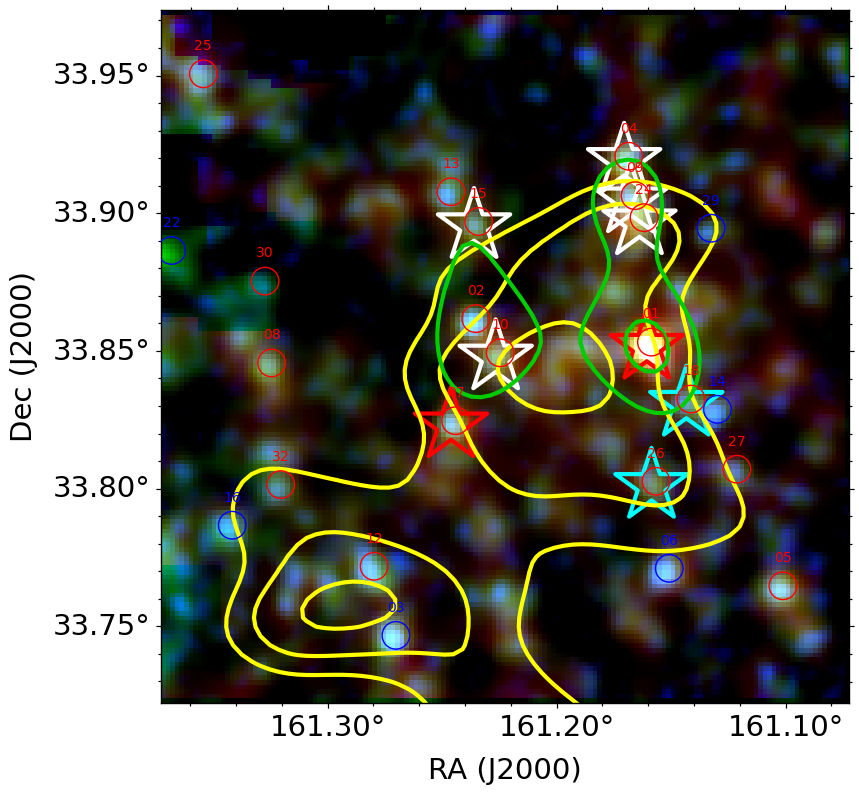}}
\;\subfloat[G088]{\includegraphics[height=5cm]{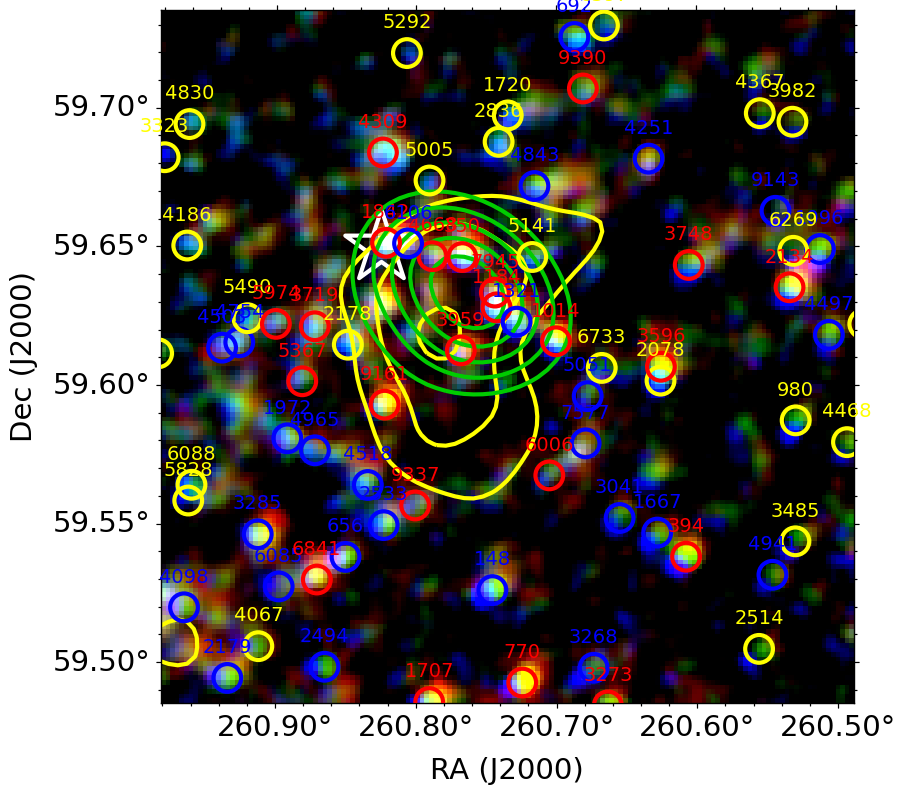}}
\;\subfloat[G059]{\includegraphics[height=5cm]{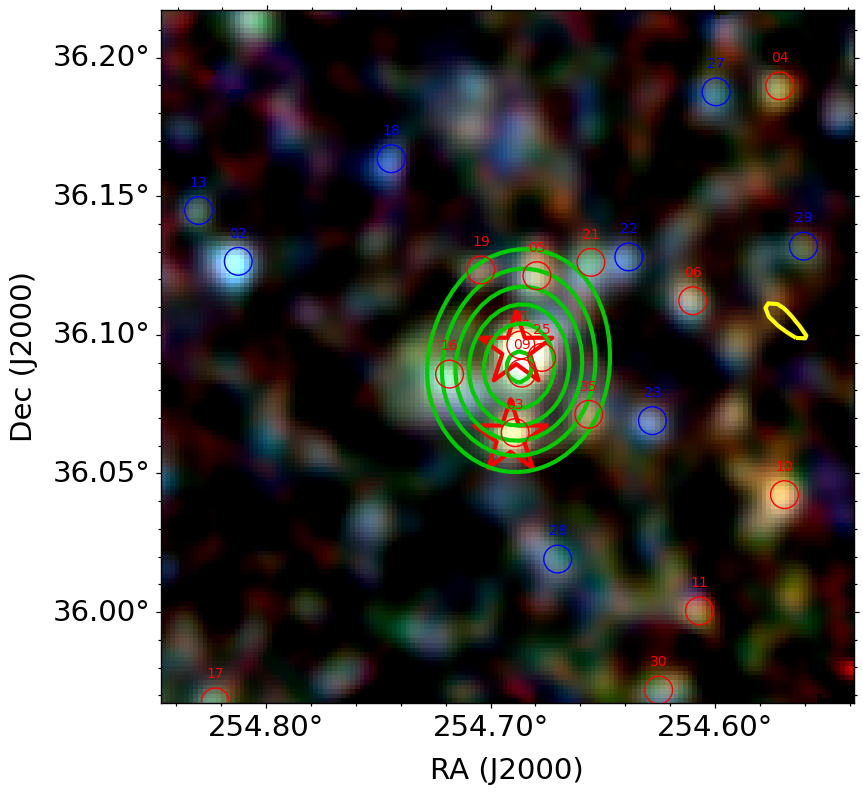}}\\
\caption{{\small 15\arcmin$\times$15\arcmin\ \herschel\ RGB (red: 500$\mu$m,
green: 350$\mu$m, blue: 250$\mu$m) maps centered on the IRAM observed PHz
fields presented here. SPIRE sources are indicated with open circles (red:
\textit{red} sub-mm colors, blue: \textit{blue} sub-mm colors, yellow: non detected in all 3
bands, thus preventing a color classification). IRAM targets are
indicated with large stars (red: CO detection at similar
redshift (i.e., ${\Delta}z$/(1${+}z$)$<$0.02), cyan: CO detected, there is only one detection or multiple
detections at different redshifts (i.e., ${\Delta}z$/(1${+}z$)$>$0.02)}, white: CO not detected). Yellow contours represent the
\planck\ {\tt red-excess} emission (50\%, 70\%, and 90\% of the maximum value).
Green contours represent the overdensity significance of \textit{red} SPIRE
sources (starting at 3$\sigma$, with steps of 1$\sigma$).}
\label{fig:her_maps} 
\end{figure*}

\begin{figure*}[h!]
\setcounter{figure}{3}
\centering
  \subfloat[G073]{\includegraphics[height=5cm]{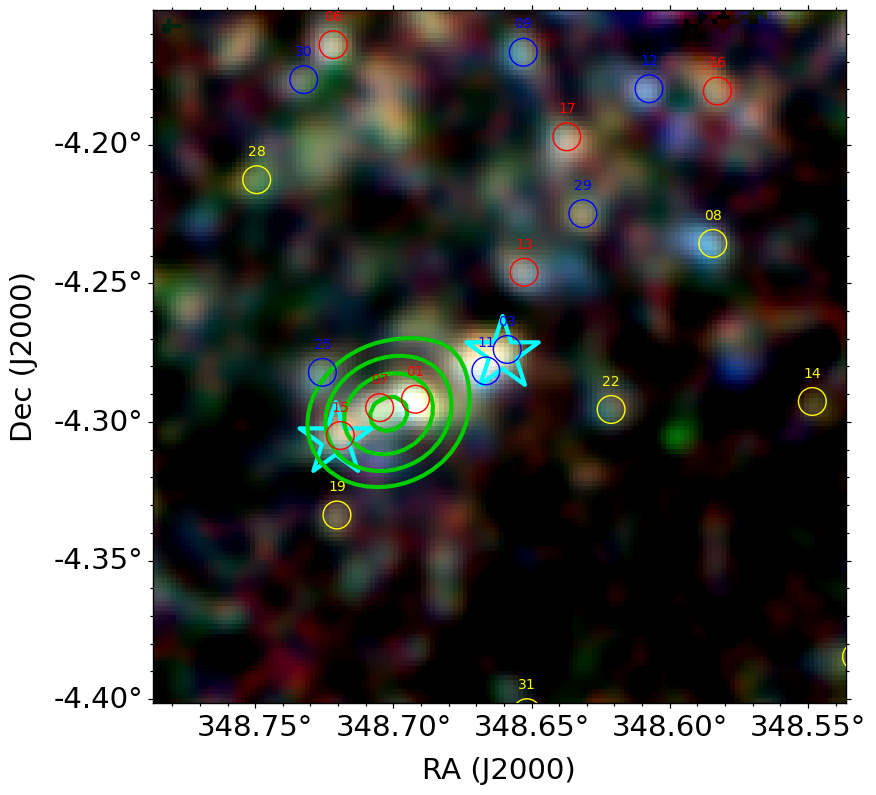}}
\;\subfloat[G124]{\includegraphics[height=5cm]{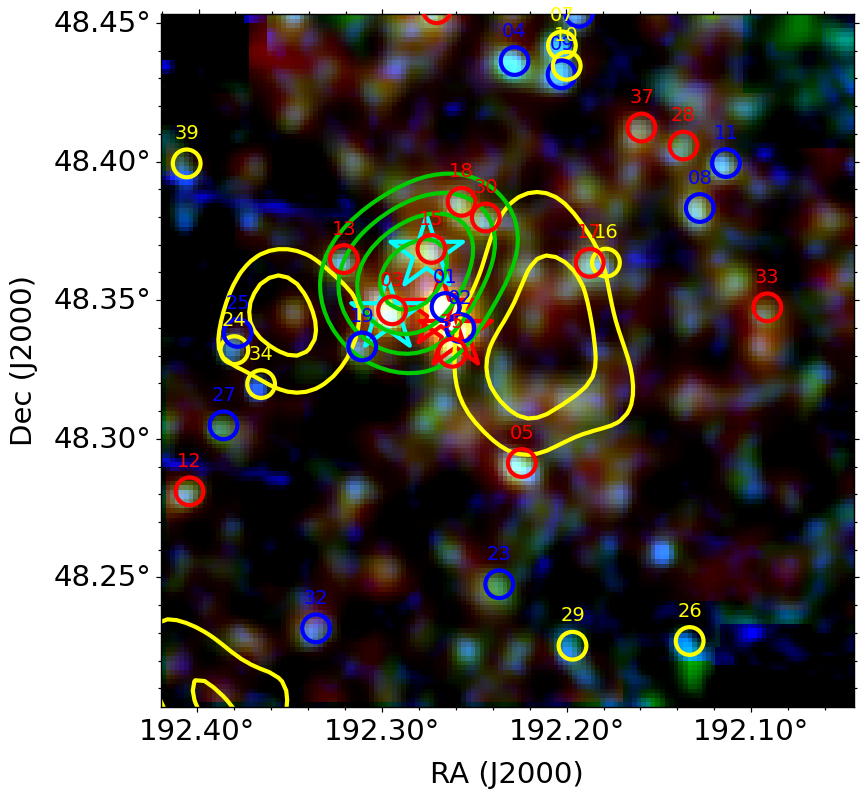}}
\;\subfloat[G072]{\includegraphics[height=5cm]{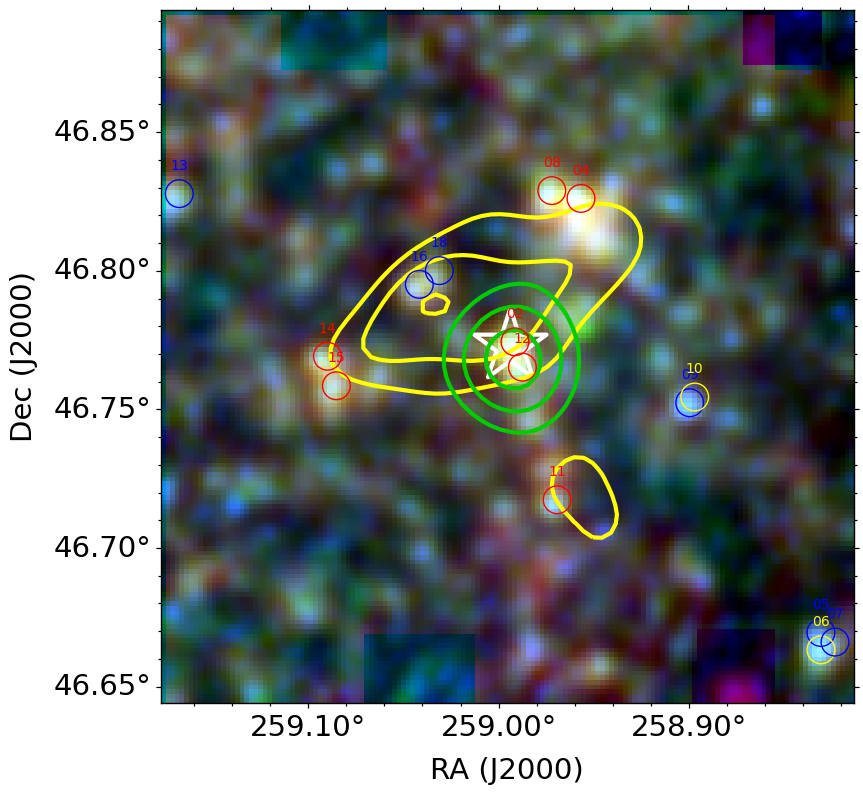}}\\
  \subfloat[G112]{\includegraphics[height=5cm]{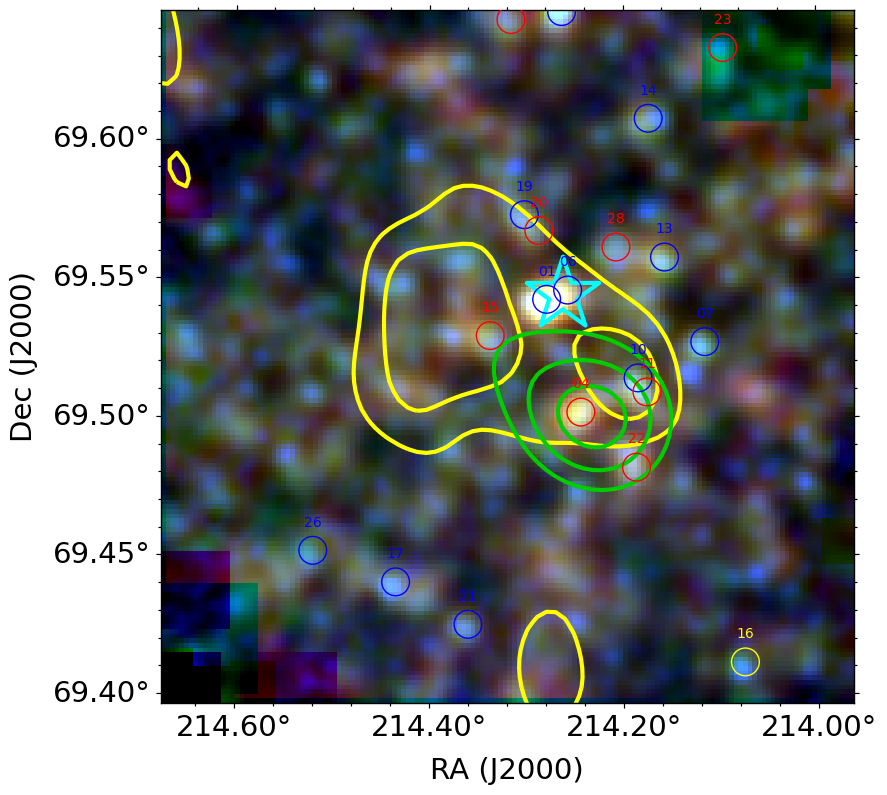}}
\;\subfloat[G143]{\includegraphics[height=5cm]{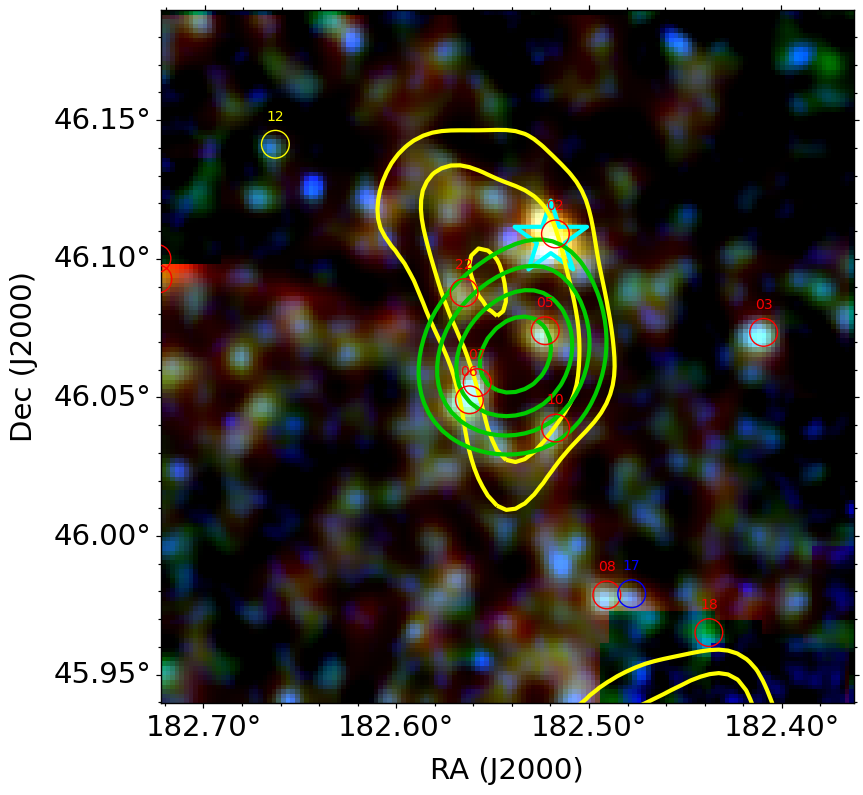}}
\;\subfloat[G131]{\includegraphics[height=5cm]{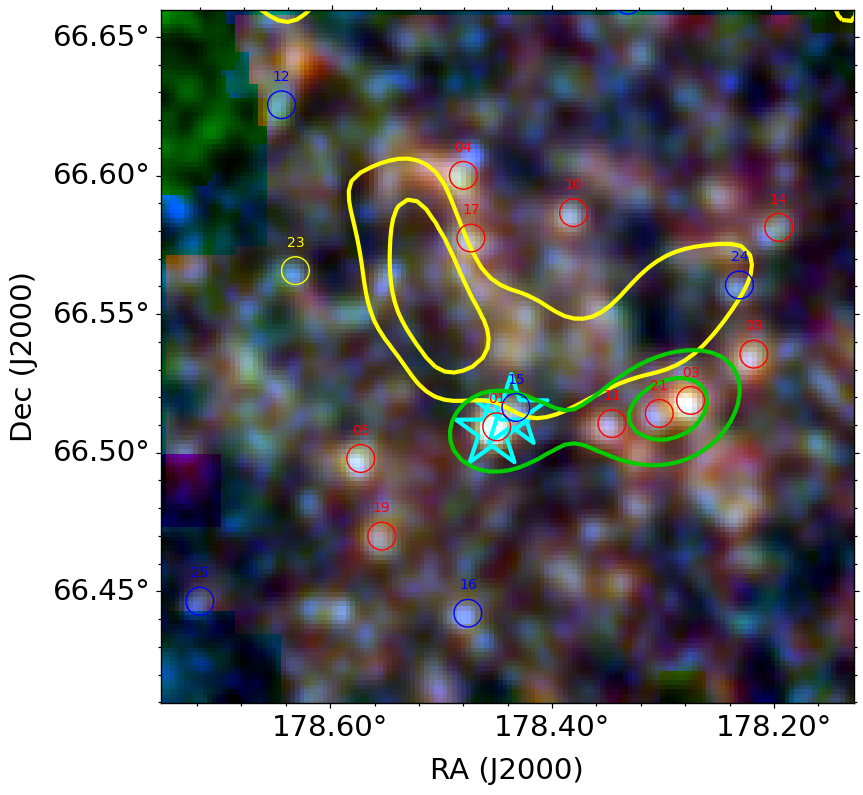}}\\
  \subfloat[G052]{\includegraphics[height=5cm]{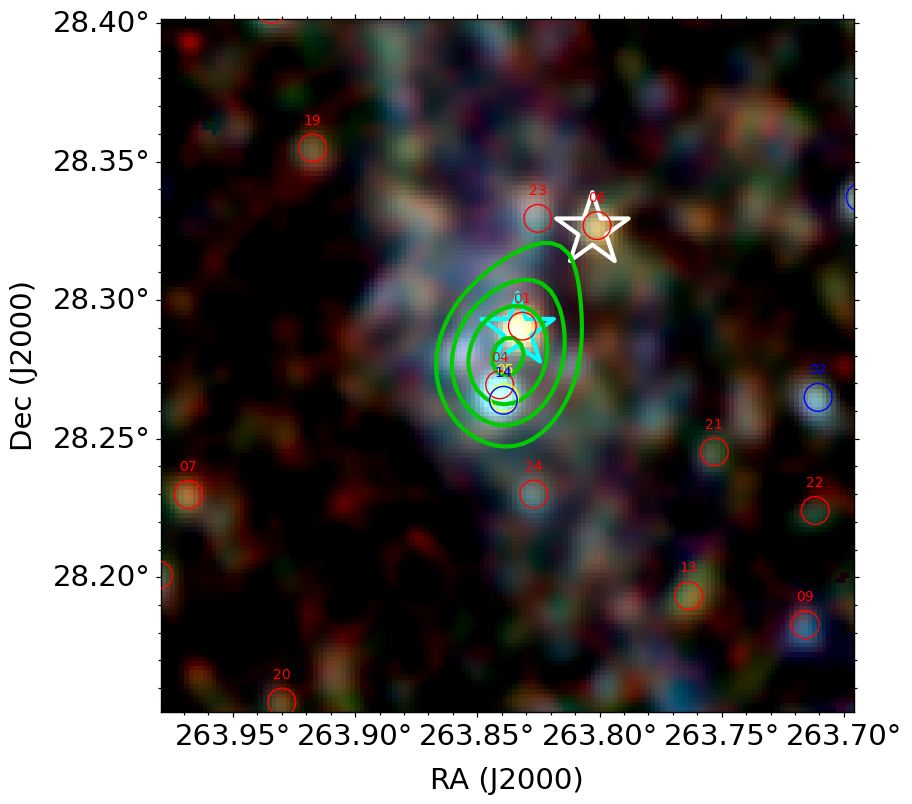}}
\;\subfloat[G068]{\includegraphics[height=5cm]{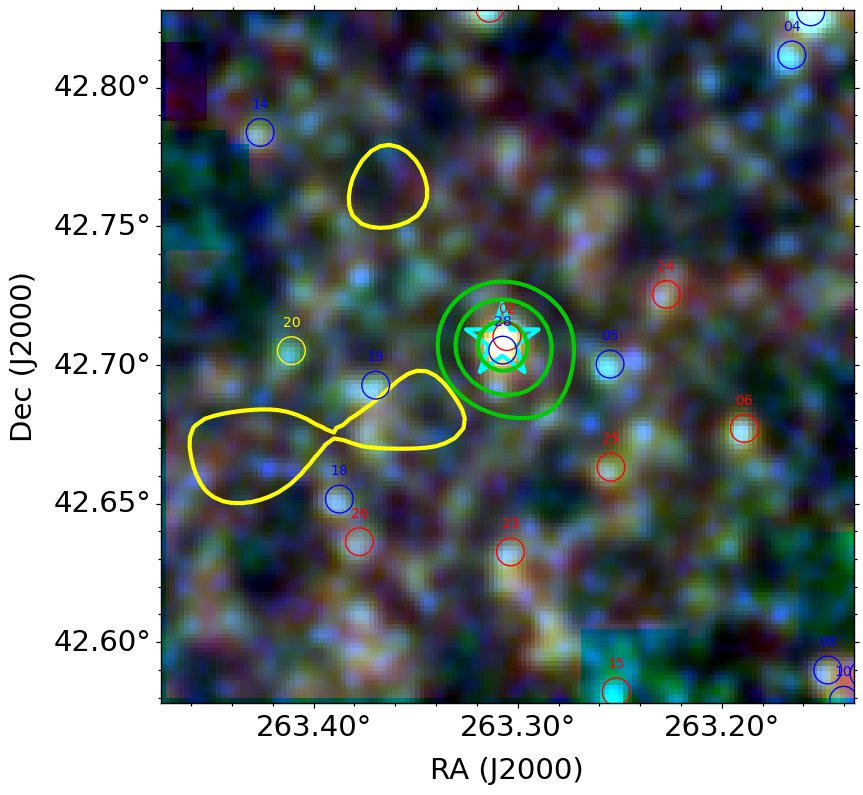}}
\;\subfloat[G063]{\includegraphics[height=5cm]{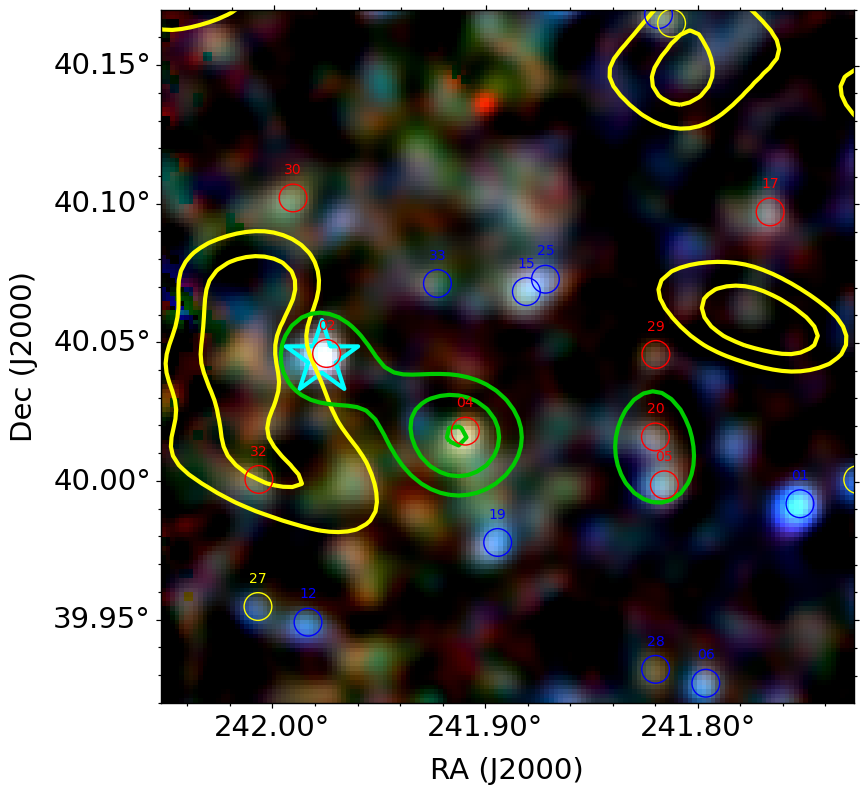}}\\
\caption{{\small  Continued.}}
\label{fig:her_maps} 
\end{figure*}

Previous works on the PHz sample have also quantified the density of
\textit{red} SPIRE and \spitzer/IRAC
sources~\citep{planck15,martinache18}.  Unlike our method, the \herschel\
overdensity was computed only in the region where the \planck\ {\tt
red-excess} signal was greater than half of the maximum value.  We did not apply the same
methodology as we do not have the {\tt red-excess} map for all selected sources,
and the low resolution of the \planck\ maps introduces a large uncertainty in
determining the location of the sub-mm emission.  The method used
by~\citet{martinache18} is instead based on the density of
\textit{red} \spitzer/IRAC sources within 1\arcmin\ from the brightest
\textit{red} \herschel\ source. 

\section{Observations and data reduction}\label{sec:emir_observations}

Observations of the selected targets were carried out using the heterodyne
receiver EMIR~\citep{carter12} on the IRAM 30-m telescope.  For the
backends, we simultaneously used the Wideband Line Multiple Autocorrelator
(WILMA, 2-MHz spectral resolution) and the fast Fourier Transform
Spectrometer (FTS200, 200-kHz resolution).  The program strategy was to
carry out a blind redshift search in the 3\,mm and 2\,mm bands for the
brightest \textit{red} SPIRE source per PHz field.  The frequency tuning was
defined to search for the CO(3--2), or the CO(4--3) lines at
$z$\,$\simeq$\,2--3 ($\nu^{\rm obs}_{\rm CO(3-2)}$\,=\,86--115\,GHz, $\nu^{\rm
obs}_{\rm CO(4-3)}$\,=\,115--154\,GHz).  Each source was observed for about
200\,min, or longer in case of a tentative detection.  In case of a line
detection, an additional observation was carried out, when possible, to look
for a second line to confirm the redshift.  In a few fields, more than one
source was observed with a tuning suited to find a line at the same
redshift.  The observed targets, project numbers, observing dates, selected
bands, and integration times are listed in Tables~\ref{tab:obs_log},
and~\ref{tab:obs_setup}.

We assume that the selected targets are smaller than the beam size at the
observing frequencies.  The FWHM of IRAM 30-m/EMIR is 27\arcsec\ at 91\,GHz,
comparable to the {\it Herschel\/}-SPIRE beam at 350\,$\mu$m\footnote{
The FWHM of the \herschel-SPIRE beams are 18\arcsec\ at 250$\mu$m,
25\arcsec\ at 350$\mu$m, and 36\arcsec\ at 500$\mu$m.}.  Observations were
performed in wobbler-switching mode with a throw of 80\arcsec--150\arcsec. 
For calibration, pointing, and focusing we used planets, and bright quasars. 
Individual scans were 28--30\,sec long, and we observed sets of 12 scans
followed by a calibration.  Data reduction was performed with the {\tt
CLASS} package in {\tt GILDAS}~\citep{gildas13}.  In a few cases, some scans
were discarded because of an unstable baseline yielding different
integration times in the two sub-bands.  A baseline, computed as
first-order polynomial of the spectrum, was subtracted from each individual
scan.  Scans were then averaged using the inverse of the square of the
individual noise level as weight, and smoothed to a velocity resolution
given by the frequency and the line strength.  We then applied the different
point source conversion factors (in the range of 5.9--6.4\,Jy/K depending on
the optics and the frequency) to convert $T^*_A$ [K] into a flux density
[Jy].  The resulting co-added spectra were inspected to look for emission
lines as multiple adjacent channels with signal higher than the rms
in the nearby channels (${\mid}{\Delta}v{\mid}{<}$2800\,\kms) and then
fitted with a single or a double Gaussian profile to measure the line
position, the integrated line flux, and the line width (FWHM). 
We consider a line detected if the its S/N is greater than 4,
but we report additional transitions with lower significance if their S/N is
at least 2. The detected lines and derived parameters are listed in
Table~\ref{tab:lines}.

\begin{table*}[!ht]
\centering
\caption{EMIR line detections\label{tab:lines}}
\setlength{\tabcolsep}{3.pt}
\begin{tabular}{lrcr c c c c c r r}
\hline \hline
 ID           &  PHz  &\herschel\ &Frequency& Line  &       Redshift       & $S_\mathrm{CO} \Delta v$ & FWHM      & rms$_{\rm line}$ & S/N & Flag\tablefootmark{a} \\
              &    ID   &  ID    &    (GHz) &       &                      &   Jy\,km\,s$^{-1}$    & km\,s$^{-1}$ & mJy &      & \\
\hline
\noalign{\vspace{2pt}}
 G176  &     57  &   01   &   92.128 & CO(3--2) &  2.75343$\pm$0.00063 &  5.57$\pm$0.37  &   482$\pm$ 38  &    0.50 & 23.2 &   1  \\
       &         &        &  153.527 & CO(5--4) &  2.75353$\pm$0.00024 &  3.26$\pm$0.52  &   225$\pm$ 34  &    1.58 &  9.3 &   1  \\
 G176  &     57  &   02   &   92.123 & CO(3--2) &  2.75363$\pm$0.00055 &  0.97$\pm$0.32  &   230$\pm$ 62  &    0.94 &  4.4 &   1  \\
       &         &        &  153.901 & CO(5--4) &  2.74441$\pm$0.00030 &  1.78$\pm$0.43  &   190$\pm$ 45  &    1.75 &  5.5 &   1  \\
 G173  &    237  &   01   &   97.209 & CO(3--2) &  2.55724$\pm$0.00035 &  0.88$\pm$0.19  &   259$\pm$ 62  &    0.51 &  6.6 &   1  \\
       &         &        &  161.962 & CO(5--4) &  2.55804$\pm$0.00027 &  0.54$\pm$0.19  &   116$\pm$ 32  &    1.22 &  3.8 &   1  \\
 G006  &    631  &   01   &   97.622 & CO(2--1) &  1.36154$\pm$0.00049 &  1.70$\pm$0.34  &   579$\pm$103  &    0.42 &  6.9 &$-$1  \\
       &         &        &   97.622 & CO(3--2) &  2.54219$\pm$0.00072 &  1.70$\pm$0.34  &   579$\pm$103  &    0.42 &  6.9 &$-$1  \\
 G237  &    712  &  962   &  113.090 & CO(3--2) &  2.05771$\pm$0.00014 &  1.52$\pm$0.29  &   134$\pm$ 27  &    1.49 &  7.6 &   1  \\
 G237  &    712  & 9741   &  145.945 & CO(4--3) &  2.15900$\pm$0.00025 &  0.63$\pm$0.20  &   149$\pm$ 50  &    0.94 &  4.7 &   1  \\
 G191  &    832  &   01   &   97.383 & CO(3--2) &  2.55089$\pm$0.00071 &  2.75$\pm$0.33  &   497$\pm$ 63  &    0.46 & 12.1 &   1  \\
       &         &        &  129.830 & CO(4--3) &  2.55035$\pm$0.00045 &  1.13$\pm$0.25  &   301$\pm$ 59  &    0.61 &  6.2 &   1  \\
       &         &        &  162.205 & CO(5--4) &  2.55271$\pm$0.00024 &  0.57$\pm$0.28  &    67$\pm$ 33  &    3.04 &  2.8 &   1  \\
 G191  &    832  &   07   &   95.430 & CO(3--2) &  2.62356$\pm$0.00032 &  2.22$\pm$0.37  &   305$\pm$ 47  &    0.87 &  8.4 &   1  \\
       &         &        &  127.262 & CO(4--3) &  2.62277$\pm$0.00040 &  2.79$\pm$0.66  &   258$\pm$ 52  &    1.87 &  5.8 &   1  \\
       &         &        &  159.033 & CO(5--4) &  2.62357$\pm$0.00042 &  2.80$\pm$0.65  &   444$\pm$ 90  &    0.95 &  6.7 &   1  \\
       &         &   07a  &   95.458 & CO(3--2) &  2.62249$\pm$0.00036 &  1.20$\pm$0.42  &   163$\pm$ 64  &    1.19 &  6.2 &   2  \\
       &         &        &  127.296 & CO(4--3) &  2.62180$\pm$0.00051 &  1.52$\pm$0.76  &   146$\pm$ 75  &    2.47 &  4.2 &   2  \\
       &         &        &  159.066 & CO(5--4) &  2.62282$\pm$0.00051 &  2.15$\pm$0.59  &   306$\pm$ 93  &    1.14 &  6.2 &   2  \\
       &         &   07b  &   95.399 & CO(3--2) &  2.62473$\pm$0.00027 &  0.96$\pm$0.42  &   122$\pm$ 55  &    1.37 &  5.7 &   2  \\
       &         &        &  127.226 & CO(4--3) &  2.62379$\pm$0.00035 &  1.30$\pm$0.78  &   109$\pm$ 64  &    2.86 &  4.2 &   2  \\
       &         &        &  158.887 & CO(5--4) &  2.62690$\pm$0.00024 &  0.91$\pm$0.34  &   103$\pm$ 35  &    1.97 &  4.5 &   2  \\
 G191  &    832  &   26   &   79.774 & CO(2--1) &  1.88989$\pm$0.00046 &  1.81$\pm$0.37  &   481$\pm$126  &    0.48 &  7.9 &   1  \\
 G059  & 124051  &   01   &  103.010 & CO(3--2) &  2.35692$\pm$0.00038 &  2.97$\pm$0.69  &   266$\pm$ 87  &    1.46 &  7.9 &   1  \\
       &         &        &  137.341 & CO(4--3) &  2.35678$\pm$0.00012 &  2.70$\pm$0.32  &   173$\pm$ 20  &    1.33 & 12.1 &   1  \\
       &         &        &  171.714 & CO(5--4) &  2.35598$\pm$0.00010 &  0.53$\pm$0.25  &    37$\pm$ 20  &    4.25 &  3.4 &   1  \\
 G059  & 124051  &   03   &  102.877 & CO(3--2) &  2.36126$\pm$0.00019 &  0.26$\pm$0.09  &    87$\pm$ 31  &    0.74 &  4.1 &   1  \\  
       &         &        &  136.903 & CO(4--3) &  2.36765$\pm$0.00023 &  0.84$\pm$0.18  &   178$\pm$ 31  &    0.74 &  6.4 &   1  \\
       &         &        &  171.372 & CO(5--4) &  2.36267$\pm$0.00060 &  1.68$\pm$1.31  &   199$\pm$224  &    3.86 &  2.2 &   1  \\
 G073  & 124052  &   03   &   90.700 & CO(2--1) &  1.54176$\pm$0.00036 &  1.00$\pm$0.31  &   285$\pm$105  &    0.71 &  4.9 &   1  \\
 G073  & 124052  &   15   &   94.859 & CO(2--1) &  1.43032$\pm$0.00012 &  1.02$\pm$0.29  &    89$\pm$ 32  &    2.08 &  5.5 &$-$1  \\
       &         &        &   94.859 & CO(3--2) &  2.64537$\pm$0.00017 &  1.02$\pm$0.29  &    89$\pm$ 32  &    2.08 &  5.5 &$-$1  \\
 G124  & 124053  &   01   &  109.669 & CO(3--2) &  2.15309$\pm$0.00026 &  2.00$\pm$0.30  &   315$\pm$ 45  &    0.67 &  9.5 &   1  \\
       &         &        &  146.203 & CO(4--3) &  2.15276$\pm$0.00029 &  1.31$\pm$0.28  &   243$\pm$ 53  &    0.81 &  6.7 &   1  \\
       &         &   01a  &  109.623 & CO(3--2) &  2.15441$\pm$0.00011 &  0.78$\pm$0.16  &    89$\pm$ 18  &    1.26 &  7.0 &   2  \\
       &         &        &  146.196 & CO(4--3) &  2.15358$\pm$0.00032 &  0.74$\pm$0.27  &   163$\pm$ 69  &    0.99 &  4.6 &   2  \\
       &         &   01b  &  109.695 & CO(3--2) &  2.15234$\pm$0.00014 &  1.15$\pm$0.23  &   148$\pm$ 38  &    0.98 &  8.0 &   2  \\
       &         &        &  146.266 & CO(4--3) &  2.15207$\pm$0.00007 &  0.51$\pm$0.20  &    51$\pm$ 22  &    1.76 &  5.7 &   2  \\
 G124  & 124053  &   02   &  109.379 & CO(3--2) &  2.16145$\pm$0.00031 &  2.59$\pm$0.36  &   430$\pm$ 69  &    0.56 & 10.8 &   1 \\
       &         &        &  145.983 & CO(4--3) &  2.15818$\pm$0.00026 &  1.75$\pm$0.47  &   167$\pm$ 36  &    2.11 &  5.0 &   1  \\
 G124  & 124053  &   03   &  154.955 & CO(5--4) &  2.71894$\pm$0.00073 &  2.02$\pm$0.55  &   408$\pm$110  &    1.01 &  5.1 &   1  \\
 G124  & 124053  &   15   &  111.975 & CO(3--2) &  2.08815$\pm$0.00024 &  2.94$\pm$0.56  &   241$\pm$ 53  &    1.55 &  7.9 &   1  \\
       &         &        &  149.564 & CO(4--3) &  2.08257$\pm$0.00013 &  1.20$\pm$0.44  &    60$\pm$ 25  &    5.03 &  4.0 &   1  \\
 G112  & 125018  &   06   &  104.197 & CO(3--2) &  2.31868$\pm$0.00073 &  1.34$\pm$0.45  &   306$\pm$120  &    0.98 &  4.5 &   1  \\
 G143  & 125026  &   02   &  160.805 & CO(5--4) &  2.58364$\pm$0.00029 &  3.06$\pm$0.68  &  269$\pm$88    &    1.41 &  8.2 &   1  \\    
 G131  & 125027  &   01   &   99.218 & CO(2--1) &  1.32355$\pm$0.00014 &  2.60$\pm$0.27  &  347$\pm$38    &    0.52 & 14.3 &   1  \\    
       &         &        &  148.849 & CO(3--2) &  1.32315$\pm$0.00018 &  2.67$\pm$0.35  &  326$\pm$38    &    0.67 & 12.8 &   1  \\    
 G131  & 125027  &   15   &   94.612 & CO(2--1) &  1.43667$\pm$0.00010 &  1.16$\pm$0.31  &   88$\pm$27    &    2.33 &  5.7 &$-$1  \\  
       &         &        &   94.612 & CO(3--2) &  2.65489$\pm$0.00014 &  1.16$\pm$0.31  &   88$\pm$27    &    2.33 &  5.7 &$-$1  \\    
 G052  & 125056  &   01   &  105.663 & CO(3--2) &  2.27263$\pm$0.00012 &  1.86$\pm$0.29  &  136$\pm$24    &    1.46 &  9.4 &   1  \\     
 G068  & 125107  &   02   &   84.803 & CO(2--1) &  1.71851$\pm$0.00052 &  3.73$\pm$0.45  &  839$\pm$94    &    0.43 & 10.4 &   1  \\    
 G063  & 125132  &   02a  &   88.798 & CO(2--1) &  1.59621$\pm$0.00041 &  1.69$\pm$0.42  &  381$\pm$115   &    0.81 &  5.7 &$-$1  \\
       &         &        &   88.798 & CO(3--2) &  2.89419$\pm$0.00061 &  1.69$\pm$0.42  &  381$\pm$115   &    0.81 &  5.7 &$-$1  \\
       &         &   02b  &  106.278 & CO(3--2) &  2.25369$\pm$0.00033 &  2.00$\pm$0.45  &  254$\pm$66    &    1.18 &  6.7 &   1  \\
\hline                                                                                                                     
\end{tabular}                                                                                                              
\tablefoot{             
\tablefoottext{a}{\small The flag is equal to 1 for secure line identifications, to $-$1 for dubious
ones, and to 2 when line parameters are obtained from double Gaussian fits.
Note that lines not securely identified (Flag\,=\,$-$1) are listed twice with the two most
likely interpretations.}
}
\end{table*}

\begin{figure} 
\centering
\includegraphics[width=9cm]{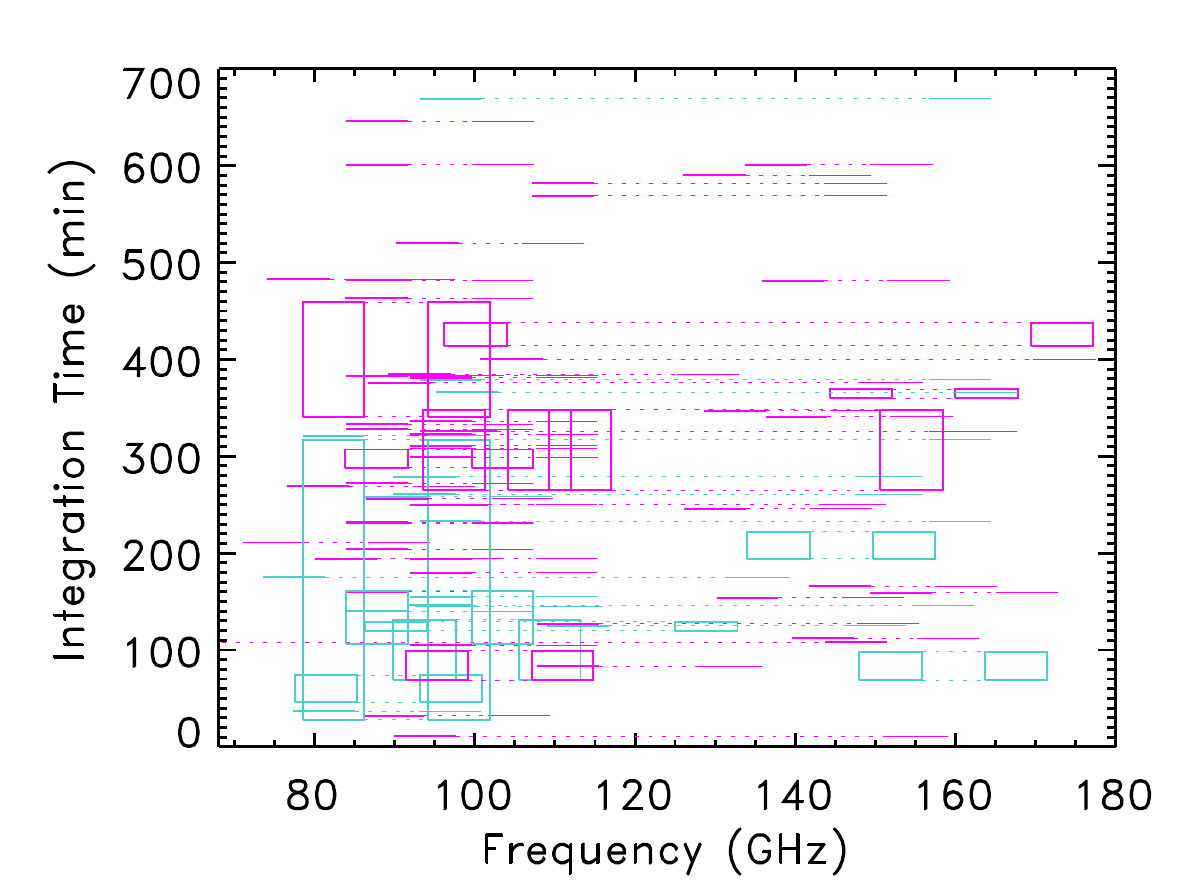}
\caption{{\small EMIR integration time per frequency band of all
observations. Boxes are shown in case some scans were not used in specific
sub-bands. Observations relative to the same target are
connected with dotted lines. The observations that yielded at least one 
CO line detection are shown in magenta, and those with no
detections are shown in turquoise.}}
\label{fig:texp_freq}
\end{figure}

In Fig.~\ref{fig:texp_freq}, we show the frequency coverage and integration
time of each target and highlight in magenta those that yielded a detection
and in turquoise those where no line was detected.  The integration
time of some observations is not fixed but covers a broad range due to the
removal of bad scans in some sub-bands, or to the combination of
observations with a slightly different tuning frequency.  These are shown as
boxes instead of straight lines in Fig.~\ref{fig:texp_freq}. The figure
shows that a line is typically detected if the integration time is longer
than 300\,min, or if the observation covers a broad frequency range.

\begin{figure} 
\centering
\includegraphics[width=9cm]{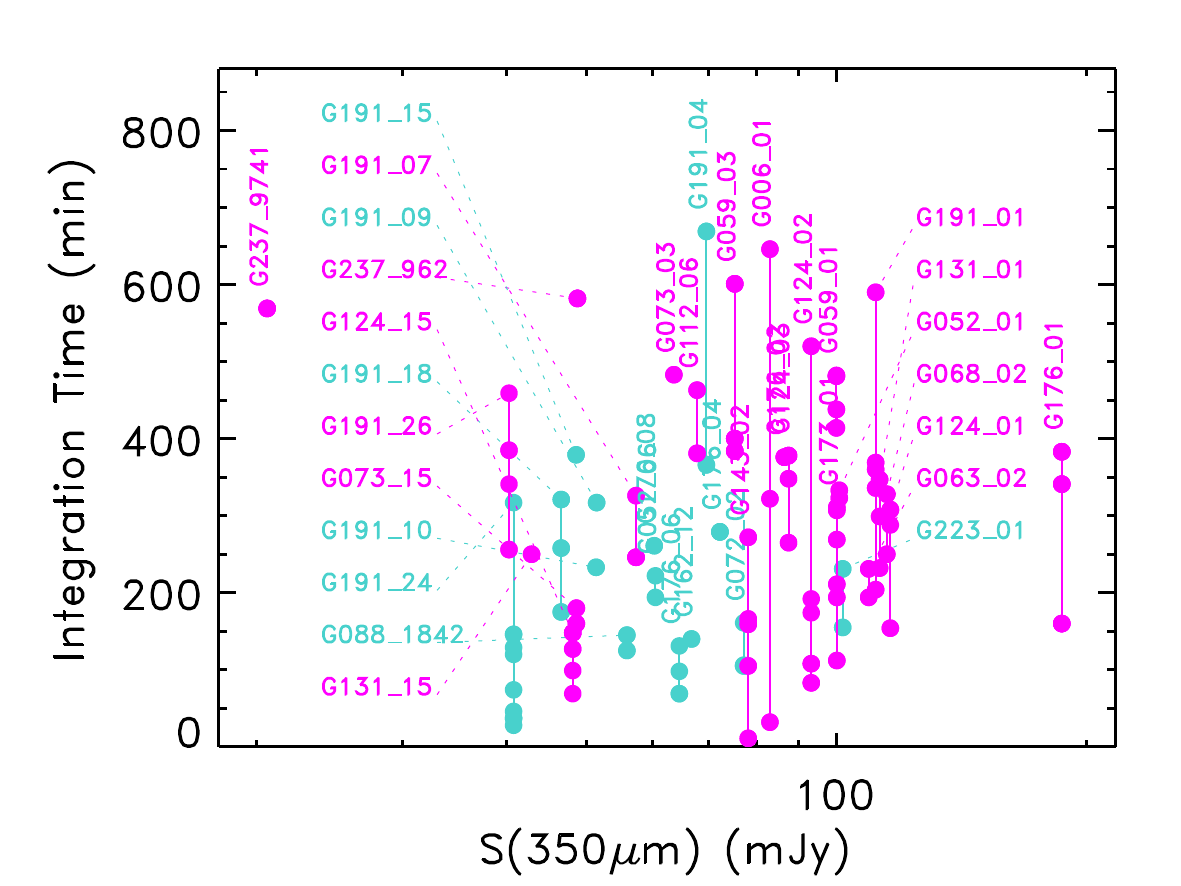}
\caption{{\small EMIR integration time of all observations per source (full
circle connected by a straight line) as a function of the source flux
density at 350$\mu$m.  The source ID is noted.  The observations that
yielded at least one CO line detection are shown in magenta, and those with
no detections are shown in turquoise.  }}
\label{fig:texp_s350}
\end{figure}

In Fig.~\ref{fig:texp_s350}, we show the range of integration times used per
target and the corresponding flux density at 350\,$\mu$m.  At a 350$\mu$m
flux density above 75\,mJy all, but two, sources (88\%) had at least one
line detected.  At fainter fluxes, a line was instead detected only in 37\%
of the cases.  Although it is plausible that no line was present in the
observed narrow frequency range, it is quite likely that, in most of the
cases, the lack of detection might be due to a combination of short
integration time, and source faintness rather than a missing line.  In
Figs.~\ref{fig:f350_histo}, and~\ref{fig:phz_colors}, we compare the
sub-mm fluxes and colors of the SPIRE sources that were CO detected with
those that were not-detected.  The latter are, on average, fainter and
slightly redder.  These properties indicate that they might be at slightly
higher redshifts, but the difference would be small with respect to the wide
redshift range of measured redshifts.  We will see later that most of the
faintest targets were selected in fields where one or two primary sources
had already been detected.  The additional targets were observed with the
goal of detecting a CO line at the same redshift.  In the event of no line
detection in $\sim$200\,min from a first quick on-site data reduction, the
observations were stopped to move onto a different target.  These additional
targets are typically 1.5--3 times fainter than the primary targets and
would have required longer integrations.

The EMIR spectra, together with the best fit Gaussian components are shown
in Fig.~\ref{fig:trp_trans}, ~\ref{fig:dbl_trans}, and~\ref{fig:single_line}.

\section{Results}\label{sec:analysis}

\subsection{Line detections and redshifts}\label{sec:lines}

\begin{figure} 
\centering
\includegraphics[width=9cm]{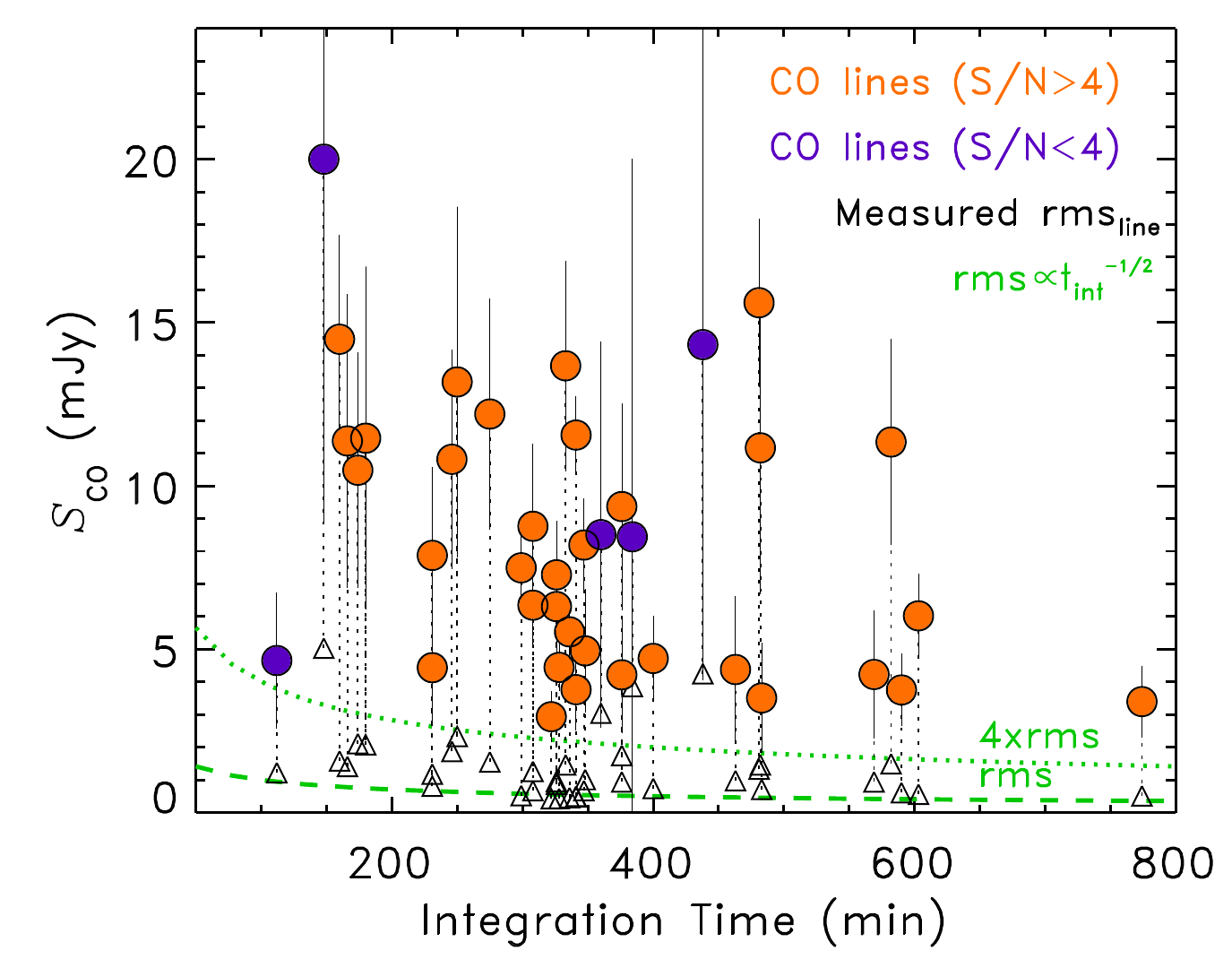}
\caption{{\small Peak flux of the detected CO lines (full orange circles if
S/N$>$4.0, and purple if S/N$\leq$4.0) as a
function of integration time.  The line rms is also shown (black triangles)
and connected to the corresponding measured flux by a dotted vertical line.
The expected trend of rms$_{\rm line}$ with the integration time is shown as
dashed green line, and the detection threshold of 4$\times$rms$_{\rm line}$
as dotted green line.  Note that rms$_{\rm line}$ varies with channel width
and depends on the line width.}}
\label{fig:sco_texp}
\end{figure}

\begin{figure} 
\centering
\includegraphics[width=9cm]{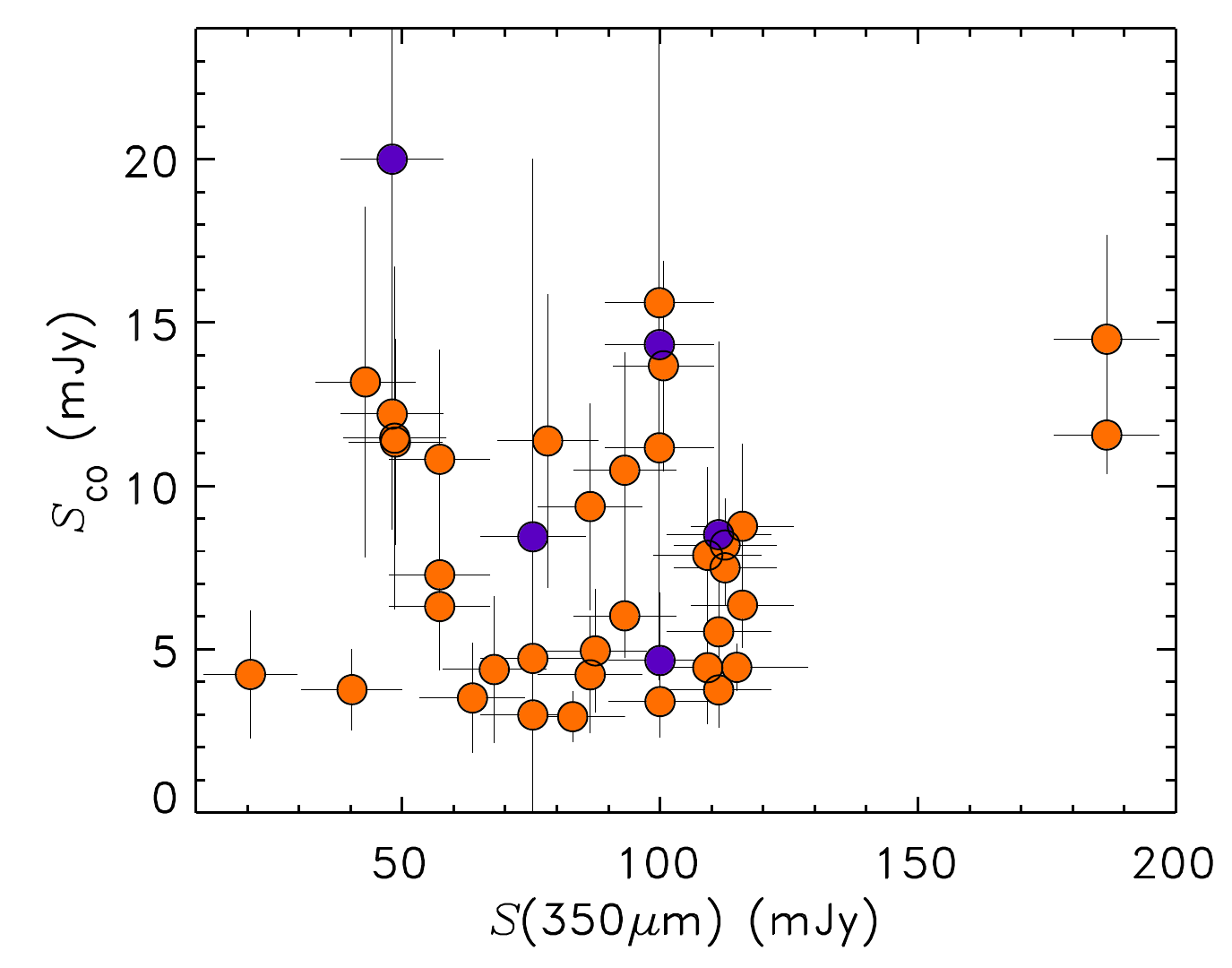}
\caption{{\small Peak flux of the detected CO lines (full orange circles if
S/N$>$4.0, and purple if S/N$\leq$4.0) as a
function of flux density at 350$\mu$m. }}
\label{fig:sco_f350}
\end{figure}

We detect 35 CO lines with signal-to-noise ratio
(S/N)\footnote{S/N\,=\,$S_{\rm CO}$/rms$_{\rm line}$, and rms$_{\rm
line}$\,=\,rms$_{\rm ch}{\times}\sqrt{\Delta v_{\rm ch}/{\rm FWHM}}$, where
rms$_{\rm ch}$ is the spectrum rms noise in mJy, and $\Delta v_{\rm ch}$ the
spectral channel width in \kms.  $\Delta v_{\rm ch}$ ranges from 24\,\kms\
to 91\,\kms, and is 37$\pm$11\,\kms, on average. greater than 4 in 24
bright} ($S_{\rm 350\mu m}>$40\,mJy, with one exception G237\,962 which has
$S_{\rm 350\mu m}{=}$21\,mJy) SPIRE sources in 14 PHz fields.  Five
additional CO lines are detected at a lower significance level,
2.2$<$S/N$\leq$4.0, but are all confirmed by at least one reliable
detection of a different transition.  The lines' transition, intensity,
width (FWHM), rms, S/N, and corresponding redshift are
listed in Table~\ref{tab:lines}.  The lines' peak fluxes as a function of
integration time, and of flux density at 350$\mu$m are shown, respectively
in Fig.~\ref{fig:sco_texp}, and Fig.~\ref{fig:sco_f350}.  The figures show a
wide range of line fluxes, independently on the integration time, and on the
sub-mm brightness.  This means that our observations were able to measure,
with the same integration time and for sources covering a limited range of
sub-mm fluxes (i.e., a factor of three in S(350$\mu$m)), CO line fluxes, and
thus gas masses spanning a factor of $\sim$5.  This large dynamic range in
CO fluxes permits us to explore the intrinsic properties that influence the gas
content in our sample.

\begin{figure} 
\centering
\includegraphics[width=9cm]{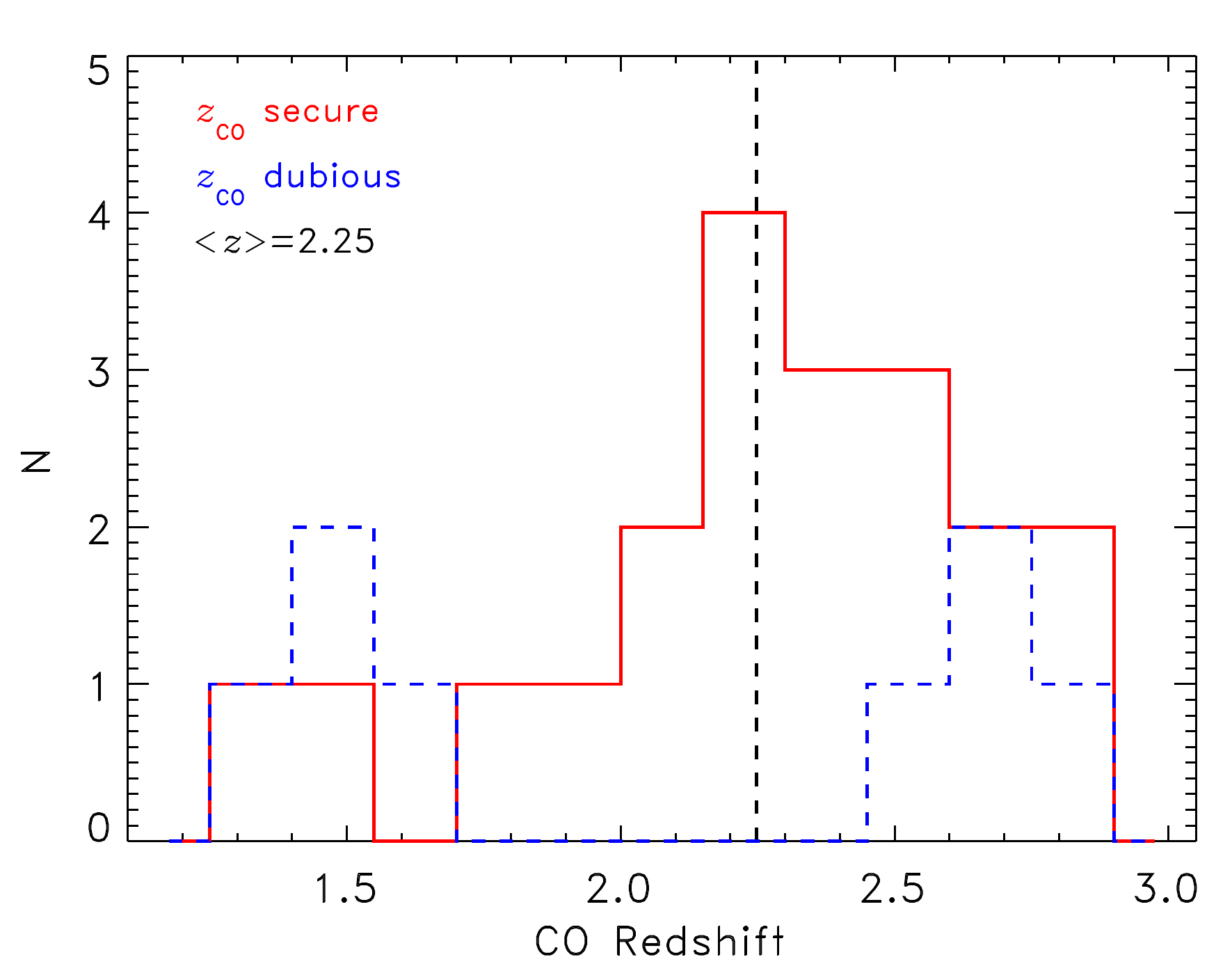} 
\caption{{\small Redshift distribution derived from secure (red) and
dubious (blue) CO line identifications of 24 PHz sources. 
The dashed vertical black line represents the mean redshift,
${<}z_{\rm CO}{>}$\,=\,2.25$\pm$0.09, obtained from the secure
identifications.}}
\label{fig:z_histo}
\end{figure}

In eleven sources we detect multiple (two in seven sources, and three in the
remaining four sources) CO transitions with consistent redshift, thus
yielding a robust redshift estimate.  A single line is instead detected in
the remaining 13 sources.  Because of the line strength and the
expected redshift range, the line identification is straightforward in ten
cases, and thus the redshift estimate is reliable as a lower or higher
redshift would be unlikely.  In four cases (G006\,01, G073\,15, G131\,15,
and G063\,02a) the line interpretation is ambiguous, we thus list the most
plausible transitions and corresponding redshifts (see Flag\,=\,$-$1 in
Table~\ref{tab:lines}).  The distribution in redshift derived from the CO
lines is shown in Fig.~\ref{fig:z_histo}.  The measured secure
redshifts span a range from 1.32 to 2.75, with an average ${<}z_{\rm
CO}{>}$\,=\,2.25$\pm$0.09. In the following analysis, we will distinguish
the parameters derived from the secure CO line identifications (red symbols
in all the following figures), and those derived from the uncertain ones
(blue symbols in all figures).

In four out of the eight fields where more than one SPIRE source was
observed, we detect CO emission from two SPIRE galaxies with consistent
redshifts (G176 at $z$\,=\,2.75, G059 at $z$\,=\,2.36, G124 at $z$\,=\,2.16,
and G191 at $z$\,$\simeq$\,2.6).  In three of these fields, we targeted more
than two sources (five in G176, nine in G191, and four in G124), but
we either detected a CO line at a different redshift, or no CO line
at all.  The number of observed, and detected sources and of those at
similar redshift in each targeted field are summarized in
Table~\ref{tab:iram_summary}. The lack of detection, in some cases, might
have been due to insufficient integration times as these additional targets
are 1.5--3 times fainter than the detected ones.  In other three fields
(G237, G073, and G131), the two detected targets were found at different
redshifts, but additional redshift measurements from
different observations in two of these fields find a significant
concentration of sources at the same redshift.  ALMA observations of G073
detect seven CO lines from five SPIRE sources at $z_{\rm
CO}{\simeq}$1.50--1.54~\citep[][;see also Hill, R.  private
comm.]{kneissl19}.
The ALMA observations detect two sources in the continuum associated with
G073\,15~\citep[ALMA IDs 11, and 12 in][]{kneissl19}, but for only one (ALMA
ID 12) there is a CO detection at $z_{\rm CO}$\,=\,1.50.  We could not
detect the expected CO line associated with such a redshift because it falls
exactly in a frequency gap of our observations, but we have a tentative
detection at another redshift that might be associated with the other
continuum source (ALMA ID 11). Optical and near-infrared spectroscopic
observations in G237 find a significant overdensity of 31 sources at
$z$\,=\,2.16--2.20~\citep{polletta21}, but most of them are not bright
enough in the sub-mm to be followed-up with the IRAM-30m telescope. In the
remaining field, G052, only one source was detected.  IRAM can
probably reveal only a few protocluster members, those with the largest gas
reservoirs~\citep[see also][]{ivison20}.  Some of these sources
might be associated with multiple galaxies that can be individually detected
only by high resolution sensitive mm observations.  The effect of
multiplicity is discussed in the next section.

\begin{table}[!ht]
\centering
\caption{IRAM observations summary\label{tab:iram_summary}}
\begin{tabular}{rc ccc} 
\hline \hline
PHz         &  \planck\ &\multicolumn{2}{c}{N. targets}    & N. at   \\
ID          &   ID      & observed & detected\tablefootmark{a} & similar $z$\tablefootmark{b} \\
\hline
      57   &   G176  &  5  & 2    & 2 \\ 
      70   &   G223  &  1  & 0    & \nodata \\ 
     237   &   G173  &  1  & 1    & \nodata \\ 
     343   &   G162  &  1  & 0    & \nodata \\ 
     631   &   G006  &  1  & 1    & \nodata \\ 
     712   &   G237\tablefootmark{c}  &  2  & 2    & 0 \\
     832   &   G191  &  9  & 3(4) & 2(3) \\ 
    1473   &   G088  &  1  & 0    & \nodata \\ 
  124051   &   G059  &  2  & 2    & 2 \\
  124052   &   G073\tablefootmark{d}  &  2  & 2    & 0 \\
  124053   &   G124  &  4  & 4(5) & 2(3) \\
  125002   &   G072  &  1  & 0    & \nodata \\
  125018   &   G112  &  1  & 1    & \nodata \\
  125026   &   G143  &  1  & 1    & \nodata \\
  125027   &   G131  &  2  & 2    & 0 \\
  125056   &   G052  &  2  & 1    & \nodata \\
  125107   &   G068  &  1  & 1    & \nodata \\
  125132   &   G063  &  1  & 1(2) & \nodata \\
\hline
  All      &         & 38  & 24(27) & 8(10) \\
\hline
\end{tabular}                                                                                                              
\tablefoot{
\tablefoottext{a}{\small We list in brackets the number of CO detections at
distinct redshifts. In three cases, we find two objects at different
redshifts associated with the same SPIRE target, G191\,07, G124\,01, and
G063\,02.}
\tablefoottext{b}{\small We consider two redshift measurements similar if
${\mid}\Delta{z}{\mid}/(1+${<}z{>}$){<}$0.02.}
\tablefoottext{c}{\small The field G237 contains a spectroscopically confirmed
protocluster at the redshift of one of the two CO-detected
sources~\citep{polletta21,koyama21}.}
\tablefoottext{d}{\small ALMA observations of both targets in G073
yielded continuum detections of five sources (three in G073\,03, and two in
G073\,15), and a CO(3--2) line detection at $z{\sim}$1.51--1.54 in three of
them~\citep[][; Hill, R.  priv.  comm.]{kneissl19}.  Our EMIR observations
detected only one CO line at such redshifts~\citep[see][]{kneissl19}, and
one at a different redshift in G073\,15.}
}
\end{table}

\subsection{Multiplicity}

The possibility that multiple sources might contribute to a single SPIRE
source emission is quite likely, especially in bright sources.  It is well
known that bright SPIRE sources suffer from multiplicity, for instance in studies
of SPIRE sources with bright density fluxes (S$_{\rm 500\mu
m}{=}$35--80\,mJy) 9--23\% of them are multiple
sources~\citep{montana21,scudder16}.

In our PHz-IRAM sample, we detect lines at two different redshifts in one
source, G063\,02. In Table~\ref{tab:CO_properties}, we report
both measurements and add the suffixes "a" and "b" to the \herschel\ ID to
differentiate them. Note that in the following analysis we adopt the full
SPIRE emission for both sources and thus, both their IR luminosity and SFR
are shown as upper limits.

There are also two SPIRE sources, G191\,07, and G124\,01, whose CO lines
have double Gaussian profiles that could be due to two sources at similar
redshifts (see Fig.~\ref{fig:dbl_trans}).
In G124\,01, both CO transitions (\COthree, and \COfour) exhibit a double Gaussian
profile. The mean difference in velocity between the two peaks (${<}z_{\rm
low}{>}$\,=\,2.1522, and ${<}z_{\rm high}{>}$\,=\,2.1540) is
${<}\delta$v{>}$\sim$251\,\kms, whereas the velocity difference
between the two transitions is 26\,\kms, and 79\,\kms, for the low, and the
high $z$ values, respectively.
In G191\,07, we detect three double peak CO lines (\COthree, \COfour, and
\COfive).  In this source, ${<}z_{\rm low}{>}$\,=\,2.62237, and ${<}z_{\rm
high}{>}$\,=\,2.62514, thus the mean velocity difference between the two
peaks is ${<}\delta$v{>}$\sim$316\,\kms, and the highest velocity
differences across the three transitions are 84\,\kms, and 257\,\kms, for the low, and
the high $z$ values, respectively.

The offset between the two peaks in both sources is larger than the
difference across the three CO transitions, strongly favoring the double
Gaussian solution in both G124\,01, and G191\,07.  In addition, the line
profile is not symmetric, with the two peaks having different intensities
and widths, favoring the presence of kinematically
distinct components suggestive of a merger system, rather than of a rotating 
disk at the origin of these double peak profiles.  Observations at high
resolution would help to distinguish these two possible scenarios.

In the only PHz field for which high spatial resolution mm observations are
available, G073, \citet{kneissl19} find from one to four ALMA-detected
objects in eight SPIRE sources.  Subsequent ALMA observations of this field,
find that {two SPIRE sources contain two galaxies at
the same redshift (Hill, priv. comm.)}.

In summary, we find evidence of multiplicity in three out of 24
sources (12\%), consistent with previous
studies~\citep{montana21,scudder16}.  The 350$\mu$m flux densities of these
three sources range from 57 to 116\,mJy, which are consistent with the peak
of the flux density distribution, and they are not the brightest examples. 
We cannot rule out the possibility that other sources might be multiple, but
the limited frequency coverage and integration times of our IRAM
observations, might have not revealed sources at different redshifts, or
with fainter CO emission.

\begin{table*}[!ht]
\centering
\caption{CO-derived properties\label{tab:CO_properties}}
\begin{tabular}{lrccrrr}
\hline \hline
 ID        & PHz ID   & Her ID  & Line    & \multicolumn{1}{c}{L$^{\prime}_\mathrm{CO}$} & \multicolumn{1}{c}{L$^{\prime}_\mathrm{CO(1-0)}$\tablefootmark{a}} & \multicolumn{1}{c}{M$_{\rm gas}$\tablefootmark{b}} \\ 
           &          &         &         & \multicolumn{2}{c}{(10$^{10}$\,K\,\kms\,pc$^2$)} &             (10$^{10}$\,\msun) \\ 
\hline
\noalign{\vspace{2pt}}
 G176    &      57  &   01   &  CO(3--2)   &   22.26$\pm$1.48    &   42.81$\pm$2.84  &    149.85$\pm$ 9.95  \\
 G176    &      57  &   02   &  CO(3--2)   &    3.88$\pm$1.28    &    7.46$\pm$2.46  &     26.10$\pm$ 8.61  \\
 G173    &     237  &   01   &  CO(3--2)   &    3.11$\pm$0.67    &    5.97$\pm$1.29  &     20.90$\pm$ 4.51  \\
 G006\tablefootmark{c}    &     631  &   01   &  CO(2--1)   &    4.36$\pm$0.87    &    5.19$\pm$1.04  &     18.18$\pm$ 3.64  \\
         &          &        &  CO(3--2)   &    5.94$\pm$1.19    &   11.42$\pm$2.28  &     39.98$\pm$ 8.00  \\
 G237    &    712   &  962   &  CO(3--2)   &    3.68$\pm$0.70    &    7.08$\pm$1.35  &     24.80$\pm$ 4.73  \\
 G237    &    712   & 9741   &  CO(4--3)   &    0.93$\pm$0.30    &    2.28$\pm$0.72  &      7.98$\pm$ 2.53  \\
 G191    &     832  &   01   &  CO(3--2)   &    9.66$\pm$1.16    &   18.58$\pm$2.23  &     65.05$\pm$ 7.81  \\
 G191    &     832  &   07   &  CO(3--2)   &    8.18$\pm$1.36    &   15.73$\pm$2.62  &     55.07$\pm$ 9.18  \\
         &          &   07a  &  CO(3--2)   &    4.42$\pm$1.55    &    8.50$\pm$2.97  &     29.75$\pm$10.41  \\
         &          &   07b  &  CO(3--2)   &    3.54$\pm$1.55    &    6.81$\pm$2.98  &     23.83$\pm$10.43  \\
 G191    &     832  &   26   &  CO(2--1)   &    8.49$\pm$1.73    &   10.10$\pm$2.07  &     35.36$\pm$ 7.23  \\
 G059    &  124051  &   01   &  CO(4--3)   &    4.66$\pm$0.55    &   11.37$\pm$1.35  &     39.80$\pm$ 4.72  \\
 G059    &  124051  &   03   &  CO(4--3)   &    1.46$\pm$0.31    &    3.57$\pm$0.76  &     12.48$\pm$ 2.67  \\
 G073    &  124052  &   03   &  CO(2--1)   &    3.24$\pm$1.00    &    3.85$\pm$1.19  &     13.49$\pm$ 4.18  \\
 G073\tablefootmark{c}    &  124052  &   15   &  CO(2--1)   &    2.87$\pm$0.82    &    3.42$\pm$0.97  &     11.97$\pm$ 3.40  \\
         &          &        &  CO(3--2)   &    3.81$\pm$1.08    &    7.33$\pm$2.08  &     25.66$\pm$ 7.29  \\
 G124    &  124053  &   01   &  CO(3--2)   &    5.25$\pm$0.79    &   10.09$\pm$1.51  &     35.33$\pm$ 5.30  \\
         &          &   01a  &  CO(3--2)   &    2.05$\pm$0.42    &    3.94$\pm$0.81  &     13.79$\pm$ 2.83  \\
         &          &   01b  &  CO(3--2)   &    3.02$\pm$0.60    &    5.80$\pm$1.16  &     20.30$\pm$ 4.06  \\
 G124    &  124053  &   02   &  CO(3--2)   &    6.84$\pm$0.95    &   13.16$\pm$1.83  &     46.06$\pm$ 6.40  \\
 G124    &  124053  &   03   &  CO(5--4)   &    2.85$\pm$0.78    &    8.90$\pm$2.42  &     31.13$\pm$ 8.48  \\
 G124    &  124053  &   15   &  CO(3--2)   &    7.31$\pm$1.39    &   14.06$\pm$2.68  &     49.22$\pm$ 9.37  \\
 G112    &  125018  &   06   &  CO(3--2)   &    4.00$\pm$1.34    &    7.69$\pm$2.58  &     26.92$\pm$ 9.04  \\
 G143    &  125026  &   02   &  CO(5--4)   &    3.96$\pm$0.88    &   12.36$\pm$2.75  &     43.28$\pm$ 9.62  \\
 G131    &  125027  &   01   &  CO(2--1)   &    6.33$\pm$0.66    &    7.53$\pm$0.78  &     26.36$\pm$ 2.74  \\
 G131\tablefootmark{c}    &  125027  &   15   &  CO(2--1)   &    3.29$\pm$0.88    &    3.92$\pm$1.05  &     13.72$\pm$ 3.67  \\
         &          &        &  CO(3--2)   &    4.36$\pm$1.17    &    8.39$\pm$2.24  &     29.36$\pm$ 7.85  \\
 G052    &  125056  &   01   &  CO(3--2)   &    5.36$\pm$0.84    &   10.31$\pm$1.61  &     36.09$\pm$ 5.63  \\
 G068    &  125107  &   02   &  CO(2--1)   &   14.73$\pm$1.78    &   17.54$\pm$2.12  &     61.39$\pm$ 7.41  \\
 G063\tablefootmark{c}    &  125132  &  02a   &  CO(2--1)   &    5.83$\pm$1.45    &    6.94$\pm$1.73  &     24.30$\pm$ 6.04  \\
         &          &        &  CO(3--2)   &    7.34$\pm$1.82    &   14.12$\pm$3.51  &     49.40$\pm$12.28  \\
         &          &  02b   &  CO(3--2)   &    5.68$\pm$1.28    &   10.93$\pm$2.46  &     38.25$\pm$ 8.61  \\
\hline                                                                                                                     
\end{tabular}                                                                                                              
\tablefoot{
\tablefoottext{a}{\small The luminosity \LpCOone\ is obtained from the \LpCO\ of the
reported CO transition and assuming the SMG brightness temperature ratios from \citet{bothwell13}.
The uncertainty introduced by this choice of brightness temperature ratio is
a factor of 1.8 on the derived \LpCOone, and gas mass.} 
\tablefoottext{b}{\small Gas masses are derived from \LpCOone\ assuming $\alpha_\mathrm{CO}$\,=\,3.5\,\msun\,pc$^{-2}$\,(K\,\kms)$^{-1}$~\citep{magdis17}.}
\tablefoottext{c}{\small The two estimates are derived from the same line,
but assuming two different transitions and relative redshifts.}
}
\end{table*}

\subsection{Do PHz fields contain high-$z$ proto-structures ?}

The main goal of the EMIR observations was to determine whether the PHz fields
contain high-$z$ structures.  Finding multiple sources at similar
redshifts~\citep[${\mid}\Delta{v}{\mid}{<}$2000\,(1+${<}z{>}$)\,\kms; see
][]{eisenhardt08} would support such a hypothesis.  

We measure redshifts for multiple Herschel sources in close projected
proximity in eight PHz fields.  In half of those fields we detect two
objects at similar redshifts. In the following, we will refer to these as
the structures' redshifts.
In two of the fields with two sources at the same redshift, G191, and G124,
two SPIRE sources have a double peak CO line implying the presence of two
sources at the same redshifts.  Thus in these two fields, we find three
sources at consistent redshift associated with two distinct SPIRE sources.

In three fields more than two SPIRE sources have been observed (four in
G124, five in G176, and nine in G191), but out of the 18 targets in this
subset, six are at the structure redshift, three are at different
redshifts, and no redshift was measured in the remaining nine.  The
number of observed, and detected sources and of those at similar redshift in
each targeted field are listed in Table~\ref{tab:iram_summary}.  The lack of
CO detection can be in part explained by insufficient exposure times. 
Indeed the primary targets that led the first CO-detections in each field
were a factor of 2--3 brighter than the secondary targets and were usually
observed with 1.5--3 times longer exposure times.  This observing strategy
was defined to find CO line detections at the same redshift as the primary
targets.  A line was successfully detected in ten out of 20
secondary targets.  In the remaining cases, we cannot rule out the
possibility that the sources are at the same redshift as the primary target
as the achieved depth might have been insufficient to detect a CO
line in those additional fainter targets.

These results, although based only on eight fields, provide support to the
hypothesis that the PHz fields contain overdensities of DSFGs at
$z{\simeq}$2--3, but also indicate that DSFGs situated along the line of
sight contribute to the \planck\ signal.  We can thus conclude that
the \planck\ selection technique is efficient in finding overdensities of
highly star-forming systems at $z{\simeq}$2, but the measured \planck\
signal is also affected by line of sight projections, as predicted
by~\citet{miller15}, and~\citet{negrello17}.

\subsection{Far-infrared properties: luminosities, dust temperatures and
SFRs}\label{sec:fir_properties}

With the CO-derived redshifts, accurate infrared luminosities can be
measured for all CO-detected SPIRE sources.  The total (8--1000\,$\mu$m) IR
luminosities (L$_{\rm IR}$), dust temperatures (T$_{\rm dust}$), and SFRs are
estimated by fitting the SPIRE data with single-temperature modified
blackbody models.  SCUBA-2 data at 850$\mu$m are also included in
two cases~\citep[G006\,01, and G068\,02; ][]{mackenzie17}.  Fits were
performed using the {\tt cmcirsed} package \citep{casey12c} and assuming the
CO-derived redshift and a dust emissivity-index $\beta$ equal to
1.8~\citep{cortese14,pokhrel16}.  The code returns also the uncertainties on
L$_{\rm IR}$ and T$_{\rm dust}$. Note that the quoted uncertainties in \LIR\ do not
account for the uncertainty in the redshift as it is negligible compared to
the uncertainties associated with the photometric points and the best fit model.  From the
IR luminosities, SFR estimates are derived assuming the relationship in
\citet{kennicutt98a}, modified for a Chabrier IMF \citep{chabrier03}, 
SFR/(M$_\odot$\,yr$^{-1}$)\,{=}\,9.5\,${\times}$\,10$^{-11}$\,L$_{\rm
IR}$/L$_\odot$.

\begin{figure*}[h!]
\centering
\includegraphics[width=14cm]{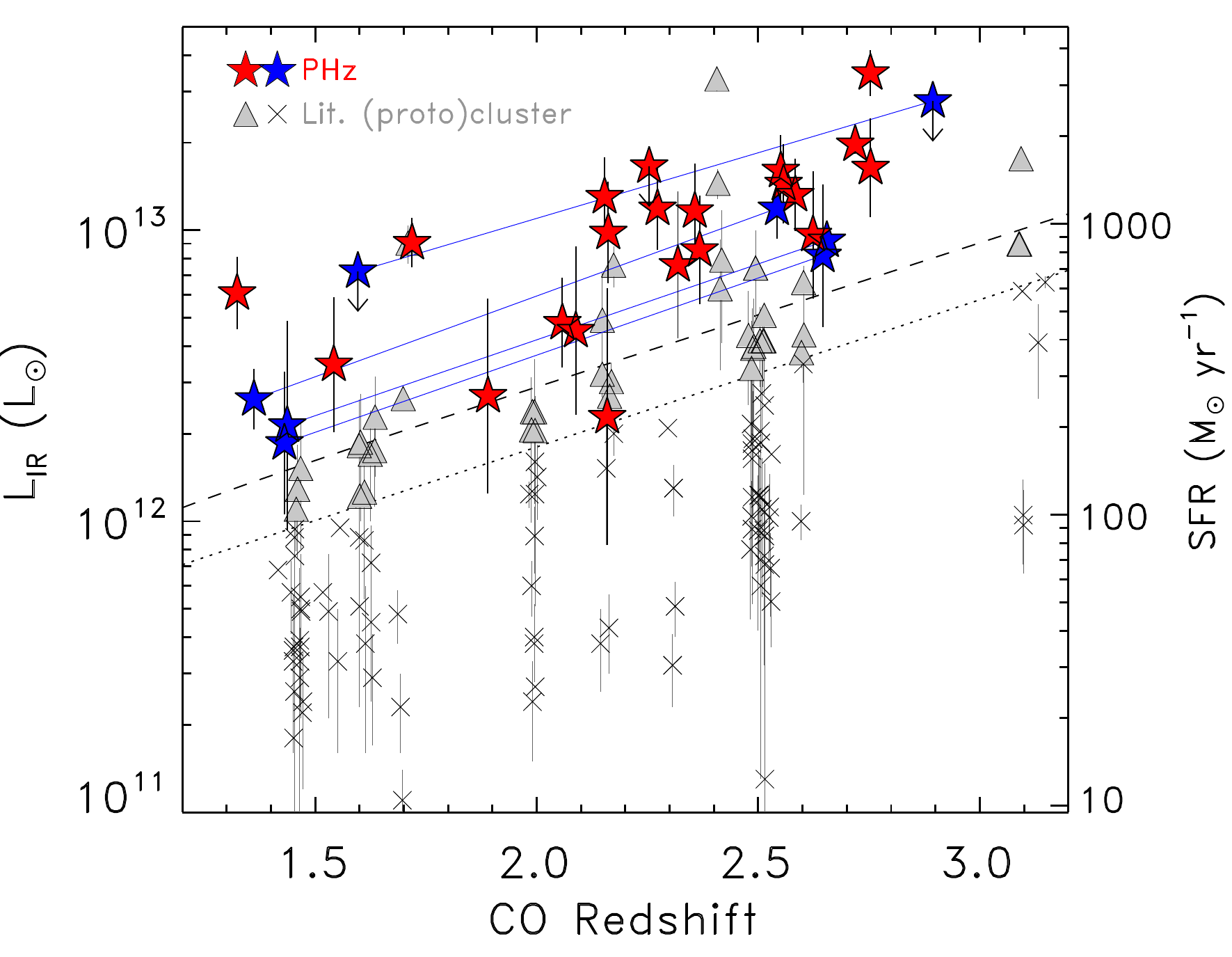}
\caption{{\small Total IR (8--1000\,$\mu$m) luminosity or SFR as a function
of redshift for the PHz-IRAM sources (full red stars for
secure line identifications, and full blue stars for uncertain line
identifications), and for 1.4${<}z{<}$3.1 cluster and protocluster members
from the literature (gray triangles and crosses).  
In case of multiple EMIR detections associated with the same
SPIRE source, an upper limit to L$_{\rm IR}$ is reported with a downward
pointing arrow. Blue stars connected with a solid blue line refer to the same source
and to two possible redshifts.
The black dashed line represents the total IR luminosity of a source with S(350$\mu$m)\,=\,40\,mJy
and T$_\mathrm{dust}$\,=\,30\,K~\citep[derived using the {\tt cmcirsed}
package; ][]{casey12c}, corresponding to the selection of the majority of
the PHz-IRAM targets.  The dotted line is the same L$_{\rm IR}$ limit scaled by $-$0.2\,dex
and used to select a sub-set of cluster and protocluster members from the
literature for comparison (gray triangles). }}
\label{fig:LIR_z}
\end{figure*}

The FIR-derived parameters L$_{\rm IR}$, T$_{\rm dust}$, and SFR are listed in
Table~\ref{tab:mm_properties}.  Infrared luminosities as a function of redshift
are shown in Fig.~\ref{fig:LIR_z}.  In case of dubious line identifications
(4 cases) we show both redshifts and IR luminosities in the figure. 
All detected targets are classified as ULIRGs \citep[L$_{\rm
IR}\,{\geq}\,10^{12}$\,L$_\odot$; ][]{sanders88a}, and 50\% are HyLIRGs
(L$_{\rm IR}\,{\geq}\,10^{13}$\,L$_\odot$) with consequently large
(${\ga}\,$100\,M$_\odot$\,yr$^{-1}$) SFRs (see Fig.~\ref{fig:LIR_z}).  The
highest SFR (${\gtrsim}\,$3000\,M$_\odot$\,yr$^{-1}$) is measured in
G176\,01, which is the highest redshift source together with G176\,02 (both
at $z_{\rm CO}\,{=}\,$2.75).  The average SFR, considering only the secure
CO identifications, is 1043$\pm$157\,\msun\,yr$^{-1}$.  The large IR
luminosities are due to the sample selection, as illustrated by the dashed
line corresponding to the IR luminosity of a source with $S_\mathrm{350\mu
m}$\,=\,40\,mJy and T$_\mathrm{dust}$\,=\,30\,K.

\begin{table*}[ht!]
\caption{Far-infrared properties\label{tab:mm_properties}}
\begin{center}
\renewcommand{\arraystretch}{1.4}
\begin{tabular}{llr cr c ccc}
\hline 
ID         & PHz ID  & Her ID &    $z$    & SFR$_{\rm IR}$          & log(L$_\mathrm{IR}$)&T$_{\rm dust}$ &$\tau_\mathrm{dep}$ & $z_\mathrm{dep}$  \\
           &         &        &           & (\msun\,yr$^{-1}$)      &       (\lsun)       &     (K)       &       (Gyr)        &                   \\
\hline                                                                                                                        
G176       &     57  &    01  &   2.7534  &   3294$^{+649}_{-542}$  &    13.54$\pm$0.08   &    34$\pm$1   &     0.45$\pm$0.09  &      2.28  \\
G176       &     57  &    02  &   2.7536  &   1557$^{+738}_{-501}$  &    13.22$\pm$0.17   &    35$\pm$3   &     0.17$\pm$0.09  &      2.56  \\
G173       &    237  &    01  &   2.5572  &   1374$^{+511}_{-372}$  &    13.16$\pm$0.14   &    30$\pm$2   &     0.15$\pm$0.06  &      2.40  \\
G006       &    631  &    01  &   1.3615  &    250$^{+ 68}_{- 53}$  &    12.42$\pm$0.10   &    20$\pm$1   &     0.73$\pm$0.23  &      1.13  \\
           &    631  &    01  &   2.5422  &   1129$^{+306}_{-241}$  &    13.08$\pm$0.10   &    30$\pm$2   &     0.35$\pm$0.11  &      2.21  \\
G237       &    712  &   962  &   2.0577  &    457$^{+194}_{-136}$  &    12.68$\pm$0.15   &    28$\pm$2   &     0.54$\pm$0.22  &      1.71  \\
G237       &    712  &  9741  &   2.1590  &    217$^{+382}_{-138}$  &    12.36$\pm$0.44   &    27$\pm$5   &     0.37$\pm$0.46  &      1.89  \\
G191       &    832  &    01  &   2.5509  &   1523$^{+499}_{-376}$  &    13.21$\pm$0.12   &    30$\pm$2   &     0.43$\pm$0.13  &      2.15  \\
G191       &    832  &    07  &   2.6236  &    919$^{+601}_{-363}$  &    12.99$\pm$0.22   &    32$\pm$3   &     0.60$\pm$0.33  &      2.07  \\
           &    832  &   07a  &   2.6225  & $<$918$^{+601}_{-363}$  & $<$12.99$\pm$0.22   &   \nodata     &  $>$0.32$\pm$0.20  &   $<$2.30  \\
           &    832  &   07b  &   2.6247  & $<$920$^{+602}_{-364}$  & $<$12.99$\pm$0.22   &   \nodata     &  $>$0.26$\pm$0.18  &   $<$2.36  \\
G191       &    832  &    26  &   1.8899  &    256$^{+299}_{-138}$  &    12.43$\pm$0.34   &    23$\pm$4   &     1.38$\pm$1.21  &      1.25  \\
G059       & 124051  &    01  &   2.3568  &   1107$^{+503}_{-346}$  &    13.07$\pm$0.16   &    30$\pm$2   &     0.36$\pm$0.14  &      2.06  \\
G059       & 124051  &    03  &   2.3677  &    813$^{+433}_{-283}$  &    12.93$\pm$0.19   &    27$\pm$2   &     0.15$\pm$0.08  &      2.23  \\
G073       & 124052  &    03  &   1.5418  &    329$^{+232}_{-136}$  &    12.54$\pm$0.23   &    26$\pm$3   &     0.41$\pm$0.26  &      1.37  \\
G073       & 124052  &    15  &   1.4303  &    177$^{+133}_{- 76}$  &    12.27$\pm$0.25   &    20$\pm$2   &     0.68$\pm$0.45  &      1.19  \\
           &         &    15  &   2.6454  &    777$^{+588}_{-335}$  &    12.91$\pm$0.25   &    30$\pm$3   &     0.33$\pm$0.22  &      2.31  \\
G124       & 124053  &    01  &   2.1531  &   1238$^{+453}_{-332}$  &    13.12$\pm$0.14   &    31$\pm$2   &     0.29$\pm$0.10  &      1.94  \\
           &         &   01a  &   2.1544  & $<$1240$^{+454}_{-332}$ & $<$13.12$\pm$0.14   &    \nodata    &  $>$0.11$\pm$0.02  &   $<$2.07  \\
           &         &   01b  &   2.1523  & $<$1237$^{+453}_{-331}$ & $<$13.12$\pm$0.14   &    \nodata    &  $>$0.16$\pm$0.03  &   $<$2.03  \\
G124       & 124053  &    02  &   2.1614  &    932$^{+458}_{-307}$  &    12.99$\pm$0.17   &    30$\pm$3   &     0.49$\pm$0.21  &      1.81  \\
G124       & 124053  &    03  &   2.7189  &   1873$^{+150}_{-139}$  &    13.30$\pm$0.03   &    38$\pm$0   &     0.17$\pm$0.05  &      2.53  \\
G124       & 124053  &    15  &   2.0882  &    429$^{+408}_{-209}$  &    12.66$\pm$0.29   &    27$\pm$4   &     1.15$\pm$0.85  &      1.44  \\
G112       & 125018  &    06  &   2.3187  &    723$^{+572}_{-319}$  &    12.88$\pm$0.25   &    30$\pm$4   &     0.37$\pm$0.26  &      2.02  \\
G143       & 125026  &    02  &   2.5836  &   1259$^{+415}_{-312}$  &    13.12$\pm$0.12   &    21$\pm$1   &     0.34$\pm$0.13  &      2.25  \\
G131       & 125027  &    01  &   1.3235  &    577$^{+193}_{-145}$  &    12.78$\pm$0.13   &    27$\pm$2   &     0.46$\pm$0.14  &      1.17  \\
G131       & 125027  &    15  &   1.4367  &    203$^{+262}_{-114}$  &    12.33$\pm$0.36   &    26$\pm$5   &     0.68$\pm$0.65  &      1.20  \\
           &         &    15  &   2.6549  &    876$^{+146}_{-125}$  &    12.97$\pm$0.07   &    38$\pm$0   &     0.34$\pm$0.10  &      2.31  \\
G052       & 125056  &    01  &   2.2726  &   1130$^{+436}_{-315}$  &    13.08$\pm$0.14   &    30$\pm$2   &     0.32$\pm$0.12  &      2.02  \\
G068       & 125107  &    02  &   1.7185  &    861$^{+186}_{-153}$  &    12.96$\pm$0.09   &    28$\pm$1   &     0.71$\pm$0.16  &      1.39  \\
G063       & 125132  &   02a  &   1.5962  & $<$686$^{+240}_{-178}$  & $<$12.86$\pm$0.13   &    \nodata    &  $>$0.35$\pm$0.09  &   $<$1.44  \\
           &         &   02a  &   2.8942  &$<$2644$^{+161}_{-152}$  & $<$13.44$\pm$0.03   &    \nodata    &  $>$0.19$\pm$0.05  &   $<$2.66  \\
           &         &   02b  &   2.2537  &$<$1578$^{+551}_{-408}$  & $<$13.22$\pm$0.13   &    \nodata    &  $>$0.24$\pm$0.06  &   $<$2.06  \\
\hline                                                                                                      
\end{tabular} \\                                                                                            
\end{center}                                                                                                
{\small The SFR$_{\rm IR}$ is derived by fitting a modified black body                                     
spectrum ($\beta$=1.8) to the SPIRE, and SCUBA, when available, photometric                                 
data points, and assuming the \citet{kennicutt98a} L$_{\rm IR}$-SFR relation
corrected for a Chabrier IMF. In case of multiple sources associated with the same
\herschel\ source, \LIR, the SFR, and $z_\mathrm{dep}$ should be considered as upper
limits, $\tau_\mathrm{dep}$ as a lower limit, and T$_{\rm dust}$ is unconstrained 
as the FIR emission is blended. }                                                                         
\end{table*}                                                                                                

In Fig.~\ref{fig:LIR_z}, we also show, for comparison, the IR luminosities
of cluster and protocluster members from the literature with
1.4${<}z{<}$3.1, and for which CO observations are available (see full list
in Table~\ref{tab:lit_samples}).  The IR luminosities from the literature
sample extend to much lower values (see also Table~\ref{tab:lit_samples})
than the PHz-IRAM sample.  For the purpose of a proper comparison, we select
only a subset of the literature sources with IR luminosities greater than
our limit $-$0.2\,dex (see dotted line in Fig.~\ref{fig:LIR_z}).  This
choice, although somewhat arbitrary, takes into account the uncertainty on
the L$_\mathrm{IR}$ values, and yields a sample size (41 sources) well suited for a
comparison.

\begin{figure}[h!]
\centering
\includegraphics[width=\linewidth]{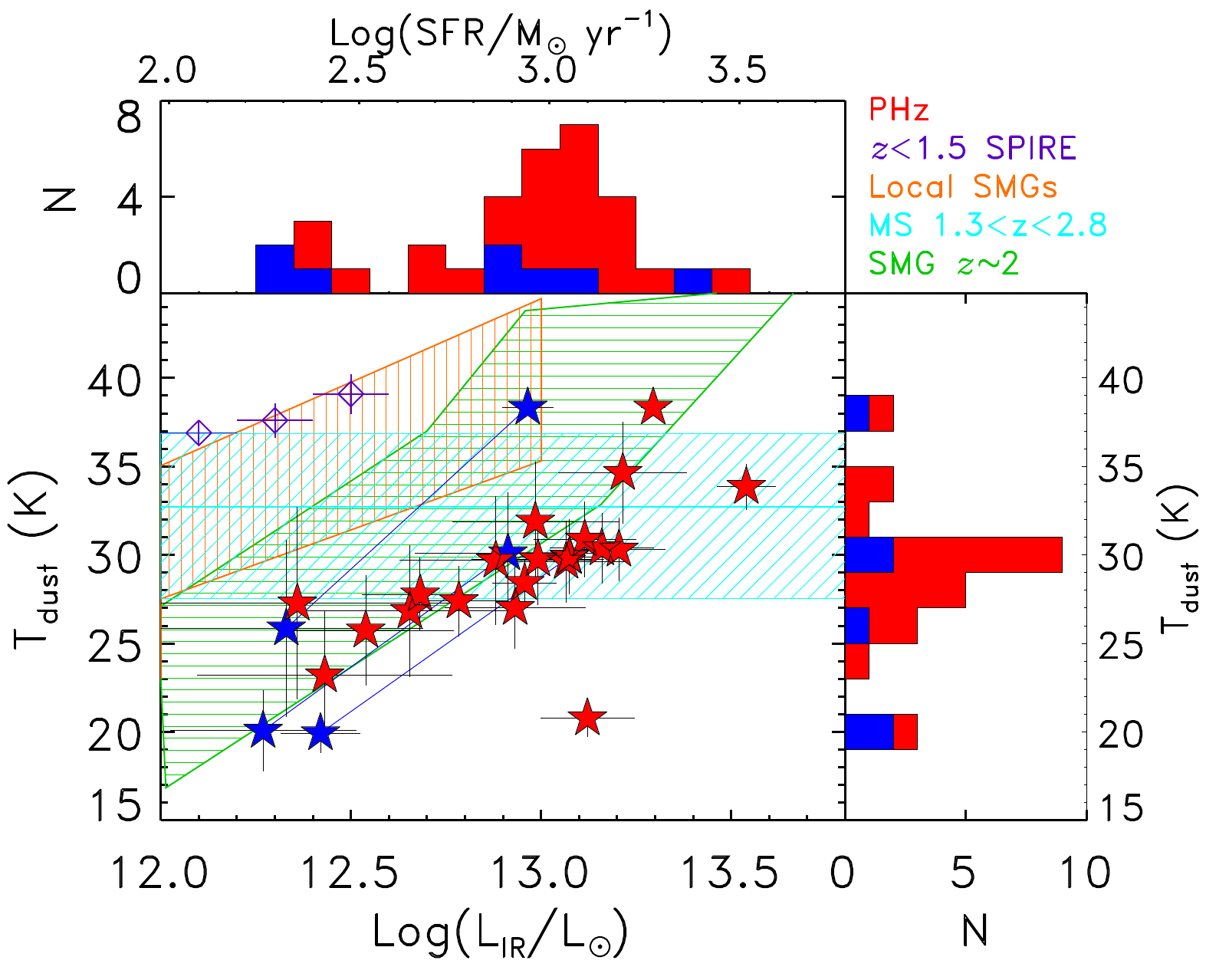}
\caption{{\small {\it Main panel:} Dust temperature as a function of the IR
(8--1000\,$\mu$m) luminosity for the PHz-IRAM sources (full stars,
red for secure CO line identifications, and blue for the uncertain ones.). The shaded orange area shows the local T$_{\rm
dust}$--\LIR\ relation derived by~\citet{chapman03}, linearly extrapolated
to 10$^{13}$\lsun. The shaded cyan area shows the range of dust
temperatures for main sequence galaxies in the redshift range of the
PHz-IRAM sources (1.3${<}z{<}$2.8) according to the relation found
by~\citet{magnelli14}, and extended to \LIR$>$10$^{13}$\,\lsun.
The shaded green area presents results for $z{\simeq}$2 SMGs
from~\citet{chapman05}. The purple diamonds represent the relation found for
$z{<}$1.5 SPIRE sources by~\citet{symeonidis13}. The {\it top}, and {\it
right} panels show the distributions of SFR (and IR luminosity) and dust
temperature, respectively, for all PHz-IRAM sources.}}
\label{fig:LIR_Tdust}
\end{figure}

In Fig.~\ref{fig:LIR_Tdust}, we show the estimated dust temperatures as a
function of IR luminosities for our sources.  The T$_{\rm
dust}$--\LIR\ relations derived for local IR-selected
galaxies~\citep{chapman03}, for $z{\simeq}$2 SMGs~\citep{chapman05}, and for
$z{<}$1.5 SPIRE sources~\citep{symeonidis13} are also shown for comparison. 
We also report the dust temperatures expected for main sequence~\citep[MS:
SFR--$\mathcal{M}$ relation of SFGs; see e.g.  ][]{speagle14} galaxies in
the redshift range of our PHz-IRAM sources (1.3${<}z{<}$2.8) based on the
relation between T$_{\rm dust}$ and the offset from the MS reported
by~\citet{magnelli14}.  In galaxies, the
intensity of the radiation field increases with lookback
time~\citep[e.g.][]{magdis12,huang14,bethermin15} along with a concurrent
rise in dust temperatures.  Indeed local galaxies have typically colder
dust than those at higher redshift~\citep[e.g., T$_{\rm
dust}{\sim}$20--55\,K in local U/LIRGs, and 25--60\,K in those at
$z{\simeq}$2--2.5, whereas normal SFGs have T$_{\rm dust}{\sim}$20\,K at
$z{\sim}$0, and T$_{\rm dust}{\gtrsim}$30\,K at
$z{\geq}$1;][]{clements18,magdis10,symeonidis18,cortese14}.

The PHz-IRAM sources, with an average $\left\langle T_{\rm
dust}\right\rangle\,{=}\,(29.2\pm0.9)$\,K exhibit dust temperatures
consistent with those observed in MS galaxies at similar redshifts, but colder 
than those typical of local SMGs and $z{<}$1.5 SPIRE sources.
Compared to the $z\,{\simeq}\,2$ SMG population, there is some overlap, but
more on the low T$_{\rm dust}$--high L$_\mathrm{IR}$ side.  Note that higher
$\beta$ values would yield lower dust temperatures, but even if we assume
$\beta$\,=\,1.5 as in~\citet{chapman05}, the observed temperature offset
will still be present.  We consider the possibility that such a difference
might be a consequence of fitting the sub-mm SED without data at
$\geq$850$\mu$m which are typically available for SMGs.  However, in the two
cases where a SCUBA2\,850$\mu$m flux measurement is available, i.e. 
G006\,01, and G068\,02, the estimated dust temperatures are similarly low. 
We thus conclude that our sources are characterized by lower dust
temperatures, by $\sim$5\,K on average, than the $z\,{\simeq}\,2$ SMG
population.

Our results on the dust temperature indicate that our PHz-IRAM sources are
similar to those observed in normal SFGs at similar redshifts.

\subsection{The CO spectral line energy distribution}\label{sec:co_sled}

The shape of the CO spectral line energy distribution (SLED) is linked to
the underlying molecular gas density and kinetic temperature~\citep[see
e.g., ][]{schirm14,daddi15,canameras18}.  Typically, the
warmer and denser the molecular gas is, the more populated the upper levels
will be with a SLED rising faster with the line frequency.
The CO SLED shape depends on the fraction of dense to total gas and can thus
indicate the star-forming mode~\citep[merger $versus$
disk;][]{daddi10b,zhang14}.  Thus studying the CO excitation properties
offers a mean to investigate what powers the elevated SFRs of our sources. 
For example, merger-driven starburst galaxies are
much more highly excited in their high-J CO transitions than disk
galaxies~\citep{weiss07,papadopoulos12}.  Lower densities and
temperatures~\citep[T$_\mathrm{kin}\sim$15--20\,K,
log(n$_\mathrm{H_2}$/cm$^{-3}$)$\sim$3.0; ][]{carilli13} yield SLEDs that
peak at lower transition values, like for instance in the Milky Way (peak at
$J_{\rm up}$\,=\,3).

\begin{figure*} 
\centering
\includegraphics[width=\linewidth]{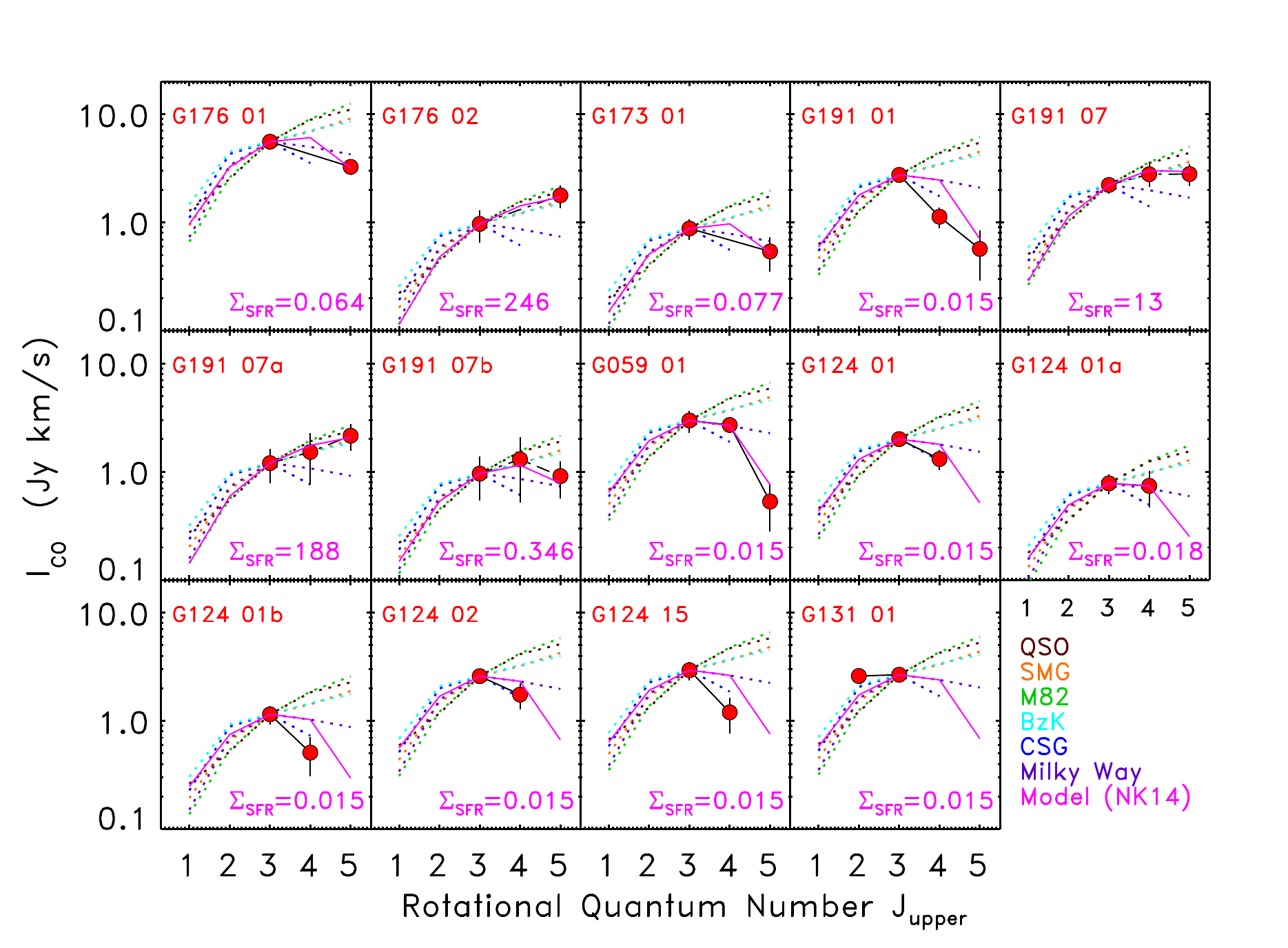}
\caption{{\small CO SLEDs of the PHz-IRAM sources (full 
red circles connected by a black solid line). 
The CO SLEDs of different types of galaxies normalized to
the observed \COthree\ intensity of each PHz-IRAM source are shown for
comparison in each panel as dotted lines: the Milky
Way~\citep[purple;][]{fixsen99}; SMGs
(orange), QSO (brown), the prototypical starburst galaxy M82 (green), and
color-selected SFGs (CSG; blue) from~\citet{carilli13}; and BzK galaxies at
$z{\sim}$1.5 (cyan) from~\citet{daddi15}. The solid magenta line in each
panel shows the SLED model from~\citet{narayanan14} obtained assuming the annotated SFR density in
\msun\,yr$^{-1}$\,kpc$^{-2}$. }}
\label{fig:co_sled}
\end{figure*}

Since we have observed, at the most, three CO transitions in the same
source, we cannot build a full CO SLED, but we can build portions of
the CO SLED and compare them with those observed in other sources from the
literature.  In this analysis we do not consider G059\,03 because only one
out of the three CO lines is well detected. In case of sources with double peak lines (G191\,07a, G191\,07b,
G124\,01a, and G124\,01b), we show the CO SLED derived from the single Gaussian fit
and for each component of the double Gaussian fit.
In Fig.~\ref{fig:co_sled}, we show portions of the CO
SLEDs of ten PHz-IRAM sources for which multiple CO transitions are
detected, compared with those typical of SMGs, quasi stellar objects (QSO),
color selected SFGs (CSGs), BzK-selected galaxies, and to that of the
starburst galaxy M82, and of the Milky Way~\citep{carilli13,bothwell13}.

\begin{figure*} 
\centering
\includegraphics[width=\linewidth]{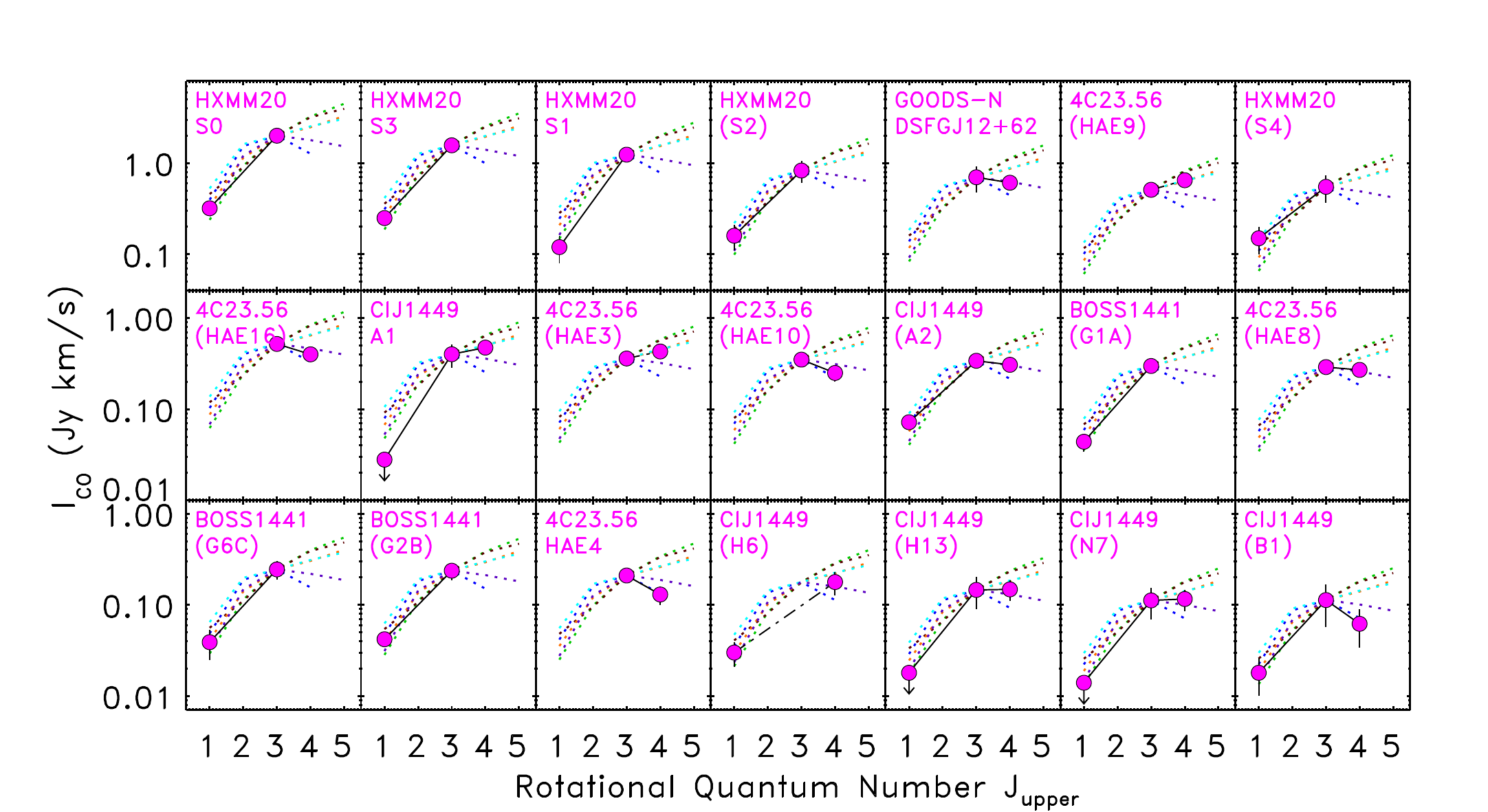}
\caption{{\small CO SLEDs of protocluster members from the
literature~\citep[full pink circles connected by black
lines;][]{coogan18,casey16,lee17,lee19,gomez19} compared with the CO
SLED of different types or sources (dotted lines of color as in
Fig.~\ref{fig:co_sled}).  Downward pointing arrows indicate upper limits.
The protocluster and source names are annotated (see Table~\ref{tab:lit_samples}). Source
names between brackets are those that do not satisfy the IR luminosity
selection criterion.}}
\label{fig:co_sled_lit}
\end{figure*}

The CO SLEDs of our sources peak, typically, at low rotational
quantum numbers, similar to the Milky Way rather than to the SMGs at similar
redshifts~\citep[$J_{\rm up}^{\rm MW}$\,=\,3, $J_{\rm up}^{\rm
SMGs}{\simeq}$6; ][]{carilli13}.  In eight sources, or 80\% of the sources
with measured line ratios, the CO SLED peaks at $J_{\rm up}$\,=\,3.  There
are only two exceptions, G191\,07 (including 07a, and 07b), and G176\,02
where the peak is at $J_{\rm up}\geq$\,4, consistent with what is observed
in SMGs at similar redshifts.  These results suggest that the molecular gas
in the most PHz-IRAM sources must be at low densities, cold
(T$\sim$10--15\,K) and at low-excitation.  To investigate whether these
properties are observed in other protoclusters, we searched for protocluster
members from the literature with multiple CO transitions.  We found 21
sources from four protoclusters~\citep[HXMM20, 4C\,23.56, BOSS\,1441,
GOODS-N, and Cl\,J1449$+$0856;][]{gomez19,lee17,lee19,casey16,coogan18}. 
Their CO SLEDs, compared with those from the same classes of galaxies used
earlier, are shown in Fig.~\ref{fig:co_sled_lit}.  The CO properties of
these sources are listed in Table~\ref{tab:lit_samples}.  For this analysis,
we include also sources that do not satisfy the IR luminosity selection
criterion applied earlier to have the largest and most complete comparison sample, but we
flag them in the Figure (see source name in bracket).

In eight cases, the available CO data are insufficient to constrain the SLED
peak because only transitions with $J_{\rm{up}}\leq$3 are available.  In the
remaining 13 cases where higher transitions are available, low excitation
SLEDs are observed in seven sources (DSFGJ123711+621331, HAE16,
HAE10, A2, HAE8, HAE4, and B1), a high excitation SLED is observed
in one case (HAE9), and a SLED with intermediate excitation properties in
the remaining five sources.  Thus, half of the members from the literature
have CO excitation properties as in normal SFGs.  Similar conclusions were
drawn by~\citet{coogan18} based on the CO SLED analysis of the
Cl\,J1449$+$0856 cluster members.

To further investigate this result and possible implications, we
resort to theoretical models.  By combining numerical simulations
with molecular line radiative transfer calculations, \citet{narayanan14}
develop a model for the physical parameters that drive variations in the CO
SLEDs of galaxies.  They find that the shape of the SLED is determined by
the gas density, temperature and optical depth distributions, and that these
quantities are correlated with a galaxy mean star formation rate surface
density ($\Sigma_{\rm SFR}$).  Based on this model, we derive the
$\Sigma_{\rm SFR}$ values that reproduce at best the observed CO SLEDs.  The
predicted SLEDs, obtained assuming the relation for unresolved
sources~\citep[see eq.~(19) and Table~3 in][]{narayanan14} are shown in
Fig.~\ref{fig:co_sled} after normalizing them at the observed \COthree\
intensity, and the derived $\Sigma_{\rm SFR}$ are reported in
Table~\ref{tab:sfr_size}.  In four PHz-IRAM sources (G191\,01, G124\,01,
G124\,01b, G124\,15, G131\,01), the measured SLED is not well reproduced by
the model.  Since, the parametrized form that links the line ratios to the
SFR surface density is valid only for $\Sigma_{\rm
SFR}\geq$1.5$\times$10$^{-2}$\,\msun\,yr$^{-1}$\,kpc$^{-2}$, and the peak of
the SLED decreases for lower $\Sigma_{\rm SFR}$, the lack of a best fit points to
$\Sigma_{\rm SFR}$ values that are lower than the minimum assumed by the
model. Such a minimum is not physically motivated, but due to the fact that
the relation in the model cannot be parameterized with the same analytical
expression at lower $\Sigma_{\rm SFR}$.

\begin{table}[!ht]
\centering
\caption{SFR and molecular gas density and extent\label{tab:sfr_size}}
\renewcommand{\arraystretch}{1.2}
\begin{tabular}{ccccc} 
\hline \hline
 PHz    &\herschel& log($\Sigma_{\rm SFR}$)       & R$_\mathrm{SFR}$\tablefootmark{a} & log($\Sigma_{\rm M_\mathrm{gas}}$) \\
 ID     &    ID   & (\msun\,yr$^{-1}$\,kpc$^{-2}$)    &       (kpc)               & (\msun\,pc$^{-2}$) \\
\hline
  G176  &     01  & $-$1.19$^{+0.28}_{-0.26}$     &    128$^{+ 61}_{- 43}$   &   \;\;1.5$^{+ 0.4}_{-0.4}$   \\ 
        &     02  &\;\;2.39$^{+1.90}_{-2.13}$     &    1.4$^{+ 19}_{-1.3}$   &   \;\;\;4.6$^{+ 2.2}_{-2.5}$ \\ 
  G173  &     01  & $-$1.12$^{+0.73}_{-0.59}$     &     76$^{+100}_{- 48}$   &   \;\;\;1.01$^{+ 1.0}_{-0.8}$\\ 
  G191  &     01  & ${<}-$1.82\tablefootmark{b}   &  $>$180                  &   $<$0.8                     \\
        &     07  &\;\;1.12$^{+1.82}_{-1.72}$     &      5$^{+ 39}_{-  4}$   &   \;\;3.9$^{+ 2.1}_{-2.0}$   \\ 
        &    07a  &\;\;2.28$^{+2.02}_{-2.28}$     &    1.2$^{+ 21}_{-1.2}$   &   \;\;4.8$^{+ 2.4}_{-2.7}$   \\ 
        &    07b  & $-$0.46$^{+2.62}_{-1.52}$     &     29$^{+187}_{-28}$    &   \;\;2.0$^{+ 3.0}_{-2.0}$   \\ 
  G059  &     01  & $-$1.82 (F)\tablefootmark{c}  &  $>$153                  &   $<$0.7                     \\ 
  G124  &     01  & ${<}-$1.82\tablefootmark{b}   &  $>$162                  &   $<$0.6                     \\
        &    01a  & $-$1.75$^{+2.61}_{-0.72}$     &    150$^{+251}_{-143}$   &   \;$-$0.3$^{+ 2.8}_{-1.0}$  \\ 
        &    01b  & ${<}-$1.82\tablefootmark{b}   &  \nodata                 &   \nodata                    \\
        &     02  & $-$1.82 (F)\tablefootmark{c}  &  $>$141                  &   $<$0.9                     \\
        &     15  & ${<}-$1.82\tablefootmark{b}   &  $>$95                   &   $<$1.2                     \\
  G131  &     01  & ${<}-$1.82\tablefootmark{b}   &  $>$111                  &   $<$0.8                     \\
\hline
\end{tabular}                                                                                                              
\tablefoot{
\tablefoottext{a}{\small R$_\mathrm{SFR}$ is derived assuming a circular
region with area derived from SFR/$\Sigma_{\rm SFR}$, and the SFRs} from
Table~\ref{tab:mm_properties}.
\tablefoottext{b}{\small An upper limit to the SFR surface density
equal to the minimum value allowed by the model~\citep[i.e,
$\Sigma_{\rm SFR}^{\rm model}{=}$0.015\msun\,yr$^{-1}$\,kpc$^{-2}$;][]{narayanan14} is assumed
when the model does not reproduce the observed CO SLED.}
\tablefoottext{c}{\small The SFR surface density is fixed 
to the minimum value allowed by the model~\citep[i.e,
$\Sigma_{\rm SFR}^{\rm model}{=}$0.015\msun\,yr$^{-1}$\,kpc$^{-2}$;][]{narayanan14}
as this provides an acceptable fit to the observed CO SLED.}
}
\end{table}

From the estimated SFR surface brightness, and the measured SFRs (see
Table~\ref{tab:mm_properties}), we can derive the size of the star-forming
region.  Making the assumption that the observed SFR is produced by the
measured molecular gas mass and that they have the same extent, we can also
derive a molecular gas density.  All these values are reported in
Table~\ref{tab:sfr_size}.  Based on the prescription between line ratios and
$\Sigma_{\rm SFR}$ parameterized by~\citet{narayanan14}, the size
of the star-forming region or molecular gas extent would be implausibly
large (R$_{\rm gas}{\gtrsim}$70\,kpc) in most of the cases.  Thus, either
the model, or the model parameters, like the gas density and temperature,
are not appropriate for our sources, or our SLEDs are not correct. 
This could be the case if dust obscuration is present as it might depress
the emission in the high transitions~\citep{papadopoulos10b}.  It is also
possible that the low transition peak of the observed SLEDs is artificially
produced by the variation of the beam size with wavelength (i.e., the
IRAM/30m main beam is 29\arcsec\ at 86\,GHz, and 16\arcsec\ at
145\,GHz)\footnote{https://publicwiki.iram.es/Iram30mEfficiencies} in case
of extended or contaminated sources, or inaccurate pointing during the
observations. Gravitational lensing can also produce a larger
magnification of the diffuse gas emission seen in the low transitions than that
coming from the compact gas traced by higher transitions, artificially
producing a low peak in the CO SLED~\citep[see][]{hezaveh12}.  This
possibility will be discussed in Sect.~\ref{sec:lensing}. To further
investigate these results it would be necessary to obtain CO observations at
higher spatial resolution.  Such observations would provide the size of the
molecular gas distribution and reveal whether multiple sources are present
in the EMIR low frequency beam enhancing the CO flux at smaller
frequencies, or whether some of our sources are affected by gravitational
lensing.

\subsection{CO luminosities}

The CO luminosities are calculated following~\citet{solomon97} as:
\begin{equation}
L'_{\rm CO} = \frac{c^2}{2 k} S_{\rm CO} \left( \Delta V\right) \nu^{-2}_{\rm obs} D^2_{\rm L} (1+z)^{-3}
\end{equation}
\noindent
where $S_{\rm CO} \left( \Delta V\right)$ is the line intensity derived from
the Gaussian fit, and $\nu_{\rm obs}$ the line observed frequency (see
Table~\ref{tab:lines}).  The derived CO luminosities are then converted to
the \COone\ luminosity, \LpCOone, using the brightness temperature ratios
measured for SMGs by~\citet{bothwell13}\footnote{\citet{bothwell13} report
the following brightness temperature ratio values,
r$_{J,J-1}$\,=\,L$^{\prime}_\mathrm{CO(J-J-1)}$/\LpCOone\,=\,0.84, 0.52,
0.41, and 0.32 for J\,=\,2, 3, 4, and 5, respectively.}.

\begin{figure}[h!]
\centering
\includegraphics[width=\linewidth]{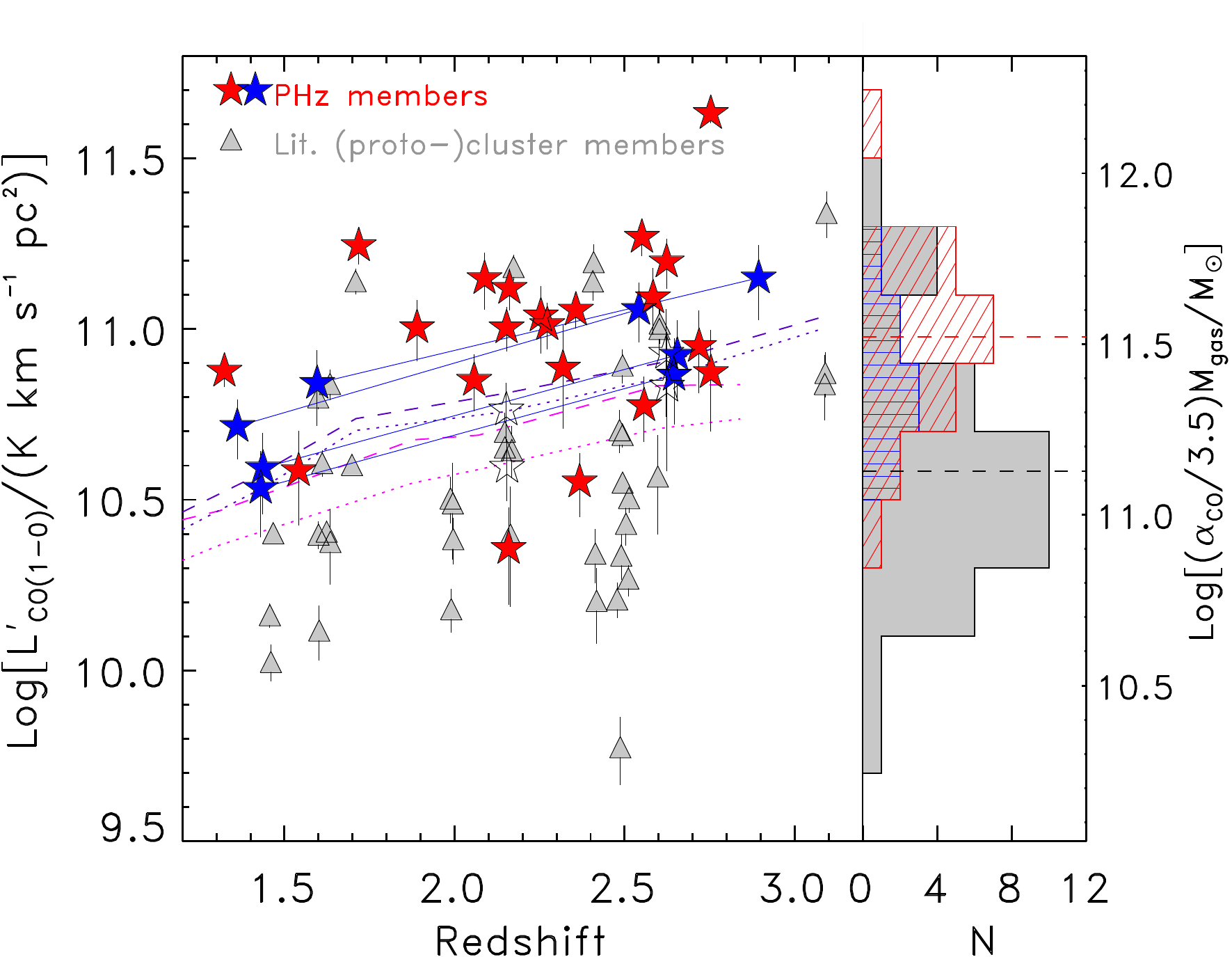}
\caption{{\small \COone\ luminosity as a function of redshift for the
PHz-IRAM sources. Symbols like in Fig~\ref{fig:LIR_z}.
The purple and magenta curves show the expected \LpCOone\ rms at,
respectively, 3\,mm (E090), and 2\,mm (E150) reached in 5\,hrs of integration time, 
in the summer (dashed curve) and in the winter (dotted curve) semesters
assuming average conditions, and a resolution of 45\,\kms.
\textit{Right panel:} CO luminosity distributions for the same samples shown
in the main panel. Horizontal dashed lines represent the mean luminosities
for the PHz-IRAM sources with secure line identifications (red), and for the literature sources (black).
The vertical y-axis on the right hand-side shows the molecular gas masses
derived from the CO(1--0) luminosity assuming $\alpha_{\rm CO}{=}$3.5.}}
\label{fig:LCO_z}
\end{figure}

Since the CO SLEDs of the PHz-IRAM sources are not always consistent
with those typical of SMGs, this choice might introduce large uncertainties
on \LpCOone\ and on the gas mass estimate.  The choice of a different SLED
would yield a \LpCOone\ value a factor of $\sim$1.8 higher (r$_{3,2}^{\rm
starburst}{/}r_{3,2}^{\rm SMG}$\,=\,0.93/0.52\,=\,1.8), or $\sim$0.5
smaller (r$_{3,2}^{\rm Milky-Way}{/}r_{3,2}^{\rm
SMG}$\,=\,0.27/0.52\,=\,0.5) when derived from \COthree.  Larger
uncertainties would result when using luminosities from CO transitions with
J$_{\rm upper}{>}$3. In case of multiple detected CO lines, we thus
choose that at the lowest transition, \COthree\ in most of the cases, to
derive \LpCOone\ and minimize the uncertainty associated with the brightness
temperature ratios.  In two cases (G059\,01, and G059\,03), we preferred a
higher transition, \COfour, because more significantly detected than
\COthree.  The derived luminosities are listed in
Table~\ref{tab:CO_properties}, and plotted as a function of redshift in
Fig.~\ref{fig:LCO_z}.  In all the figures where \LpCOone\ is shown,
we report a unique value per source in case of secure line identification,
and two values in case the line is not univocally identified.

For comparison, we include the \COone\ luminosities
from cluster and protocluster members from the literature at 1.4${<}z{<}$3.1
after applying the L$_\mathrm{IR}$ cut (see Table~\ref{tab:lit_samples}). 
The distributions of \COone\ luminosities for the PHz-IRAM sources and the
literature sample are shown on the right hand panel of Fig.~\ref{fig:LCO_z}. 
The PHz \COone\ luminosities are, on average, 0.4\,dex higher than those of
the literature sources. From \LpCOone, it is possible to
derive the molecular gas mass through a conversion factor,
$\alpha_\mathrm{CO}$, that depends on the physical properties of the gas. 
The $\alpha_\mathrm{CO}$ for normal SFGs at high-$z$ is typically
3.5~\citep{magdis17}.  The molecular gas masses derived from \LpCOone\
assuming $\alpha_{\rm CO}$\,=\,3.5 (see Sect.~\ref{sec:mol_masses}), are
shown on the right hand axis in Fig.~\ref{fig:LCO_z}.  The PHz sources are
among the most luminous and, thus with the largest gas reservoirs,
CO-detected galaxies found in overdense regions at 1.4${<}z{<}$3.1.  This
result can be in part explained by the selection bias in favor of the
brightest sub-mm sources targeted and detected per field, as illustrated by
the dashed and dotted curves shown in the Figure.  On the other hand, there
must also be an intrinsic property associated with the extreme luminosities
and gas masses found in our sources because of the wide range of \COone\
luminosities observed at fixed IR luminosity.

\subsection{PHz-IRAM sources: normal SFGs or starbursts ?}\label{sec:ms_sb}

\begin{figure*}[h!]
\centering
\includegraphics[width=14cm]{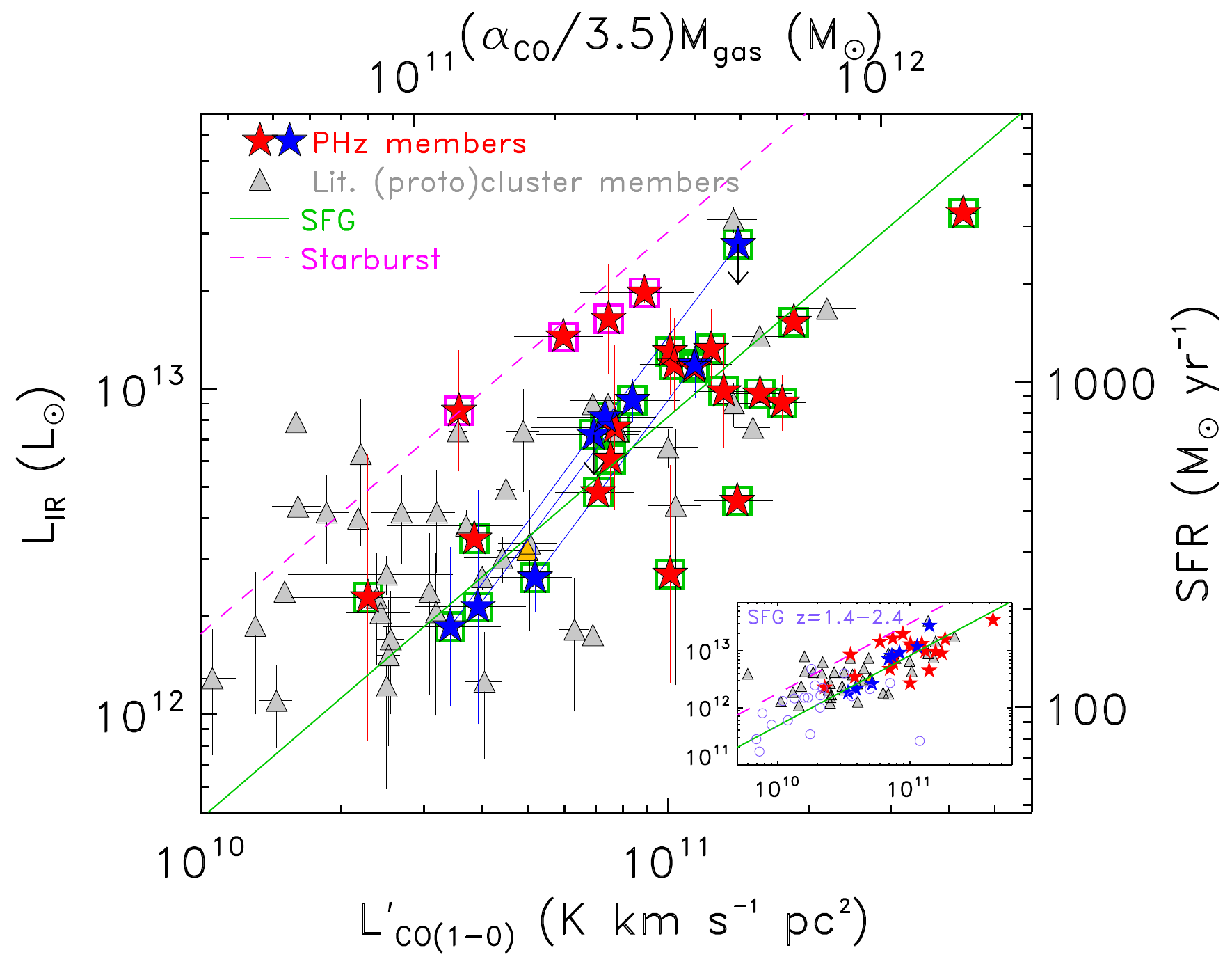}
\caption{{\small Total IR (8--1000\,$\mu$m) luminosity or SFR as a function
of \COone\ luminosity for the PHz-IRAM sources, and for 1.4${<}z{<}$3.1 cluster
and protocluster members from the literature (same symbols as in
Fig.~\ref{fig:LCO_z}). The protocluster massive spiral
HAE229~\citep{dannerbauer17} is shown as full gold triangle.  The dashed
magenta line and solid green lines represent, respectively, the average
relations for normal SFGs and starburst galaxies~\citep{sargent14}.  Open
magenta or green squares are overplotted over the PHz sources classified as
starburst, or normal SFGs based on the agreement with the drawn relations. 
In the \textit{inset}, we show the same graph and data, but extended to
lower luminosities and include a sample of CO-detected normal SFGs at
$z$\,=\,1.4--2.4~\citep[lilac open circles;][]{daddi10b,tacconi13}.}}
\label{fig:LIR_LCO}
\end{figure*}

Star-forming galaxies follow a relation between the \COone\ luminosity and
L$_\mathrm{IR}$ that describes, in observable terms, the relationship
between the luminosity due to star formation and the total gas content. 
This relation is different for normal SFGs and starburst galaxies, with the
latter having \LpCOone\ smaller by 0.46\,dex, on average, at fixed IR
luminosity than normal SFGs~\citep[as formulated by
][]{sargent14}\footnote{\citet{sargent14} find
log(\LpCOone/(K\,\kms\,pc$^2$))\,=\,0.54$\pm$0.02$+$
(0.81$\pm$0.03)$\times$log(L$_{\rm IR}$/\lsun) for normal SFGs.}.  In
Fig.~\ref{fig:LIR_LCO}, we show the relations for SFGs and starburst
galaxies, and the values derived for our PHz-IRAM sources and for the
cluster and protocluster galaxies drawn from the literature.  Both the
PHz-IRAM sources, and those drawn from the literature agree with the scaling
relations, although with a large scatter.  Indeed at fixed IR luminosity,
\LpCOone\ can vary by up to a factor of five in our sample, compared to the
factor of almost three that separates normal SFGs from starburst galaxies. 
The large scatter means that the SFE varies across the sample, but some of
the scatter might be also due to a multiplicity effect.  The IR luminosity
might be overestimated because derived assuming that all the SPIRE flux is
emitted by the CO source.  The same might be true for the CO luminosity if
multiple sources at the same redshift are present.

Based on the comparison with the two relations, we classify our sources as
normal SFGs if log(\LpCOone)$>$log(\LpCOone)$^\mathrm{SFG}{-}$0.3, where
\LpCOone$^\mathrm{SFG}$ is the expected \LpCOone\ assuming the SFG
relation~\citep{sargent14} at the observed L$_\mathrm{IR}$, and starburst if
log(\LpCOone)$<$log(\LpCOone)$^\mathrm{SFG}{-}$0.3.  Normal SFGs and
starbursts are identified in the figure with, respectively, green and
magenta open squares.  Most of our sources (16/20 sources or 80\%
of those with secure CO identification) are consistent with or below
the normal SFGs relation, and four (20\%) are in the starburst
locus.  Those classified as starburst (G176\,02, G173\,01, G059\,03, and
G124\,03) are all at high redshifts ($z{>}$2.3), but other sources at
similarly high redshift are classified as normal SFGs.  If we consider the
sources with double peak lines, G191\,07, and G124\,01, as made of two
separate galaxies, we find that in one case (G124\,01), both galaxies (01a,
and 01b) would be classified as starbursts.  This is, however, the case if
we do not deblend the SPIRE flux.  If the IR luminosity is scaled by their
contribution to the total CO luminosity their values would be consistent
with the SFG relation.  The starburst galaxies represent 20$\pm$10\%
(4 out of 20 sources) of the PHz-IRAM sources with secure line
identifications.  If we interpret this fraction as the starburst phase
timescale in the PHz-IRAM sample, we can infer that such a phase
lasts $\sim$20\% of the galaxy lifetime.  Thus, during most of
their lifetime the bright PHz sources behave like normal SFGs.

Among the sources from the literature, a larger fraction, i.e.  32\% (13 out
of 41 sources) falls in the starburst locus compared to our sample, but the
majority is on the normal SFG relation. It is interesting to point
out that, although the PHz-IRAM SFGs agree with the normal SFG relation,
their luminosities are systematically larger than measured in typical SFGs
at similar redshifts.  In the inset of Fig.~\ref{fig:LIR_LCO}, we show the
luminosities of the sub-sample of normal SFGs at $z$\,=\,1.4--2.4 used
by~\citet{sargent14} to derive the normal SFG relation.  This includes 14
galaxies from the PHIBSS sample~\citep{tacconi13}, and six BzK
galaxies~~\citep{daddi10b}.  Our PHz-IRAM sources occupy the high luminosity
tail of the luminosity distribution observed in the normal SFG sample, with
average values of IR and CO luminosities being eight, and five
times larger than those relative to the normal SFG sample, respectively. 
With respect to the protocluster SFG members drawn from the literature, the
PHz-IRAM sources exhibit IR and CO luminosities that are, on average, only a
factor of two higher.  A factor of two is not significant considering the
scatter in the luminosity distributions\footnote{The mean \LpCOone, and IR
luminosities of the SFGs in the PHz-IRAM sample, of the SFG protocluster
members drawn from the literature, and of the normal SFGs at
$z$\,=\,1.4--2.4 are, respectively,
${<}$log(\LpCOone/(K\,\kms\,pc$^2$))${>}$\,=\,11.0$\pm$0.1, 10.7$\pm$0.1,
and 10.3$\pm$0.1, and ${<}$log(L$_\mathrm{IR}$/\lsun)${>}$\,=\,12.9$\pm$0.1,
12.6$\pm$0.1, and 12.0$\pm$0.1.}.  Thus, our selection of bright red sub-mm
galaxies yields mostly normal SFGs, but with larger IR and CO luminosities,
and thus SFRs and gas masses, than typically observed in normal SFGs, and in
other protocluster members at similar redshifts.

\subsubsection{CO excitation and L$_\mathrm{IR}$--\LpCOone\ relation}

In Sect.~\ref{sec:co_sled}, we were able to identify the sources in
our sample with highly excited CO, as expected in starburst galaxies.  It is
thus interesting to compare the classification based on the
L$_\mathrm{IR}$--\LpCOone\ relation, with the CO excitation level.  Among
the four sources classified as starbursts, the CO SLED has been measured for
only G176\,02, and G173\,01.  The CO SLED of G176\,02 (peak at $J_{\rm
up}$\,=\,5) is consistent with a starburst classification, G173\,01 has,
instead, a line ratio I$_{\rm CO(5-4)}{/}$I$_{\rm
CO(3-2)}$\,=\,0.61$\pm$0.25, consistent with the MW CO SLED (i.e., I$_{\rm
CO(5-4)}{/}$I$_{\rm CO(3-2)}$\,=\,0.82), and much lower than observed in
starburst galaxies (e.g., I$_{\rm CO(5-4)}{/}$I$_{\rm CO(3-2)}$\,=\,2.24 in
the prototypical starburst M82).  For the other two starburst galaxies, the
CO SLED is not constrained.  Conversely, the only other source with a highly
excited CO SLED, G191\,07, is consistent with the normal SFG
L$_\mathrm{IR}$--\LpCOone\ relation.  In summary, out of the ten sources for
which the CO SLED is measured, one is classified starburst and has a highly
excited CO SLED (G176\,02), eight are classified normal SFGs and have a CO SLED
consistent with low excitation, and one is classified starburst, but is not
highly excited (G173\,01).  We remind the reader that the IR luminosity of
our sources might be overestimated in case of multiplicity affecting the
measured \herschel\ emission, and that in such a case the true \LIR\ could
be lower moving a source towards the normal SFG relation.  We can thus
conclude that the classification based on the L$_\mathrm{IR}$--\LpCOone\ diagram
well matches the gas excitation level.

\subsection{Molecular gas masses and depletion times}\label{sec:mol_masses}

Gas masses depend on the assumed CO SLED as this provides the intensity of
the \COone\ line from higher transitions, and on the CO--H$_{\mathrm 2}$
conversion factor ($\alpha_{\mathrm{CO}}$\,=\,M$_{\rm H_{\rm
2}}$/L$^{\prime}_\mathrm{CO}$) to convert the CO gas mass into H$_{\mathrm
2}$ mass~\citep[see review by][]{bolatto13}.  As discussed earlier, the
brightness ratios adopted for our sources to derive the \COone\ luminosity
from that at higher transitions, are those observed in
SMGs~\citep{bothwell13}, even if the CO SLEDs of our sources seem to differ
from those.  On the other hand, assuming brightness ratios typical of normal
SFGs like the Milky Way would yield \COone\ luminosities, and gas masses
$\sim$2 times larger~\citep{carilli13}.  Since our values are already higher
than those typically observed in other $z{\sim}$2 galaxies, we prefer to use
the standard SMG ratios that are also commonly adopted in the literature,
and more suited for a comparison with those samples.

Regarding the CO--H$_{\mathrm 2}$ conversion factor, in normal SFGs such as
the Milky Way, $\alpha_\mathrm{CO}$ is $\sim$4.36 (including the
contribution of Helium to the molecular gas mass), and for starburst
galaxies and mergers, it is typically $\sim$0.8~\citep{bolatto13}.  Studies of
the molecular gas at high redshifts have shown that such a dichotomy breaks
down and that $\alpha_\mathrm{CO}$ covers a broad and continuous range of
values between $\sim$0.2 and
$\sim$10\,\msun\,pc$^{-2}$\,(K\,\kms)$^{-1}$~\citep{tacconi08,casey14}.  The
$\alpha_{\rm CO}$ value decreases with the offset from the MS relation that links a galaxy SFR with its stellar mass~\citep[see eq. 
2 in ][]{castignani20}.  Thus, galaxies above the MS, like the starburst
galaxies, have smaller $\alpha_{\rm CO}$ values.  

The $\alpha_{\rm CO}$ value depends also on the galaxy gas
metallicity~\citep{genzel12,inoue21}.
Since we do not have estimates of stellar mass nor of gas-phase metallicity
for our sources, we assume a solar metallicity for all sources and adopt
$\alpha_\mathrm{CO}$\,=\,3.5\,\msun\,pc$^{-2}$\,(K\,\kms)$^{-1}$ as reported
in~\citet{magdis17} for normal SFGs with solar metallicity at $z{\sim}$2. 
For the PHz classified as starbursts based on their position in the
\LpCOone--L$_\mathrm{IR}$ diagram (see previous section), we assume
the same $\alpha_\mathrm{CO}$ value for consistency within the PHz-IRAM
sample.  The estimated gas masses assuming
$\alpha_\mathrm{CO}$\,=\,3.5\,\msun\,pc$^{-2}$\,(K\,\kms)$^{-1}$ are listed
in Table~\ref{tab:CO_properties}.

\begin{figure} 
\centering
\includegraphics[width=\linewidth]{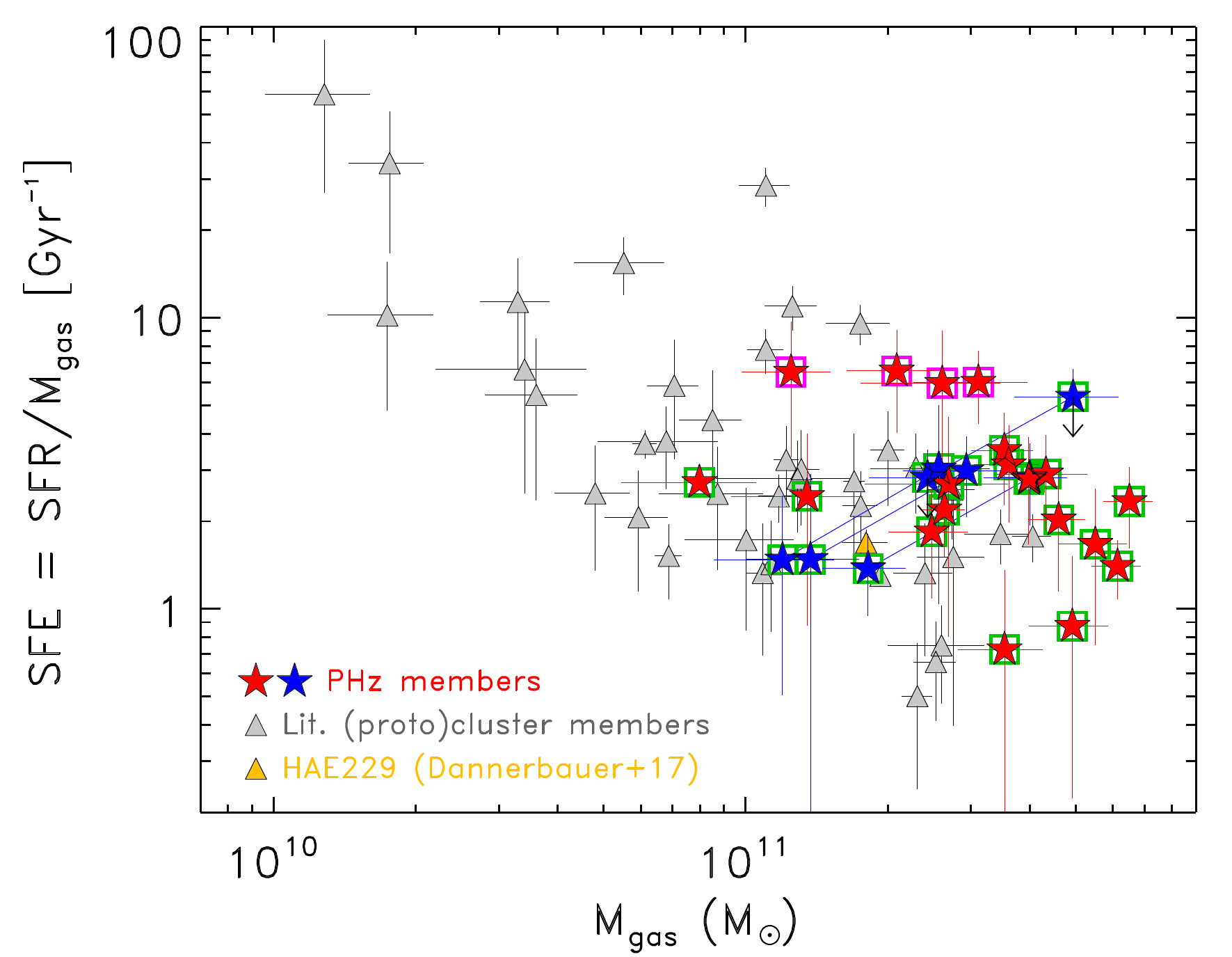}
\caption{{\small Star formation efficiency (SFE) of the PHz-IRAM sources as
a function of gas mass. Symbols as in Fig.~\ref{fig:LIR_LCO}.}}
\label{fig:sfe_mgas}
\end{figure}

From the gas masses, we can compute the star formation efficiency (SFE),
given by the ratio between the SFR, and the molecular gas mass.  The SFE
shows how efficiently the molecular gas mass is converted into stars.  The
estimated SFEs for our sources as a function of gas mass are compared with
those from the literature in Fig.~\ref{fig:sfe_mgas}.

The PHz-IRAM sources exhibit a broad range of SFEs and gas masses,
both spanning 1\,dex.  The broad range of SFEs is probably a reflection of
the extent in redshifts, SFRs, and gas masses observed in our sample,
although there is no clear correlation between the SFE and those quantities. 
The only noticeable features are that the sources with the highest SFE
values are all classified as starburst, and the least efficient sources are
among those with the largest gas masses (M$_{\rm
gas}{>}$3$\times$10$^{11}$\,\msun). 

\begin{figure} 
\centering
\includegraphics[width=\linewidth]{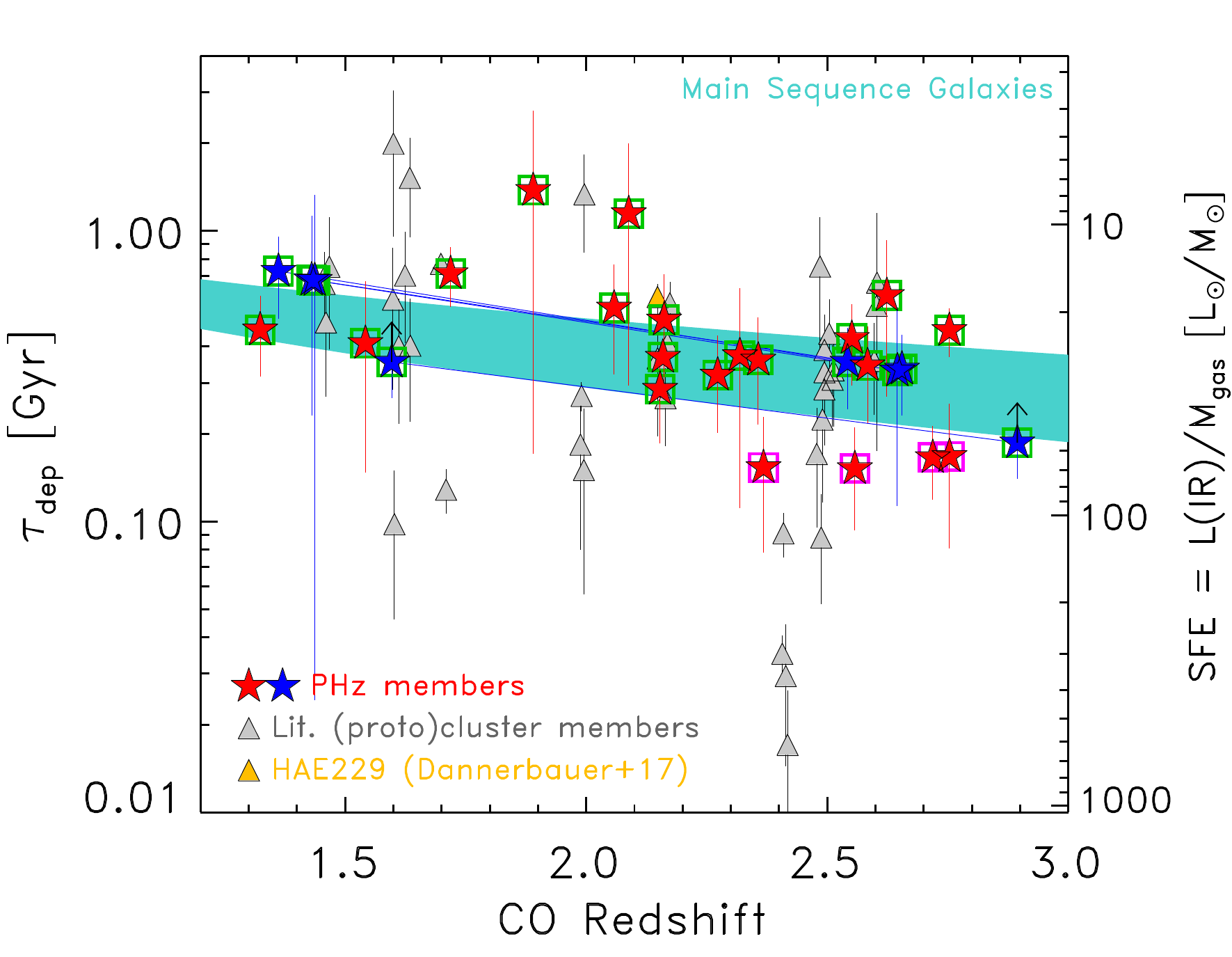} \caption{{\small
Estimated depletion times of the PHz-IRAM sources as a function of redshift. 
Symbols as in Fig.~\ref{fig:LIR_LCO}.  The turquoise region is the locus of
normal SFGs~\citep{saintonge13}.}}
\label{fig:tdepl_z}
\end{figure}

The inverse of the SFE, defined as the molecular gas depletion time,
$\tau_\mathrm{dep}{\equiv}$M$_\mathrm{gas}$/SFR, represents the
time-scale required to exhaust the molecular gas mass with the current SFR. 
Normal SFGs at high redshifts ($z{\gtrsim}$1.5) are characterized by
$\tau_\mathrm{dep}{\gtrsim}$0.45\,Gyr~\citep{tacconi13,saintonge13},
slightly decreasing at higher
redshifts~\citep[1.5$\times$(1${+}z$)$^{-1.5}{<}\tau_{\rm
dep}{<}$1.5$\times$(1${+}z$)$^{-1}$; ][]{saintonge13}.  In
Fig.~\ref{fig:tdepl_z}, we show the estimated gas depletion times of our
PHz-IRAM sources (see values in Table~\ref{tab:mm_properties}).  We compare
them with the selected protocluster sample from the literature, and with
the aforementioned evolutionary track for normal
SFGs~\citep{saintonge13}.

The average depletion time of the PHz-IRAM sources is ${<}\tau_{\rm
dep}{>}$\,=\,0.47$\pm$0.07\,Gyr.  Similar depletion times are observed in
the protoclusters drawn from the literature.  The large uncertainties
associated with these estimates do not allow us to analyze any trend with
the redshift. Indeed, they are all consistent with the average value, and
with the normal SFGs' evolutionary track.

\begin{figure} 
\centering
\includegraphics[width=\linewidth]{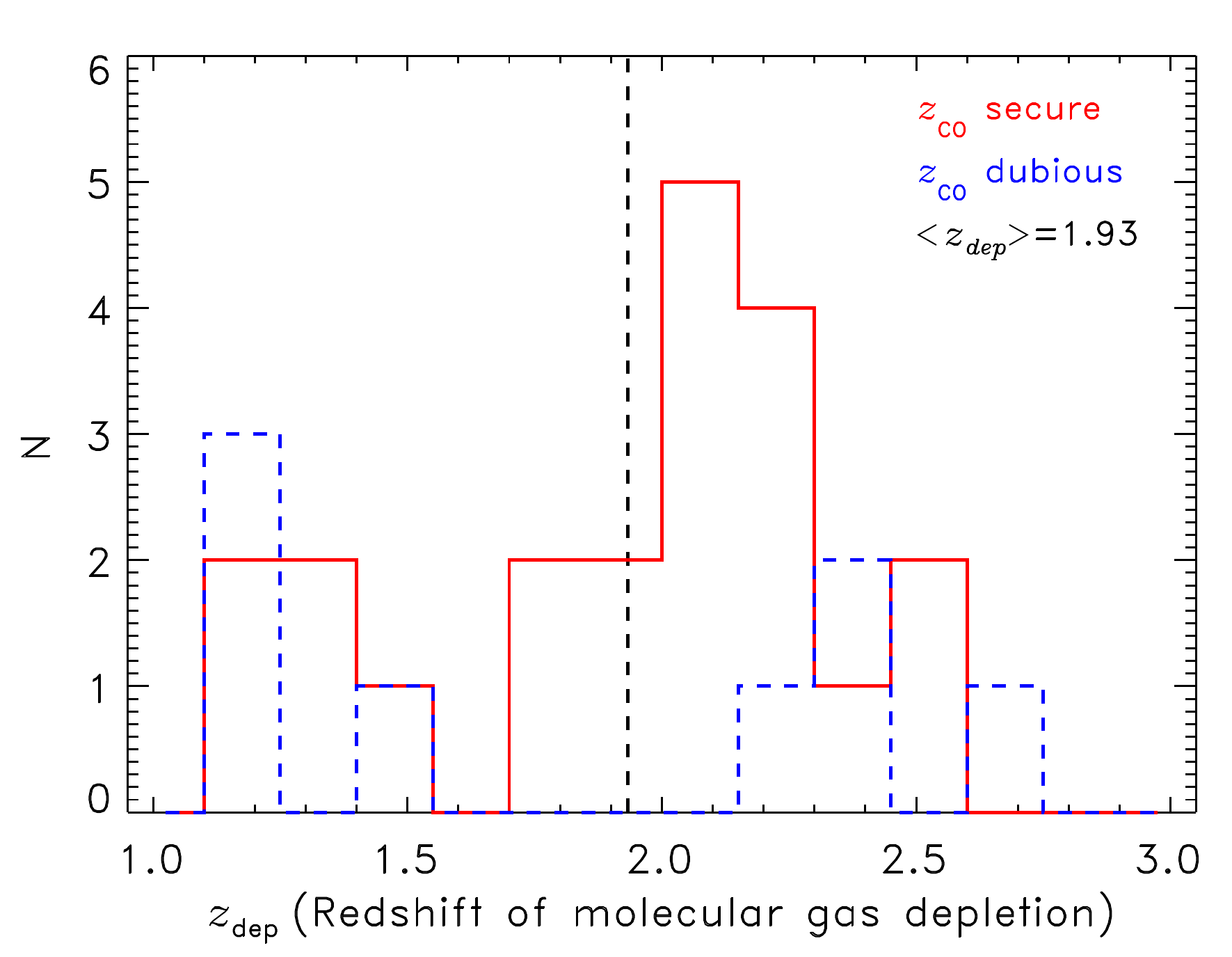}
\caption{{\small Estimated redshift by which the molecular gas would be
depleted ($z_\mathrm{dep}$) assuming the current SFR and 100\% conversion of gas into stars with
no refueling (red solid line: sources with secure redshift, blue dashed line: sources
with uncertain redshift). The vertical black dashed line represents the mean $z_\mathrm{dep}$
value derived from the sources with secure redshifts}.}
\label{fig:zdepl_histo}
\end{figure}

Assuming the current SFR and 100\% conversion of gas into stars with no
refueling, we can estimate by which redshift the molecular gas would be
fully consumed.  The estimated depletion redshifts of the sources
with a secure redshift, listed in Table~\ref{tab:mm_properties} and shown in
Fig.~\ref{fig:zdepl_histo}, range from 1.17 to 2.56 and have a mean value of
1.93$\pm$0.34.  These redshifts are consistent with the epoch of build up of
the passive galaxy population in clusters~\citep[e.g., ][]{pallero19}.

\subsection{CO kinematics}

\begin{figure}[h!]
\centering
\includegraphics[width=\linewidth]{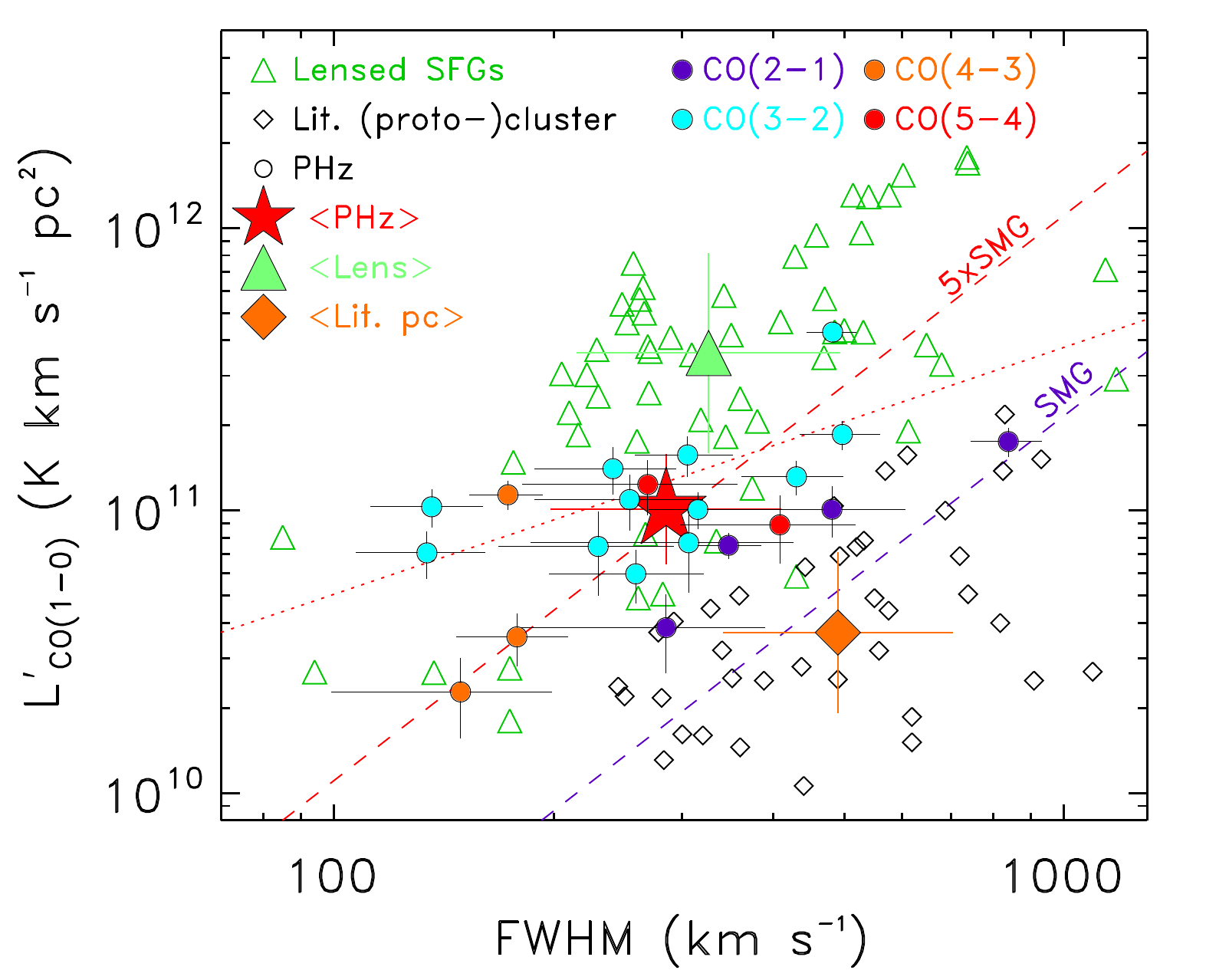}
\caption{{\small \LpCOone\ as a function of FWHM derived from the
lowest transition of each PHz-IRAM source with secure redshift
(full circles: purple for \COtwo, cyan for
\COthree, orange for \COfour, and red for \COfive).  The large red star
represents the median value of all PHz-IRAM sources.  
Black diamonds represent CO-detected cluster and
protocluster galaxies at 1.4${<}z{<}$3.1 and
log(L$_\mathrm{IR}$/\lsun)$>$11.44$+$0.5${\times}z$ from the
literature, and the large filled orange diamond the corresponding median
value.  Green open triangles represent the observed values for lensed SFGs
at 1.4${<}z{<}$3.1~\citep{harris12,saintonge13,dessauges15,canameras15,aravena16,yang17,harrington18,solimano21},
and the large full green triangle the corresponding median value.  The
dashed purple line is the best fit relation for unlensed high-$z$ SMGs
from~\citet{bothwell13}, and the red dashed line is the same relation but
multiplied by a factor of 5 to match the PHz-IRAM median value.  The
dotted red line is the best linear fit to the PHz-IRAM values.}}
\label{fig:LCO_FWHM}
\end{figure}

The CO line emission traces the kinematics of the potential well in which a
galaxy's molecular gas lies, modulo any inclination or dispersion effects. 
In particular, the CO line width is sensitive to both the dynamical mass and
inclination effects, while the luminosity (\LpCOone) is sensitive to the
total molecular gas mass (see Sect.~\ref{sec:mol_masses}).  In
Fig.~\ref{fig:LCO_FWHM}, we show the \COone\ luminosity derived from the
lowest transition line for each target with secure redshift as a function of
line width. For each line, we report the measured width, while the
luminosity is converted to the \COone\ luminosity using the line ratios
from~\citet{bothwell13}.  We also show for comparison the relation between
the \COone\ luminosity and FWHM derived for unlensed CO-detected SMGs
by~\citet{bothwell13}, the values measured in several cluster and
protocluster members at 1.4${<}z{<}$3.1 from the literature (see
Table~\ref{tab:lit_samples}), and those measured in lensed SFGs at
1.4${<}z{<}$3.0 from the HATLAS, GEMS, SPT, and PCCS samples, and in
intermediate mass or low L$_\mathrm{IR}$ lensed
SFGs~\citep{harris12,saintonge13,dessauges15,canameras15,aravena16,yang17,harrington18,solimano21}.
We also report the median values of each subsample\footnote{The median line widths and
luminosities of the PHz-IRAM sources are
median(FWHM)\,=\,305$^{+140}_{-96}$\,\kms, and
median(\LpCOone)\,=\,(10.1$^{+6.0}_{-3.8}{)}{\times}$10$^{10}$\,K\,\kms\,pc$^{2}$,
for the protocluster members from the literature
median(FWHM)\,=\,490$^{+214}_{-149}$\,\kms, and
median(\LpCOone)\,=\,(3.7$^{+3.4}_{-1.8}{)}{\times}$10$^{10}$\,K\,\kms\,pc$^{2}$,
and for the lensed SFGs median(FWHM)\,=\,326$^{+168}_{-111}$,\kms, and
median(\LpCOone)\,=\,(36$^{+46}_{-20}{)}{\times}$10$^{10}$\,K\,\kms\,pc$^{2}$.}. 

The PHz-IRAM sources show a relatively flat distribution of \LpCOone\ with
line widths spanning a wide range (FWHM$\sim$134--839\,\kms).  They
do not follow the relation between \LpCOone\ and FWHM observed in the SMGs
and in the protocluster members from the
literature~\citep[\LpCOone$\propto$FWHM$^2$; see dashed purple line in
Fig.~\ref{fig:LCO_FWHM}; ][]{bothwell13}.  Lensed SMGs are known for
exhibiting a similar flat trend~\citep{aravena16,yang17}.  Such a behavior
has been interpreted as lensing magnification acting on a population with
intrinsically steep number counts and detected above a constant CO flux
threshold~\citep{harris12}.  Note that our lens compilation, that includes
lensed SFGs with low IR luminosities~\citep{dessauges15}, at intermediate
masses~\citep{solimano21}, and with extreme IR
luminosities~\citep{harrington18} produces a different distribution than the
flat one reported in some previous works.  An analysis of the \LpCOone--FWHM
relation in lensed SFGs is beyond the scope of this work, but understanding
the origin of such a flat trend might help to interpret our sample as well,
and vice versa, even if our sources are not affected by lensing.  Another
striking difference compared to the protocluster sample from the literature is the
significant fraction (24\%) of sources with narrow
(FWHM$<$200\,\kms) lines.

To explain the observed linewidths, we consider the parameters that
influence them, the gas mass density profile, extent, and
velocity dispersion.  The relation between \LpCOone\ and the linewidth (FWHM) can be written as: 
\begin{equation} 
\LpCOone = \frac{C({\rm FWHM}/2.35)^2R}{\alpha_\mathrm{CO}G} 
\label{eq:co_kin}
\end{equation} 
where $C$ is a constant parameterizing the kinematics of the galaxy that can
assume values from $\leq$1 to $\geq$5, with the latter being the case of a
uniform sphere and lower values for disk distributions~\citep[see
e.g.,][]{erb06}, $G$ is the gravitational constant, and $R$ is the radius of
the CO(1--0) emission region~\citep[e.g.  7\,kpc; ][]{ivison11,bothwell13}. 
High-$z$ SMGs follow this relation assuming $\alpha_\mathrm{CO}$\,=\,1.36,
$R$\,=\,7\,kpc, and $C$\,=\,2.  Their scatter around such a relation
is small~\citep[$\Delta$\LpCOone/\LpCOone\,=\,0.38; ][]{bothwell13}
implying compact sizes and regular gas motions.  The PHz sources are, on
average, on a relation shifted towards larger luminosities by a factor of 5
(see red dashed line in Fig.~\ref{fig:LCO_FWHM}), implying larger $C$, and
$R$ values, and thus more random motions or large inclinations in case of
disks, and larger sizes.

Taking into account the different $\alpha_{\rm CO}$ factors assumed for our
sources and for the SMGs, we can write $C^{PHz}{\times}R^{PHz}$ as
5${\times}C^{SMG}{\times}R^{SMG}{\times}(\alpha^{PHz}_\mathrm{CO}/\alpha^{SMG}_{\mathrm
CO}$).  Replacing $C^{SMG}$, $R^{SMG}$, and the the different $\alpha_{\rm CO}$
factors with the appropriate values, we obtain $C^{PHz}{\times}R^{PHz}$\,=\,180.
Assuming the highest $C$ value, that is 5, valid for a uniform sphere,
$R^{PHz}$ would be $\sim$36\,kpc.  The derived $R^{PHz}$ value is
plausible, but unusually large, compared to the sizes of the molecular gas
reservoirs in high-redshift galaxies~\citep[i.e.,
$<$25\,kpc;][]{riechers11,sharon16,keating20}.  Interestingly, an
extended ($R{\simeq}$20\,kpc) molecular gas distribution has been reported
in another protocluster galaxy, HAE229~\citep{dannerbauer17}, but as
a rotating disk.  

It is important to point out that the PHz-IRAM sources do not lie on the
expected relation, even if scaled by a factor of 5.  The relation
that best describes our sample has a slope of 0.87$\pm$0.07 (i.e.,
log(\LpCOone)$\propto$(0.87$\pm$0.07)$\times$log(FWHM); see red dotted line
in Fig.~\ref{fig:LCO_FWHM}), rather than a slope of 2 as expected for a rotating
disk (see eq.~\ref{eq:co_kin}).  The different slope implies that the simple
rotating disk model is inadequate to explain the observed kinematics.  A
rotating disk is also disfavored by the large $C^{PHz}{\times}R^{PHz}$
factor required to fit the relation in eq.~\ref{eq:co_kin}.
Interestingly, the peculiarity of the PHz sources in the
\LpCOone--FWHM plane is observed for all transitions, thus we can rule out
different extents and kinematics for gas with different excitations
as the origin of the flat distribution.  We thus conclude that the molecular
gas in our sources must be characterized by kinematics that are not
typically observed in SMGs at $z{\sim}$2.

A possible explanation is that we are seeing an extended gas
component that the large beam of the IRAM-30 telescope is able to reveal.  A
significant fraction of such a gas component might not be in equilibrium,
but cold dynamically, prompting an interesting question on its origin and
extended nature.  Another possible explanation, suggested by the location of
our sources in between lensed and unlensed sources in
Fig.~\ref{fig:LCO_FWHM}, it is that they might be moderately lensed, with
significant variations source-by-source. This possibility will be discussed
further in Sect.~\ref{sec:lensing}.

\section{Discussion}\label{sec:discussion}

\subsection{The molecular gas properties of the PHz-IRAM sources}\label{sec:PHz_nature}

The properties of the PHz-IRAM sources derived from their sub-mm and
CO emission suggest that the majority contain large and extended molecular
gas reservoirs at low density and excitation.  These findings are surprising
considering their elevated SFRs, more often observed in galaxies with
compact and highly excited molecular gas distributions like merger-driven
starburst galaxies.  It is also possible that the sensitivity of the
IRAM-30m telescope to extended gas distribution might have allowed us to
detect such a diffuse component.  Although rarely, widespread star formation
across an extended (FWHM$\gtrsim$10\,kpc) disk of molecular gas has been
found in other SMGs at $z{\gtrsim}$2, including protocluster
members~\citep{ivison11,hodge12,calistro18,dannerbauer17}.
Rotating disks associated with starbursting activity triggered by an
interaction might not be necessarily rotationally supported~\citep[see e.g. 
][]{hodge12,ivison13}. Such disks might exhibit broad CO line profiles (i.e.,
700-1200\,\kms), due to the contributions from multiple CO components. 
We do not have CO observations at high spatial resolution to
carry out a proper dynamical analysis of our sources' gas reservoirs, but the
relatively narrow CO linewidths, and the CO spectral ladder favor a scenario
in which the molecular gas is cold and extended, as it might be expected in
case of cold accretion, rather than in a rotating disks or in a compact merger.

An alternative, although unlikely, explanation to the observed properties
might be the presence of large-scale molecular gas in the
circumgalactic medium, as observed in some other
protoclusters~\citep{emonts16,ginolfi17}, and also in isolated
quasars~\citep[see][]{cicone21}.  It is also possible that these massive,
extended, and poorly excited molecular gas reservoirs are due to the
combined emission of multiple galaxies at similar redshifts (with the low
excitation caused by the decreasing beam size in EMIR at increasing
frequencies).  This scenario would imply the presence of multiple sources at
the same redshift within the EMIR beam, as found for example in the
protocluster ClJ1001 where ALMA detects four CO emitters at the same
redshift, all within a region of 20\arcsec$\times$20\arcsec\ and associated
with a single bright ($S_{\rm 350\mu m}$\,=\,77$\pm$6\,mJy) SPIRE
source~\citep{wang16}.  Another scenario that will be discussed in
Sect.~\ref{sec:lensing}, it is that the observed properties of the PHz-IRAM
sources might be due to gravitational lensing affecting at different levels
some of our sources.

\subsection{What powers the observed SFRs?}

The finding that the majority of the selected targets are normal SFGs in
spite of their large IR luminosities and SFRs was unexpected as our
selection was thought to find the most starbursting protoclusters.  This
result is in part due to the strong evolution of the ULIRG population and to
their increasing contribution to the IR luminosity cosmic density with
redshift~\citep{goto10,magnelli13,magnelli19}.  Ultra-luminous IR galaxies
are more common at high redshifts than in the local Universe.  This is
analogous to the idea that ULIRGs at $z{\sim}$0 do not lie on the
star-forming MS, while those at high-$z$ do~\citep{rodighiero11}.  The
physical origin of ULIRGs in the local Universe is often mergers and
interactions~\citep{sanders96,taniguchi98}.  Both merger events, and the
availability of cold gas increase with
redshift~\citep{lopez13,romano21,walter20}.  Our findings support the idea
that the majority of ULIRGs in overdensities at $z{\sim}$2 are powered by
cold gas accretion, rather than merger events.  Massive dark matter halos,
where protocluster cores breed, are expected to host large amounts of cold
gas.  The availability of large amounts of cold gas might boost the number
of galaxies with large SFRs as found in the PHz fields.  The predominance of
normal SFGs, in terms of their position relative to the star-forming
main sequence, is observed in most known protoclusters at $z{\sim}$2--3,
even those with extreme total SFRs~\citep[for a compilation see Fig.~20
in][]{polletta21}.

It would be interesting to measure the baryonic fraction of our
sources, as this would provide constraints on the mass of their dark matter
haloes and thus on their association with protocluster cores.  A massive dark
matter halo would help retaining the measured large gas reservoirs.  
Measuring their halo masses from the molecular gas kinematics and
morphology, and their stellar masses would provide key constraints on the
gas dynamical properties, and on the nature of our sources.

\subsection{The evolutionary stage of the PHz-IRAM sources}\label{sec:evolution}

Based on the amount of molecular gas and the current SFR, we estimated the
timescale ($\tau_\mathrm{dep}$) and redshift at which the gas will be
exhausted in our sources (see Sect.~\ref{sec:mol_masses}).  We can also
estimate how long these galaxies have spent in building their stellar mass. 
In Sect.~\ref{sec:fir_properties}, we show that our sources' dust
temperatures cover a similar range as the $z{\sim}$2 normal SFGs analyzed
by~\citet{magnelli14}.  For such a sample, \citet{magnelli14} finds a strong
relation between the dust temperature, and the specific SFRs
(sSFR\,=\,SFR/$\mathcal{M}$), that is T$_{\rm
dust}{=}98{\times}(1+z)^{-0.065}+6.9{\times}log({\rm sSFR})$.
Assuming such a relation and the average redshift and
dust temperature of our PHz-IRAM sources, we derive, an average
${<}log$(sSFR)${>}$\,=\,$-$8.92$\pm$0.15.

Assuming the derived average $<$sSFR$>$, and the measured $<$SFR$>$ (i.e.;
1043$\pm$157\,\msun\,yr$^{-1}$), the
estimated average stellar mass of our sources would be ${<}{\mathcal
M}{>}{\simeq}$(8.8$\pm$3.5)$\times$10$^{11}$\,\msun.  Our galaxies would thus be
already quite massive.  Assuming the predicted sSFR, the time necessary to
build such stellar mass would be, on average, $\tau_{\rm
buildup}$\,=\,1/$<$sSFR${>}{\simeq}$0.8$^{+0.4}_{-0.2}$\,Gyr.  The derived
timescale is longer than the average depletion time estimated in
Sect.~\ref{sec:mol_masses}, ${<}\tau_\mathrm{dep}{>}{=}$0.47$\pm$0.07\,Gyr,
implying that these galaxies have undergone most of their star-formation
episode (started, on average, at $z{\simeq}$3.0, and supposed to end by
$z{\simeq}$1.9).  These estimates assume a box-shape star formation history
which is implausible.  In SFGs, star formation is expected to gradually
increase with time (e.g.  SFR${\propto}e^{a{\rm t}}$, where $a$ is a
constant that describes how quickly the SFR increases with time).  It is
thus more likely that star formation activity started earlier than
$z{\simeq}$3.0.  Such an early formation epoch is typical of galaxies
located in dense environments, rather than in the field~\citep[see
][]{thomas05}.  This result is consistent with the expectation that these
galaxies trace overdense environments.

Based on the estimated SFRs and large reservoirs of molecular gas,
our sources could consume all their molecular gas in $\sim$0.5\,Gyr, unless
some quenching mechanisms is activated.  According to this scenario and
assuming the estimated average stellar mass, our sources could reach stellar
masses $\mathcal M{\geq}$10$^{12}$\,\msun, which are extremely rare.  We can
thus expect that their star formation will end before exhausting the
available molecular gas. Their star formation activity could be halted by
environment-related processes, like starvation or ram-pressure
stripping~\citep{hayashi17,wang18,foltz18}, or by internal feedback, like
the injection of energy from a powerful AGN.  These possibilities could be
explored through a morphological and kinematical study of their molecular
gas.  To reveal the presence of AGN activity, observations in the X-rays or
in the radio would be crucial as they can reveal buried AGN and powerful
jets.  A certain number of PHz fields have been observed by LOFAR as part of
the LoTSS survey~\citep{shimwell17,shimwell22}, and although a radio
counterpart is detected in 25 PHz-IRAM sources, none of them seems to host
a radio-loud AGN, ruling out the possibility that a radio jet might eject or
heat the available molecular gas and halt the star formation. This
analysis will be presented in future works.

\subsection{Gravitational lensing effect}\label{sec:lensing}

Some of the properties observed in the PHz-IRAM sample might be
explained by an effect of moderate gravitational lensing.  In the entire PHz
parent sample, a dozen of strongly lensed SPIRE sources, called GEMs, has
been confirmed~\citep{canameras15}.  Compared to the PHz-IRAM sources, the
GEMs are brighter (i.e., $S_{\rm 350\mu m}{>}$200\,mJy), and are easily
discernible in the \herschel\ images as single bright sources.  It is
however possible that some of the brightest PHz-IRAM sources, for example
those with $S_{\rm 350\mu m}{>}$100\,mJy, might be also lensed, but with
less magnification factors than the GEMs sources and thus less
straightforward to identify as lensed systems.  Lensed DSFGs with sub-mm
flux densities consistent with ours have indeed been found~\citep{weiss13}. 
Interestingly, it is predicted that at intermediate magnifications (e.g.,
$\mu{<}$10), the diffuse emission might be magnified by a larger factor than
the compact component~\citep{hezaveh12}.  Such an effect could significantly
bias molecular lines ratios and explain the CO SLEDs dominated by gas at low
transitions observed in our sample.  Lensing might also explain the high
SFRs and gas masses, and the low dust temperatures observed in the PHz-IRAM
sample.  Additional observations, in particular at high spatial resolution,
would be necessary to determine whether the PHz-IRAM sources are
gravitationally-lensed galaxies.  Our selection of the brightest red
\herschel\ sources in the PHz fields might have biased our sample towards
this unusual association of moderately magnified galaxies and overdense
regions. 

\subsection{Comparison with simulations}

\begin{figure}[h!]
\centering
\includegraphics[width=\linewidth]{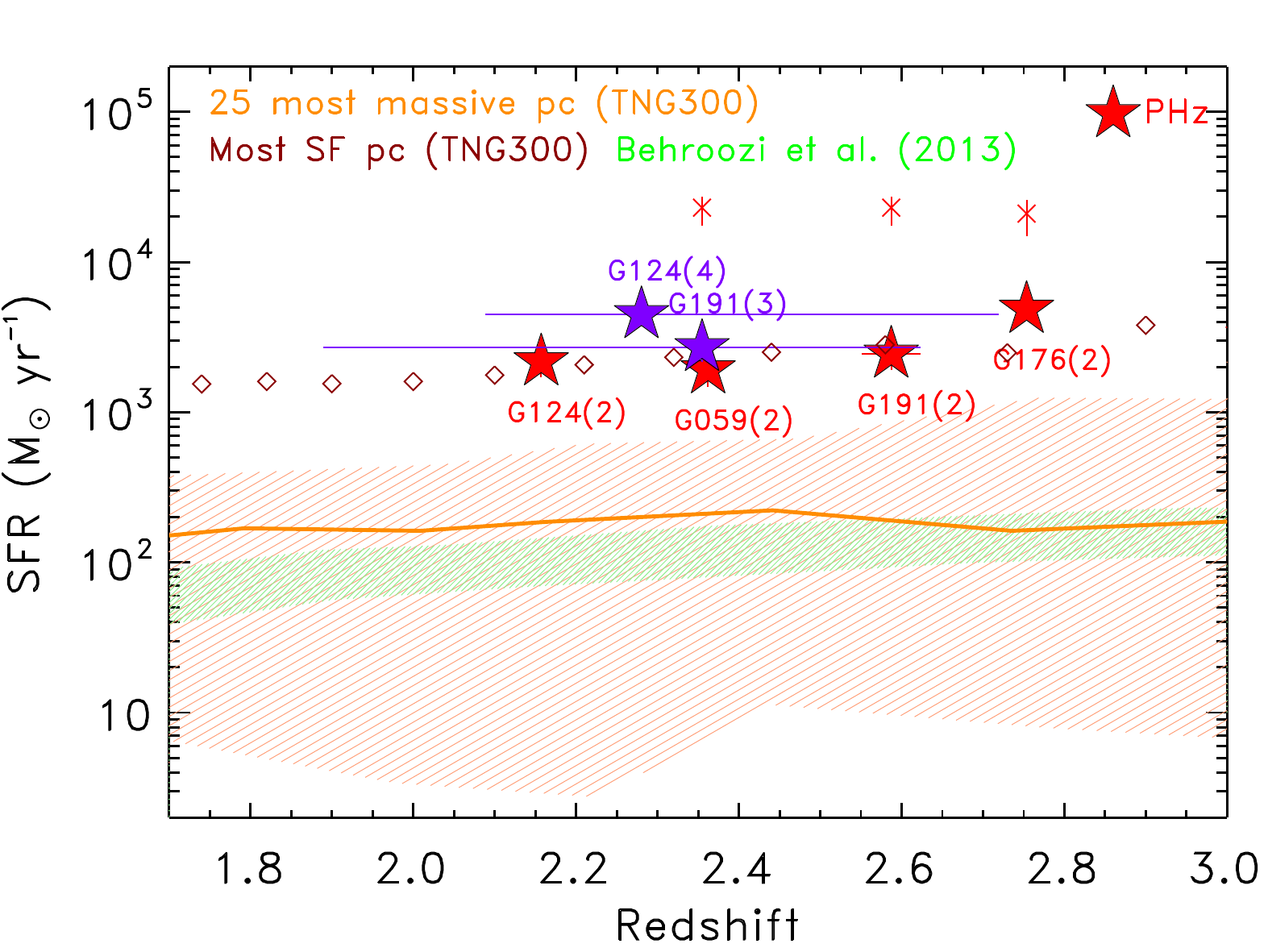}
\caption{{\small Total SFR as a function of redshift of the PHz fields with
multiple CO-detected sources at the same redshift (red full stars), and at
all redshifts (purple full stars) compared with the total SFRs of simulated
protoclusters.  Horizontal bars represent the range of redshifts of all
sources considered for the computation of the total SFR (see number in
parenthesis).  The PHz field names are annotated. The total SFR derived from
the \planck\ fluxes and assuming a dust temperature of 30\,K are shown as red crosses (these are available only for
the the subset of fields that are in the official PHz list). The predictions from the
empirical models of~\citet{behroozi13} are shown as green hatched region,
those from the protocluster that will become the 25 most massive $z$\,=\,0
clusters in the TNG300 simulation as salmon hatched area), the median SFR of
the five most massive TNG300 simulated clusters as salmon thick
line~\citep[adapted from ][]{lim21}, and the most star-forming protocluster
at 1.3${\leq}z{\leq}$3 from the TNG300 simulations as brown open
diamonds~\citep[Gouin, priv.  comm.; ][]{nelson19}.}}
\label{fig:sim_SFR}
\end{figure}

To further assess the protocluster nature of our PHz sources, we can compare
the results we have obtained in this study with the predictions derived from
large volume simulations like IllustrisTNG300~\citep{nelson19}.  Previous
studies on SMG overdensities claim that even a large overdensity of SMGs may
not probe massive clusters in formation~\citep[see e.g.,
][]{blain04,chapman09,casey16}.  This conclusion is supported by
large-volume, semi-analytic simulations that find very few massive
structures at $z{\lesssim}$2.5 containing more than one
SMG~\citep{miller15}.  Here, we revisit this assessment by examining the
SFR distribution in simulated high-$z$ structures, and the number of member
galaxies with predicted SFRs as large as those measured in the PHz-IRAM
sources.  For this analysis, we consider only the fields where we have
evidence of multiple sources at the same redshift.  This is the case of four
fields (G176, G191, G059, and G124) where two SPIRE sources are detected in
CO at similar redshifts.  In Fig.~\ref{fig:sim_SFR}, we show the total SFRs
of the four selected PHz fields, derived as the sum of the single members
SFRs.  In the G191, and G124 fields, CO detections for two additional
sources are available, but at a different redshift.  For these two fields,
we also show the total SFR, computed by considering these additional
sources, and the redshift range (see purple stars in
Fig.~\ref{fig:sim_SFR}).  For the fields that are in the official PHz
catalog, we also show the total SFR as derived from the \planck\ fluxes
assuming a dust temperature of 30\,K, and after correcting them for a
Chabrier IMF (SFR$^{Chabrier}$\,=\,SFR$^{Salpeter}$/1.8).  The predicted SFR
derived from the empirical model in~\citet{behroozi13}, and the TNG300
simulations~\citep{nelson19} are also shown.  For the latter we consider the
protoclusters that will become the 25 most massive $z$\,=\,0 clusters, and
the median SFR of the five most massive TNG300 simulated
clusters~\citep{lim21}.  We also consider the most star-forming structures
at various redshifts between $z$\,=\,1.3 and 3 from the IllustrisTNG
simulation (Gouin et al., in prep.).  Their total SFRs are computed as the
sum of the SFR of all galaxies within a 5\arcmin\ diameter\ aperture, and a
redshift interval corresponding to
${\mid}\Delta{v}{\mid}{<}$2000$\times$(1+${<}z{>}$) (see Gouin et al., in
prep.  for specific details on this procedure).  Our fields exhibit total
SFRs that are much higher thank those measured in the most massive
protoclusters, but consistent with the most star-forming ones, although the
latter are obtained by adding the SFRs of about 25 star-forming members,
while in our cases we consider only two SPIRE sources.  Furthermore, the
highest SFRs in the simulated members reach $\sim$200\,\msun\,yr$^{-1}$ at
the most, while our members' SFRs range from $\sim$800 to
3300\,\msun\,yr$^{-1}$.  Large volume hydrodynamic cosmological simulations
like Magneticum show that the progenitors of the most massive clusters at
$z{\simeq}$0 are not necessarily those with the most intense star formation
activity during the assembling phase (Remus et al., in prep.\footnote{See
Remus,~R.-S.'s presentation at the Galaxy Cluster Formation II (GCF 2021) -
https://www.youtube.com/watch?v=iDVtDaElSLc.}).  It is beyond our scope to
explain such a finding, but it is interesting to consider the implications,
even in a simplistic way.  The cause might be written in these galaxies
genes, for example these extreme star formers with massive gas reservoirs
might have higher baryonic fractions than galaxies with more typical SFRs,
and thus occupy less massive dark matter haloes than expected. 
Alternatively, it might be their lifestyle, too much intrinsic growth might
limit that of their environment through a sort of feedback process.  To
reconcile our results with the simulations, our PHz-IRAM sources will have
to separate out into multiple galaxies with smaller SFRs, or simulations
need to include phases of extreme star formation activity.  A more detailed
comparison between the PHz sources and simulations will be carried in a
future work (Gouin et al., in prep.), but to really benefit from this
comparison we need to acquire more information on our sources, and in
particular assess whether they suffer from multiplicity, and determine their
baryonic fraction.  High resolution sub-mm/mm observations are crucial to
answer these questions.

Compared to the SFRs derived from \planck~\citep{planck16}, shown as
red crosses in Fig.~\ref{fig:sim_SFR}, our estimates are about 8--17 times
lower, even when considering the CO-detected sources at different redshifts. 
This discrepancy has been solved in at least one
field~\citep[G237;][]{polletta21}, where a detailed comparison between
\planck\ and \herschel\ fluxes integrated over the same area has
been carried out using deep \herschel\ observations from the HerMES
survey~\citep{oliver12}.  The higher \planck\ SFRs are due to the large
number of sub-mm sources within the \planck\ beam
(FWHM\,=\,4.6\,arcmin at 350\um), typically $\gtrsim$10
\herschel-detected sources, but most of them are along the line of sight of
the protocluster and fainter than the IRAM-30m telescope can detect (see
Sect.~\ref{sec:spire_density} and Fig.~\ref{fig:her_maps}).

\section{Summary and conclusions}\label{sec:conclusion}

We report on IRAM-30m/EMIR observations of 38 \herschel\ sources chosen as
the brightest \textit{red} sub-mm sources in 18 \planck-selected (PHz)
fields.  These fields are considered protocluster candidates at
$z{\simeq}$2--4 hosting overdensities of DSFGs.  We detect 40 CO
lines on a total of 24 bright ($S_{\rm 350\mu m}{>}$40\,mJy)
\herschel\ sources in 14 PHz fields.  For the sources with no CO detection
we cannot place any constraints on their redshift, or CO luminosity
as the observations cover a limited frequency range, typically 16\,GHz in
bandwidth, and the exposure times were often too short for their brightness. 
For the 24 CO-detected sources we estimate SFRs, and dust temperatures from
fitting the sub-mm SED with a modified black-body.  We also measure CO
lines' intensity, luminosity, width, and in ten cases the CO SLED.  We
then derive molecular gas masses, depletion times, and star formation
efficiencies and classify the sources as normal SFGs, or starbursts based on
their position with respect to the L$_\mathrm{IR}$--\LpCOone\
relation~\citep{sargent14}.

The PHz-IRAM sample is characterized by the following average properties,
${<}z_\mathrm{CO}{>}$\,=\,2.25$\pm$0.09,
${<}SFR{>}$\,=\,1043$\pm$157\,\msun\,yr$^{-1}$,
$<$T$_\mathrm{dust}{>}$\,=\,29.2$\pm$0.9\,K,
$<$FWHM$>$\,=\,(317$\pm$133)\,\kms,
$<$M$_\mathrm{gas}{>}$\,=\,(4.0$\pm$0.7)$\times$10$^{11}$\,\msun, and
${<}\tau_\mathrm{dep}{>}{=}$0.47$\pm$0.07\,Gyr.  We compare the distributions of
these quantities with those of cluster and protocluster members drawn from the
literature with similar IR luminosities, and redshifts
(1.4${\leq}z{\leq}$3.1) and with the scaling relations describing normal
SFGs, and starbursts at $z\sim$2. 

The main results are:
\begin{enumerate}

\item The majority (80\%) of the PHz-IRAM sources are consistent
with the L$_\mathrm{IR}$--\LpCOone\ relation typical of normal SFGs, but are
characterized by SFRs, and gas masses that are, respectively, eight and five
times, on average, higher than those typical of normal SFGs at similar redshifts.
This result is not simply due to a selection effect that favors bright
sub-mm sources, as we observe a wide range of CO luminosities at fixed IR
luminosity.

\item We find evidence of multiplicity in three out of 24 (12$\pm$7\%)
PHz-IRAM sources through the detection of two CO lines at inconsistent
redshifts in one case, and of double peak line profiles in other two
cases.  This fraction is consistent with those found in previous studies of
bright SPIRE sources~\citep[9--23\%; ][]{montana21,scudder16}.  However, the
derived properties suggest that multiplicity might affect a larger fraction
of sources than our IRAM-30m observations are able to reveal.

\item  We find two or three \herschel\ sources in close projected
proximity at similar redshifts in four of the PHz fields where multiple
\herschel\ sources were observed, supporting the idea that the PHz fields
contain protoclusters at $z{\simeq}$2--3.  However, we also detect sources
situated along the line of sight that contribute to the measured \planck\
flux, as predicted by~\citet{miller15}, and~\citet{negrello17}.

\item The CO SLED of eight out of ten PHz-IRAM sources with multiple
CO transitions peaks at a low rotational number ($J_\mathrm{up}$\,=\,3),
implying that a significant fraction of the molecular gas must be at low
densities, cold, and at low-excitation.

\item The CO line widths of the PHz-IRAM sample are, on average, smaller
than typically observed in field SMGs, and cluster and protocluster sources
from the literature.  This difference is explained by an extended
gas distribution to which the IRAM 30m-telescope is particularly sensitive.

\item About 20\% of the PHz-IRAM sample is undergoing a starburst phase.
Based on this result, we infer that these galaxies spend most of their
lifetime as normal SFGs, and exhibit a starburst phase only during 20\%
of their lifetime. 

\item Compared to other protocluster members from the literature selected
above a similar L$_{\rm IR}$ limit, and across a similar redshift
range (1.4{<}z{<}3.1), we find that our sources are characterized, on
average, by slightly larger SFRs, and gas masses, but consistent SFEs, gas
depletion timescales and excitations.  

\item We compare the SFRs of our PHz-IRAM sources with those predicted by
state-of-the-art simulations for the most massive and most star-forming
protoclusters~\citep[][Gouin, in prep.]{nelson19,behroozi13}.  Although,
our total SFRs are consistent with the simulated ones, there is a
significant discrepancy in terms of number of sources contributing to the
total SFR, typically 2--3 in our fields and 25 in the simulated
protoclusters, and in the typical SFR of the galaxy members, the observed
ones being about ten times larger than in the simulations.  
\end{enumerate}

Overall the \planck-\herschel-IRAM selection reveals a class of SFGs that
follows the scaling relations typical of normal SFGs at $z{\sim}$2, but with
larger SFRs, and molecular gas masses than typically found in these
galaxies.  The peculiarity of our sources resides in their relatively low
dust temperatures, low CO excitations, and relatively narrow CO
line widths.  These properties suggest that the majority of our
sources are powered by a secular steady-state mechanism, rather than by
merger events. Multiplicity and some level of magnification due to
gravitational lensing might also play a role in producing the observed properties.

Millimeter and CO observations at high spatial resolution would be
necessary to assess whether the sub-mm, and CO fluxes are affected by
multiplicity, or gravitational lensing, and determine whether the molecular
gas is associated with a merger, a large massive disk, or extended
emission.  Such observations can also provide dynamical masses
from the CO gas kinematics and morphology, and thus halo masses, and
baryonic fractions.  These estimates are fundamental to further characterize
the environment of these galaxies.  Finally, they are instrumental to
identify the optical-NIR counterparts to the sub-mm emission, and thus
obtain an accurate determination of the stellar mass by modeling the
optical-NIR spectral energy distribution (SED), and of the SFR (by
deblending the SPIRE flux as done in~\citet{kneissl19}).

The EMIR observational campaign, with a detection rate of $\sim$78\%
(14 PHz fields with a redshift estimate out of 18), was successful in
providing an initial redshift guess for the overdensities in the
PHz fields. Such information is crucial for an effective follow-up
campaign at shorter wavelengths.  Spectroscopic measurements through CO
lines in the mm and in the rest-frame optical range are both necessary to
fully study these structures as they typically reveal different galaxy
types.  The former are indeed more sensitive to heavily obscured galaxies
that are often out of reach in the optical rest-frame, and the latter can
reveal SFGs that are difficult to detect at mm wavelengths because not
sufficiently massive in dust and molecular gas.  Combining the results from
these two wavelength windows is essential, but not straightforward.  With
sensitive facilities such as ALMA, JWST, and future near-IR spectrographs,
these complementary datasets will become quickly accessible for many galaxy
members, enabling the study of the stellar component, and of both the
ionized and molecular gas at similar angular resolution.  Such a step
forward in observational studies is essential to test simulations and
improve our theoretical understanding on structure assembly and galaxy
evolution.

\begin{acknowledgements}                 
We kindly thank the second referee for a careful reading and for many
suggestions that improved the clarity of the paper, and for replying in due
time after the long-sought report from the first referee.
We are very grateful to the IRAM staff for supporting us during the
observations, for their warm hospitality, and for providing the data in the
archive.
We thank C.~Gouin for providing the data on the most star-forming simulated
IllustrisTNG300 protoclusters prior publication.
We thank C.~Gouin, N.~Aghanim, and B.~Garilli for useful discussions, and C.~Casey for making the
{\tt cmcirsed} package available and for clarifying its usage.
We thank T. Perdereau for his assistance during an observing run at
IRAM-30m.
MP thanks the IRAP Institute for providing support and hospitality during
her frequent visits.
BLF gratefully acknowledges support from the Universit\'{e} de Paris-Saclay.  
This work is based on observations carried out under project numbers 107--14,
219--14, 197--15, 077--16, 186--16, 171--17, 085--19, 213--19, 178--20, and
082--21 with the IRAM 30m telescope.  IRAM is
supported by INSU/CNRS (France), MPG (Germany) and IGN (Spain).
The research leading to these results has received funding from the European
Union's Horizon 2020 research and innovation program under grant agreement
No 730562 [RadioNet].
The work is based on observations obtained with Planck
http://www.esa.int/Planck, an ESA science mission with instruments and
contributions directly funded by ESA Member States, NASA, and Canada.
%
This research has made use of data from HerMES project
(http://hermes.sussex.ac.uk/). HerMES is a Herschel Key Programme utilising
Guaranteed Time from the SPIRE instrument team, ESAC scientists and a mission
scientist.
%
The HerMES data was accessed through the Herschel Database in Marseille
(HeDaM - http://hedam.lam.fr) operated by CeSAM and hosted by the Laboratoire
d'Astrophysique de Marseille.
{\em Software:}  This research made use of astropy, a community developed
core Python package for astronomy~\citep{astropy}, of APLpy, an open-source
plotting package for Python~\citep{aplpy}, of topcat~\citep{topcat}, and of
the HEALPix software~\citep{gorski05}.
\end{acknowledgements}                                                                                                     
                                                                                                                           
%
%
                                                                                                                           

\clearpage

\appendix
\appendixpage

\section{Tables: IRAM Observations}

The list of observed targets, the log of the IRAM observations, and the instrument
set-up are listed in Tables~\ref{tab:targets}, \ref{tab:obs_log}, and
\ref{tab:obs_setup}.

\begin{sidewaystable*}[!ht]
\centering
\caption{List of observed targets\label{tab:targets}}
\begin{tabular}{rlcccrrrr}
\hline \hline
     PHz  & PHz                   & \herschel\ &  $\alpha$ [\herschel] & $\delta$ [\herschel]  & $S_\mathrm {250\mu m}$ & $S_\mathrm{350 \mu m}$ & $S_\mathrm{500 mu m}$ & $S_\mathrm{850 mu m}$ \\
     ID   & name                  &     ID &hr:min:sec  & \deg:\arcmin:\arcsec &              mJy      &          mJy           &           mJy           &     mJy \\ 
\hline
\noalign{\vspace{2pt}}
      57  &     PHz\,G176.60$+$59.01   &   01  &  10:37:07.85 &\;\;41:25:32.24 &  165.9$\pm$11.3 &  186.6$\pm$10.3 &  134.6$\pm$12.2  & \nodata  \\
          &                            &   02  &  10:36:52.66 &\;\;41:24:10.61 &   78.2$\pm$10.6 &   86.4$\pm$10.1 &   58.9$\pm$12.0  & \nodata  \\
          &                            &   04  &  10:37:05.51 &\;\;41:27:31.44 &   81.2$\pm$10.6 &   72.3$\pm$10.0 &   44.2$\pm$11.8  & 12.5$\pm$3.0   \\
          &                            &   06  &  10:36:46.32 &\;\;41:24:42.30 &   64.4$\pm$11.1 &   64.6$\pm$10.0 &   43.1$\pm$11.8  & \nodata  \\
          &                            &   08  &  10:37:03.43 &\;\;41:32:51.86 &   37.9$\pm$10.6 &   60.3$\pm$10.1 &   41.0$\pm$11.8  & \nodata  \\
      70  &     PHz\,G223.87$+$41.22   &   01  &  09:37:14.05 &\;\;10:00:06.42 &   70.2$\pm$10.8 &  101.6$\pm$10.1 &  100.7$\pm$12.0  & 16.2$\pm$1.9   \\
     237  &     PHz\,G173.93$+$56.97   &   01  &  10:28:40.91 &\;\;43:24:05.26 &   80.0$\pm$10.4 &  100.0$\pm$10.1 &   75.7$\pm$11.3  & \nodata  \\
     343  &     PHz\,G162.14$-$59.25   &   12  &  02:07:04.34 & $-$02:15:07.70 &   37.9$\pm$11.1 &   45.4$\pm$10.4 &   65.3$\pm$11.8  & \nodata  \\
     631  &     PHz\,G006.06$+$61.78   &   01  &  14:33:46.98 &\;\;12:12:57.08 &   59.8$\pm$10.7 &   83.1$\pm$10.1 &   85.7$\pm$11.8  & 16.0$\pm$2.8   \\
     712  &     PHz\,G237.01$+$42.50   &  962  &  10:01:42.33 &\;\;02:18:35.62 &   46.6$\pm$ 6.7 &   48.7$\pm$ 9.1 &   38.6$\pm$ 8.6  & \nodata  \\
          &                            & 9741  &  10:01:52.18 &\;\;02:19:00.95 &   19.6$\pm$ 6.7 &   20.6$\pm$ 9.1 &   18.9$\pm$ 8.5  & \nodata  \\
     832  &     PHz\,G191.24$+$62.04   &   01  &  10:44:38.53 &\;\;33:51:05.88 &   89.4$\pm$10.5 &  111.4$\pm$10.2 &   85.0$\pm$11.3  & \nodata  \\
          &                            &   04  &  10:44:40.90 &\;\;33:55:08.88 &   50.7$\pm$10.5 &   69.6$\pm$ 9.9 &   55.4$\pm$11.4  & \nodata  \\
          &                            &   07  &  10:44:59.07 &\;\;33:49:23.14 &   52.8$\pm$10.5 &   57.3$\pm$ 9.8 &   47.2$\pm$11.2  & \nodata  \\
          &                            &   09  &  10:44:40.24 &\;\;33:54:17.36 &   51.2$\pm$10.4 &   51.4$\pm$10.0 &   37.1$\pm$11.3  & \nodata  \\
          &                            &   10  &  10:44:54.39 &\;\;33:50:51.89 &   38.5$\pm$10.6 &   51.3$\pm$ 9.8 &   39.2$\pm$11.6  & \nodata  \\
          &                            &   15  &  10:44:56.66 &\;\;33:53:43.58 &   40.0$\pm$10.4 &   48.5$\pm$10.3 &   34.0$\pm$11.3  & \nodata  \\
          &                            &   18  &  10:44:34.38 &\;\;33:49:51.37 &   45.3$\pm$10.4 &   46.6$\pm$10.0 &   30.6$\pm$11.4  & \nodata  \\
          &                            &   24  &  10:44:39.30 &\;\;33:53:48.59 &   28.8$\pm$10.2 &   40.8$\pm$ 9.8 &   29.5$\pm$11.3  & \nodata  \\
          &                            &   26  &  10:44:38.04 &\;\;33:48:03.95 &   29.8$\pm$10.4 &   40.3$\pm$ 9.8 &   34.6$\pm$11.3  & \nodata  \\
    1473  &    PHz\,G088.39$+$34.26    & 1842  &  17:23:17.95 &\;\;59:38:59.39 &   47.3$\pm$ 5.6 &   55.9$\pm$ 6.0 &   50.0$\pm$ 5.9  & \nodata  \\
  124051  &   PLCK\,DU\,G059.1$+$37.4  &   01  &  16:58:45.26 &\;\;36:05:41.32 &   76.5$\pm$10.7 &   99.9$\pm$10.5 &   61.4$\pm$11.9  & \nodata  \\
          &                            &   03  &  16:58:45.87 &\;\;36:03:47.32 &   54.3$\pm$10.7 &   75.3$\pm$10.3 &   62.8$\pm$11.8  & \nodata  \\
  124052  &   PLCK\,DU\,G073.4$-$57.5  &   03  &  23:14:38.52 & $-$04:16:31.93 &   66.0$\pm$10.5 &   63.6$\pm$10.3 &   38.1$\pm$11.3  & \nodata  \\
          &                            &   15  &  23:14:53.00 & $-$04:18:23.66 &   43.3$\pm$10.6 &   48.6$\pm$10.0 &   46.9$\pm$11.7  & \nodata  \\
  124053  &   PLCK\,DU\,G124.1$+$68.8  &   01  &  12:49:04.34 &\;\;48:20:45.73 &  110.5$\pm$10.7 &  116.0$\pm$10.0 &   66.5$\pm$12.3  & \nodata  \\
          &                            &   02  &  12:49:02.42 &\;\;48:20:18.33 &   81.8$\pm$10.8 &   93.1$\pm$10.0 &   53.4$\pm$12.3  & \nodata  \\
          &                            &   03  &  12:49:11.31 &\;\;48:20:41.34 &   97.6$\pm$11.3 &   87.5$\pm$10.0 &   53.2$\pm$12.3  & \nodata  \\
          &                            &   15  &  12:49:06.20 &\;\;48:22:01.10 &   42.0$\pm$10.9 &   48.1$\pm$10.0 &   39.0$\pm$12.4  & \nodata  \\
  125002  &   PLCK\,HZ\,G072.8$+$35.4  &   02  &  17:15:58.52 &\;\;46:46:22.32 &   60.7$\pm$10.5 &   77.2$\pm$ 9.7 &   59.1$\pm$11.0  & \nodata  \\
  125018  &   PLCK\,HZ\,G112.4$+$45.8  &   06  &  14:17:02.88 &\;\;69:32:38.66 &   51.4$\pm$10.3 &   67.9$\pm$10.1 &   36.0$\pm$11.8  & \nodata  \\
  125026  &   PLCK\,HZ\,G143.6$+$69.4  &   02  &  12:10:04.74 &\;\;46:06:26.79 &   44.0$\pm$10.8 &   78.2$\pm$ 9.8 &  140.2$\pm$11.7  & \nodata  \\
  125027  &   PLCK\,HZ\,G131.8$+$49.6  &   01  &  11:53:48.98 &\;\;66:30:28.82 &  152.6$\pm$10.5 &  112.6$\pm$10.0 &   67.7$\pm$11.6  & \nodata  \\
          &                            &   15  &  11:53:44.83 &\;\;66:30:53.96 &   47.2$\pm$10.4 &   42.9$\pm$ 9.7 &   23.7$\pm$11.6  & \nodata  \\
  125056  &   PLCK\,DU\,G052.2$+$28.1  &   01  &  17:35:20.04 &\;\;28:17:20.46 &   89.7$\pm$11.0 &  100.7$\pm$ 9.8 &   69.2$\pm$11.3  & \nodata  \\ 
          &                            &   06  &  17:35:12.69 &\;\;28:19:31.00 &   51.1$\pm$11.0 &   60.5$\pm$ 9.9 &   49.5$\pm$11.2  & \nodata  \\
  125107  &   PLCK\,HZ\,G068.3$+$31.9  &   02  &  17:33:13.82 &\;\;42:42:31.31 &  132.0$\pm$10.5 &  114.9$\pm$13.9 &   76.0$\pm$11.5  & 18.8$\pm$2.8   \\
  125132  &   PLCK\,G63.7$+$47.7       &   02  &  16:07:54.48 &\;\;40:02:40.00 &  126.2$\pm$10.5 &  109.2$\pm$10.5 &   70.6$\pm$11.9  & \nodata  \\ 
\hline
\end{tabular}
\end{sidewaystable*}

\clearpage
\onecolumn

\begin{landscape}
\begin{longtable}{rr cc c cl cl}
\caption{EMIR observations log\label{tab:obs_log}}\\
\hline\hline
    PHz   &  \herschel\ &  $\alpha$ [EMIR]& $\delta$ [EMIR]  & Program & \multicolumn{2}{c}{E090}   & \multicolumn{2}{c}{E150}   \\
    ID    &  ID         & hr:min:sec & \deg:\arcmin:\arcsec &      & $\tau$\tablefootmark{a} & Date [year/month/day] & $\tau$\tablefootmark{a} & Date [year/month/day] \\ 
\hline
\endfirsthead
\caption{continued.}\\
\hline\hline
   PHz   &  \herschel\  &  RA [EMIR]& DEC [EMIR]  & Program & \multicolumn{2}{c}{E090}   & \multicolumn{2}{c}{E150}   \\
    ID    &  ID         & hr:min:sec & \deg:\arcmin:\arcsec &      & $\tau$\tablefootmark{a} & Date [year/month/day] & $\tau$\tablefootmark{a} & Date [year/month/day] \\ 
\hline
\endhead
\hline
\endfoot
      57  &   01  &  10:37:07 &\;\;41:25:32 &   186-16 &       0.21 &    2017/01/21-24            &     0.25  &   2017/01/24-24             \\ 
          &   02  &  10:36:52 &\;\;41:24:10 &   186-16 &       1.08 &    2017/05/24-25            &     1.08  &   2017/05/24-25             \\ 
          &   04  &  10:37:05 &\;\;41:27:31 &   085-19 &       0.31 &    2019/05/29               &     0.31  &   2019/05/29                \\ 
          &   06  &  10:36:46 &\;\;41:24:41 &   085-19 &       0.57 &    2019/05/28-29            &     0.57  &   2019/05/28-29             \\ 
          &   08  &  10:37:02 &\;\;41:32:50 &   085-19 &       0.30 &    2019/05/29-30            &     0.30  &   2019/05/29-30             \\ 
      70  &   01  &  09:37:14 &\;\;10:00:06 &   197-15 &       0.25 &    2015/12/26-28            &           &                             \\ 
     237  &   01  &  10:28:40 &\;\;43:24:05 &   197-15 &       0.29 &    2015/12/23-25            &           &                             \\ 
          &       &  10:28:40 &\;\;43:24:05 &   077-16 &       0.33 &    2016/06/14-16            &           &                             \\ 
          &       &  10:28:40 &\;\;43:24:05 &   085-19 &            &                             &     0.20  &    2019/05/31               \\ 
     343  &   12  &  02:07:04 & $-$02:15:06 &   219-14 &       0.32 &    2015/01/13-15            &           &                             \\ 
     631  &   01  &  14:33:46 &\;\;12:12:57 &   186-16 &       0.16 &    2017/01/18-24            &           &                             \\ 
          &       &  14:33:45 &\;\;12:12:56 &   085-19 &       0.18 &    2019/05/31-06/01         &           &                             \\ 
     712  &  962  &  10:01:42 &\;\;02:18:35 &   178-20 &       0.23 &    2020/12/06 \& 2021/03/03-04 &  0.23  &   2020/12/06 \& 2021/03/03-04  \\
          & 9741  &  10:01:52 &\;\;02:19:00 &   178-20 &       0.27 &    2021/03/04-05            &     0.27  &   2021/03/04-05             \\
     832  &   01  &  10:44:38 &\;\;33:51:05 &   197-15 &       0.25 &    2015/12/25,26,28         &     0.28  &   2015/12/28 \& 2016/04/25-26 \\
          &       &  10:44:38 &\;\;33:51:05 &   082-21 &            &                             &     0.68  &   2021/07/27-28             \\
          &   04  &  10:44:40 &\;\;33:55:08 &   186-16 &       0.45 &    2017/05/25-26            &     0.45  &   2017/05/25-26             \\ 
          &       &  10:44:40 &\;\;33:55:07 &   171-17 &       0.16 &    2017/12/07,10            &     0.16  &   2017/12/07 \& 2018/03/23  \\ 
          &   07  &  10:44:59 &\;\;33:49:22 &   171-17 &       0.11 &    2017/12/06               &     0.11  &   2017/12/06                \\ 
          &       &  10:44:59 &\;\;33:49:23 &   082-21 &            &                             &     0.56 &    2021/07/38,30             \\
          &   09  &  10:44:40 &\;\;33:54:17 &   171-17 &       0.27 &    2017/12/07,08,10         &     0.27  &   2017/12/07,08,10           \\
          &   10  &  10:44:54 &\;\;33:50:51 &   171-17 &       0.37 &    2017/12/08               &     0.37  &   2017/12/08                \\ 
          &   15  &  10:44:55 &\;\;33:53:43 &   171-17 &       0.65 &    2017/12/09               &     0.65  &   2017/12/09                \\ 
          &   18  &  10:44:34 &\;\;33:49:50 &   213-19 &       0.49 &    2020/05/03 \& 2020/07/06 &     0.58  &   2020/07/05-06             \\ 
          &   24  &  10:44:38 &\;\;33:53:47 &   213-19 &       0.43 &    2020/05/01 \& 2020/07/06 &     0.55  &   2020/07/05-06             \\ 
          &   26  &  10:44:37 &\;\;33:48:02 &   213-19 &       0.35 &    2020/05/02-03 \& 2020/07/04 &  0.37  &   2020/05/04 \& 2020/07/04  \\ 
    1473  & 1842  &  17:23:17 &\;\;59:38:59 &   171-17 &       0.26 &    2017/12/09,10,12         &     0.08  &   2017/12/12                \\ 
  124051  &   01  &  16:58:45 &\;\;36:05:40 &   107-14 &       0.51 &    2014/08/02-03            &     0.33  &   2014/08/03-04             \\ 
          &       &  16:58:45 &\;\;36:05:41 &   082-21 &       0.31 &    2021/07/31-08/01         &     0.31  &   2021/07/31-08/01          \\ 
          &   03  &  16:58:45 &\;\;36:03:47 &   186-16 &       0.24 &    2017/01/18-23            &     0.14  &   2017/01/21-23             \\ 
          &       &  16:58:45 &\;\;36:03:47 &   082-21 &       0.57 &    2021/08/02,27,29         &     0.57  &   2021/08/02,27,29          \\ 
  124052  &   03  &  23:14:38 & $-$04:16:31 &   077-16 &       0.27 &    2016/09/13-15            &           &                             \\ 
          & 15\tablefootmark{b}&23:14:52 & $-$04:18:22 &   219-14 &       0.40 &    2015/03/19-23            &     0.36  &   2015/03/23                \\ 
  124053  &   01  &  12:49:04 &\;\;48:20:45 &   197-15 &       0.32 &    2015/12/23-25            &     0.35  &   2015/12/25                \\ 
          &   02  &  12:49:02 &\;\;48:20:18 &   077-16 &       0.39 &    2016/09/14-15            &           &                             \\ 
          &       &  12:49:02 &\;\;48:20:18 &   082-21 &       0.33 &    2021/07/31-08/01         &     0.44  &   2021/07/28,31-08/01       \\ 
          &   03  &  12:49:11 &\;\;48:20:40 &   085-19 &       0.46 &    2019/05/28-29            &     0.46  &   2019/05/28-29             \\ 
          &       &  12:49:11	&\;\;48:20:40 &   082-21 &       0.77 &    2021/08/27,29            &     0.68  &   2021/08/27,29             \\
          &   15  &  12:49:05 &\;\;48:22:00 &   213-19 &       0.37 &    2020/05/01-03            &           &                             \\ 
          &       &  12:49:05 &\;\;48:22:00 &   082-21 &       0.50 &    2021/08/02,27            &     0.50  &   2021/08/02,27            \\ 
  125002  &   02  &  17:15:58 &\;\;46:46:22 &   219-14 &       0.34 &    2015/01/17 \&  2015/03/20,23,30 &           &                             \\
  125018  &   06  &  14:17:05 &\;\;69:32:34 &   107-14 &       0.49 &    2014/08/06-10            &           &                             \\ 
  125026  &   02  &  12:10:04 &\;\;46:06:30 &   107-14 &       0.50 &    2014/08/08-10            &           &                             \\ 
          &       &  12:10:03 &\;\;46:06:26 &   171-17 &       0.16 &    2017/12/10,12            &     0.08  &   2017/12/12                \\ 
          &       &  12:10:03 &\;\;46:06:26 &   085-19 &            &                             &     0.33  &   2019/05/30-31             \\ 
  125027  &   01  &  11:53:48 &\;\;66:30:28 &   077-16 &       0.31 &    2016/06/14-17            &           &                             \\
          &       &  11:53:48 &\;\;66:30:28 &   077-16 &       0.12 &    2016/09/14               &     0.27  &   2016/09/14                \\   
          &   15  &  11:53:44 &\;\;66:30:53 &   186-16 &       0.36 &    2017/05/25-26            &     0.36  &   2017/05/25-26             \\
  125056  &   01  &  17:35:20 &\;\;28:17:20 &   219-14 &       0.21 &  2015/01/14-17 \& 2015/03/23 &           &                             \\
          &   06  &  17:35:11 &\;\;28:19:30 &   085-19 &            &                             &     0.35  &   2019/05/30-31             \\
  125107  &   02  &  17:33:13 &\;\;42:42:31 &   077-16 &       0.24 &    2016/06/15-17            &           &                             \\
  125132  &   02  &  16:07:54 &\;\;40:02:39 &   077-16 &       0.41 &    2016/06/14-17            &           &                             \\
\hline                                                                                                                     
\end{longtable}                                                                                                                                                 
\tablefoot{
\tablefoottext{a}{\small $\tau$ is the maximum opacity at 225\,GHz due to the atmosphere
measured during the observations.}
\tablefoottext{b}{\small This source was observed in the E230 band, rather than E150.}
}
\end{landscape}                                                                                                               

\clearpage
\twocolumn
\begin{table*}[!ht]

\centering
\caption{EMIR frequency coverage and integration time\label{tab:obs_setup}}
\begin{tabular}{rc c c r}
\hline \hline
    PHz   &  \herschel\ & Program &  Frequency & Time\tablefootmark{a}  \\
    ID    &  ID         &         &    (GHz)   &  (min) \\
\hline
\noalign{\vspace{2pt}}
      57  &   01  &   186-16 &  83.9--91.7  \&  99.6--103.6   & 383       \\
          &       &          &  91.9--99.7  \& 107.6--115.4   & 341       \\
          &       &          & 136.3--144.0 \& 151.9--159.7   & 160       \\
      57  &   02  &   186-16 &  86.7--94.5  \& 148.2--156.0   & 376       \\
      57  &   04  &   085-19 &  89.8--97.6  \& 148.0--155.8   & 279       \\
      57  &   06  &   085-19 &  89.8--97.6  \& 105.5--113.2   & 69--98--131 \\
          &       &          & 148.0--155.8 \& 163.7--171.4   & 69--98     \\ 
      57  &   08  &   085-19 &  89.8--97.6  \& 148.0--155.8   & 261       \\
      70  &   01  &   197-15 &  83.9--91.7  \&  99.6--107.4   & 231       \\ 
          &       &          &  91.9--99.7  \& 107.6--115.4   & 155       \\ 
     237  &   01  &   197-15 &  83.9--91.7  \&  99.6--107.3   & 307       \\ 
          &       &          &  91.9--99.7  \& 107.6--115.3   & 311       \\ 
          &       &   077-16 &  71.0--78.8  \&  86.7--94.4    & 211       \\
          &       &          &  76.5--84.3  \&  92.3--100.0   & 269       \\ 
          &       &          &  80.0--87.8  \&  95.7--103.4   & 194       \\
          &       &   085-19 & 139.6--147.4 \& 155.2--163.0   & 112       \\ 
     343  &   12  &   219-14 &  83.9--91.7  \&  99.6--107.3   & 140       \\
     631  &   01  &   186-16 &  83.9--91.7  \&  99.6--107.4   & 646       \\
          &       &          &  91.9--99.7  \& 107.6--115.4   & 322       \\
          &       &   085-19 &  86.0--93.7  \& 101.7--109.4   &  32       \\
     712  &  962  &   178-20 & 107.1--114.9 \& 143.6--151.4   & 582       \\
     712  & 9741  &   178-20 & 107.1--114.9 \& 143.6--151.4   & 569       \\
     832  &   01  &   197-15 &  83.9--91.7  \&  99.6--107.3   & 204       \\
          &       &          &  91.9--99.7  \& 107.6--115.3   & 336       \\
          &       &          & 126.0--133.8 \& 141.7--149.5   & 590       \\
          &       &   082-21 & 144.3--152.1 \& 160.0--167.8   & 360       \\
     832  &   04  &   186-16 &  95.1-102.9  \& 160.0-167.8    & 366       \\
          &       &   171-17 &  93.1--100.8 \& 156.7--164.4   & 669       \\
     832  &   07  &   171-17 &  93.1--102.8 \& 156.7--167.7   & 326       \\
          &       &   082-21 & 126.1--133.9 \& 141.8--149.6   & 246       \\
     832  &   09  &   171-17 &  93.1--100.8 \& 156.7--164.4   & 317       \\ 
     832  &   10  &   171-17 &  93.1--100.8 \& 156.7--164.4   & 233       \\
     832  &   15  &   171-17 &  93.1--100.8 \& 156.7--164.4   & 379       \\
     832  &   18  &   213-19 &  78.5--86.2  \&  94.2--101.9   & 321       \\
          &       &          &  86.2--94.0  \& 101.9--109.6   & 258       \\
          &       &          &  73.5--81.3  \& 131.5--139.3   & 175       \\ 
     832  &   24  &   213-19 &  78.5--86.2  \&  94.2--101.9   & 116--28--287--317 \\ 
          &       &          &  77.5--85.3  \&  93.2--100.9   & 74--46--74--74    \\
          &       &          &  77.3--85.0  \&  93.0--100.8   &  37       \\
          &       &          &  86.3--94.0  \& 125.0--132.8   & 129--120--129--129 \\
          &       &          &  91.8--99.5  \& 154.6--162.3   & 146       \\
     832  &   26  &   213-19 &  78.5--86.2  \&  94.2--101.9   & 341--459--418--453  \\
          &       &          &  86.4--94.2  \& 102.1--109.8   & 256       \\
          &       &          &  89.2--97.0  \& 125.3--133.0   & 385       \\
\hline                                                                                                                     
\end{tabular}                                                                                                              
\tablefoot{
\tablefoottext{a}{\small Multiple values of integration time are reported in
case some scans had to be removed because of artifacts present in the data.}
}
\end{table*}

\begin{table*}[!ht]
\setcounter{table}{2}
\centering
\caption{continued}
\begin{tabular}{rc c c r}
\hline \hline
    PHz   &  \herschel\ & Program &  Frequency &  Time\tablefootmark{a}  \\
    ID    &  ID         &         &    (GHz)   &  (min) \\
\hline
\noalign{\vspace{2pt}}
    1473  & 1842  &   171-17 & 109.0--116.7 \& 146.1--153.8  &  125   \\
          &       &          &  92.5--100.2 \& 108.2--115.9  &  145   \\
  124051  &   01  &   107-14 &  83.9--91.7  \&  99.6--107.3  &  482   \\
          &       &          & 135.8--143.6 \& 151.5--159.3  &  481   \\
          &       &   082-21 &  96.2--104.0 \& 169.4--177.2  &  414--438   \\
  124051  &   03  &   186-16 &  83.9--91.7  \&  99.6--107.4  &  601   \\
          &       &          & 133.7--141.5 \& 149.4--157.2  &  400   \\
          &       &   082-21 & 100.7--108.5 \& 169.7--177.5  &  384  \\
  124052  &   03  &   077-16 &  74.1--81.9  \&  89.8--97.5   &  483   \\
  124052  &   15  &   219-14 &  83.9--91.7  \&  99.6--107.4  &  160   \\
          &       &          &  91.9--99.7  \& 107.6--115.3  &  180   \\
  124053  &   01  &   197-15 &  83.8--91.7  \&  99.6--107.3  &288--307  \\
          &       &          &  91.9--99.7  \& 107.6--115.3  & 308    \\
          &       &          & 130.2--137.9 \& 145.8--153.6  & 154    \\
  124053  &   02  &   077-16 &  90.2--98.0  \& 105.9--113.6  & 520    \\
          &       &   082-21 & 107.7--115.5                  & 83     \\  
          &       &   082-21 & 128.0--135.8                  & 108    \\
          &       &   082-21 & 143.7--151.5                  & 174--192 \\ 
  124053  &   03  &   085-19 & 104.2--112.0 \& 150.6--158.4  & 265--348 \\
          &       &   082-21 &  93.5--101.3 \& 109.2--117.0  & 378    \\
  124053  &   15  &   213-19 &  91.4--99.1  \& 107.1--114.8  &69--99--127 \\
          &       &   082-21 & 107.7--115.5 \& 147.7--155.5  & 148    \\
  125002  &   02  &   219-14 &  83.9--91.7  \&  99.6--107.3  & 106--161 \\
          &       &          &  91.9--99.6  \& 107.6--115.3  & 105    \\
  125018  &   06  &   107-14 &  83.8--91.7  \&  99.6--107.3  & 463    \\
          &       &          &  91.9--99.6  \& 107.6--115.3  & 381    \\
  125026  &   02  &   107-14 &  83.9--91.7  \&  99.6--107.3  & 272    \\
          &       &   171-17 &  91.9--99.7  \& 107.6--115.3  & 105    \\
          &       &          &  89.9--97.6  \& 151.4--159.1  &  11    \\
          &       &   085-19 & 141.7--149.5 \& 157.4--165.2  & 166    \\
          &       &          & 149.4--157.1 \& 165.1--172.8  & 159    \\
  125027  &   01  &   077-16 &  83.9--91.7  \&  99.6--107.3  & 232    \\
          &       &          &  91.9--99.7  \& 107.6--115.4  & 299    \\
          &       &   077-16 & 128.6--136.4 \& 144.3--152.0  & 347    \\
  125027  &   15  &   186-16 &  93.9--101.7 \& 143.5--151.3  & 250    \\
  125056  &   01  &   219-14 &  83.9--91.7  \&  99.6--107.3  & 333    \\
          &       &          &  91.9--99.7  \& 107.6--115.4  & 323    \\
  125056  &   06  &   085-19 & 134.0--141.8 \& 149.7--157.4  & 194--222 \\
  125107  &   02  &   077-16 &  83.9--91.7  \&  99.6--107.3  & 328    \\
          &       &          &  91.9--99.7  \& 107.6--115.3  & 250    \\
  125132  &   02  &   077-16 &  83.9--91.7  \&  99.6--107.3  & 231    \\
          &       &          &  91.9--99.7  \& 107.6--115.3  & 194    \\
\hline                                                                                                                     
\end{tabular}                                                                                                              
\tablefoot{
\tablefoottext{a}{\small Multiple values of integration time are reported in case some scans had to be removed because of artifacts present in the data.}
}
\end{table*}

\section{EMIR spectra}

In this section, we show the spectra of the detected CO lines listed in Table~\ref{tab:lines}. In
Fig.~\ref{fig:trp_trans}, and~\ref{fig:dbl_trans}, we show the sources where multiple transitions
have been detected, and in Fig.~\ref{fig:single_line} those with only one CO 
line detection. The CO line transition is annotated in the
corresponding panel only in case of multiple line detections per source.

\begin{figure*} 
\centering
\includegraphics[width=9cm]{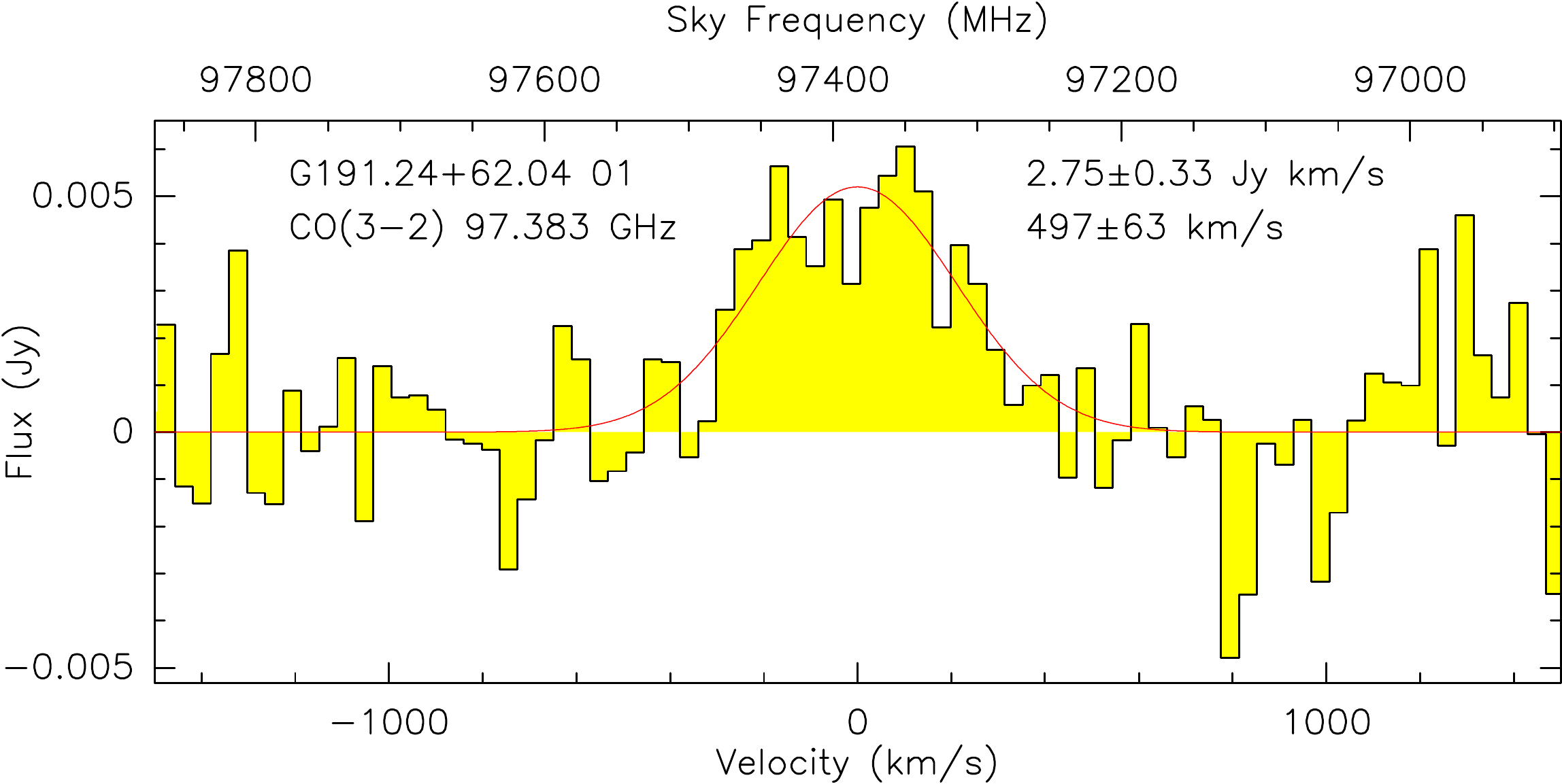}
\includegraphics[width=9cm]{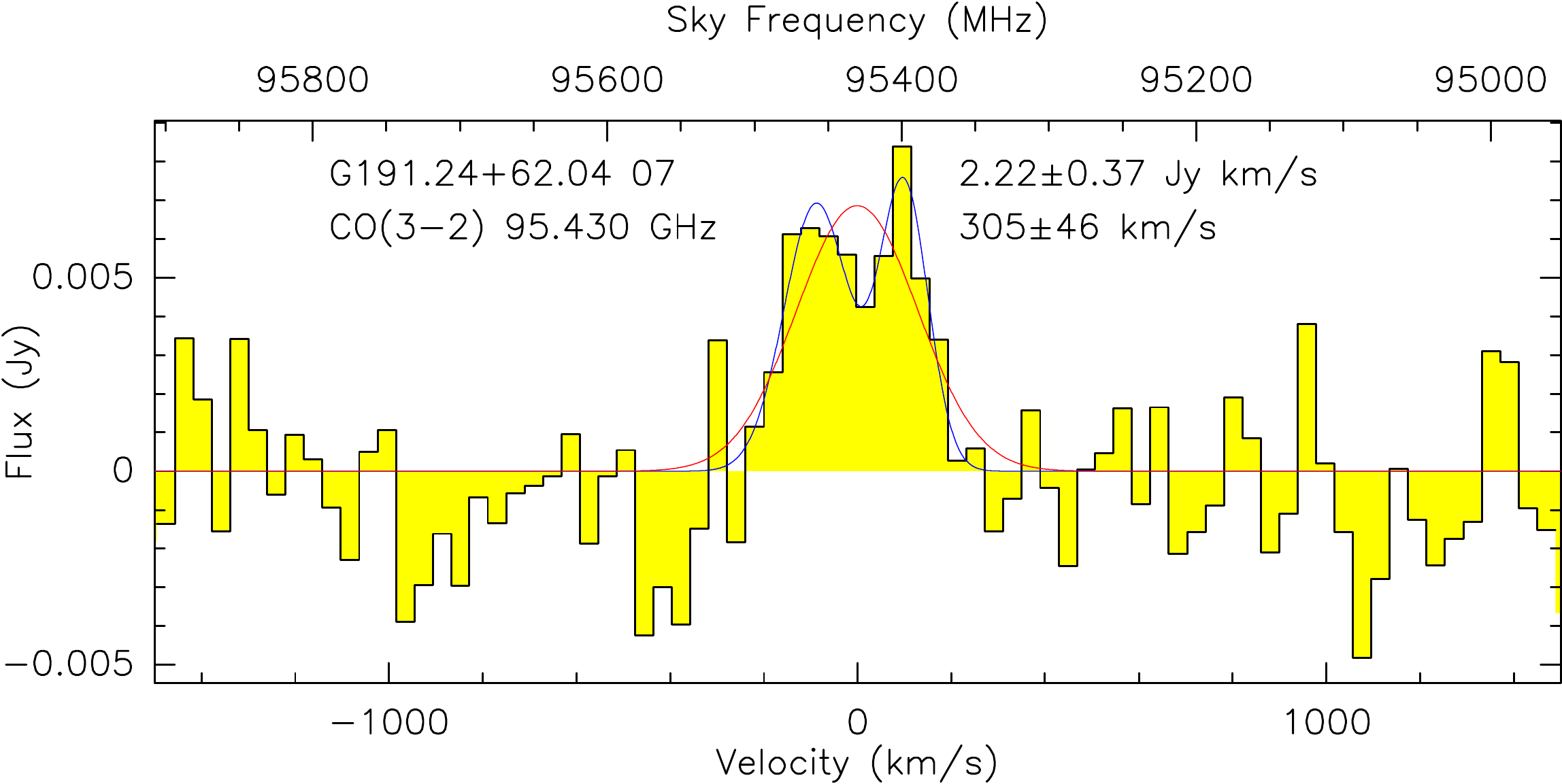}
\includegraphics[width=9cm]{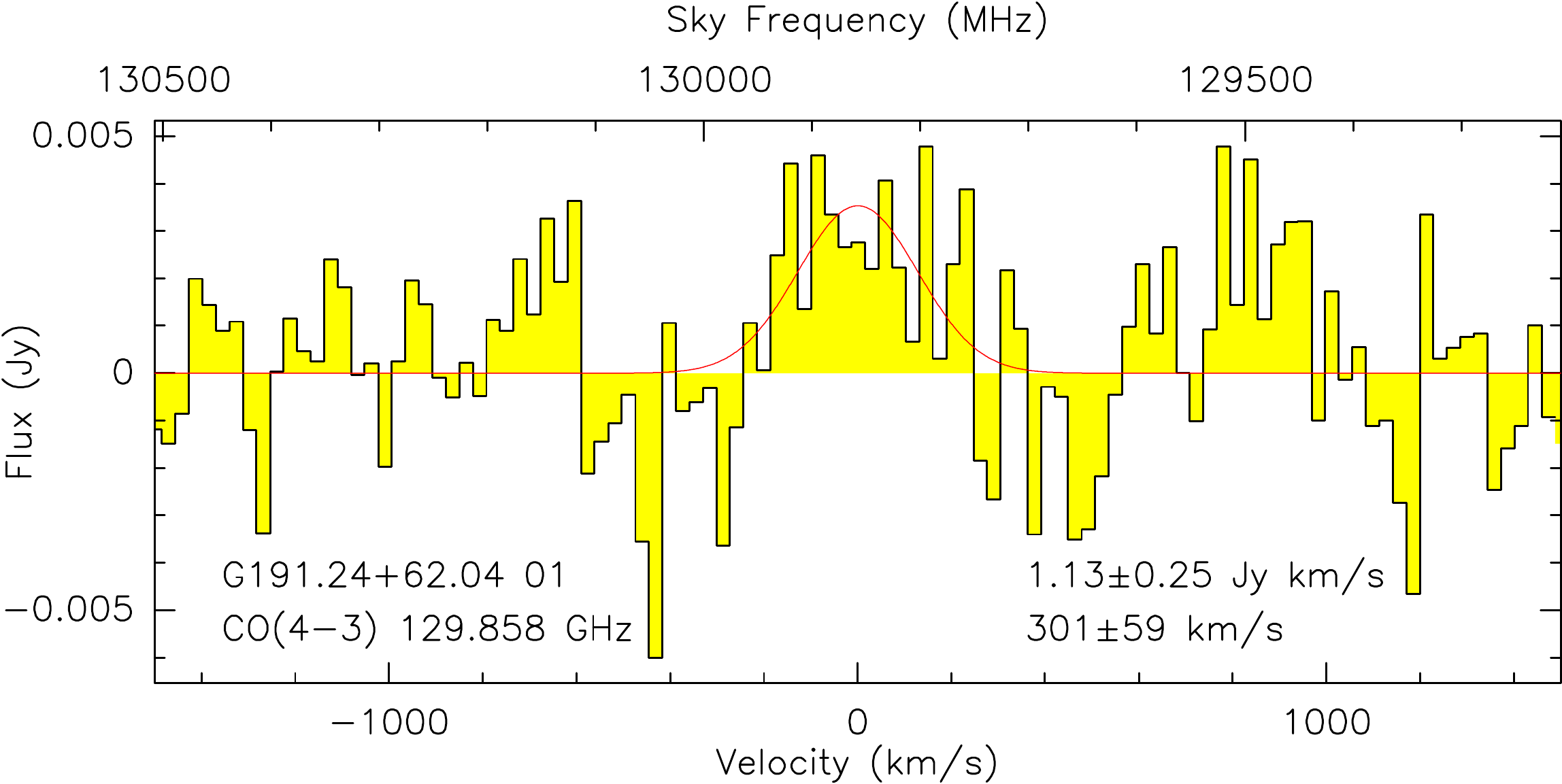}
\includegraphics[width=9cm]{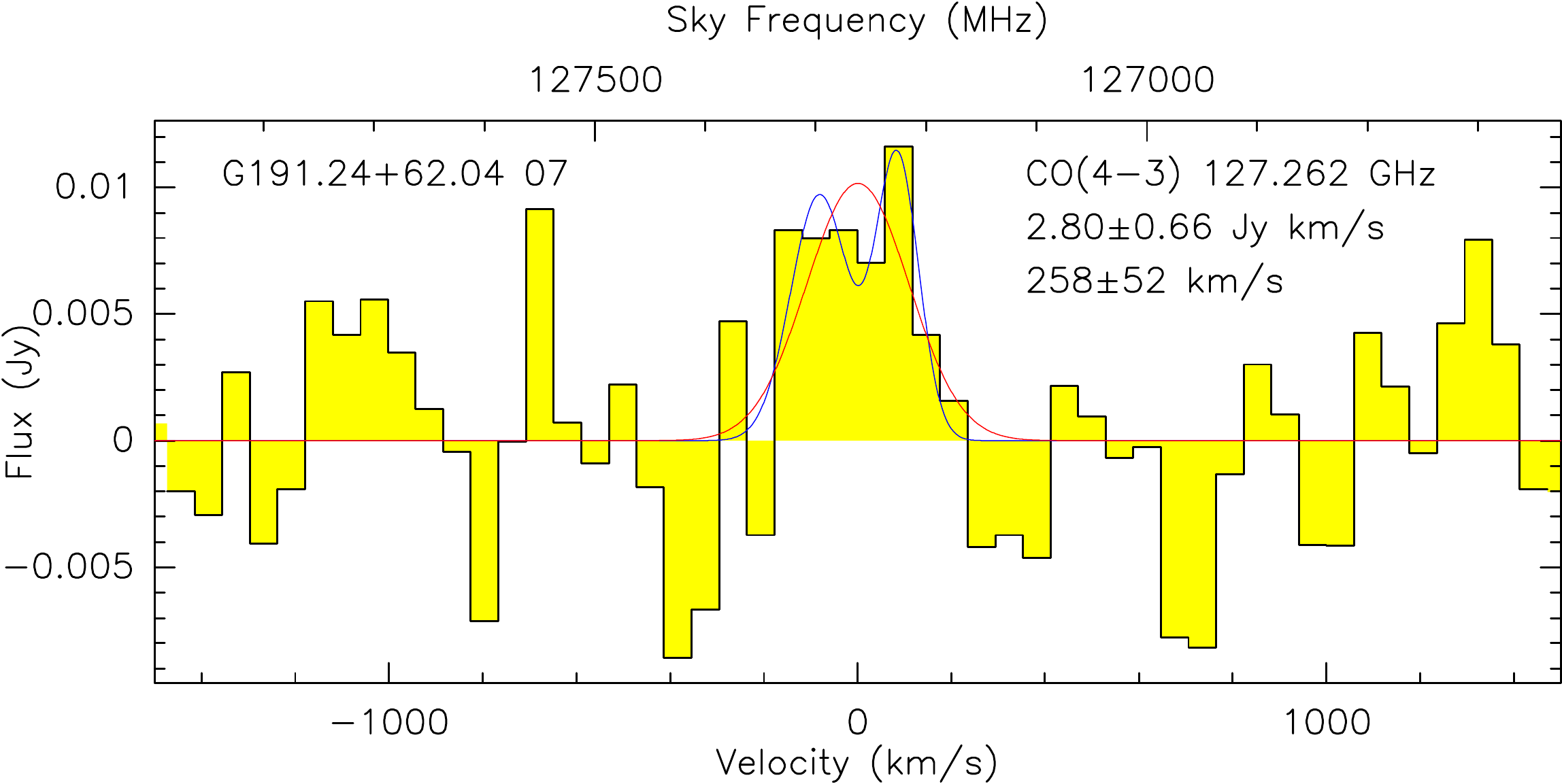}
\includegraphics[width=9cm]{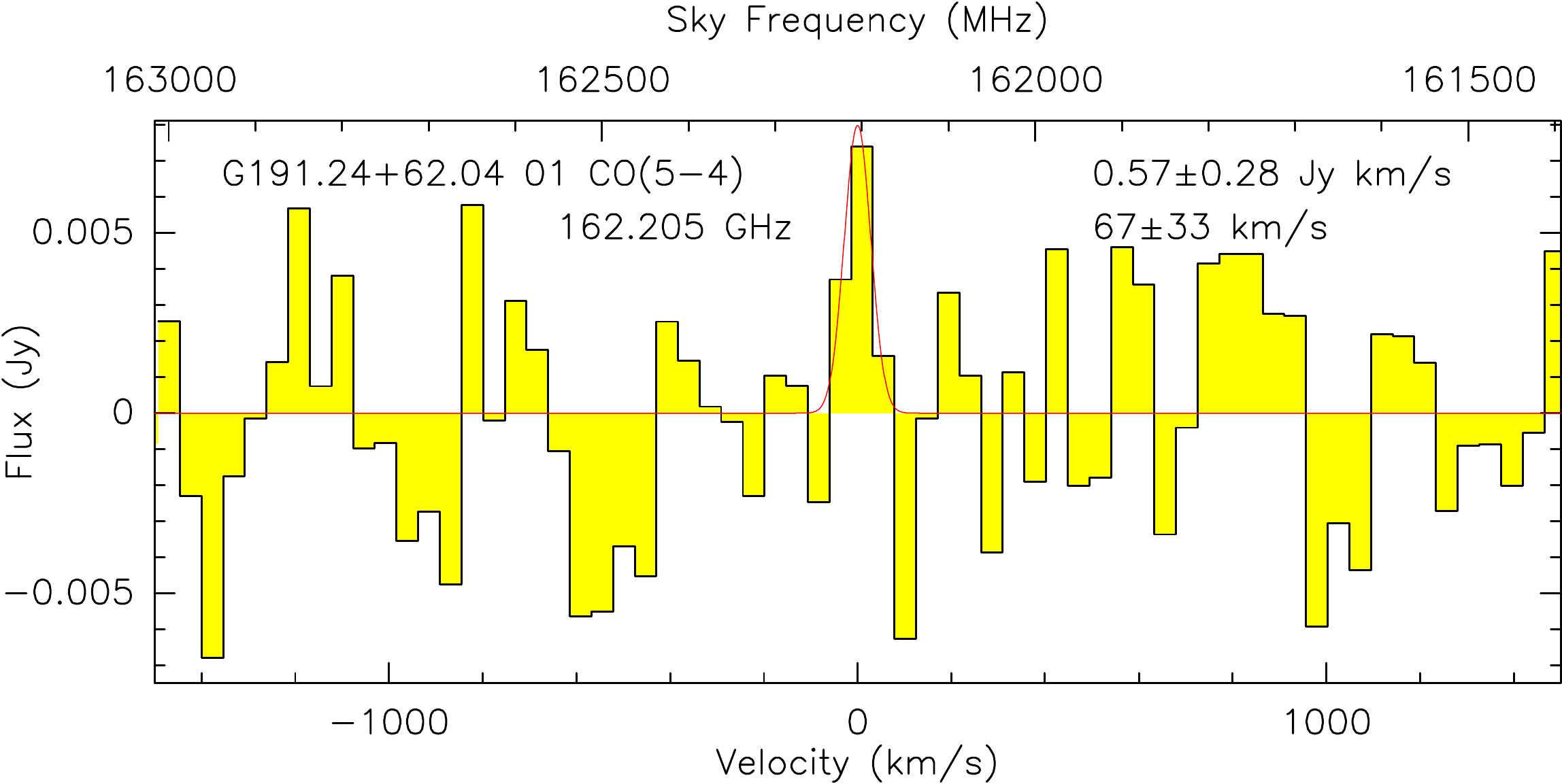}
\includegraphics[width=9cm]{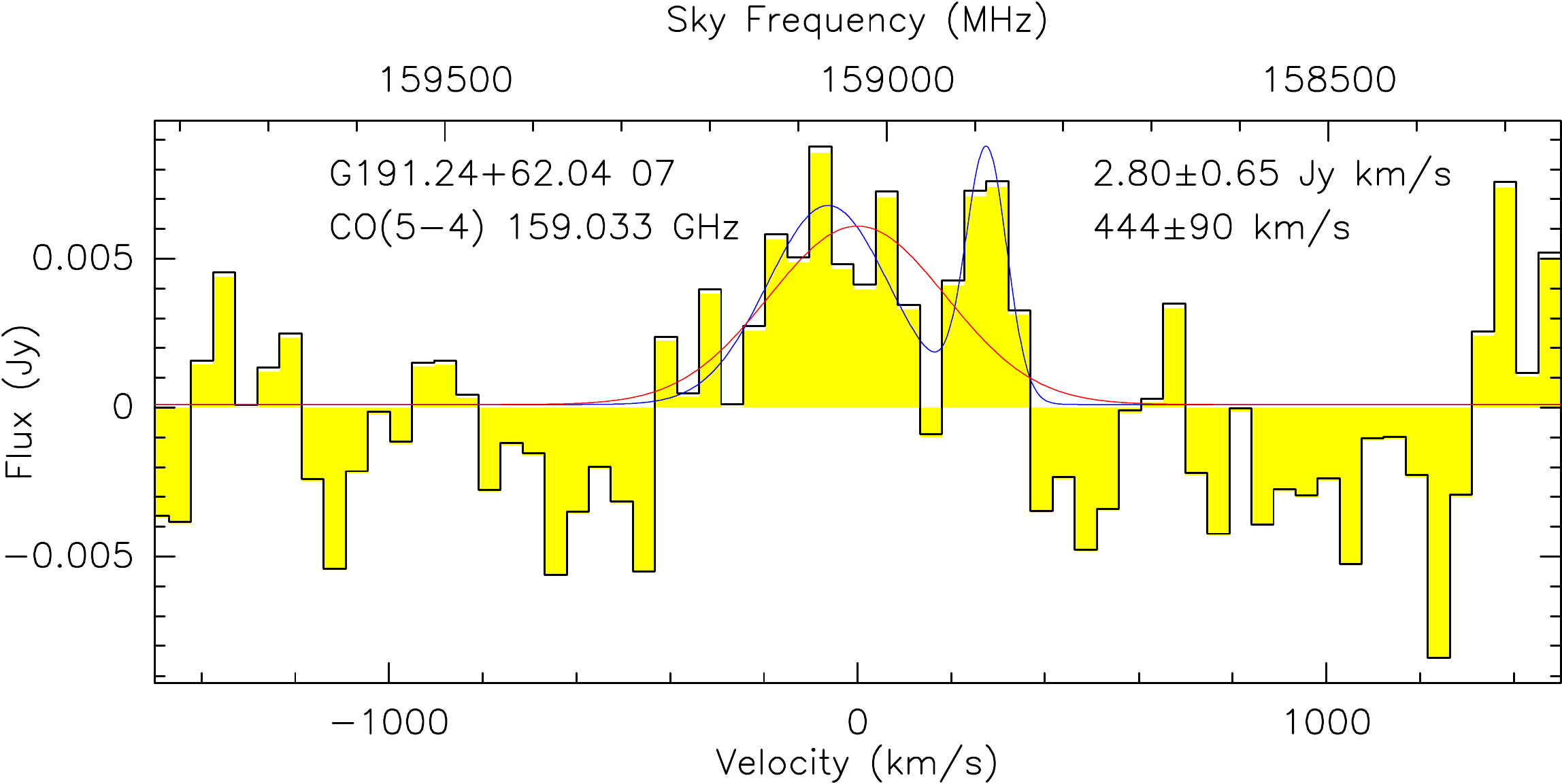}
\caption{{\small Emission line detections in the continuum-subtracted EMIR
spectra of the PHz-IRAM sources where three CO transitions have been
observed.  The fit to each line is overlaid with a red curve for single
Gaussian fits, or with a blue curve for double Gaussian fits.  Source name, line
transition and observed }central frequency, intensity, and width are annotated in each panel.
All left panels refer to one source, and all right panels to another.}
\label{fig:trp_trans}
\end{figure*}

\begin{figure*} 
\setcounter{figure}{0}
\centering
\includegraphics[width=9cm]{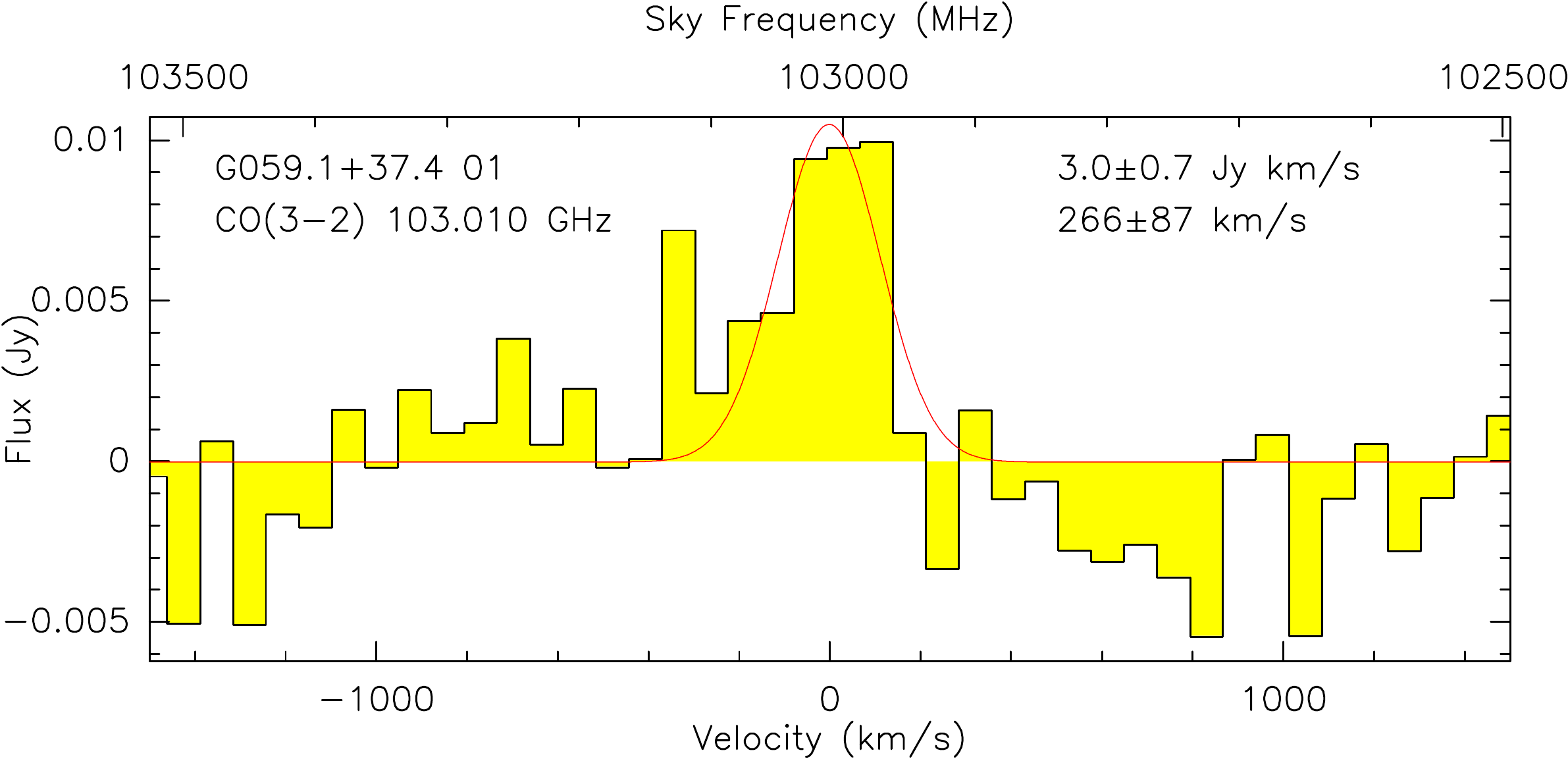}
\includegraphics[width=9cm]{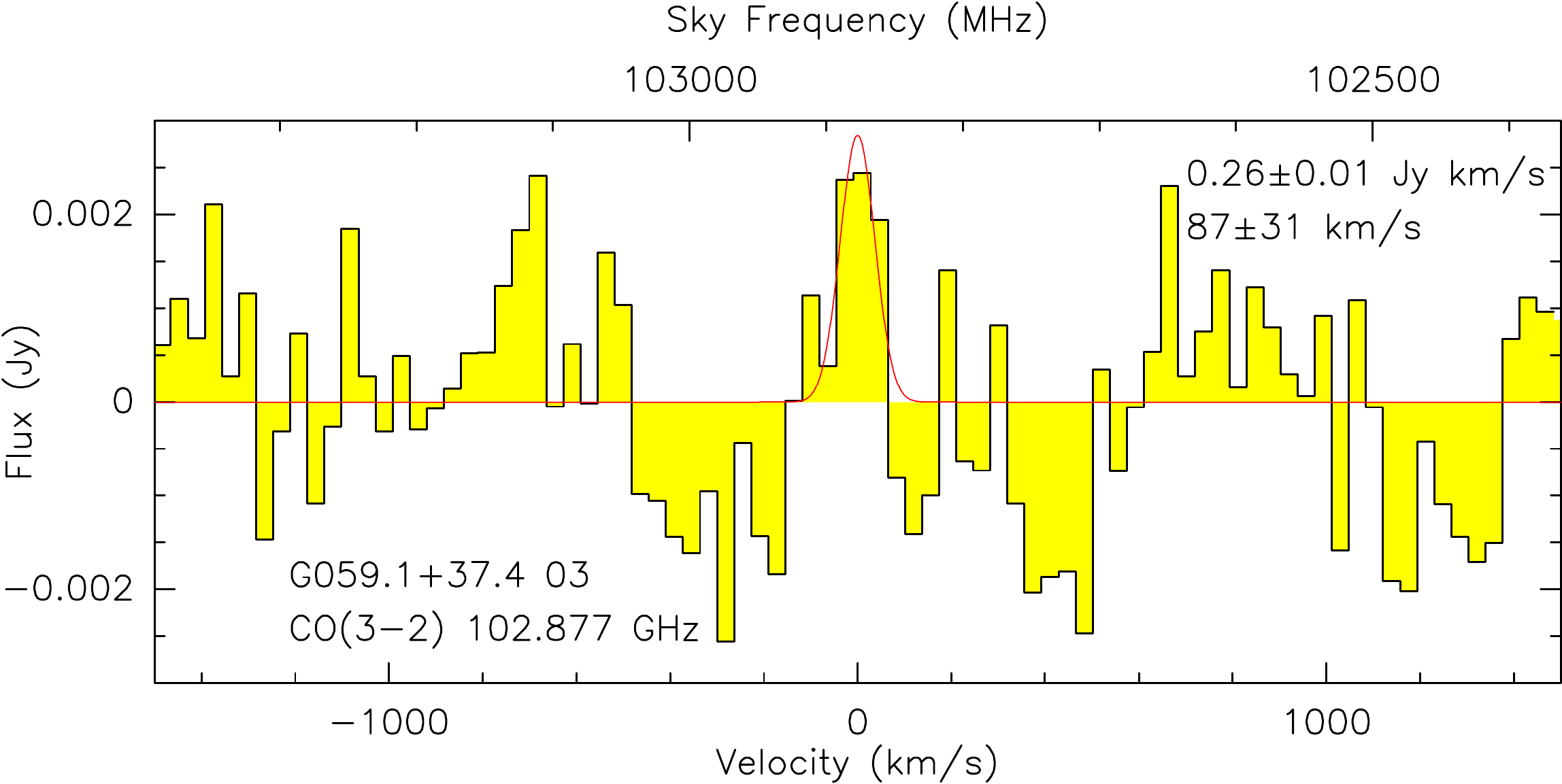}
\includegraphics[width=9cm]{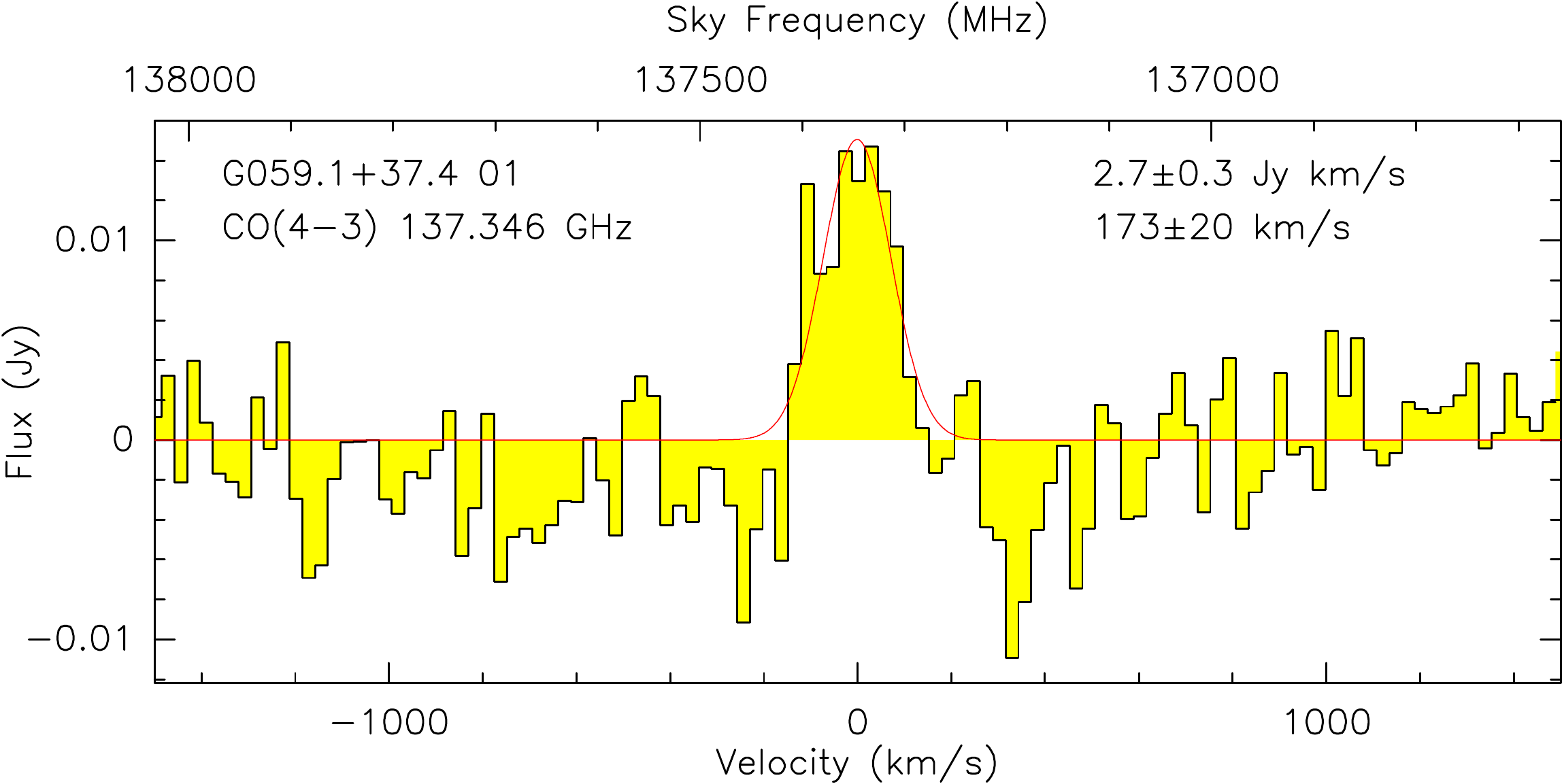}
\includegraphics[width=9cm]{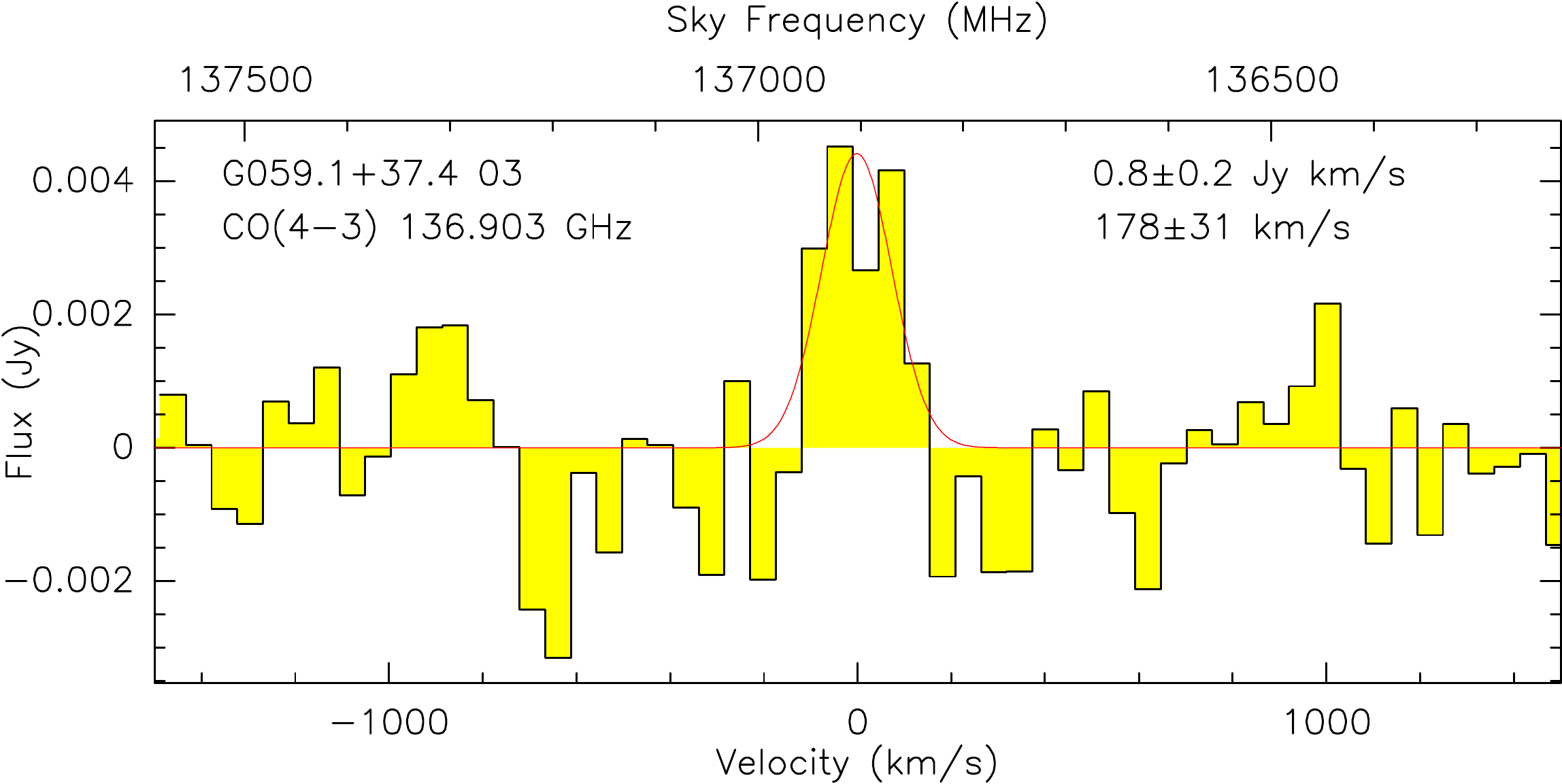}
\includegraphics[width=9cm]{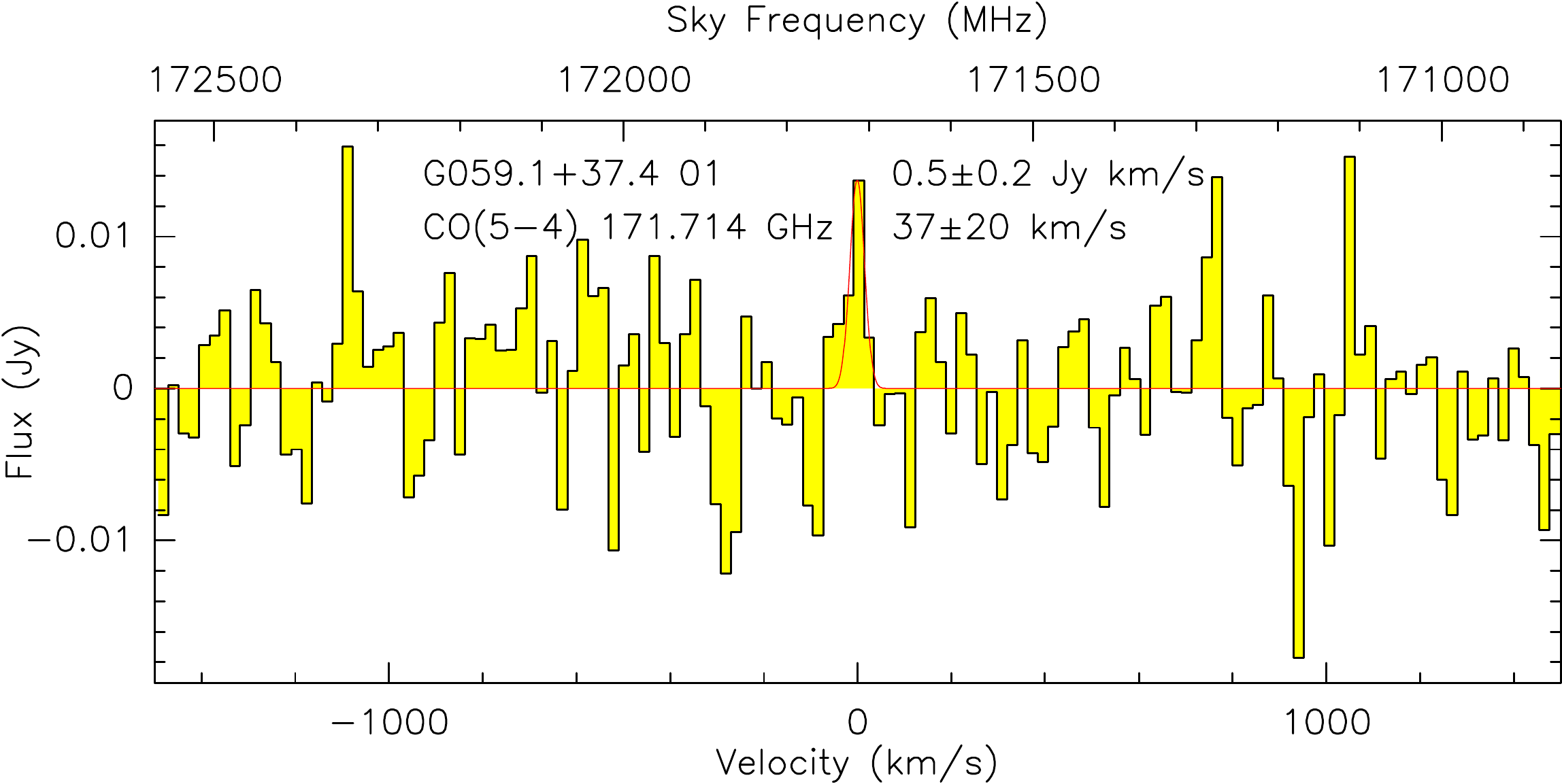}
\includegraphics[width=9cm]{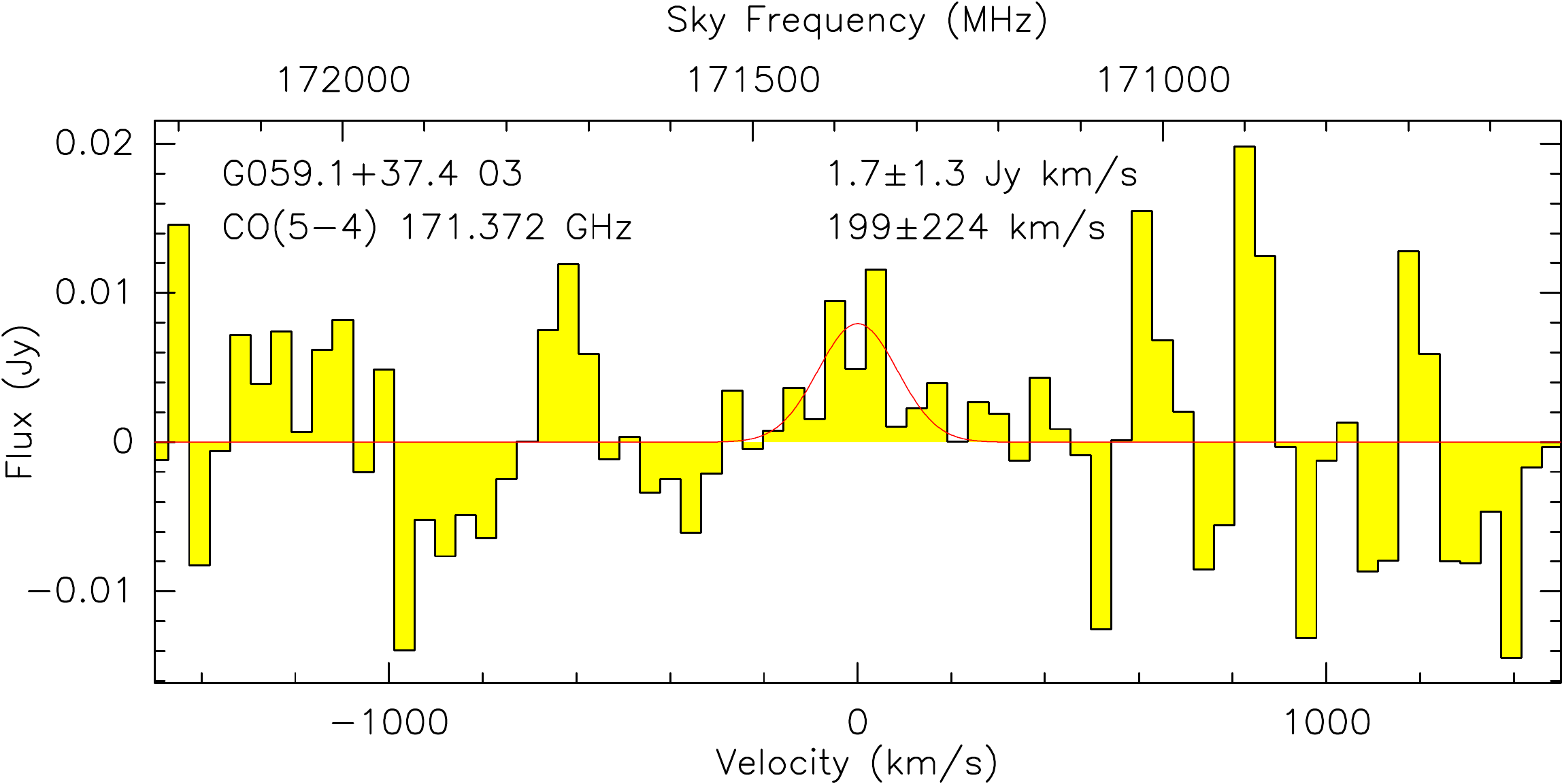}
\caption{{\small Continued.}}
\label{fig:trp_trans}
\end{figure*}

\begin{figure*} 
\centering
\includegraphics[width=9cm]{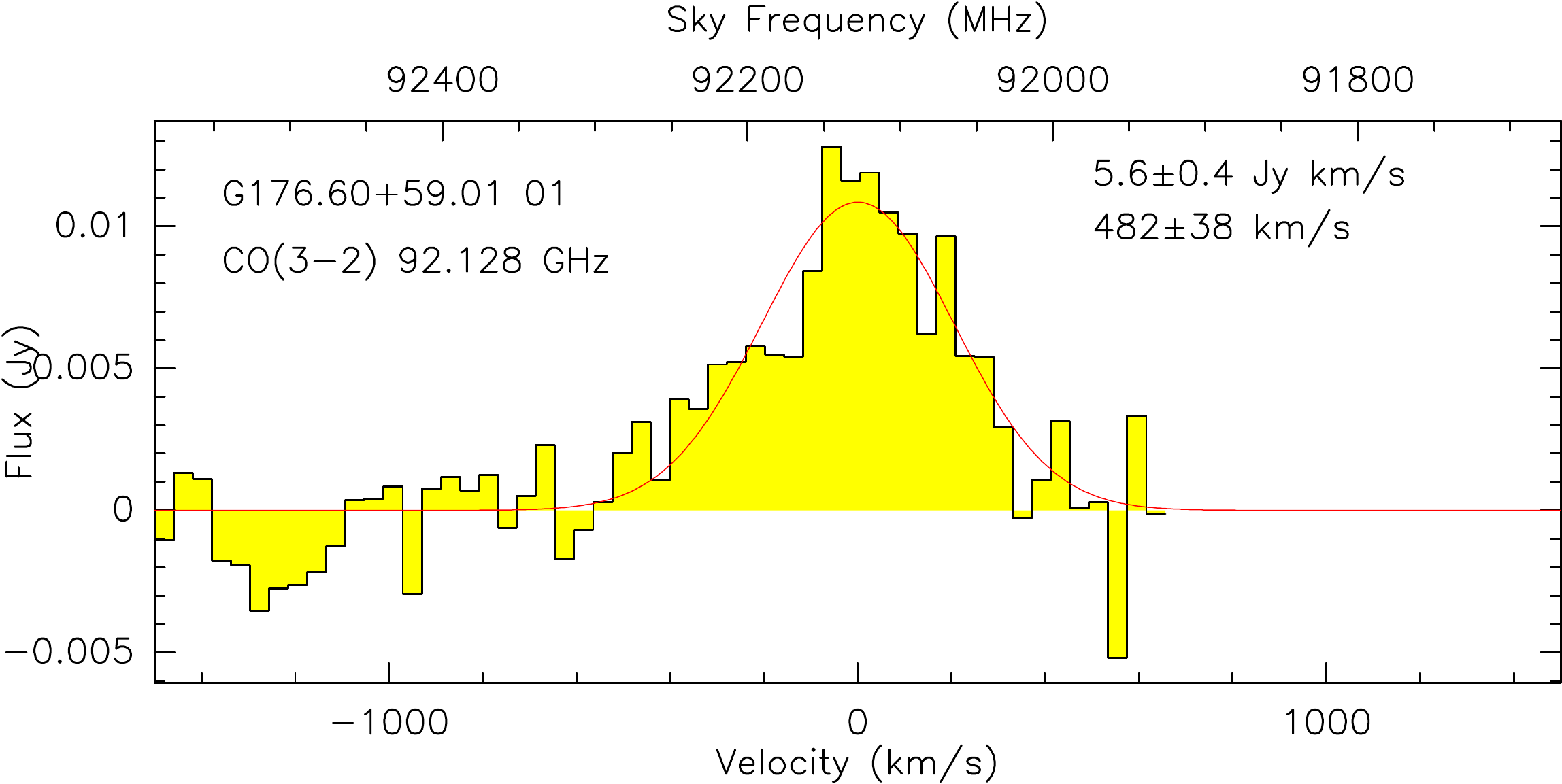}
\includegraphics[width=9cm]{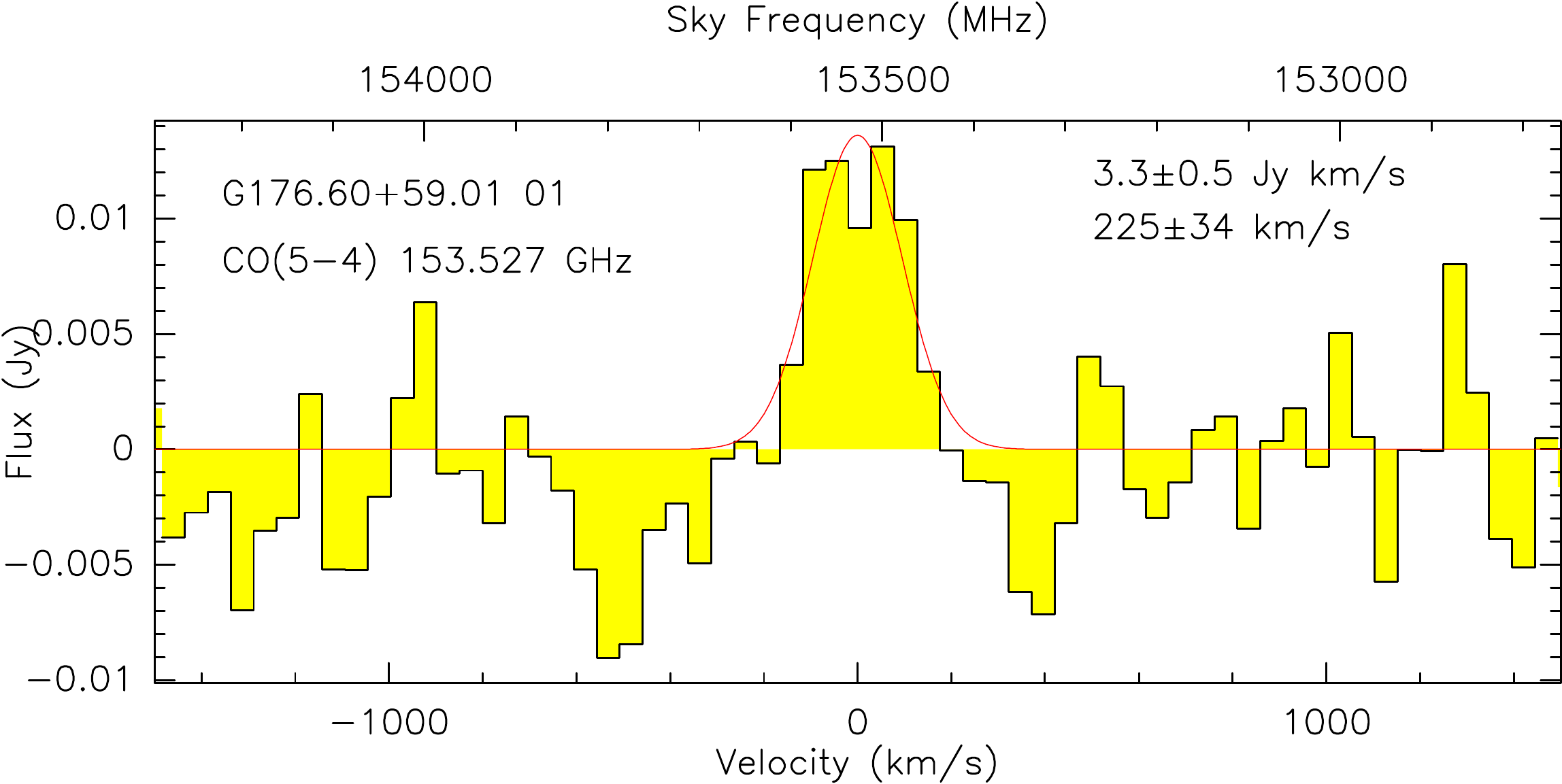}
\includegraphics[width=9cm]{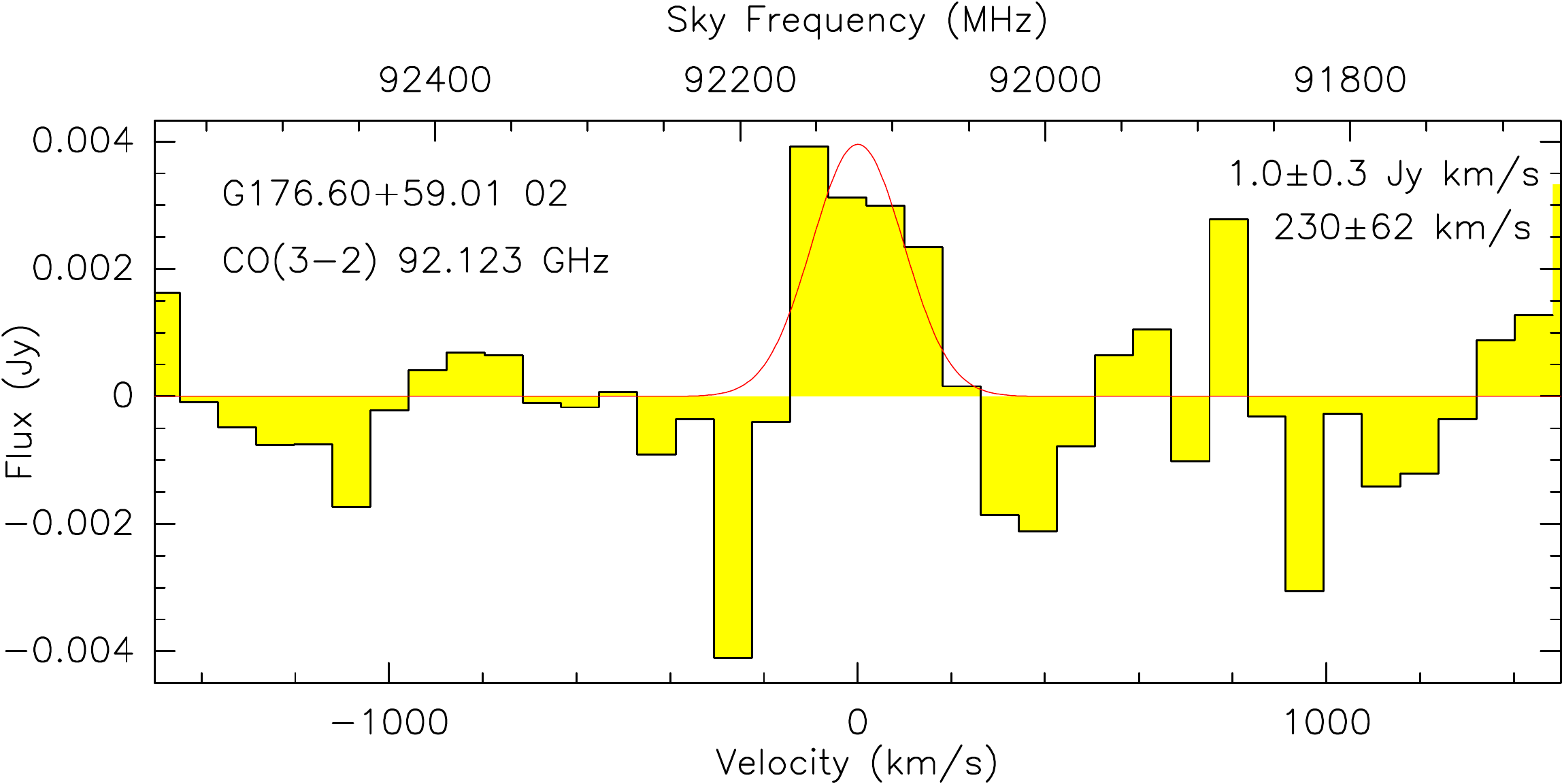}
\includegraphics[width=9cm]{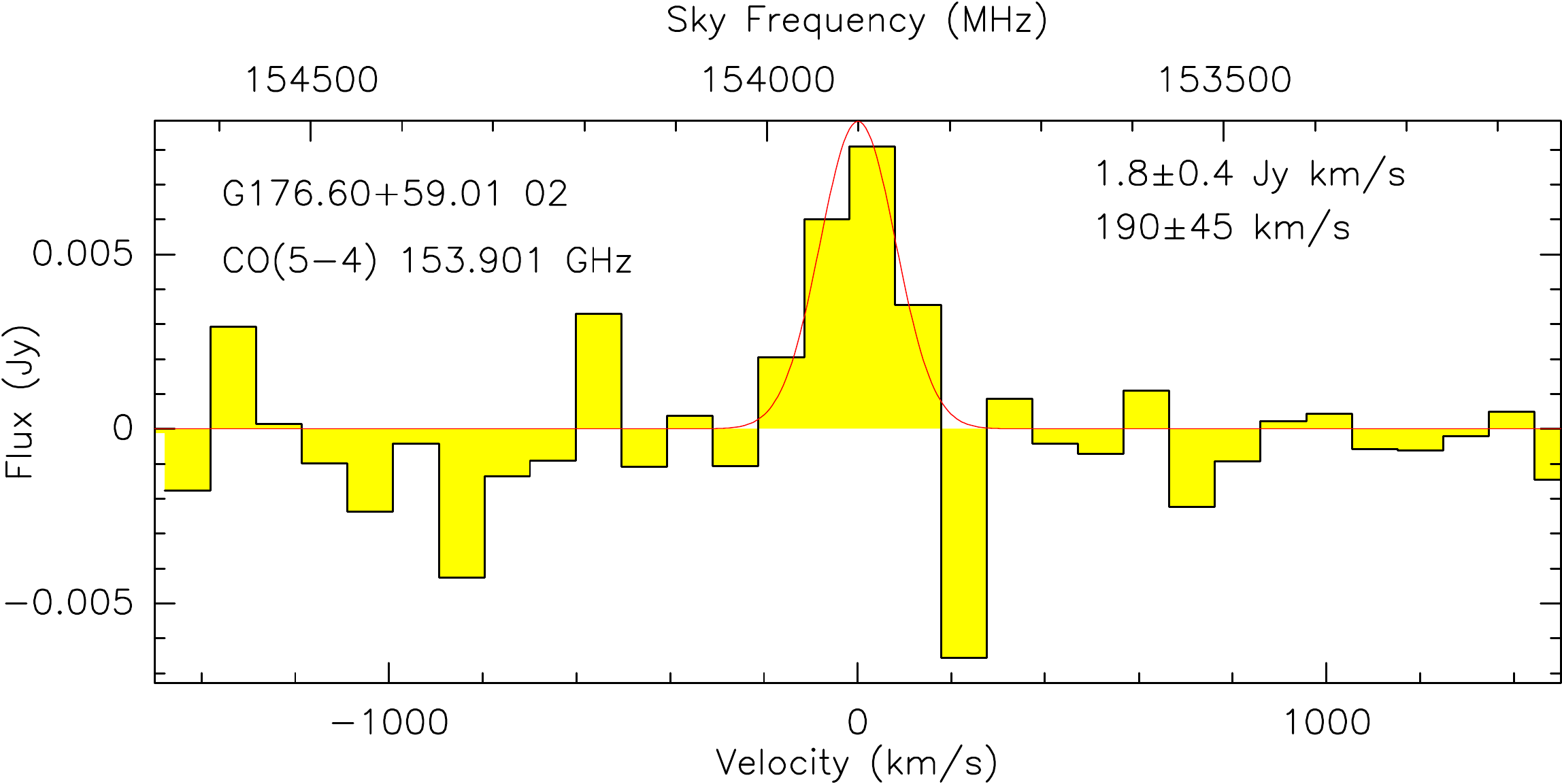}
\includegraphics[width=9cm]{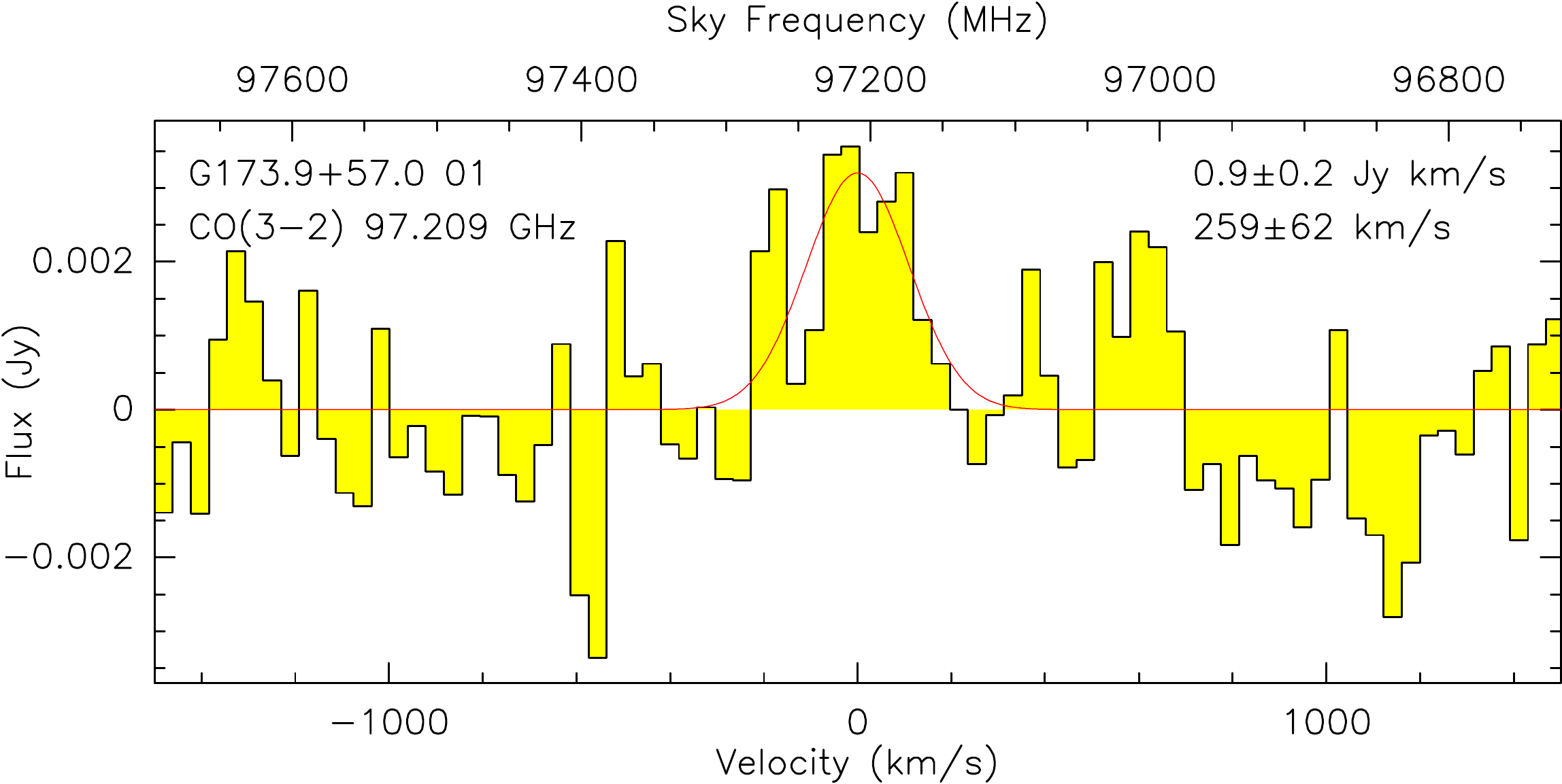}
\includegraphics[width=9cm]{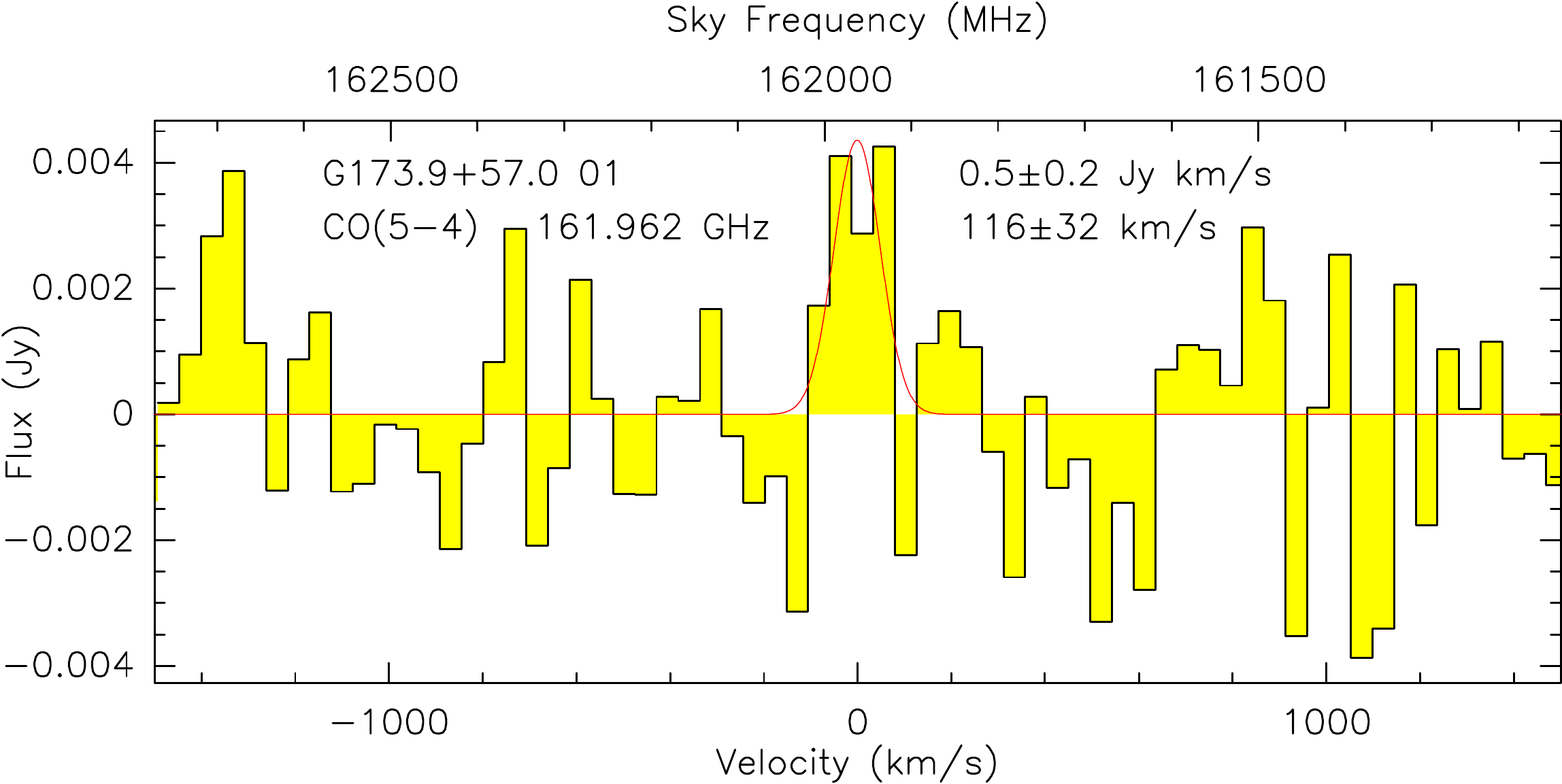}
\includegraphics[width=9cm]{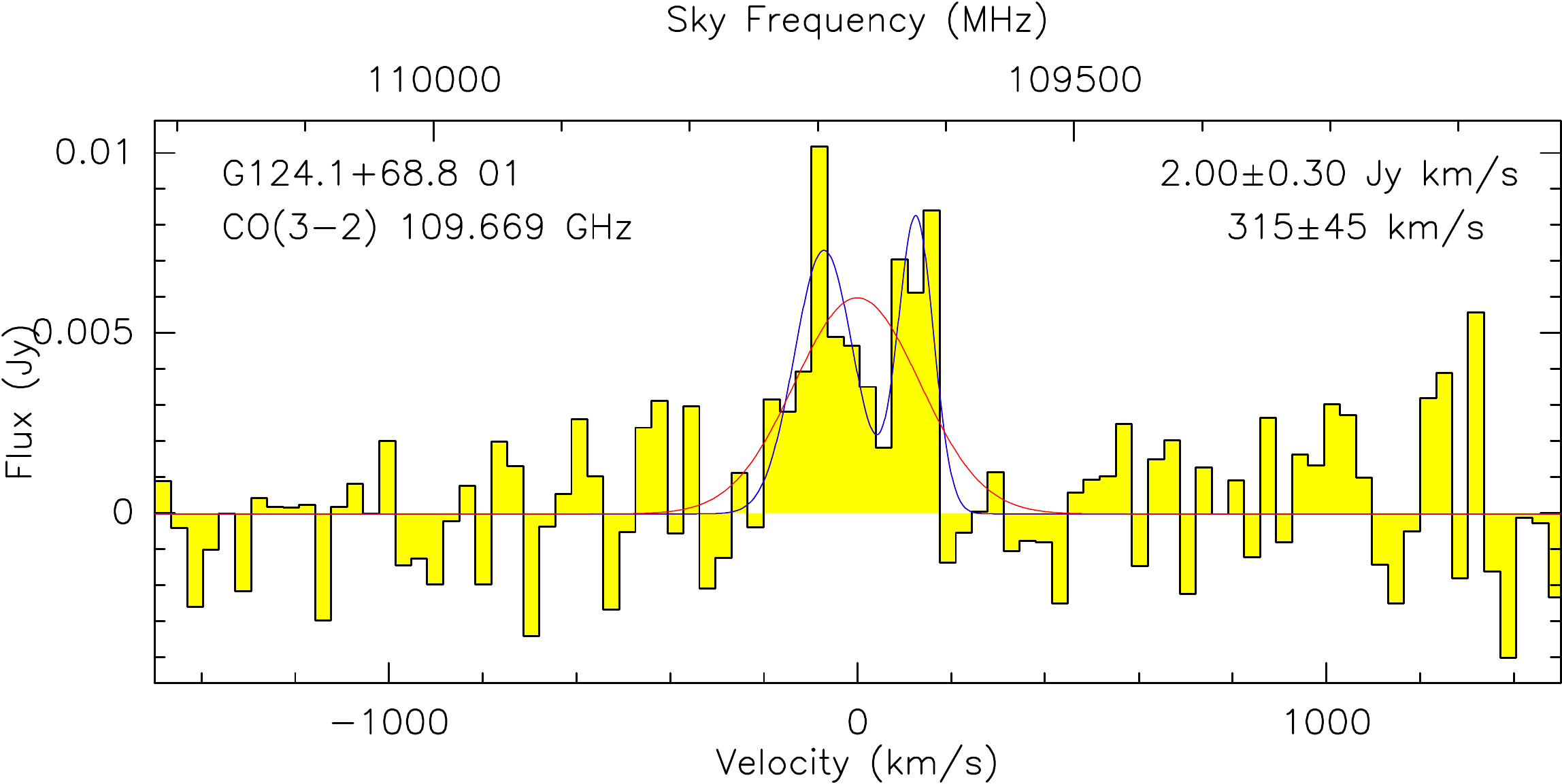}
\includegraphics[width=9cm]{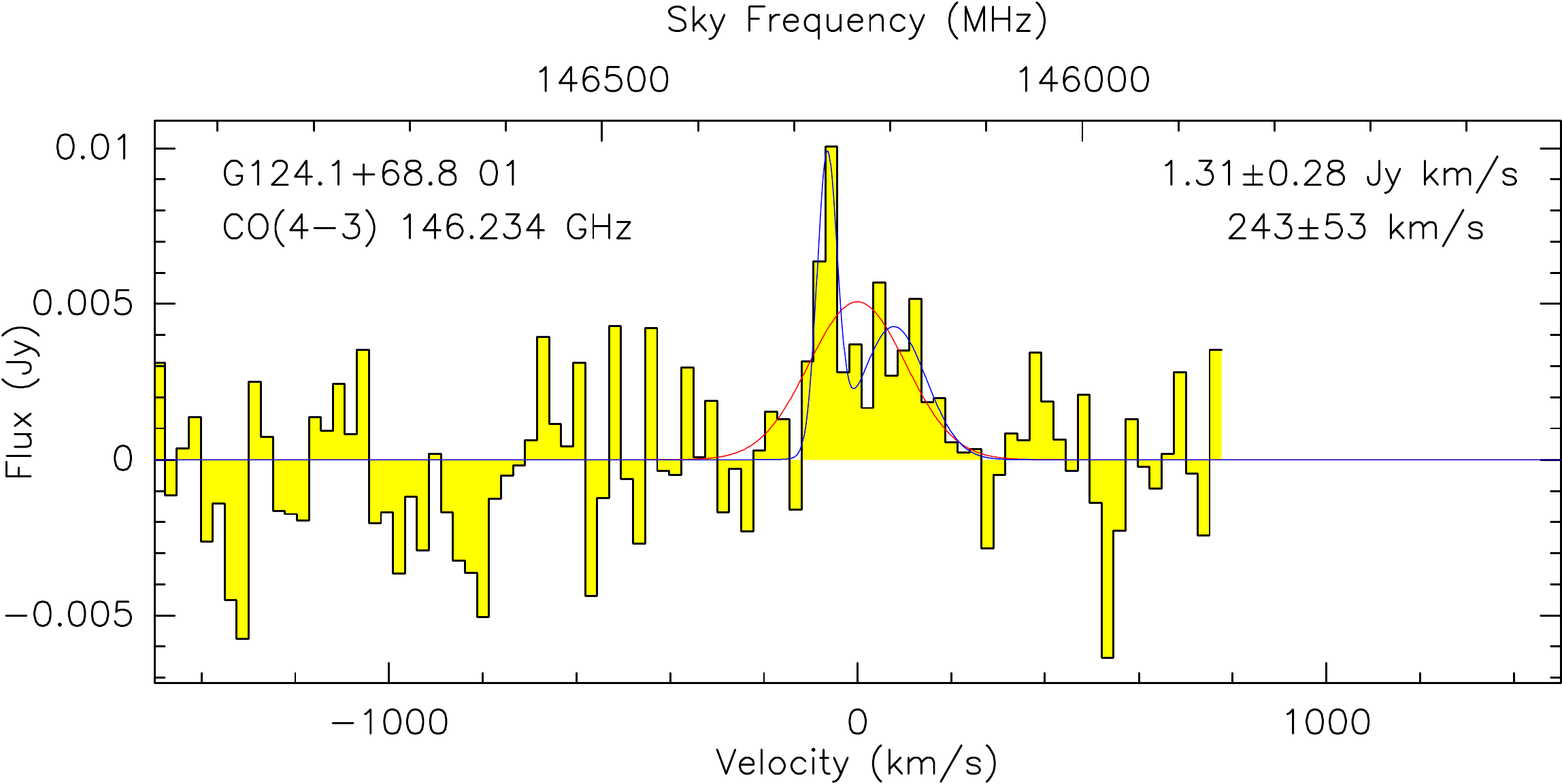}
\caption{{\small Emission line detections in the continum-subtracted EMIR
spectra of the PHz-IRAM sources where two CO transitions have been
detected.  The fit to each line is overlaid with a red curve for single
Gaussian fits, or with a blue curve for double Gaussian fits.  Source name, line
transition and observed central frequency, intensity, and width are annotated in each panel.
Each row corresponds to a specific source.}}
\label{fig:dbl_trans}
\end{figure*}
\begin{figure*} 
\setcounter{figure}{1}
\centering
\includegraphics[width=9cm]{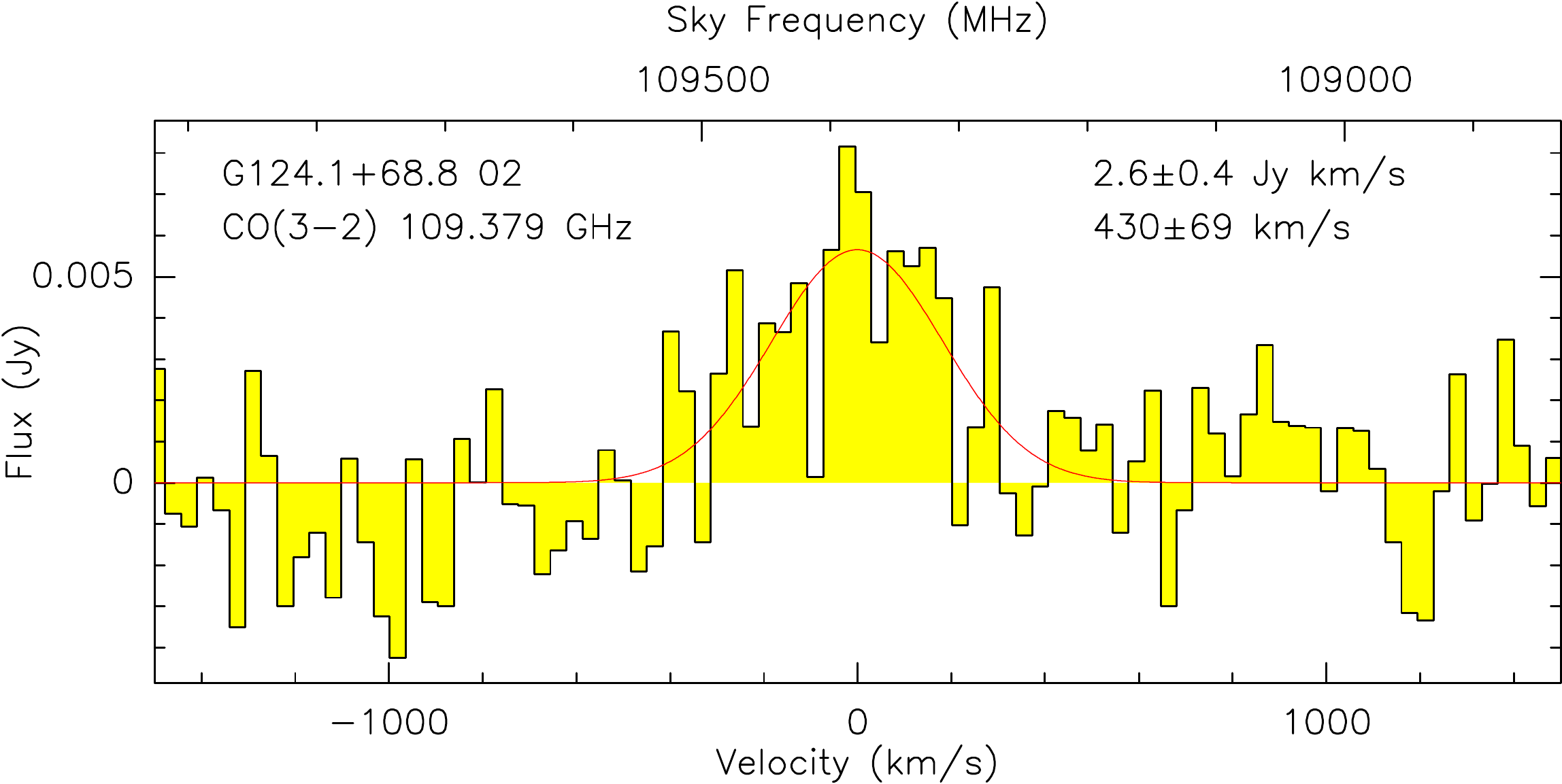}
\includegraphics[width=9cm]{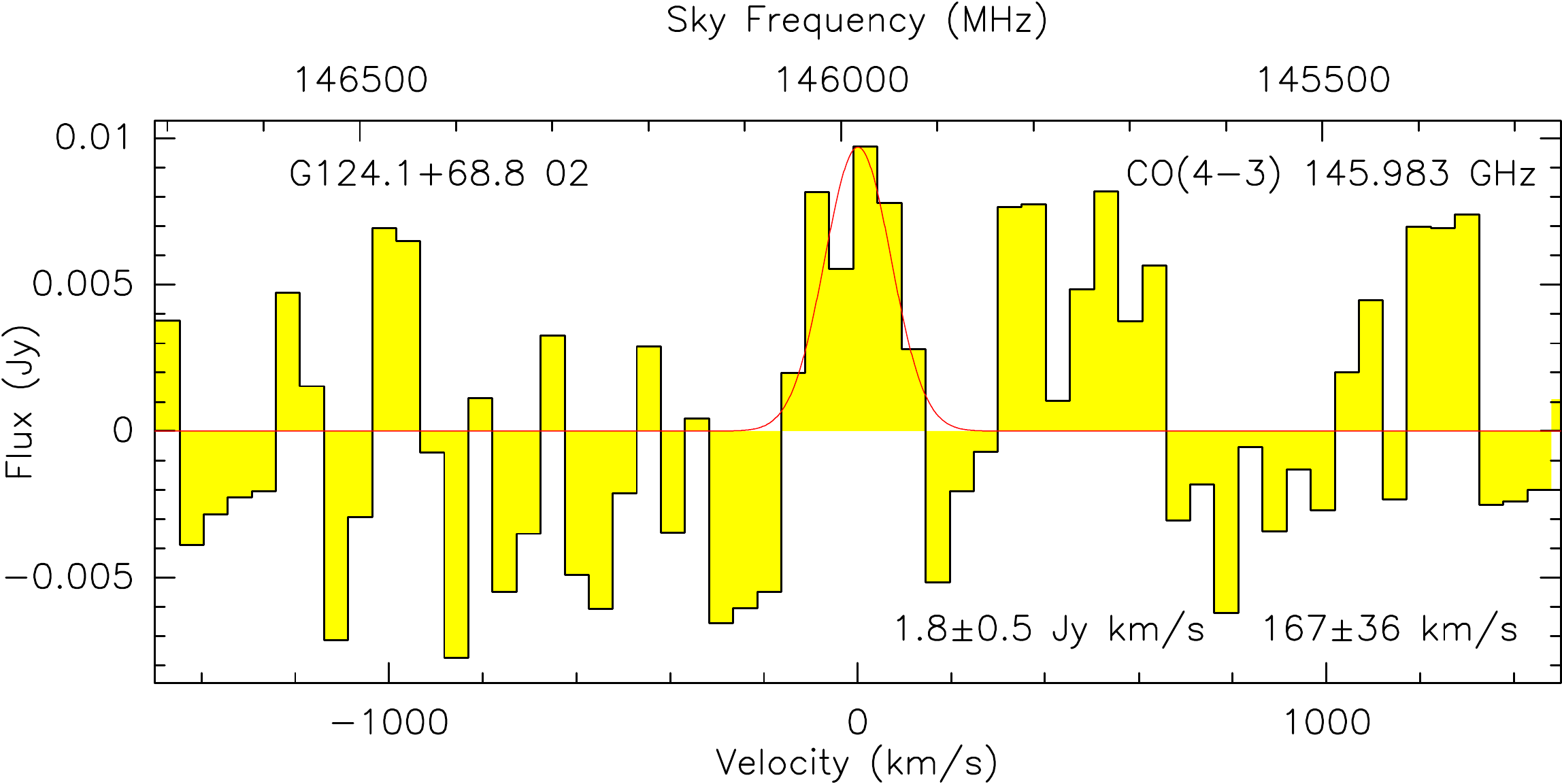}
\includegraphics[width=9cm]{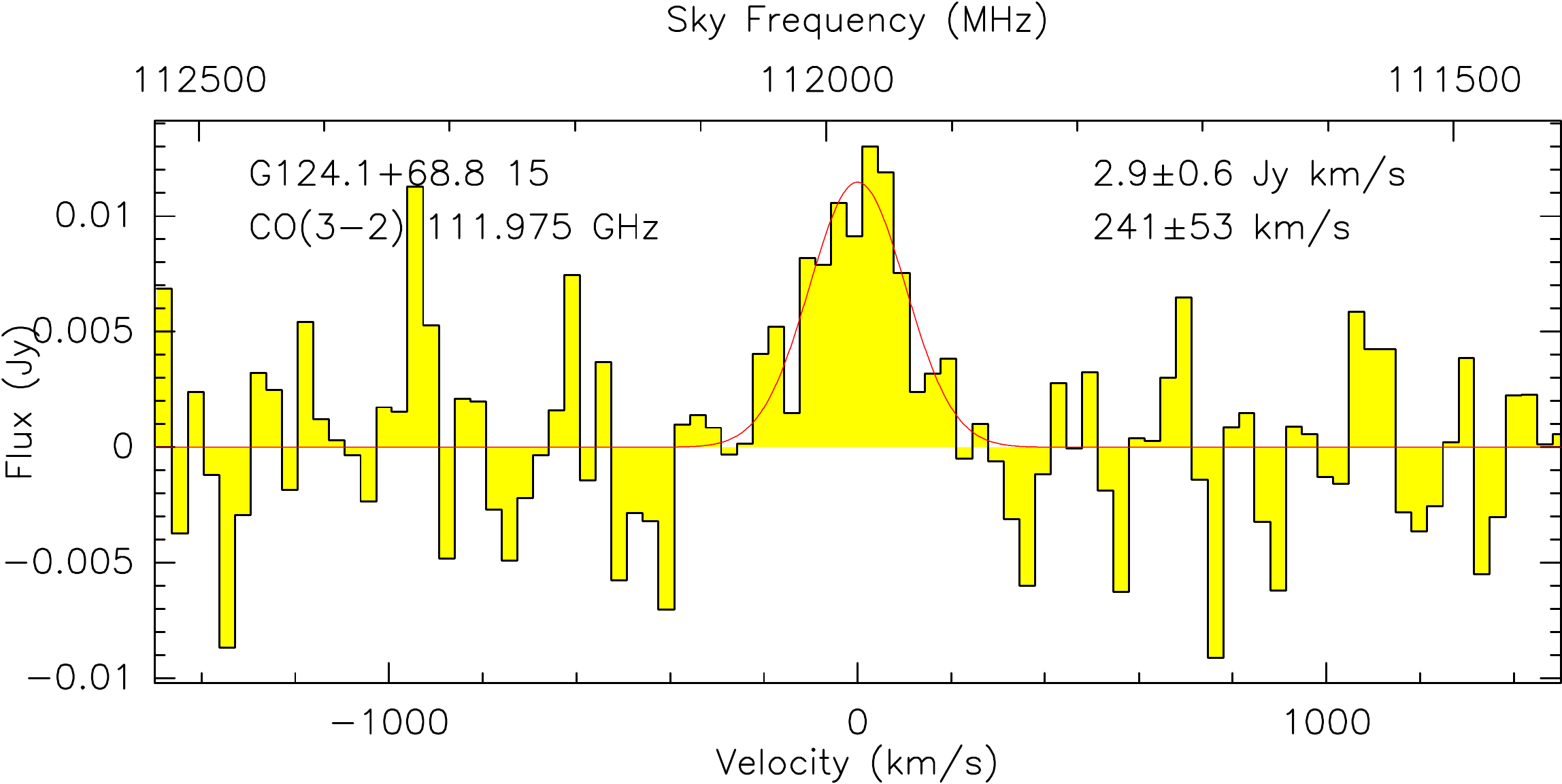}
\includegraphics[width=9cm]{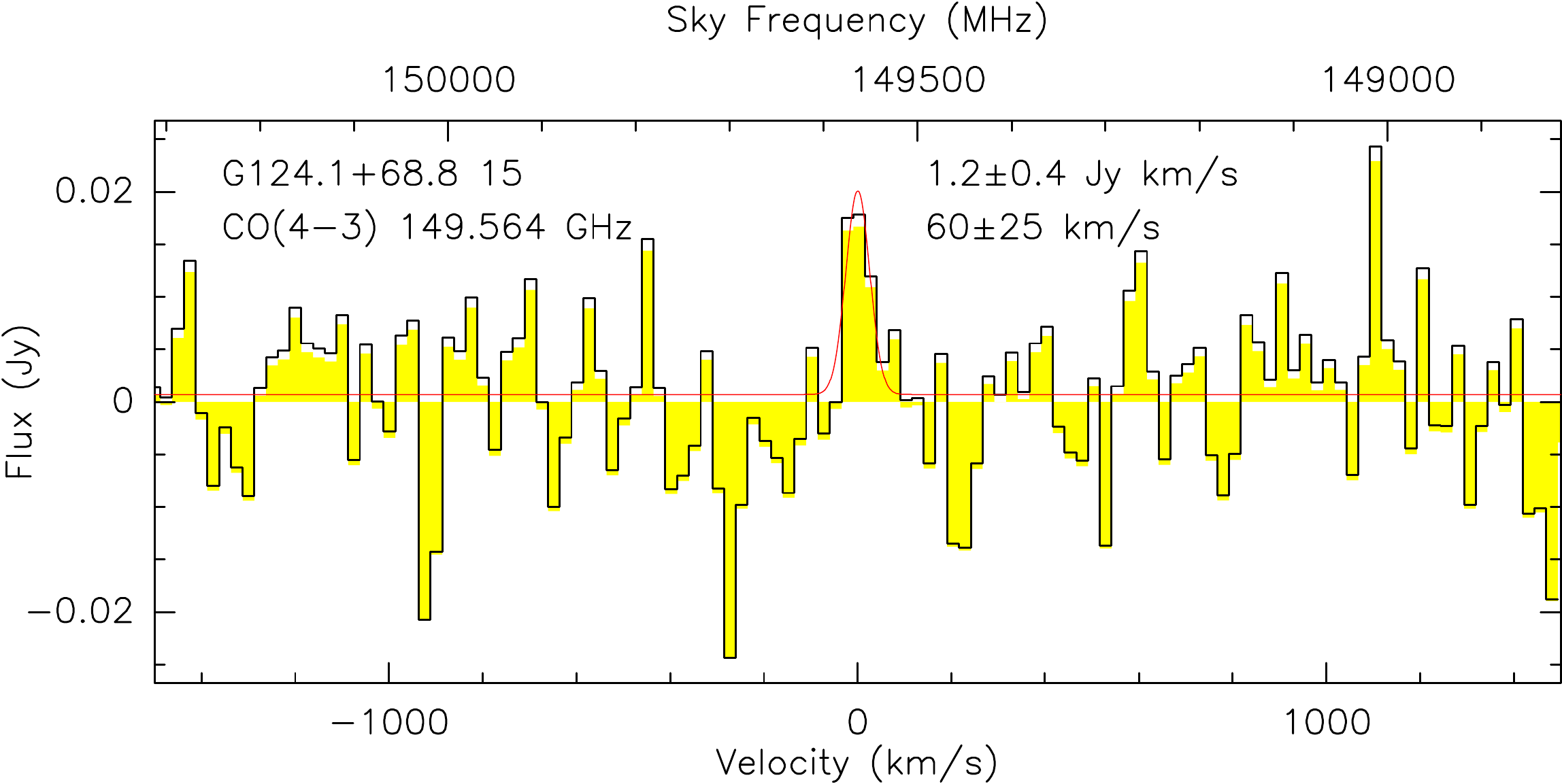}
\includegraphics[width=9cm]{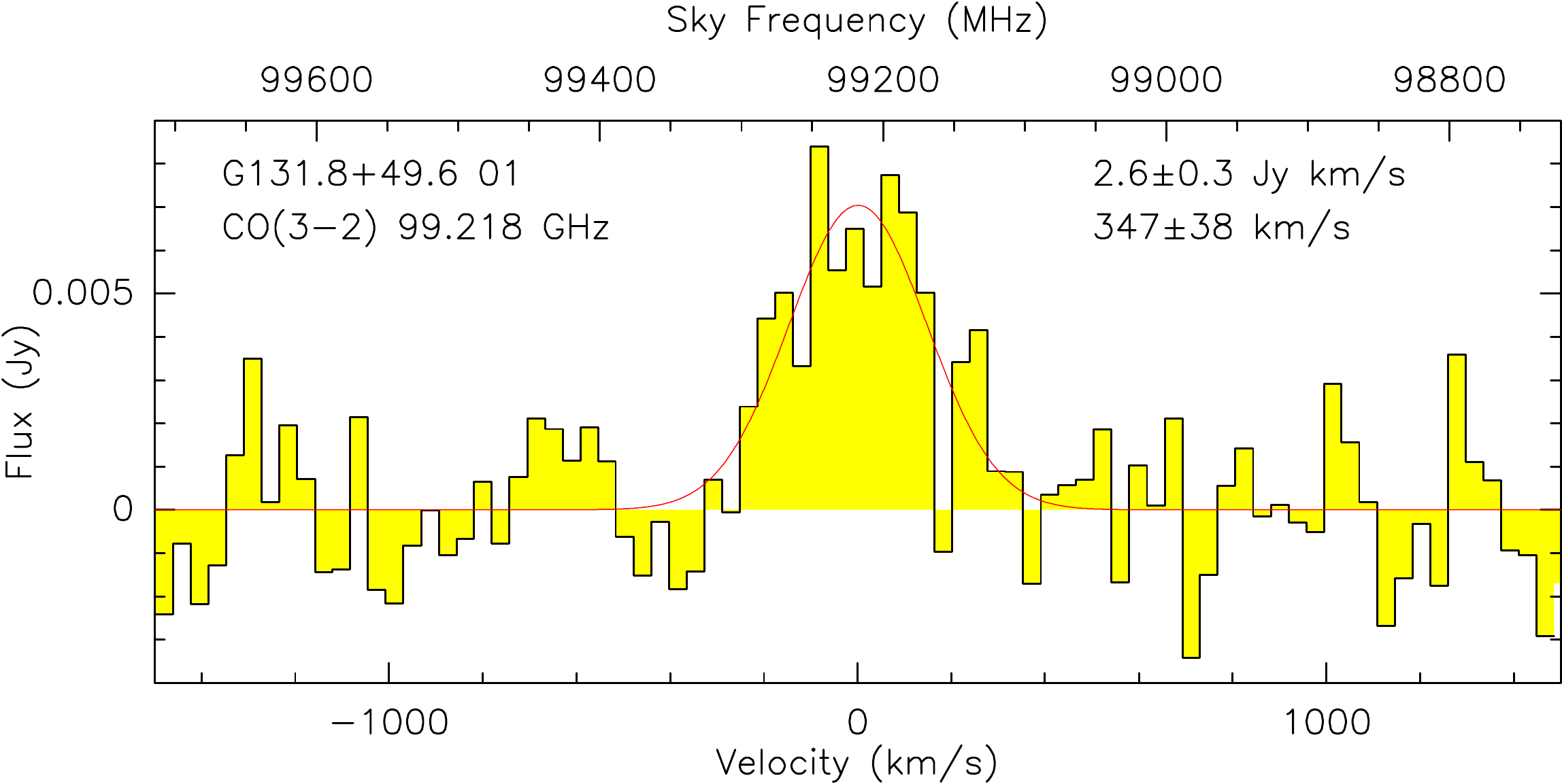}
\includegraphics[width=9cm]{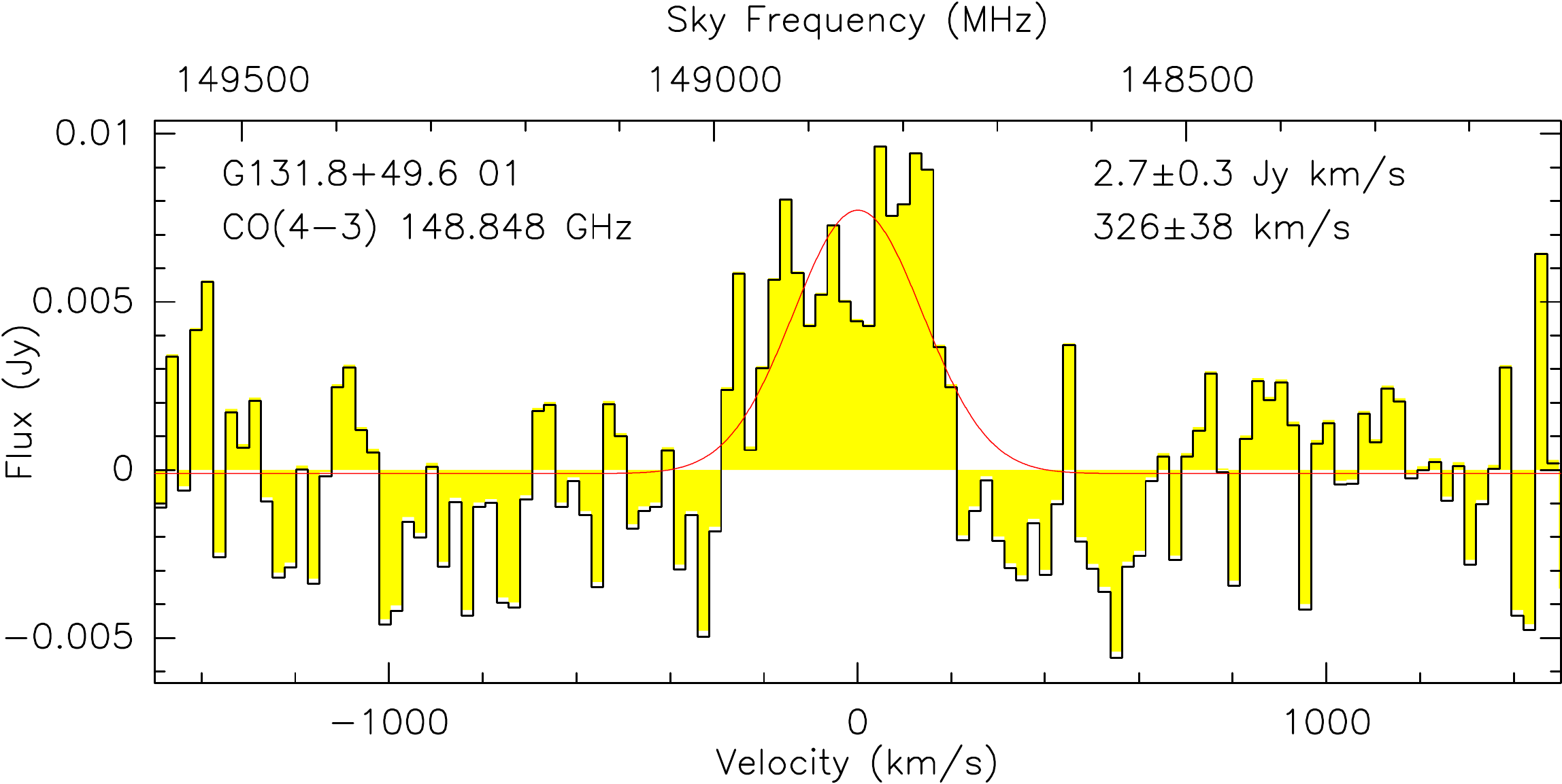}
\caption{{\small Continued.}}
\end{figure*}

\begin{figure*} 
\centering
\includegraphics[width=9cm]{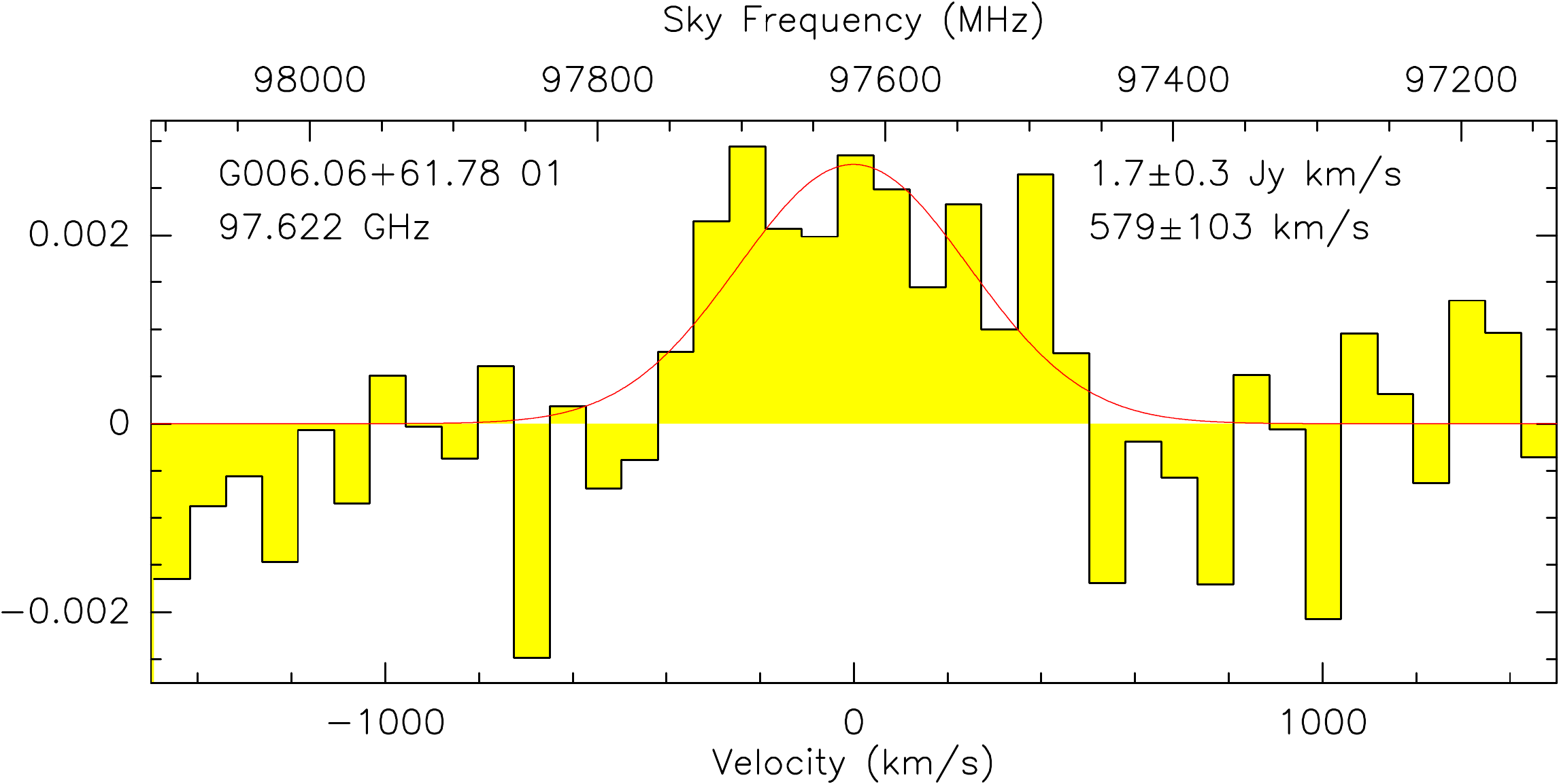}
\includegraphics[width=9cm]{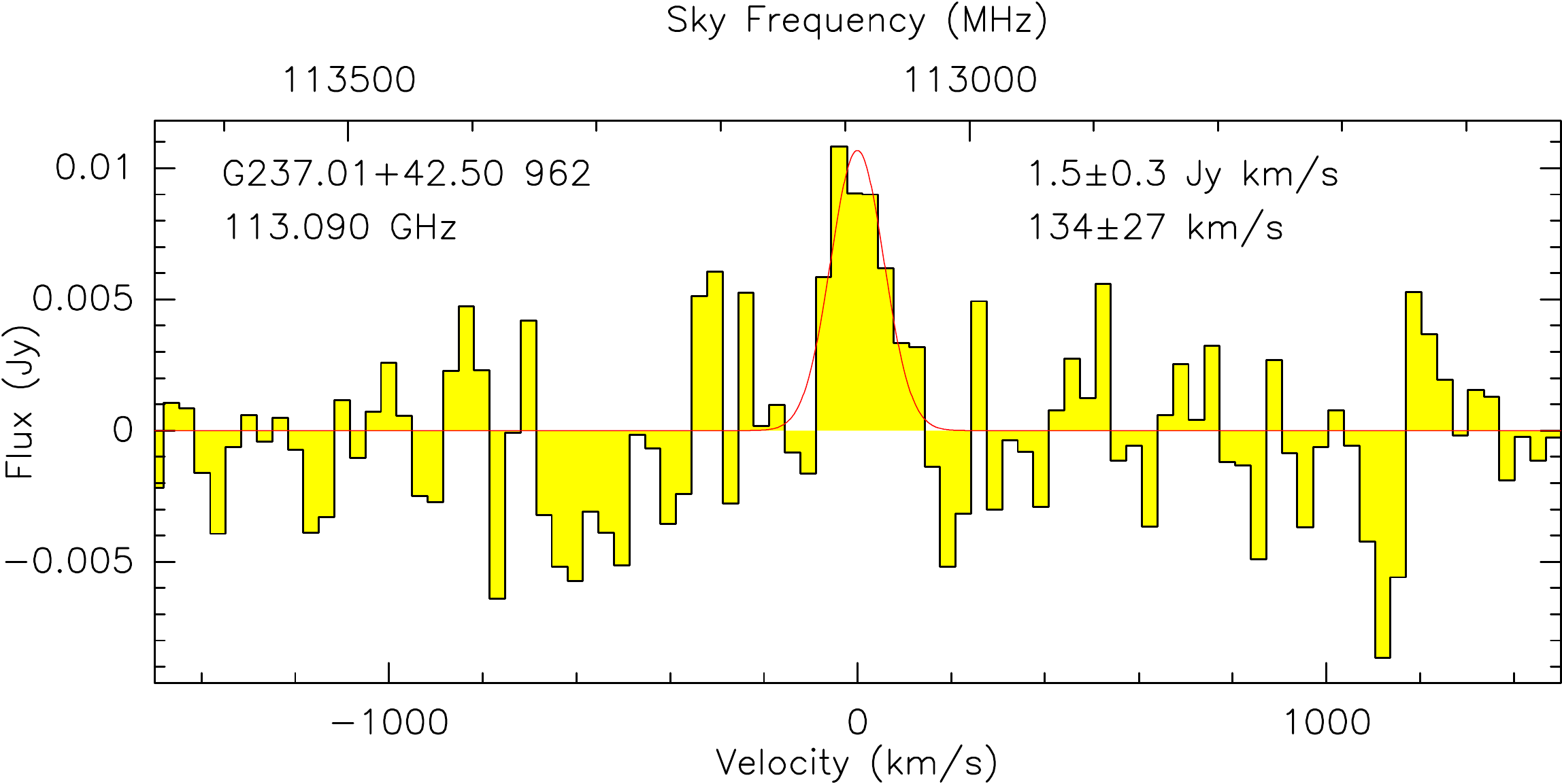}
\includegraphics[width=9cm]{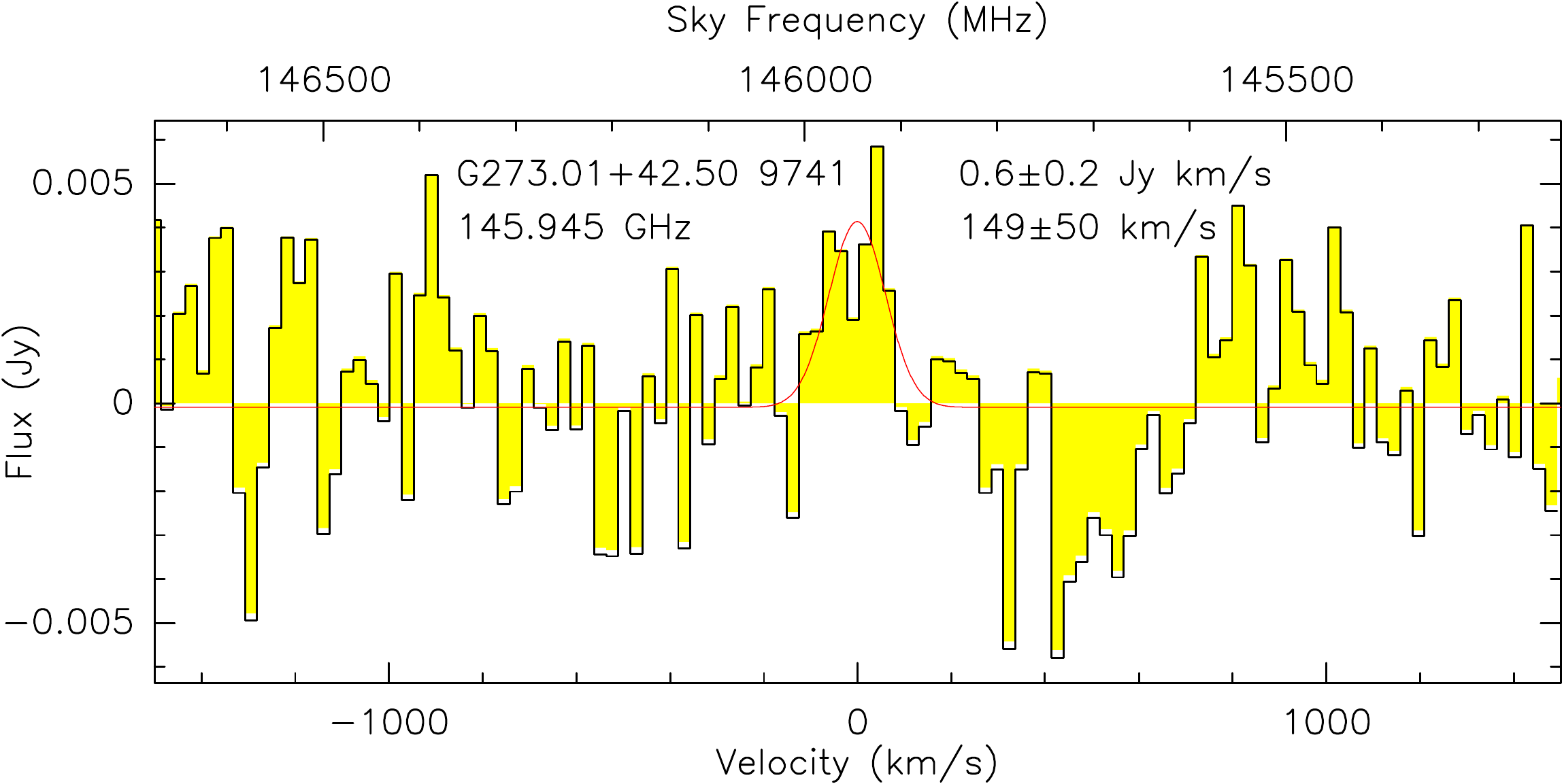}
\includegraphics[width=9cm]{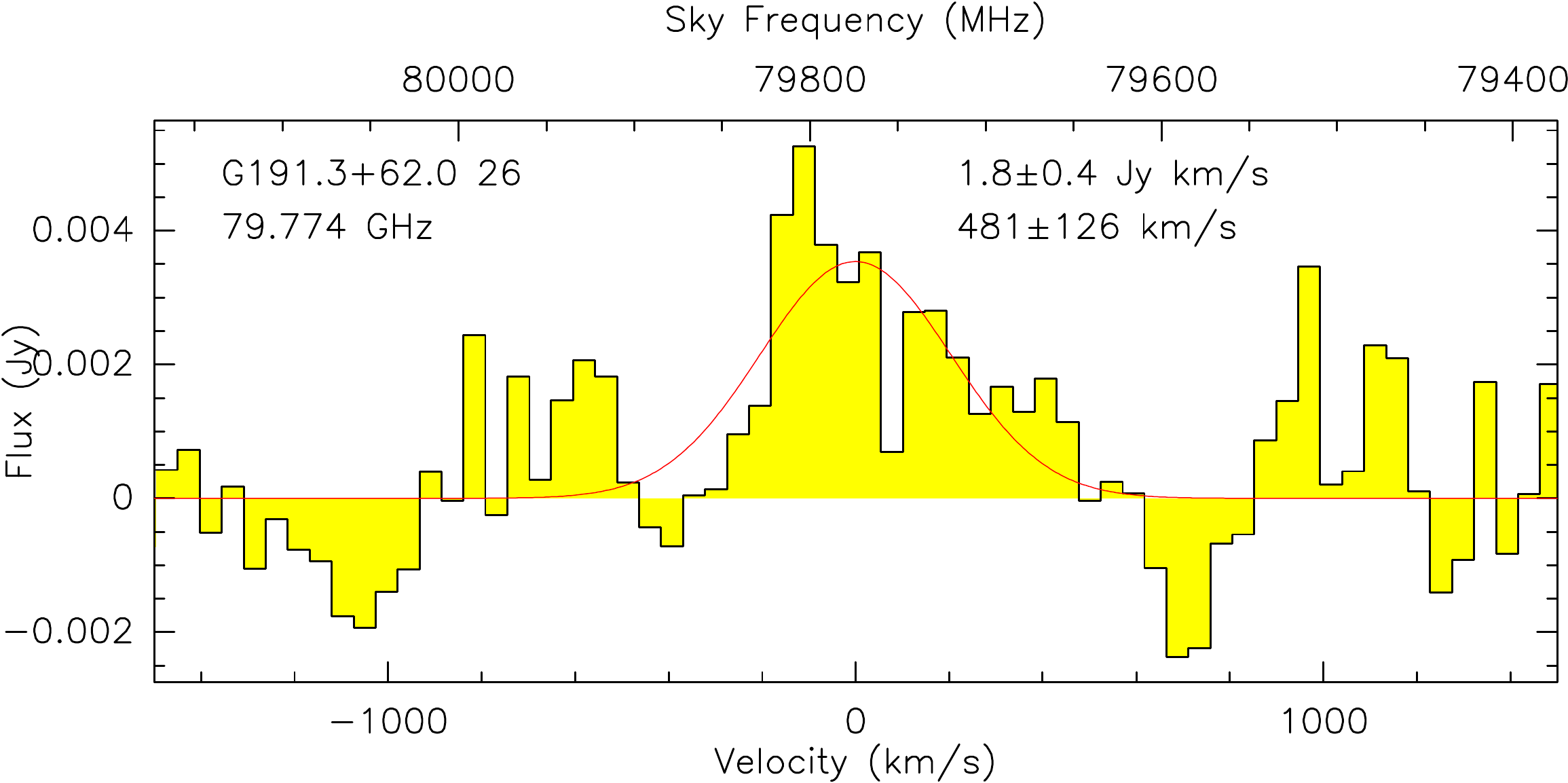}
\includegraphics[width=9cm]{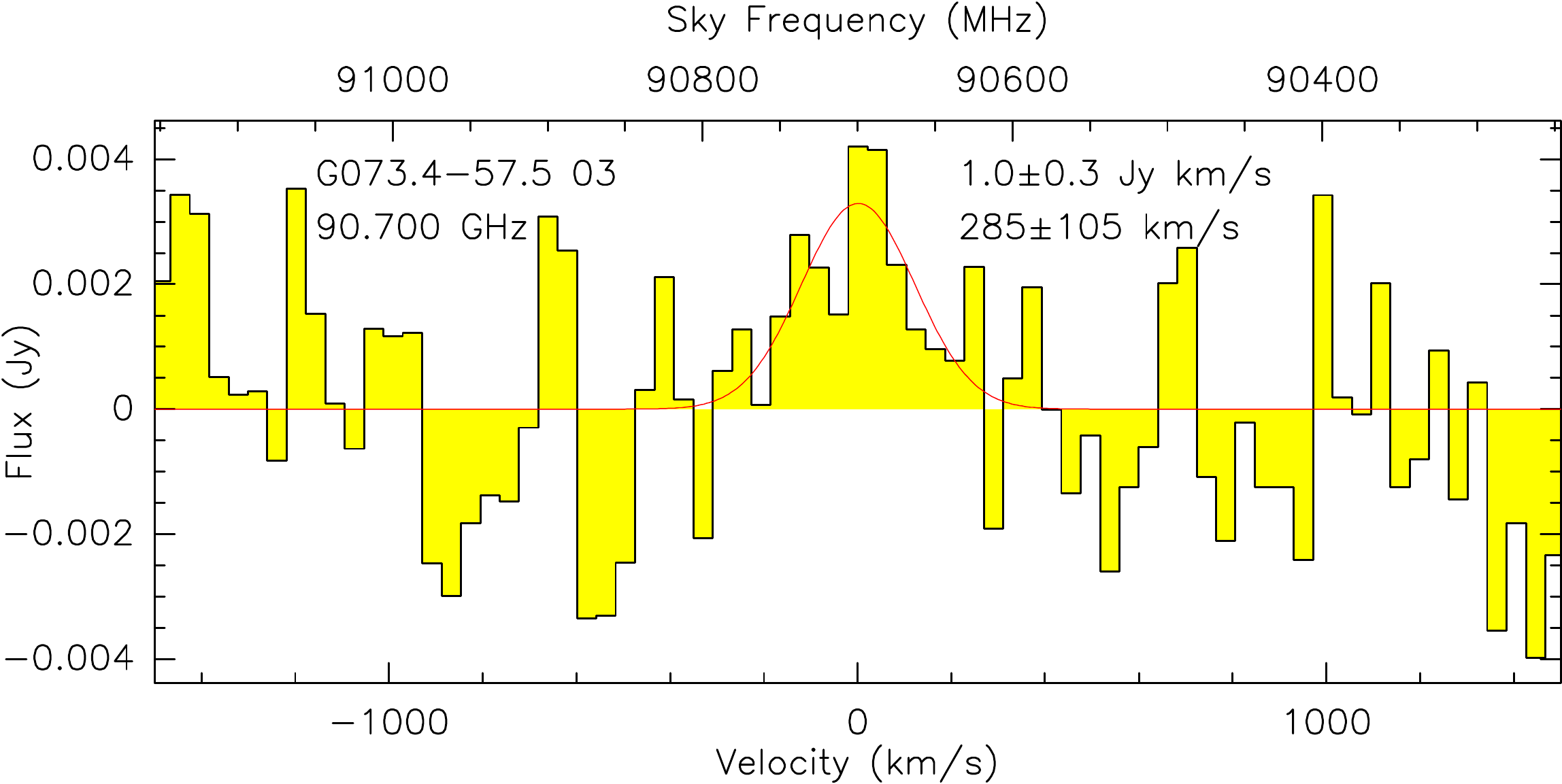}
\includegraphics[width=9cm]{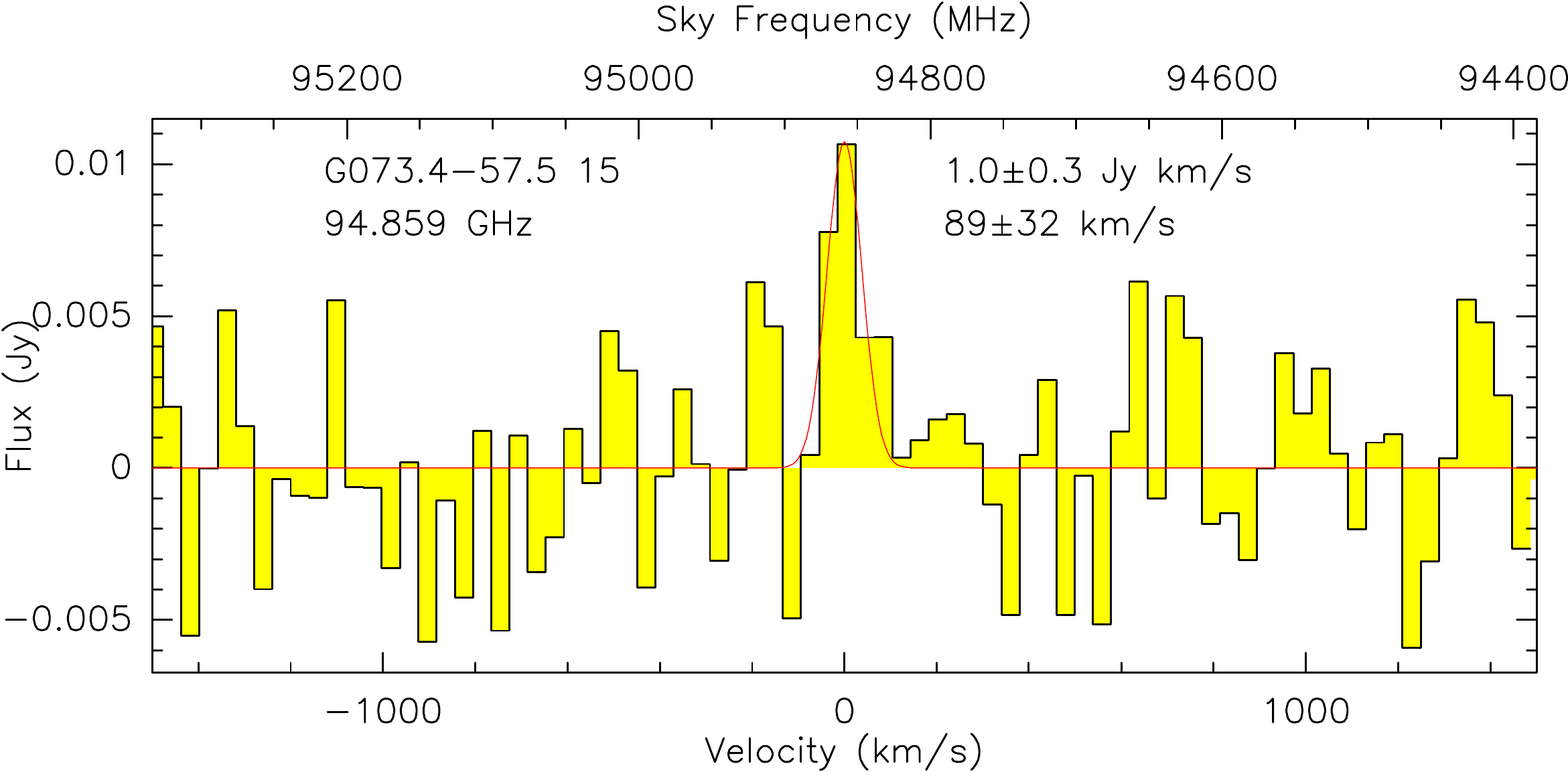}
\includegraphics[width=9cm]{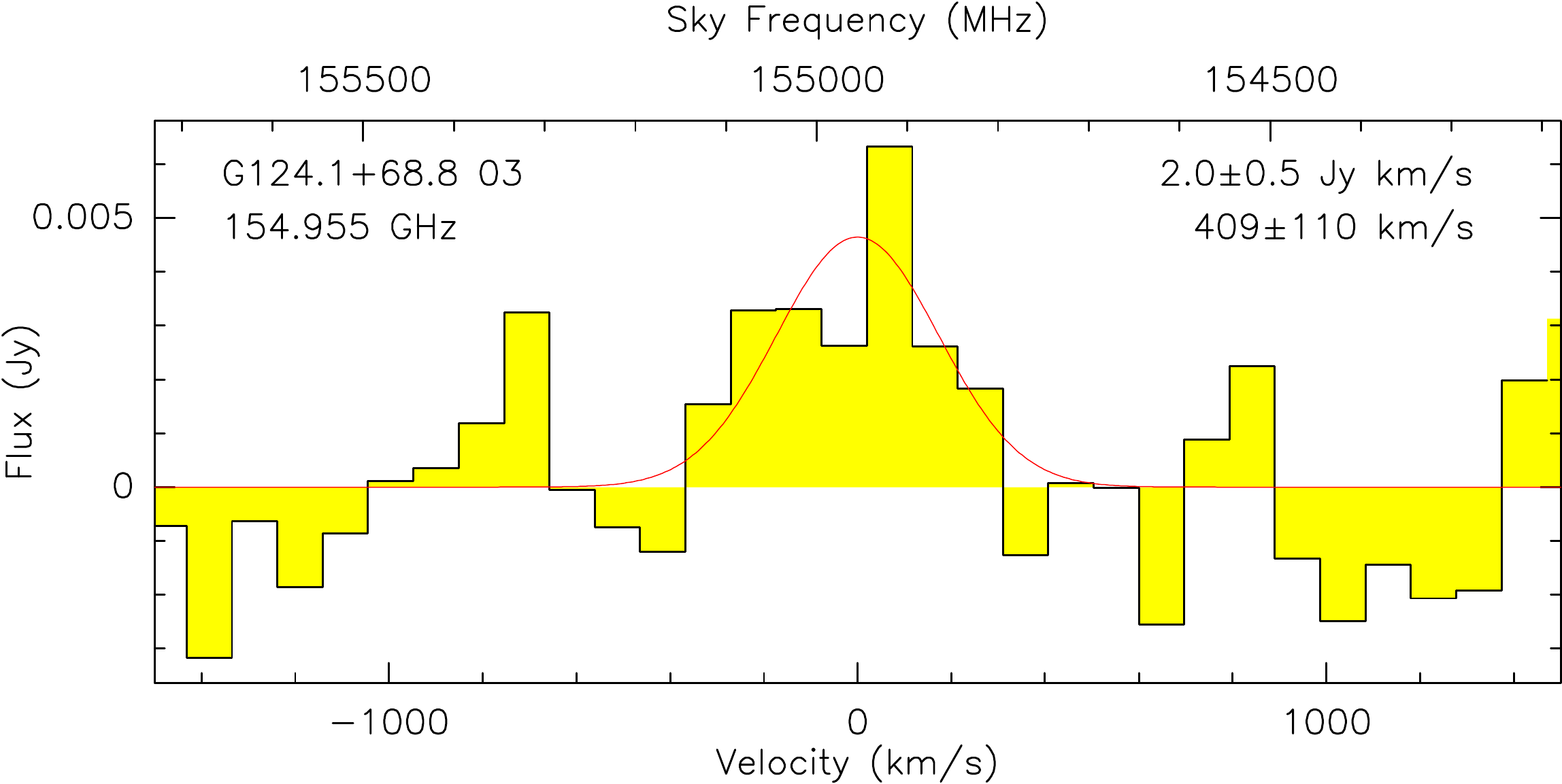}
\includegraphics[width=9cm]{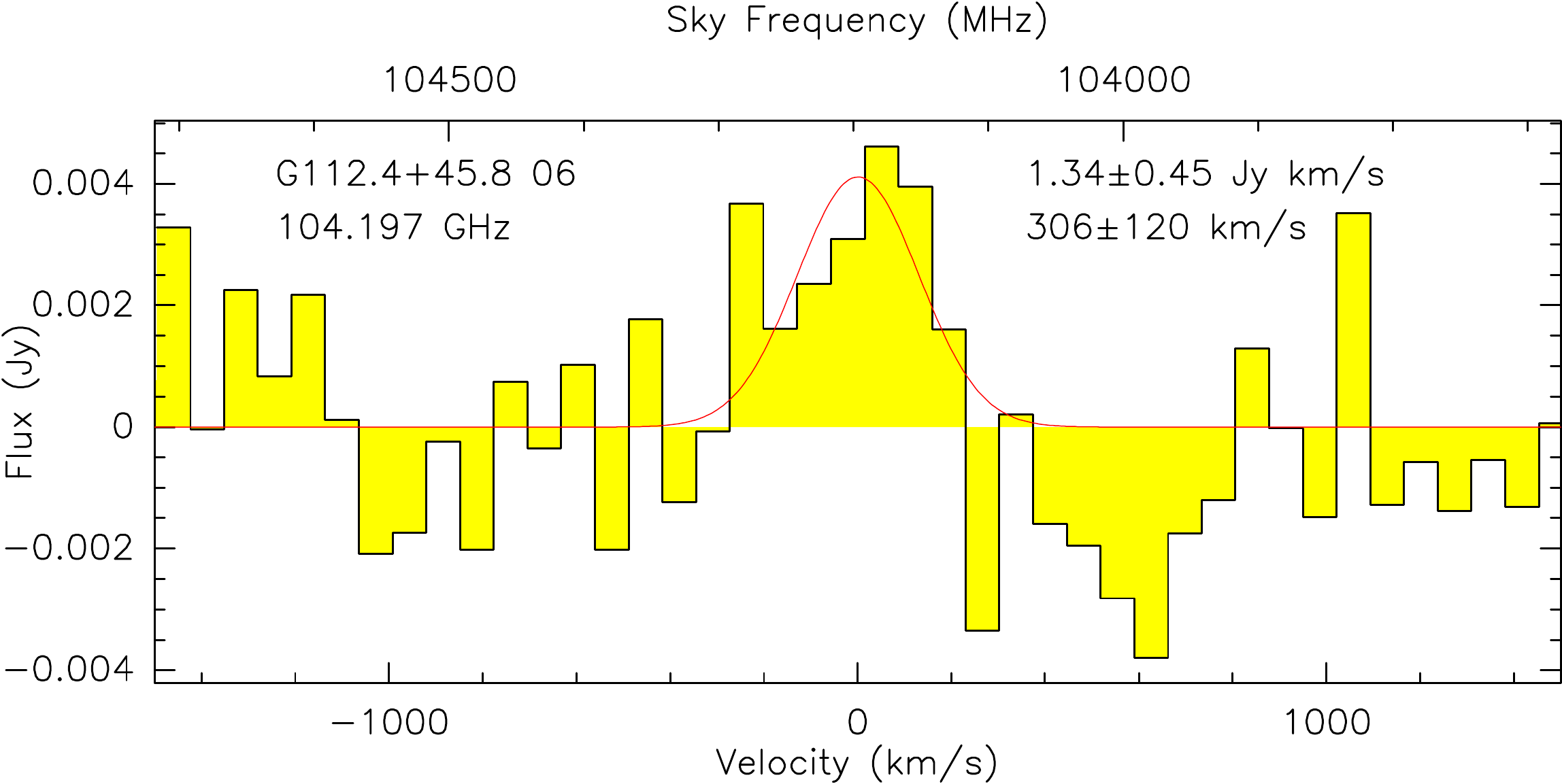}
\caption{{\small Emission line detections in the continum-subtracted EMIR
spectra of the PHz-IRAM sources where a single CO line has been detected. 
The Gaussian fit to each line is overlaid with a red curve.  Source name,
line observed central frequency, intensity, and width are annotated in each panel.}}
\label{fig:single_line}
\end{figure*}

\begin{figure*} 
\setcounter{figure}{2}
\centering
\includegraphics[width=9cm]{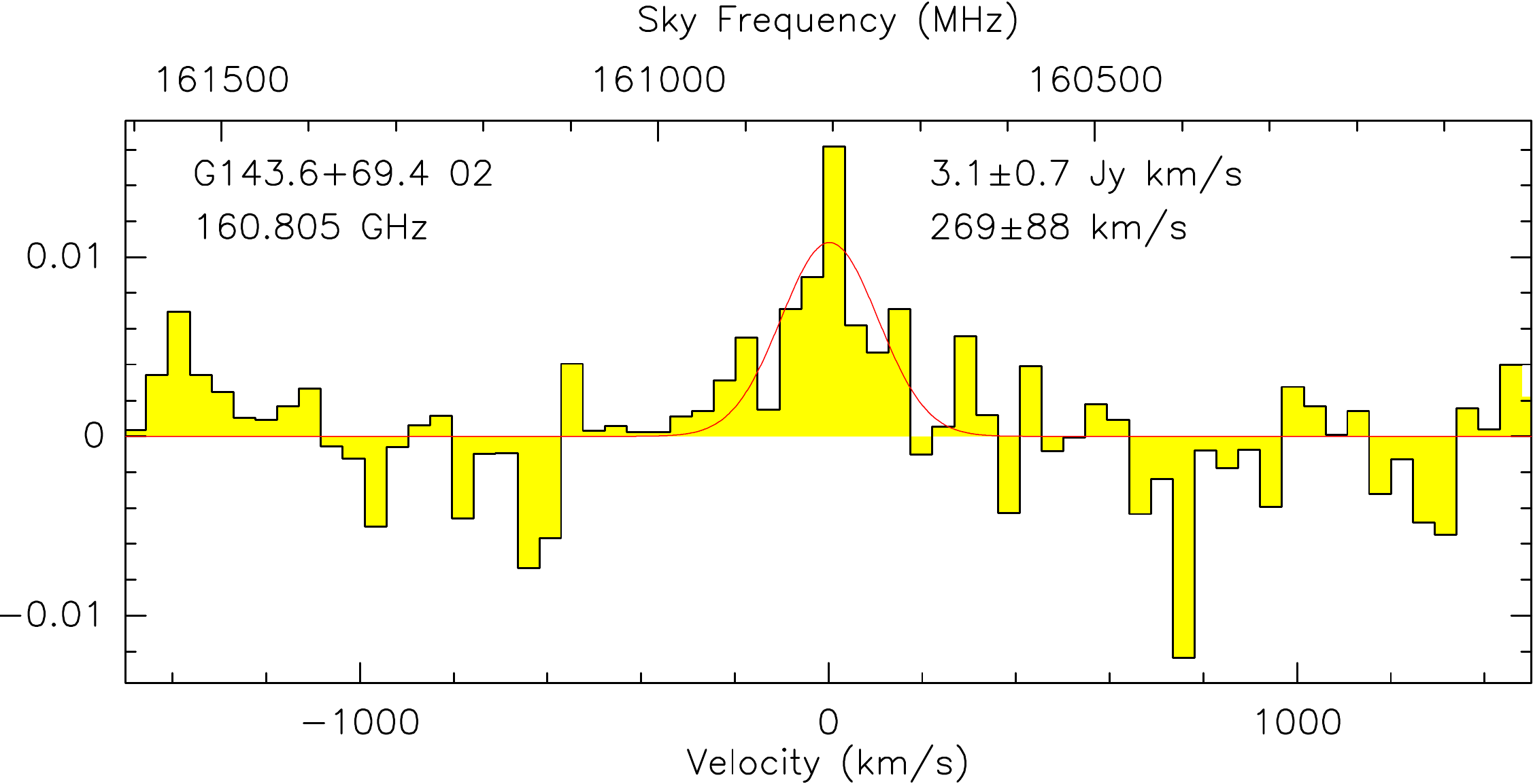}
\includegraphics[width=9cm]{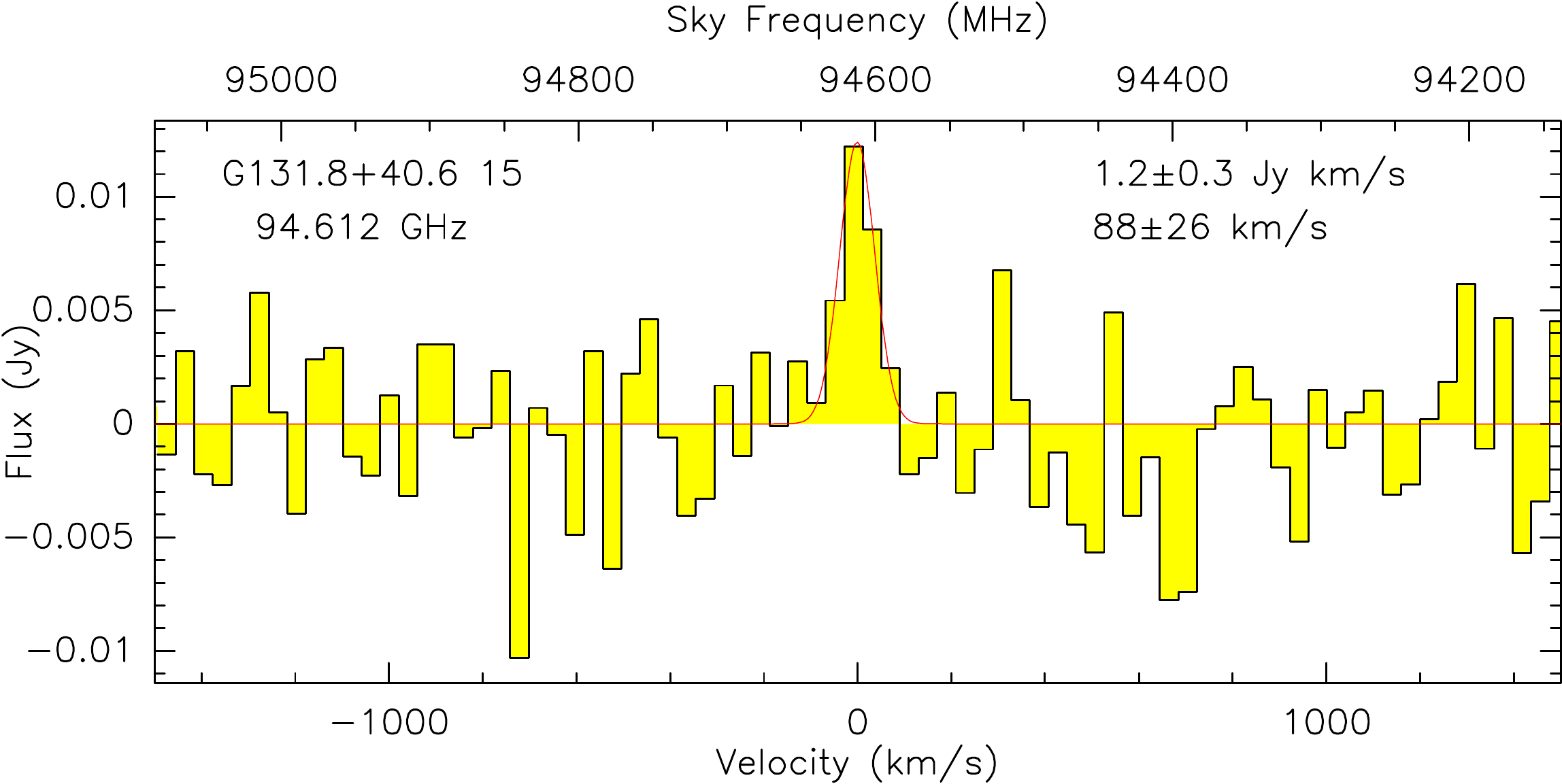}
\includegraphics[width=9cm]{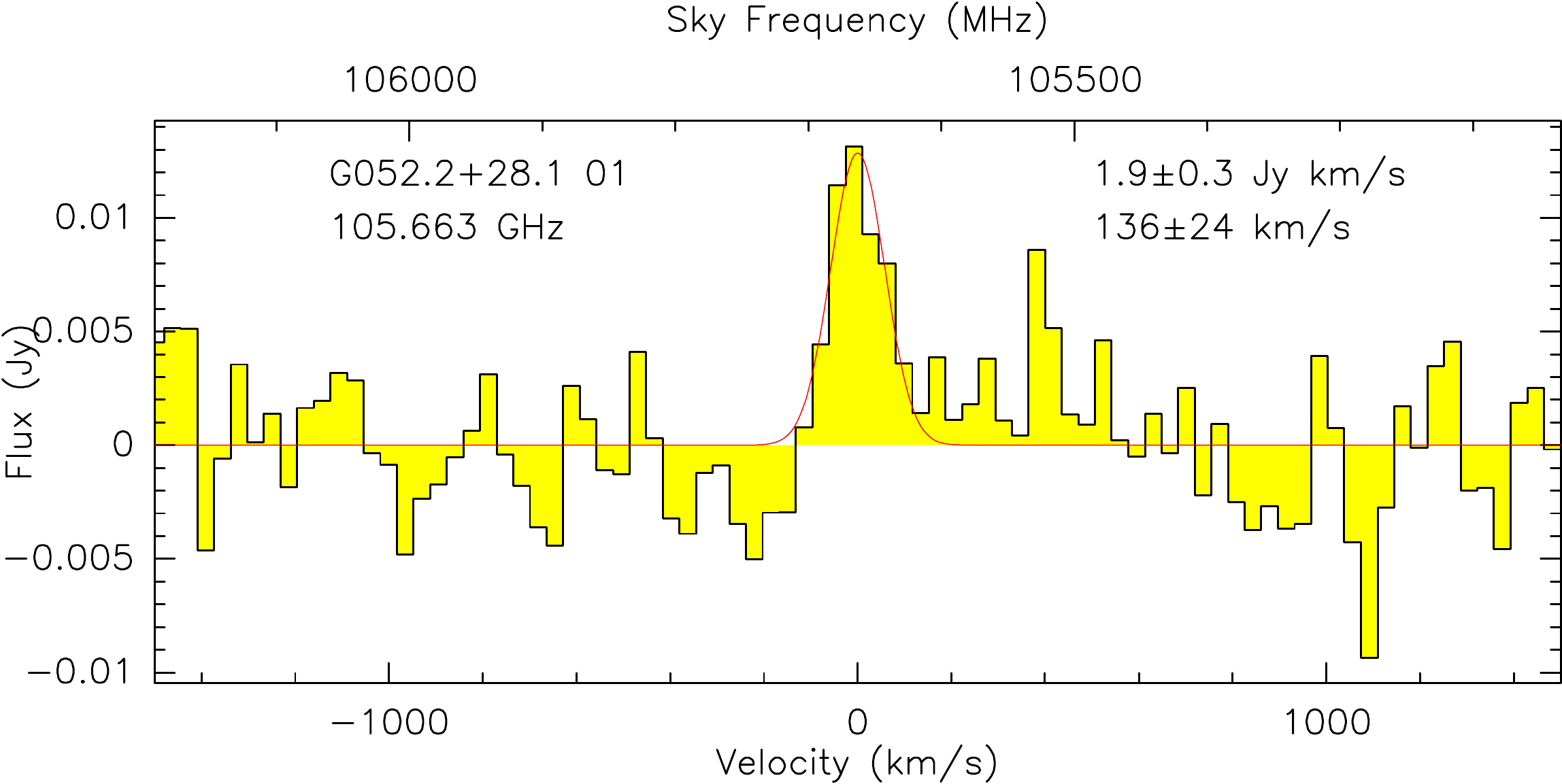}
\includegraphics[width=9cm]{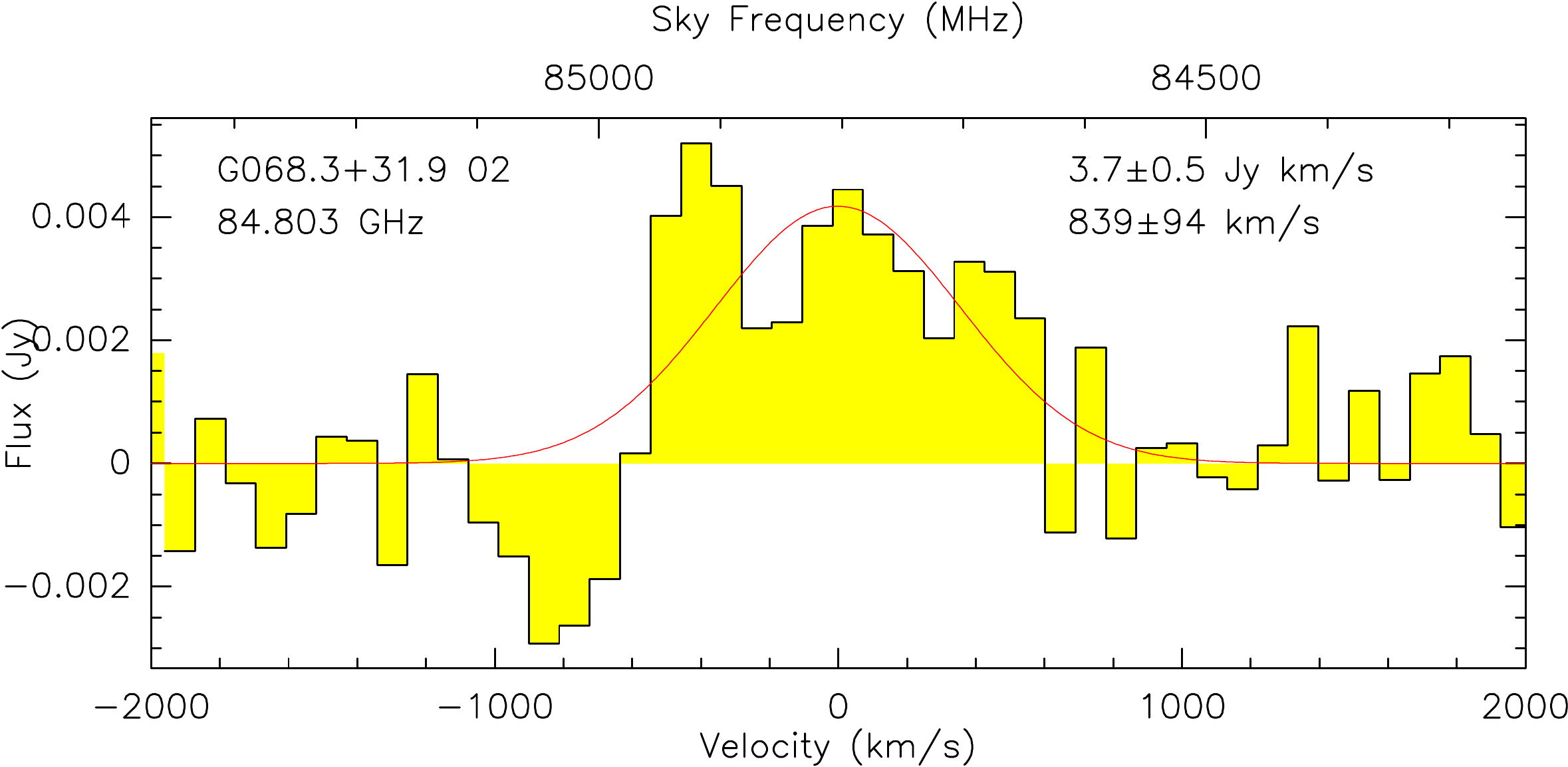}
\includegraphics[width=9cm]{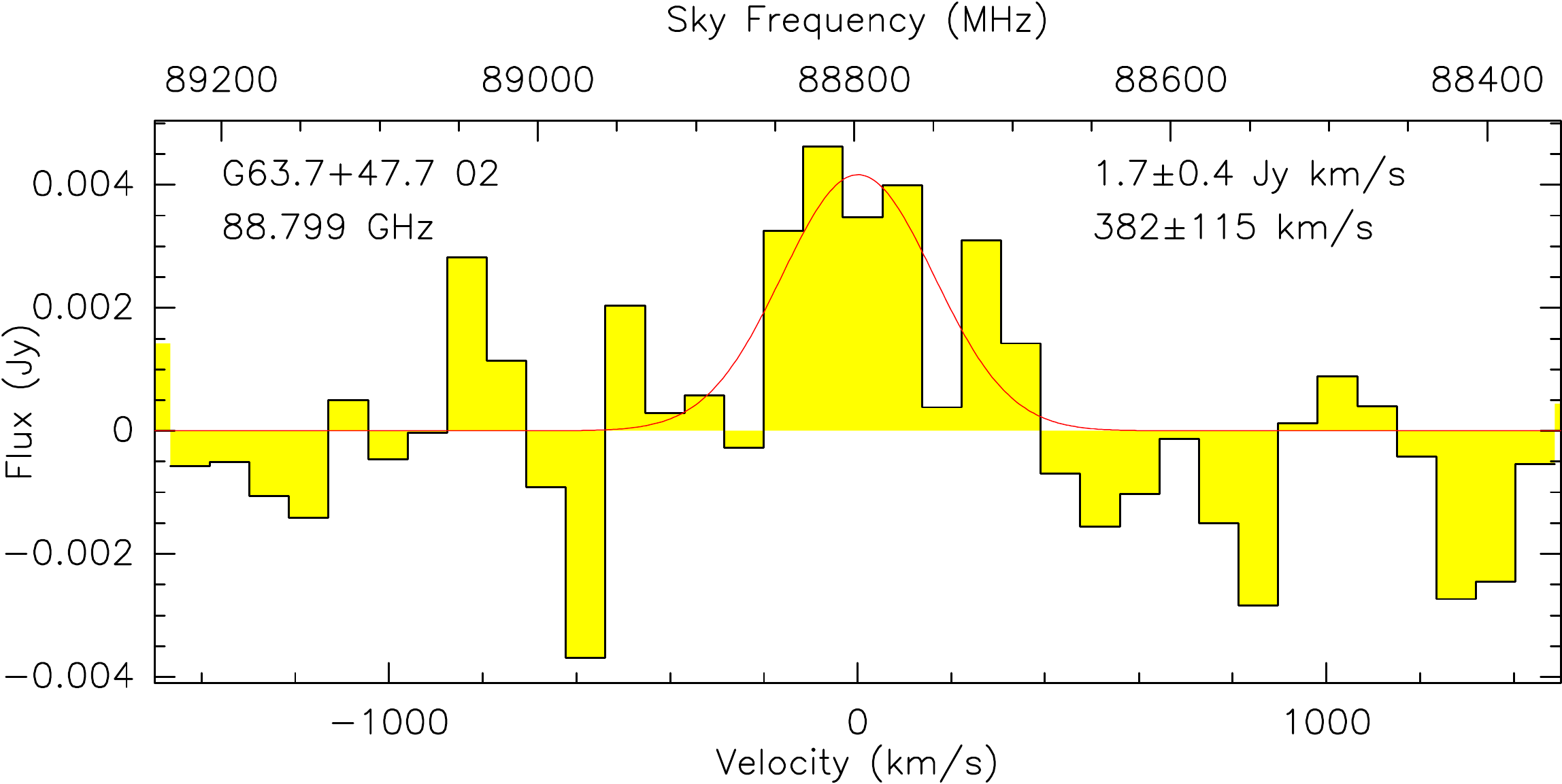}
\includegraphics[width=9cm]{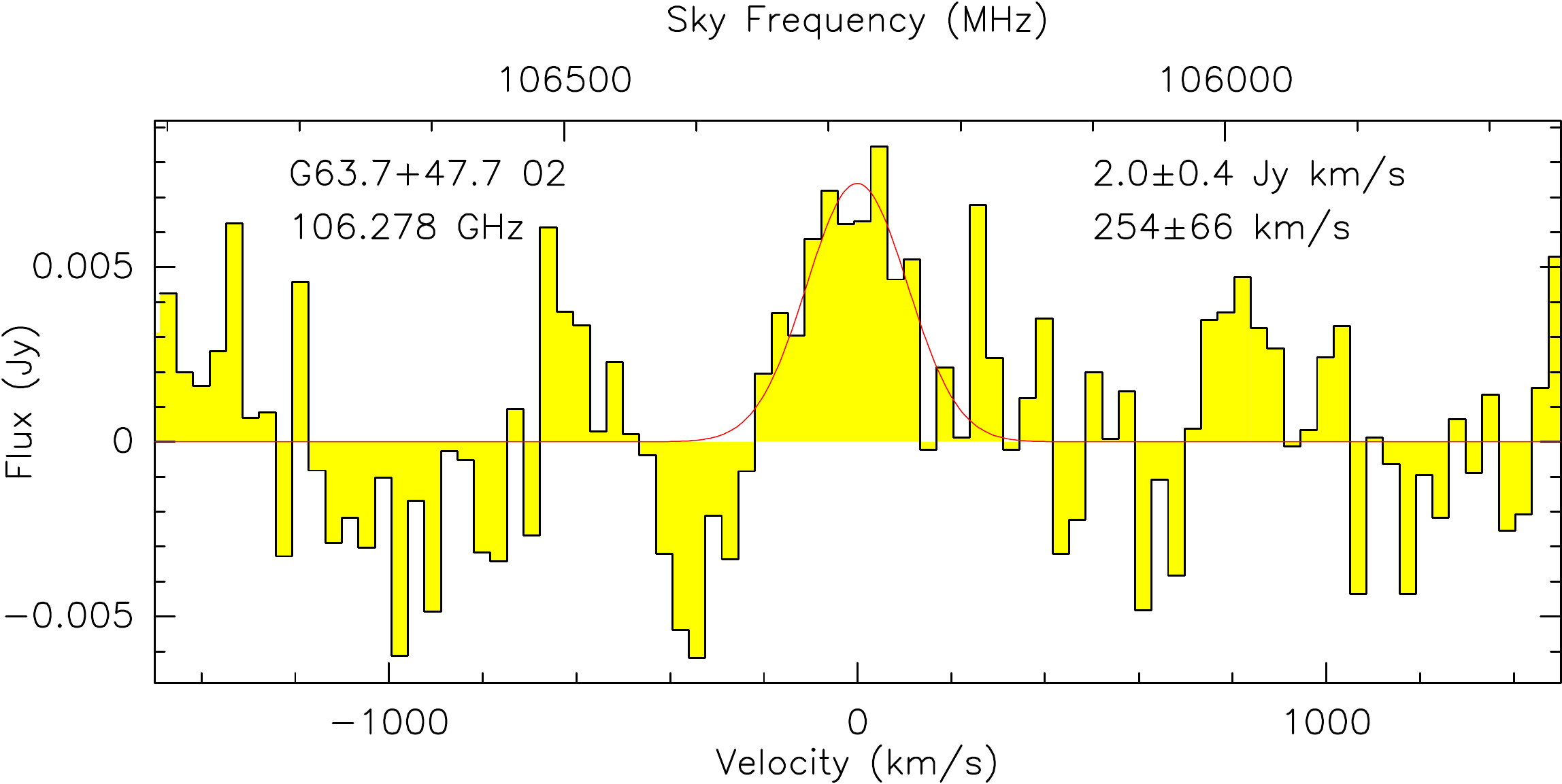}
\caption{{\small Continued.}}
\end{figure*}
                                                                                                                           
\section{Literature samples}

In order to compare the properties of the PHz-IRAM sources with other
CO-detected sources found in overdense regions, we compiled published
datasets of cluster and protocluster galaxies at 1.4${<}z{<}$3.1 that have
been observed in CO, and for which a SFR or the total IR luminosity are
available.  We list the name of the protocluster they belong, their CO
redshift, when available, and line transition, intensity and width in
Table~\ref{tab:lit_samples}.  Note that, for the purpose of this work we
selected a subsample with IR luminosities above the L$_\mathrm{IR}$ limit of
the PHz-IRAM sample minus 0.2dex, that is
log(L$_\mathrm{IR}$)$>$11.44$+$0.5${\times}z$ (see details in
Sect.~\ref{sec:fir_properties}, and Fig.~\ref{fig:LIR_z}, and sources
flagged with $b$ in the Table).  For the comparison, we converted the CO
luminosities into \LpCOone\ using the brightness temperature ratios
from~\citet{bothwell13}, as done for the PHz-IRAM sample.  Note that in a
few cases the IR luminosity was derived from the SFR using the relation
L$_\mathrm{IR}$/\lsun\,=\,SFR/(9.5$\times$10$^{-11}$) with SFR in
\msun\,yr$^{-1}$, assuming a~\citet{chabrier03} IMF, even if the SFR was
derived from other observables, like the SED or the \halpha\ luminosity.

\clearpage
\onecolumn
\begin{landscape}
\begin{longtable}{lccccccccccr}
\caption{List of protocluster and cluster galaxies from the literature\label{tab:lit_samples}}\\
\hline\hline
     Member  &$z_\mathrm{CO}$& Line  &$S_\mathrm{CO}{\Delta}v$&     FWHM      &  Tel. &  L$^{\prime}_\mathrm{CO}$\tablefootmark{a}            &      L$_\mathrm{IR}$          & $\alpha_\mathrm{CO}$ & M$_{\rm gas}$ &  Reference   \\
     Name    &            &       & (Jy\,km\,s$^{-1}$)  & (km\,s$^{-1}$)&       & (10$^{10}$\,K\,\kms\,pc$^2$)  &  (10$^{12}$\lsun)   &               & (10$^{10}$\msun) & \\
\hline
\endfirsthead
\caption{continued.}\\
\hline\hline
     Member  &$z_\mathrm{CO}$& Line  &$S_\mathrm{CO}{\Delta}v$&     FWHM      &  Tel. &  L$^{\prime}_\mathrm{CO}$\tablefootmark{a}            &      L$_\mathrm{IR}$          & $\alpha_\mathrm{CO}$ & M$_{\rm gas}$ &    Reference   \\
     Name    &            &       & (Jy\,km\,s$^{-1}$)  & (km\,s$^{-1}$)&       & (10$^{10}$\,K\,\kms\,pc$^2$)  &  (10$^{12}$\lsun)   &               & (10$^{10}$\msun) &  \\
\hline
\endhead
\hline
\endfoot
\multicolumn{11}{c}{Protoclusters} \\
\hline
\multicolumn{11}{c}{J1030$+$0524 ($z$\,=\,1.694)} \\
                         a0\tablefootmark{b}   &     1.6984   &   2--1   &      0.88$\pm$ 0.06  &     818$\pm$144  &    ALMA   &      3.36$\pm$ 0.02   &      2.64$\pm$0.22   &    4.30   &     19.30$\pm$ 0.10   &    1  \\
                         a1                    &     1.6966   &   2--1   &      0.07$\pm$ 0.02  &     133$\pm$ 38  &    ALMA   &      0.26$\pm$ 0.07   &      0.11$\pm$0.03   &    4.30   &      1.50$\pm$ 0.40   &    1  \\
                         a2                    &     1.6925   &   2--1   &      0.12$\pm$ 0.03  &     447$\pm$ 89  &    ALMA   &      0.47$\pm$ 0.11   &      0.23$\pm$0.07   &    4.30   &      2.70$\pm$ 0.60   &    1  \\
                         a3                    &     1.6864   &   2--1   &      0.22$\pm$ 0.04  &     304$\pm$182  &    ALMA   &      0.84$\pm$ 0.14   &      0.48$\pm$0.10   &    4.30   &      4.80$\pm$ 0.75   &    1  \\
\multicolumn{11}{c}{GOODS-N ($z$\,=\,1.990)} \\
       DSFG\,J123618+621550\tablefootmark{b}   &     1.996    &   4--3   &      0.77$\pm$ 0.12  &     \nodata\     &    PdBI   &     0.995$\pm$0.155   &      2.05$\pm$0.59   &    1.00   &      26.0$\pm$  6.0   &    2,3  \\
       DSFG\,J123618+621550                    &     2.001    &   4--3   &      0.50$\pm$ 0.08  &     \nodata\     &    PdBI   &     0.649$\pm$0.104   &      1.42$\pm$0.41   &    1.00   &      18.0$\pm$  4.0   &    2,3  \\
       DSFG\,J123711+621331\tablefootmark{b}   &     1.988    &   4--3   &      1.02$\pm$ 0.16  &     558$\pm$121  &    PdBI   &     1.308$\pm$0.205   &      2.06$\pm$1.07   &    1.00   &       3.6$\pm$  0.8   &    3,4  \\
       DSFG\,J123711+621331                    &     1.996    &   4--3   &      0.61$\pm$ 0.08  &     318$\pm$ 86  &    PdBI   &     0.788$\pm$0.103   &      1.24$\pm$0.64   &    1.00   &       2.2$\pm$  0.5   &    3,4  \\ 
       DSFG\,J123711+621331\tablefootmark{b}   &     1.995    &   3--2   &      0.70$\pm$ 0.22  &     \nodata\     &    PdBI   &     1.606$\pm$0.505   &      2.38$\pm$1.23   &    1.00   &       3.4$\pm$  1.2   &    3,4  \\ 
       DSFG\,J123712+621322                    &     1.996    &   3--2   &      1.20$\pm$ 0.40  &     350$\pm$100  &    PdBI   &     2.756$\pm$0.919   &      1.60$\pm$1.09   &    1.00   &       1.9$\pm$  0.5   &    3,5  \\
       DSFG\,J123632+620800                    &     1.994    &   3--2   &      1.80$\pm$ 0.50  &     310$\pm$110  &    PdBI   &     4.127$\pm$1.146   &      0.38$\pm$0.92   &    1.00   &       8.9$\pm$  2.8   &    3,5  \\
\multicolumn{11}{c}{MRC1138$-$262 ($z$\,=\,2.160)} \\
                     HAE229\tablefootmark{b}   &     2.1478   &   1--0   &      0.22$\pm$ 0.02  &     359$\pm$ 34  &    ATCA   &      5.00$\pm$ 0.70   &      3.20            &    4.00   &     18.00$\pm$ 2.00   &     6  \\
                    1138.42                    &     2.163    &   3--2   &      0.19$\pm$ 0.04  &     185          &    ALMA   &      0.51$\pm$ 0.11   &      0.43$\pm$0.13   &   10.81   &     16.22$\pm$ 3.24   &     7  \\
                    1138.48                    &     2.157    &   3--2   &      0.42$\pm$ 0.03  &     232          &    ALMA   &      1.12$\pm$ 0.08   &      1.52$\pm$0.69   &    5.55   &      9.77$\pm$ 0.56   &     7  \\
                    1138.54\tablefootmark{b}   &     2.148    &   3--2   &      0.88$\pm$ 0.04  &     328          &    ALMA   &      2.34$\pm$ 0.11   &      4.91$\pm$2.25   &    4.41   &     16.98$\pm$ 0.88   &     7  \\
                    1138.56                    &     2.144    &   3--2   &      0.20$\pm$ 0.04  &     224          &    ALMA   &      0.53$\pm$ 0.11   &      0.38$\pm$0.12   &    9.64   &     12.02$\pm$ 1.93   &     7  \\
\multicolumn{11}{c}{HELAISS02 ($z$\,=\,2.171)} \\
              HELAISS02\,S0\tablefootmark{b}   &     2.17350  &   3--2   &      3.13$\pm$ 0.30  &     931$\pm$ 58  &    ALMA   &      7.90$\pm$ 0.70   &      7.59$\pm$1.23   &    3.50   &     40.60$\pm$ 3.85   &     8  \\
              HELAISS02\,S1\tablefootmark{b}   &     2.16434  &   3--2   &      0.53$\pm$ 0.21  &     910$\pm$230  &    ALMA   &      1.30$\pm$ 0.50   &      2.69$\pm$0.37   &    9.70   &      6.79$\pm$ 1.94   &     8  \\
              HELAISS02\,S2\tablefootmark{b}   &     2.16850  &   3--2   &      0.91$\pm$ 0.16  &     575$\pm$ 78  &    ALMA   &      2.30$\pm$ 0.40   &      3.02$\pm$0.49   &    5.60   &     11.76$\pm$ 1.12   &     8  \\
              HELAISS02\,S3                    &     2.17397  &   3--2   &      0.51$\pm$ 0.19  &     610$\pm$160  &    ALMA   &      1.30$\pm$ 0.50   &      2.00$\pm$0.32   &    6.70   &      6.70$\pm$ 1.34   &     8  \\
\multicolumn{11}{c}{HS\,1700$+$64 ($z$\,=\,2.30)} \\
                      DRG55                    &     2.296    &   3--2   &      1.14$\pm$ 0.31  &     680$\pm$141  &    PdBI   &      3.60$\pm$1.00   &      2.09             &    4.36   &      32.0 $\pm$ 8.7   &     9  \\
\multicolumn{11}{c}{BOSS\,1441 ($z$\,=\,2.317)} \\
                        G1A                    &    2.3088    &   3--2   &     0.298$\pm$0.044  &     180$\pm$30   &   NOEMA   &     7.1$\pm$1.1       &      1.30$\pm$0.26   & \nodata   &       \nodata         &    10  \\
                        G1A                    &    2.3088    &   1--0   &     0.044$\pm$0.010  &     170$\pm$35   &     VLA   &     1.2$\pm$0.3       &      1.30$\pm$0.26   &    3.6    &       4.3$\pm$1.1     &    11  \\
                        G2B                    &    2.3123    &   3--2   &     0.237$\pm$0.051  &     370$\pm$90   &   NOEMA   &     6.7$\pm$1.4       &      0.51$\pm$0.11   & \nodata   &       \nodata         &    10  \\
                        G2B                    &    2.3123    &   1--0   &     0.042$\pm$0.006  &      85$\pm$20   &     VLA   &     1.1$\pm$0.2       &      0.51$\pm$0.11   &    3.6    &       4.0$\pm$0.7     &    11  \\ 
                        G6C                    &    2.3067    &   3--2   &     0.245$\pm$0.056  &     280$\pm$70   &   NOEMA   &     5.4$\pm$1.2       &      0.32$\pm$0.09   & \nodata   &       \nodata         &    10  \\ 
                        G6C                    &    2.3067    &   1--0   &     0.039$\pm$0.014  &     230$\pm$55   &     VLA   &     1.0$\pm$0.4       &      0.32$\pm$0.09   &    3.6    &       3.6$\pm$1.4     &    11  \\
\multicolumn{11}{c}{HATLAS\,J084933 ($z$\,=\,2.410)} \\
                 J084933\,W\tablefootmark{b}   &     2.4066   &   1--0   &      0.49$\pm$ 0.06  &     825$\pm$115  &    JVLA   &     13.80$\pm$ 1.70   &     33.10$\pm$3.00   &    0.80   &     11.04$\pm$ 1.36   &   12   \\
                 J084933\,T\tablefootmark{b}   &     2.4090   &   1--0   &      0.56$\pm$ 0.07  &     610$\pm$ 55  &    JVLA   &     15.70$\pm$ 2.00   &     14.50$\pm$1.70   &    0.80   &     12.56$\pm$ 1.60   &   12   \\
                 J084933\,M\tablefootmark{b}   &     2.4176   &   1--0   &      0.06$\pm$ 0.01  &     320$\pm$ 70  &    JVLA   &      1.60$\pm$ 0.40   &      7.90$\pm$3.80   &    0.80   &      1.28$\pm$ 0.32   &   12   \\
                 J084933\,C\tablefootmark{b}   &     2.4138   &   1--0   &      0.08$\pm$ 0.01  &     250$\pm$100  &    JVLA   &      2.20$\pm$ 0.40   &      6.30$\pm$3.00   &    0.80   &      1.76$\pm$ 0.32   &   12   \\
\multicolumn{11}{c}{4C\,23.56 ($z$\,=\,2.490)} \\
                     4C23.3                    &     2.488    &   3--2   &      0.45$\pm$ 0.08  &     482          &    ALMA   &      1.54$\pm$ 0.27   &      1.92$\pm$0.88   &     4.6   &     12.88$\pm$ 2.29   &    7   \\
                     4C23.4\tablefootmark{b}   &     2.490    &   3--2   &      0.33$\pm$ 0.05  &     281          &    ALMA   &      1.13$\pm$ 0.17   &      3.99$\pm$1.83   &     4.2   &      8.51$\pm$ 1.29   &    7   \\
                     4C23.8                    &     2.486    &   3--2   &      0.50$\pm$ 0.05  &     197          &    ALMA   &      1.71$\pm$ 0.17   &      1.75$\pm$0.80   &     4.8   &     14.79$\pm$ 1.48   &    7   \\
                     4C23.9\tablefootmark{b}   &     2.485    &   3--2   &      0.77$\pm$ 0.11  &     739          &    ALMA   &      2.63$\pm$ 0.38   &      3.35$\pm$1.54   &     5.1   &     23.99$\pm$ 3.43   &    7   \\
                    4C23.10                    &     2.485    &   3--2   &      0.47$\pm$ 0.09  &     303          &    ALMA   &      1.61$\pm$ 0.31   &      1.04$\pm$0.34   &     5.5   &     15.85$\pm$ 3.03   &    7   \\
                    4C23.16                    &     2.484    &   3--2   &      0.70$\pm$ 0.12  &     544          &    ALMA   &      2.39$\pm$ 0.41   &      2.16$\pm$0.99   &     5.6   &     23.99$\pm$ 4.11   &    7   \\
                       HAE3                    &     2.4861   &   3--2   &      0.35$\pm$ 0.06  &     500          &    ALMA   &      1.20$\pm$ 0.20   &      1.85$\pm$0.82   &    4.71   &     10.36$\pm$ 1.88   &   13   \\
                       HAE4\tablefootmark{b}   &     2.4780   &   3--2   &      0.25$\pm$ 0.03  &     300          &    ALMA   &      0.84$\pm$ 0.10   &      4.35$\pm$1.84   &    4.41   &      7.06$\pm$ 0.88   &   13   \\
                       HAE5\tablefootmark{b}   &     2.4873   &   3--2   &      0.09$\pm$ 0.02  &     100          &    ALMA   &      0.31$\pm$ 0.07   &      3.93$\pm$1.47   &    5.48   &      3.29$\pm$ 0.55   &   13   \\
                       HAE8                    &     2.4861   &   3--2   &      0.26$\pm$ 0.03  &     300          &    ALMA   &      0.90$\pm$ 0.10   &      1.64$\pm$0.66   &    5.19   &      8.82$\pm$ 1.04   &   13   \\
                       HAE9                    &     2.4861   &   3--2   &      0.54$\pm$ 0.06  &    1000          &    ALMA   &      1.85$\pm$ 0.20   &      0.94$\pm$0.42   &    5.35   &     18.19$\pm$ 2.14   &   13   \\
                      HAE10                    &     2.4861   &   3--2   &      0.36$\pm$ 0.06  &     500          &    ALMA   &      1.24$\pm$ 0.20   &      1.21$\pm$0.49   &    5.72   &     13.16$\pm$ 2.29   &   13   \\
                      HAE16                    &     2.4826   &   3--2   &      0.49$\pm$ 0.07  &     600          &    ALMA   &      1.68$\pm$ 0.24   &      0.80$\pm$0.34   &    5.94   &     18.41$\pm$ 2.38   &   13   \\
\multicolumn{11}{c}{CL\,J1001 ($z$\,=\,2.510)\tablefootmark{c} } \\
                         S6                    &     2.49391  &   1--0   &     0.078$\pm$0.006  &     519$\pm$ 71  &    ALMA   &      2.24$\pm$ 0.20   &      \nodata         &    3.5    &      7.84$\pm$ 0.70   &    8  \\
                         S8                    &     2.50400  &   1--0   &     0.025$\pm$0.009  &     370$\pm$100  &    ALMA   &      0.72$\pm$ 0.30   &      \nodata         &    3.5    &      2.52$\pm$ 1.05   &    8  \\
                        S12                    &     2.51225  &   1--0   &     0.031$\pm$0.013  &     720$\pm$230  &    ALMA   &      0.91$\pm$ 0.40   &      \nodata         &    3.5    &      3.19$\pm$ 1.40   &    8  \\
                    3\,(S5)                    &     2.514    &   1--0   &     \nodata\         &     280$\pm$120  &    JVLA   &      0.60$\pm$ 0.10   &      0.76$\pm$0.44   &    4.08   &      2.45$\pm$ 0.41   &   14  \\
                          4                    &     2.501    &   1--0   &     \nodata\         &     690$\pm$160  &    JVLA   &      0.60$\pm$ 0.10   &      1.23$\pm$0.55   &    4.08   &      2.45$\pm$ 0.41   &   14  \\
                   5\,(S11)                    &     2.508    &   1--0   &     \nodata\         &     240$\pm$ 40  &    JVLA   &      2.70$\pm$ 0.40   &      2.75$\pm$0.96   &    4.08   &     11.02$\pm$ 1.63   &   14  \\
                          6\tablefootmark{b}   &     2.494    &   1--0   &     \nodata\         &     550$\pm$ 40  &    JVLA   &      4.90$\pm$ 0.40   &      7.41$\pm$2.59   &    4.09   &     20.04$\pm$ 1.64   &   14  \\
                   7\,(S10)                    &     2.505    &   1--0   &     \nodata\         &     680$\pm$160  &    JVLA   &      2.80$\pm$ 0.80   &      1.20$\pm$0.62   &    4.09   &     11.45$\pm$ 3.68   &   14  \\
                          8\tablefootmark{b}   &     2.513    &   1--0   &     \nodata\         &     340$\pm$ 20  &    JVLA   &      3.20$\pm$ 0.30   &      4.17$\pm$1.45   &    4.10   &     13.12$\pm$ 1.23   &   14  \\
                          9                    &     2.500    &   1--0   &     \nodata\         &      90$\pm$ 40  &    JVLA   &      0.50$\pm$ 0.10   &      0.93$\pm$0.51   &    4.10   &      2.05$\pm$ 0.41   &   14  \\
                         10                    &     2.506    &   1--0   &     \nodata\         &     690$\pm$370  &    JVLA   &      1.00$\pm$ 0.30   &      0.60$\pm$0.47   &    4.10   &      4.10$\pm$ 1.23   &   14  \\
                         11                    &     2.506    &   1--0   &     \nodata\         &     600$\pm$190  &    JVLA   &     12.70$\pm$ 4.20   &      1.86$\pm$0.83   &    4.10   &     52.07$\pm$17.22   &   14  \\
                   12\,(S7)                    &     2.515    &   1--0   &     \nodata\         &      90$\pm$ 30  &    JVLA   &      0.30$\pm$ 0.10   &      0.13$\pm$0.07   &    4.11   &      1.23$\pm$ 0.41   &   14  \\
                   13\,(S9)                    &     2.505    &   1--0   &     \nodata\         &     530$\pm$ 80  &    JVLA   &      2.00$\pm$ 0.30   &      2.04$\pm$1.19   &    4.11   &      8.22$\pm$ 1.23   &   14  \\
                         14                    &     2.515    &   1--0   &     \nodata\         &     140$\pm$ 30  &    JVLA   &     13.40$\pm$ 2.70   &      0.89$\pm$0.69   &    4.12   &     54.80$\pm$11.12   &   14  \\
                  ID1\,(S0)\tablefootmark{b}   &     2.494    &   1--0   &    0.1220$\pm$0.0064  &    486$\pm$ 36  &    ALMA   &      3.55$\pm$ 0.16   &      7.41$\pm$2.25   &    6.5    &     22.96$\pm$ 0.74   &   15  \\
                  ID1\,(S0)\tablefootmark{b}   &     2.494    &   3--2   &    1.2610$\pm$0.1430  &    532$\pm$ 78  &    ALMA   &      4.07$\pm$ 0.47   &      7.41$\pm$2.25   &    6.5    &      0.00$\pm$ 0.00   &   15  \\
                        ID2                    &     2.496    &   1--0   &    0.0390$\pm$0.0039  &    532$\pm$ 54  &    ALMA   &      1.12$\pm$ 0.13   &      \nodata         &    6.5    &      7.35$\pm$ 0.45   &   15  \\
                  ID3\,(S3)                    &     2.503    &   1--0   &    0.0463$\pm$0.0069  &   1271$\pm$143  &    ALMA   &      1.35$\pm$ 0.22   &      1.23$\pm$0.46   &    6.5    &      8.76$\pm$ 0.81   &   15  \\                     
                  ID4\,(S2)\tablefootmark{b}   &     2.504    &   1--0   &    0.0926$\pm$0.0126  &   1095$\pm$135  &    ALMA   &      2.69$\pm$ 0.37   &      4.17$\pm$1.27   &    6.5    &     17.53$\pm$ 1.47   &   15  \\                     
                        ID5                    &     2.507    &   1--0   &    0.0374$\pm$0.0051  &    172$\pm$ 71  &     VLA   &      1.10$\pm$ 0.15   &      \nodata         &    6.5    &      7.10$\pm$ 0.59   &   15  \\                     
                  ID6\,(S4)                    &     2.509    &   1--0   &    0.0500$\pm$0.0070  &    981$\pm$145  &    ALMA   &      1.45$\pm$ 0.23   &      1.10$\pm$0.55   &    6.5    &      9.50$\pm$ 0.82   &   15  \\                     
                  ID7\,(S1)\tablefootmark{b}   &     2.511    &   1--0   &    0.0641$\pm$0.0064  &    619$\pm$ 56  &    ALMA   &      1.86$\pm$ 0.21   &      4.17$\pm$1.27   &    6.5    &     12.19$\pm$ 0.75   &   15  \\
\multicolumn{11}{c}{USS\,1558$-$003 ($z$\,=\,2.510)} \\
                   bHAE-191                    &     2.5145   &   1--0   &     0.052$\pm$0.008  &     251          &    JVLA   &      1.50$\pm$ 0.00   &      2.50            &    0.80   &      1.20$\pm$ 0.00   &   16   \\
                   rHAE-193                    &     2.5134   &   1--0   &     0.096$\pm$0.015  &     437          &    JVLA   &      2.80$\pm$ 0.00   &      5.10            &    0.80   &      2.24$\pm$ 0.00   &   16   \\
                   rHAE-213                    &     2.5300   &   1--0   &     0.026$\pm$0.007  &     294          &    JVLA   &      0.80$\pm$ 0.00   &      1.70$\pm$1.70   &    0.80   &      0.64$\pm$ 0.00   &   16   \\
                    1558.43                    &     2.528    &   3--2   &      0.26$\pm$ 0.04  &     698          &    ALMA   &      0.92$\pm$ 0.14   &      0.69$\pm$0.22   &    4.18   &      6.92$\pm$ 1.06   &    7   \\
                    1558.54                    &     2.515    &   3--2   &      0.23$\pm$ 0.02  &     242          &    ALMA   &      0.80$\pm$ 0.07   &      0.71$\pm$0.21   &    6.94   &     10.00$\pm$ 0.87   &    7   \\
                    1558.59                    &     2.513    &   3--2   &      0.49$\pm$ 0.02  &     420          &    ALMA   &      1.71$\pm$ 0.07   &      0.93$\pm$0.27   &    4.92   &     15.14$\pm$ 0.62   &    7   \\
                    1558.64                    &     2.529    &   3--2   &      0.21$\pm$ 0.04  &     721          &    ALMA   &      0.74$\pm$ 0.14   &      0.53$\pm$0.16   &   26.03   &     34.67$\pm$ 6.60   &    7   \\
                    1558.73                    &     2.526    &   3--2   &      0.19$\pm$ 0.03  &     261          &    ALMA   &      0.67$\pm$ 0.11   &      1.12$\pm$0.33   &    6.74   &      8.13$\pm$ 1.28   &    7   \\
                   1558.137                    &     2.525    &   3--2   &      0.16$\pm$ 0.02  &     264          &    ALMA   &      0.56$\pm$ 0.07   &      1.03$\pm$0.31   &    5.71   &      5.75$\pm$ 0.72   &    7   \\
\multicolumn{11}{c}{HXMM20 ($z$\,=\,2.602)} \\
                 HXMM20\,S0\tablefootmark{b}   &     2.60226  &   1--0   &      0.32$\pm$ 0.05  &     688$\pm$ 81  &    ALMA   &     10.00$\pm$ 1.60   &      6.61$\pm$0.92   &    2.90   &     34.67$\pm$ 5.61   &    8  \\
                 HXMM20\,S1\tablefootmark{b}   &     2.59757  &   1--0   &      0.12$\pm$ 0.04  &     278$\pm$ 63  &    ALMA   &      3.70$\pm$ 1.20   &      3.80$\pm$0.44   &    3.70   &     12.88$\pm$ 4.23   &    8  \\
                 HXMM20\,S2                    &     2.60201  &   1--0   &      0.16$\pm$ 0.05  &     490$\pm$110  &    ALMA   &      5.00$\pm$ 1.60   &      3.47$\pm$0.48   &    2.40   &     17.38$\pm$ 5.70   &    8  \\
                 HXMM20\,S3\tablefootmark{b}   &     2.60315  &   3--2   &      1.57$\pm$ 0.21  &     484$\pm$ 45  &    ALMA   &      5.40$\pm$ 0.70   &      4.37$\pm$3.14   &    1.20   &     27.54$\pm$ 4.46   &    8  \\
                 HXMM20\,S4                    &     2.59682  &   1--0   &      0.15$\pm$ 0.05  &     590$\pm$230  &    ALMA   &      4.70$\pm$ 1.50   &      1.00$\pm$0.14   &    0.80   &     16.22$\pm$ 5.32   &    8  \\
\multicolumn{11}{c}{B3\,J2330 ($z$\,=\,3.090)} \\
     JVLA\,J233024$+$392708\tablefootmark{b}   &     3.0884   &   1--0   &     0.162$\pm$0.034  &     720$\pm$170  &     VLA   &      6.90$\pm$ 1.50   &      8.97$\pm$0.47   &    0.80   &      5.52$\pm$ 1.20   &   17  \\
     JVLA\,J233024$+$392708\tablefootmark{b}   &     3.0901   &   4--3   &      1.12$\pm$ 0.17  &     520$\pm$110  &     VLA   &      3.05$\pm$ 0.46   &      8.97$\pm$0.47   &    0.80   &      5.52$\pm$ 1.20   &   17  \\
     JVLA\,J233024$+$392708\tablefootmark{b}   &     3.0934   &   4--3   &      3.29$\pm$ 0.51  &     830$\pm$100  &     VLA   &      8.97$\pm$ 1.39   &     17.61$\pm$0.49   &    0.80   &     17.50$\pm$ 2.71   &   17  \\
\multicolumn{11}{c}{SSA22 ($z$\,=\,3.090)} \\
      J221735.15$+$001537.3                    &     3.09630  &   3--2   &      0.70$\pm$ 0.20  &     560$\pm$110  &    PdBI   &      6.30$\pm$ 1.60   &      6.17            &    1.00   &      6.30$\pm$ 2.20   &    5  \\
          ALMA1+ALMA2+ALMA5                    &      3.0982  &   4--3   &      0.44$\pm$ 0.03  &     322$\pm$ 29  &    ALMA   &     1.203$\pm$0.082   &      1.05$\pm$0.34   &    3.60   &       6.8$\pm$  1.5   &   18  \\
                      ALMA3                    &      3.0989  &   4--3   &      0.12$\pm$ 0.02  &     316$\pm$ 65  &    ALMA   &     0.328$\pm$0.055   &      0.97$\pm$0.31   &    3.60   &       1.7$\pm$  0.5   &   18  \\
\multicolumn{11}{c}{GOODSN ($z$\,=\,3.140)} \\
                  GN\,CL\,2                    &     3.148    &   5--4   &      1.59$\pm$ 0.07  &     369$\pm$ 23  &   NOEMA   &      2.93$\pm$ 0.14   &      6.63$\pm$0.32   &    1.36   &     13.00$\pm$ 2.00   &   19  \\
                  GN\,CL\,3                    &     3.132    &   5--4   &      1.50$\pm$ 0.10  &     500$\pm$ 41  &   NOEMA   &      2.75$\pm$ 0.18   &      4.11$\pm$1.47   &    1.36   &     12.00$\pm$ 2.00   &   19  \\
\hline
\multicolumn{11}{c}{Clusters}\\                                                                                                                                                                                                                        
\hline
\multicolumn{11}{c}{7C\,1756$+$6520 ($z$\,=\,1.420)} \\
                  AGN\,1317                    &     1.4161   &   2--1   &      0.52$\pm$ 0.06  &     254$\pm$ 33  &    PdBI   &      1.36$\pm$0.15   &      0.68             &     0.8   &       1.30$\pm$0.14   &   20   \\
\multicolumn{11}{c}{XCS\,J2215 ($z$\,=\,1.460)} \\
             XCS\,J2215\,03                    &     1.453    &   2--1   &      0.50$\pm$ 0.10  &     530$\pm$ 90  &    ALMA   &      1.46$\pm$0.29   &      0.88 $\pm$0.33   &     3.5   &      6.08 $\pm$1.21   &   21   \\
             XCS\,J2215\,06                    &     1.454    &   2--1   &      0.25$\pm$ 0.08  &     130$\pm$ 30  &    ALMA   &      0.73$\pm$0.23   &      0.94 $\pm$0.22   &     3.5   &      3.04 $\pm$0.96   &   21   \\
             XCS\,J2215\,07                    &     1.450    &   2--1   &      0.19$\pm$ 0.09  &     190$\pm$120  &    ALMA   &      0.55$\pm$0.26   &      0.36 $\pm$0.20   &     3.5   &      2.29 $\pm$1.08   &   21   \\
             XCS\,J2215\,08                    &     1.466    &   2--1   &      0.60$\pm$ 0.10  &     460$\pm$ 90  &    ALMA   &      1.79$\pm$0.30   &      0.33 $\pm$0.10   &     3.5   &      7.46 $\pm$1.25   &   21   \\
             XCS\,J2215\,11                    &     1.467    &   2--1   &      0.40$\pm$ 0.10  &     390$\pm$ 90  &    ALMA   &      1.19$\pm$0.30   &      0.55 $\pm$0.22   &     3.5   &      4.96 $\pm$1.25   &   21   \\
             XCS\,J2215\,13                    &     1.472    &   2--1   &      0.30$\pm$ 0.10  &     370$\pm$ 90  &    ALMA   &      0.90$\pm$0.30   &      0.24 $\pm$0.12   &     3.5   &      3.75 $\pm$1.25   &   21   \\
                 ALMA.B3.01                    &     1.466    &   2--1   &      \nodata\        &     370$\pm$ 20  &    ALMA   &      1.99$\pm$0.11   &      0.37 $\pm$0.37   &    6.09   &      4.26 $\pm$0.61   &   22   \\
                 ALMA.B3.02                    &     1.450    &   2--1   &      \nodata\        &     310$\pm$ 80  &    ALMA   &      0.45$\pm$0.09   &      0.33 $\pm$0.16   &    5.03   &      2.52 $\pm$0.50   &   22   \\
                 ALMA.B3.03                    &     1.453    &   2--1   &      \nodata\        &     490$\pm$ 30  &    ALMA   &      2.16$\pm$0.12   &      0.26 $\pm$0.22   &    4.29   &      3.00 $\pm$0.43   &   22   \\
                 ALMA.B3.04                    &     1.466    &   2--1   &      \nodata\        &     480$\pm$110  &    ALMA   &      0.62$\pm$0.12   &      0.06 $\pm$0.03   &    5.78   &      6.36 $\pm$0.58   &   22   \\
                 ALMA.B3.05                    &     1.467    &   2--1   &      \nodata\        &     250$\pm$ 60  &    ALMA   &      0.47$\pm$0.09   &      0.50 $\pm$0.27   &    4.35   &      5.65 $\pm$0.44   &   22   \\
                 ALMA.B3.06\tablefootmark{b}   &     1.467    &   2--1   &      \nodata\        &     490$\pm$ 40  &    ALMA   &      2.12$\pm$0.12   &      1.52 $\pm$0.72   &    4.18   &     10.87 $\pm$0.84   &   22   \\
                 ALMA.B3.07                    &     1.452    &   2--1   &      \nodata\        &     480$\pm$ 70  &    ALMA   &      1.10$\pm$0.12   &      0.37 $\pm$0.51   &    4.88   &      8.30 $\pm$0.49   &   22   \\
                 ALMA.B3.08\tablefootmark{b}   &     1.457    &   2--1   &      \nodata\        &     360$\pm$ 40  &    ALMA   &      1.22$\pm$0.10   &      1.10 $\pm$0.31   &    4.58   &      6.87 $\pm$0.46   &   22   \\
                 ALMA.B3.09                    &     1.468    &   2--1   &      \nodata\        &     350$\pm$ 70  &    ALMA   &      0.68$\pm$0.10   &      0.49 $\pm$0.28   &    5.41   &      5.95 $\pm$0.54   &   22   \\
                 ALMA.B3.10                    &     1.454    &   2--1   &      \nodata\        &     270$\pm$ 20  &    ALMA   &      1.40$\pm$0.09   &      0.76 $\pm$0.29   &    4.98   &      6.47 $\pm$0.50   &   22   \\
                 ALMA.B3.11                    &     1.451    &   2--1   &      \nodata\        &     530$\pm$100  &    ALMA   &      0.96$\pm$0.12   &      0.18 $\pm$0.30   &    5.15   &      8.76 $\pm$0.52   &   22   \\
                 ALMA.B3.12                    &     1.445    &   2--1   &      \nodata\        &     210$\pm$ 30  &    ALMA   &      0.56$\pm$0.08   &      0.57 $\pm$0.16   &    4.92   &      3.94 $\pm$0.49   &   22   \\
                 ALMA.B3.13                    &     1.471    &   2--1   &      \nodata\        &     520$\pm$ 70  &    ALMA   &      1.06$\pm$0.12   &      0.22 $\pm$0.58   &    4.36   &     10.46 $\pm$0.44   &   22   \\
                 ALMA.B3.14                    &     1.451    &   2--1   &      \nodata\        &     480$\pm$100  &    ALMA   &      0.62$\pm$0.11   &      0.03             &    4.16   &     10.40 $\pm$0.83   &   22   \\
                 ALMA.B3.15                    &     1.465    &   2--1   &      \nodata\        &     520$\pm$ 90  &    ALMA   &      1.09$\pm$0.12   &      0.29 $\pm$0.10   &    5.50   &      3.30 $\pm$0.55   &   22   \\
                 ALMA.B3.16                    &     1.465    &   2--1   &      \nodata\        &     590$\pm$ 80  &    ALMA   &      1.36$\pm$0.13   &      0.39 $\pm$0.30   &    4.21   &      3.37 $\pm$0.42   &   22   \\
                 ALMA.B3.17\tablefootmark{b}   &     1.460    &   2--1   &      \nodata\        &     440$\pm$ 80  &    ALMA   &      0.89$\pm$0.11   &      1.29 $\pm$0.54   &    4.56   &      5.93 $\pm$0.91   &   22   \\
\multicolumn{11}{c}{COSMOS1002$+$0134 ($z$\,=\,1.550)} \\
                      51613                    &     1.517    &   1--0   &      0.20$\pm$ 0.05  &     200$\pm$ 80  &    JVLA   &      2.42$\pm$0.58   &      0.57             &    3.60   &      8.64 $\pm$2.16   &   23   \\
                      51858                    &     1.556    &   1--0   &      0.10$\pm$ 0.03  &     360$\pm$220  &    JVLA   &      1.26$\pm$0.38   &      0.95             &    3.60   &      4.32 $\pm$1.44   &   23   \\
                      51207                    &     1.530    &   1--0   &     0.085$\pm$0.021  &     300          &    JVLA   &      1.03$\pm$0.25   &      0.49 $\pm$0.28   &    3.60   &      3.70 $\pm$0.90   &   23   \\
                      51380                    &     1.551    &   1--0   &     0.090$\pm$0.021  &     300          &    JVLA   &      1.12$\pm$0.26   &      0.33 $\pm$0.17   &    3.60   &      4.00 $\pm$0.94   &   23   \\
\multicolumn{11}{c}{SpARCS\,J022546$-$035517 ($z$\,=\,1.590)} \\
                J0225$-$371\tablefootmark{b}   &     1.599    &   2--1   &      1.26$\pm$ 0.10  &     442$\pm$ 39  &    ALMA   &      5.30$\pm$0.40   &      1.82 $\pm$0.80   &    4.36   &     10.03 $\pm$2.62   &   24   \\
                J0225$-$460\tablefootmark{b}   &     1.600    &   2--1   &      0.50$\pm$ 0.05  &     388$\pm$ 44  &    ALMA   &      2.10$\pm$0.20   &      1.22 $\pm$0.63   &    4.36   &     23.11 $\pm$1.74   &   24   \\
                J0225$-$281\tablefootmark{b}   &     1.611    &   2--1   &      0.80$\pm$ 0.08  &     292$\pm$ 34  &    ALMA   &      3.40$\pm$0.30   &      1.26 $\pm$0.53   &    4.36   &      4.80 $\pm$0.87   &   24   \\
                J0225$-$541                    &     1.611    &   2--1   &      1.12$\pm$ 0.26  &     341$\pm$ 89  &    ALMA   &      4.80$\pm$1.10   &      0.86 $\pm$0.32   &    4.36   &      9.16 $\pm$0.87   &   24   \\
                J0225$-$429\tablefootmark{b}   &     1.602    &   2--1   &      0.25$\pm$ 0.05  &     283$\pm$ 65  &    ALMA   &      1.10$\pm$0.20   &      1.87 $\pm$0.87   &    4.36   &      1.74 $\pm$0.44   &   24   \\
                J0225$-$407                    &     1.599    &   2--1   &      0.26$\pm$ 0.04  &     290$\pm$ 57  &    ALMA   &      1.10$\pm$0.20   &      0.88 $\pm$0.29   &    4.36   &      4.80 $\pm$0.87   &   24   \\
                J0225$-$324                    &     1.600    &   2--1   &      0.09$\pm$ 0.02  &     193$\pm$ 61  &    ALMA   &      0.40$\pm$0.10   &      0.51 $\pm$0.28   &    4.36   &     14.82 $\pm$1.31   &   24   \\
                J0225$-$303                    &     1.596    &   2--1   &      0.55$\pm$ 0.15  &     687$\pm$222  &    ALMA   &      2.30$\pm$0.60   &      0.03 $\pm$0.03   &    4.36   &     20.93 $\pm$4.80   &   24   \\
\multicolumn{11}{c}{SpARCS\,J033057$-$284300 ($z$\,=\,1.613)} \\
                 J0330$-$57                    &     1.613    &   2--1   &      0.31$\pm$ 0.13  &     155$\pm$ 40  &    ALMA   &      1.40$\pm$0.60   &      0.38 $\pm$0.22   &    4.36   &      6.10 $\pm$2.62   &   25   \\
\multicolumn{11}{c}{SpARCS\,J022426$-$032330 ($z$\,=\,1.630)} \\
               J0224$-$3656                    &     1.626    &   2--1   &      0.30$\pm$ 0.06  &     539$\pm$113  &    ALMA   &      1.30$\pm$0.30   &      0.45 $\pm$0.21   &    4.36   &      5.67 $\pm$1.31   &   25   \\
                J0224$-$159\tablefootmark{b}   &     1.635    &   2--1   &      0.46$\pm$ 0.11  &     245$\pm$ 68  &    ALMA   &      2.00$\pm$0.50   &      2.28 $\pm$0.86   &    4.36   &      8.72 $\pm$2.18   &   25   \\
          J0224$-$3680/3624                    &     1.626    &   2--1   &      1.07$\pm$ 0.19  &     776$\pm$192  &    ALMA   &      4.70$\pm$0.80   &      0.72 $\pm$0.25   &    4.36   &     20.49 $\pm$3.49   &   25   \\
            J0224$-$396/424\tablefootmark{b}   &     1.634    &   2--1   &      1.32$\pm$ 0.12  &     493$\pm$ 53  &    ALMA   &      5.80$\pm$0.60   &      1.75 $\pm$0.63   &    4.36   &     25.29 $\pm$2.62   &   25   \\
\multicolumn{11}{c}{CLG\,J0218$-$0510 ($z$\,=\,1.630)} \\
                      30169                    &     1.629    &   1--0   &      0.06$\pm$ 0.01  &     836          &     VLA   &      0.76$\pm$0.18   &      0.29 $\pm$0.12   &    4.36   &      3.49 $\pm$0.44   &   26   \\
                      30545\tablefootmark{b}   &     1.624    &   1--0   &      0.19$\pm$ 0.01  &     351$\pm$ 12  &     VLA   &      2.55$\pm$0.18   &      1.70 $\pm$0.70   &    4.36   &     11.34 $\pm$0.44   &   26   \\
\multicolumn{11}{c}{SpARCS\,1049$+$56 ($z$\,=\,1.710)} \\
     SpARCS\,1049$+$56\,BCG\tablefootmark{b}   &     1.7091   &   2--1   &      3.60$\pm$ 0.30  &     569$\pm$ 63  &     LMT   &     11.60$\pm$1.00   &      9.03 $\pm$1.37   &    0.80   &     11.04 $\pm$0.96   &   27   \\
\multicolumn{11}{c}{Cl\,J1449$+$0856 ($z$\,=\,1.990)} \\
                         A1\tablefootmark{b}   &     1.9902   &   4--3   &    474.00$\pm$67.00  &     619$\pm$111  &    ALMA   &      0.62$\pm$0.09   &      2.38 $\pm$0.23   &    3.6    &      6.12 $\pm$0.36   &   28   \\
                         A2                    &     1.9951   &   4--3   &    307.00$\pm$54.00  &     387$\pm$ 89  &    ALMA   &      0.40$\pm$0.07   &      0.89 $\pm$0.23   &    3.5    &      3.42 $\pm$0.60   &   29   \\
                        H13                    &     1.9944   &   4--3   &    148.00$\pm$38.00  &     343$\pm$ 93  &    ALMA   &      0.19$\pm$0.05   &      0.40 $\pm$0.12   &    4.5    &      1.78 $\pm$1.78   &   29   \\
                         H6                    &     1.9832   &   4--3   &    178.00$\pm$50.00  &     541$\pm$162  &    ALMA   &      0.23$\pm$0.06   &      1.24 $\pm$0.13   &    4.2    &      2.75 $\pm$0.90   &   29   \\
                         N7                    &     1.9965   &   4--3   &    116.00$\pm$31.00  &     302$\pm$115  &    ALMA   &      0.15$\pm$0.04   &      0.27             &    4.2    &      1.26 $\pm$1.26   &   29   \\
                         B1                    &     1.9883   &   4--3   &     62.00$\pm$28.00  &     533$\pm$213  &    ALMA   &      0.08$\pm$0.04   &      0.60 $\pm$0.13   &    3.9    &      1.51 $\pm$0.76   &   29   \\
                         H3                    &     1.9903   &   4--3   &     70.00$\pm$24.00  &     252$\pm$ 83  &    ALMA   &      0.09$\pm$0.03   &      0.24 $\pm$0.09   &    4.7    &      1.35 $\pm$1.35   &   29   \\
\end{longtable}                                                                                                                                                 
\tablefoot{                
\tablefoottext{a}{\small The CO luminosity refers to the listed
transition.  Gas masses are obtained by multiplying \LpCOone\ by the
reported $\alpha_\mathrm{CO}$ value.  \LpCOone\ is obtained from the measured \LpCO\ and the
brightness temperature ratios from~\citet{bothwell13}.}
\tablefoottext{b}{\small Source selected for the comparison with the
PHz-IRAM sample because CO detected and with
log(L$_\mathrm{IR}$)$>$11.44$+$0.5${\times}z$.}
\tablefoottext{c}{\small The ALMA data of the galaxy members in CL\,J1001 have been
analyzed by both~\citet{gomez19}, and~\citet{champagne21}. In case of
objects in common, the measurements from the latter work were reported, but
the identifier from~\citet{gomez19} is added in parenthesis (e.g. S\#). In case of
an object that was also observed with the JVLA
by~\citet{wang18}, we report the JVLA data and add the identifier from~\citet{gomez19}
in parenthesis.}           

\tablebib{\small
 (1) \citet{damato20};
 (2) \citet{bothwell10};
 (3) \citet{casey16};                                                                                    
 (4) \citet{casey11};
 (5) \citet{bothwell13};
 (6) \citet{dannnerbauer17};                                                                                       
 (7) \citet{tadaki19};
 (8) \citet{gomez19};
 (9) \citet{chapman15}
(10) \citet{li21};
(11) \citet{emonts19};
(12) \citet{ivison13};
(13) \citet{lee17};
(14) \citet{wang18};
(15) \citet{champagne21};
(16) \citet{tadaki14};
(17) \citet{ivison12};
(18) \citet{umehata21};
(19) \citet{jones21};
(20) \citet{casasola13};
(21) \citet{stach17};
(22) \citet{hayashi17};
(23) \citet{aravena12};
(24) \citet{noble19};
(25) \citet{noble17};
(26) \citet{rudnick17};
(27) \citet{webb17};
(28) \citet{gobat11};
(29) \citet{coogan18}.}       
}
\end{landscape}            

\clearpage
\twocolumn

\section{The impact of the CO--H$_\mathrm{2}$ conversion factor}\label{sec:alpha_co}

\begin{figure}[h!]
\centering
\includegraphics[width=\linewidth]{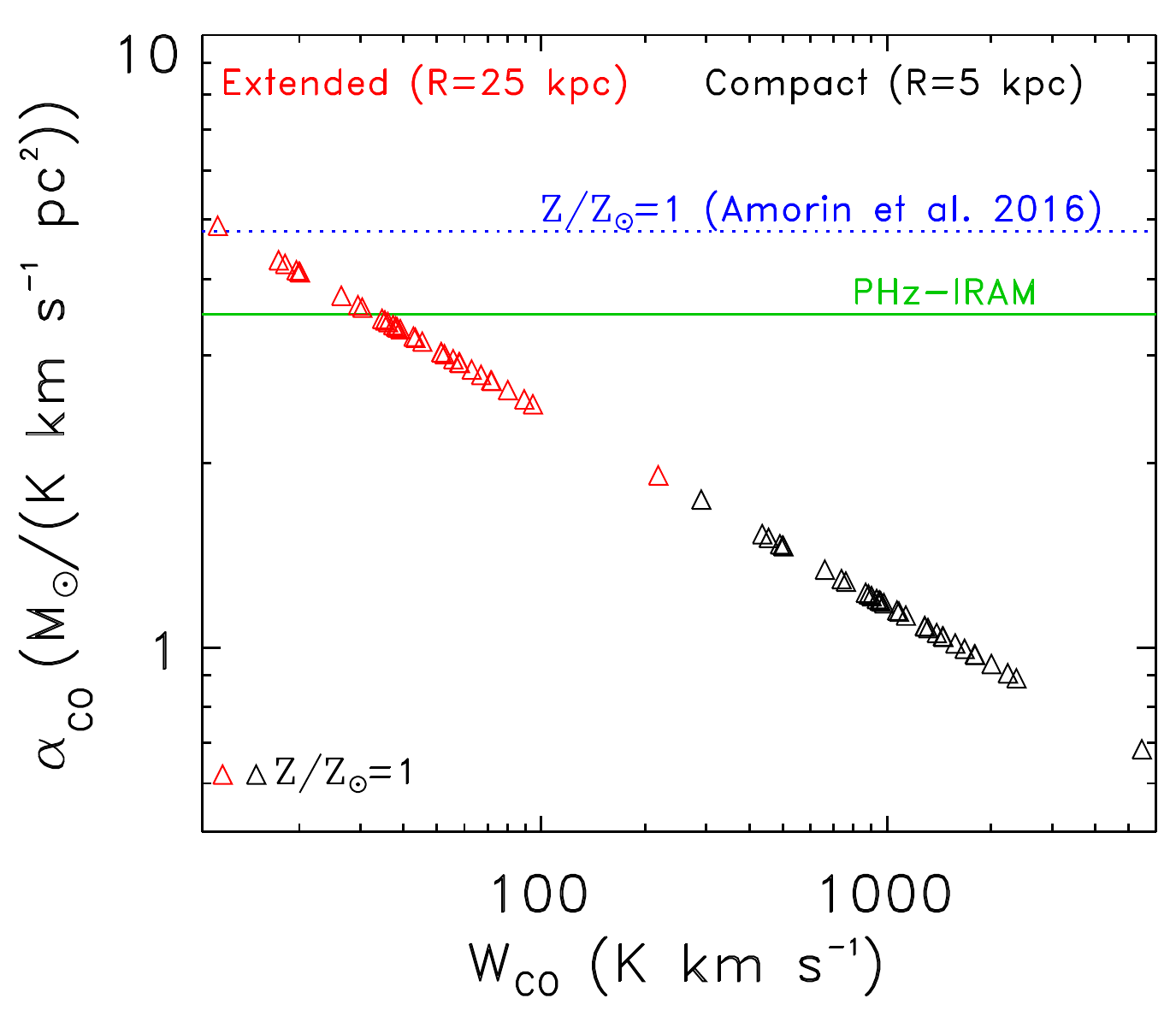}
\caption{{\small Predicted $\alpha_\mathrm{CO}$ conversion factors (open
triangles) by the functional form derived in~\citet{narayanan12} in case of
solar gas-phase metallicity and compact (radius of 5\,kpc; red), or extended
(radius of 25\,kpc; black) molecular gas distributions, as a function of CO
surface brightness given by the ratio between \LpCOone\ and the molecular
gas surface.  We also show the value adopted in this work
($\alpha_\mathrm{CO}$\,=\,3.5\,\msun\,pc$^{-2}$\,(K\,\kms)$^{-1}$; horizontal
solid green line), and the predicted value derived from the empirical relation
derived by~\citet{amorin16} assuming solar metallicity
($\alpha_\mathrm{CO}$\,=\,4.8\,\msun\,pc$^{-2}$\,(K\,\kms)$^{-1}$; horizontal
dotted blue line).}}
\label{fig:alpco_wco}
\end{figure}

Using simulations, \citet{narayanan12} showed that the $\alpha_\mathrm{CO}$
conversion factor covers a wide range of values depending on several
physical parameters, and in particular, on the molecular gas surface
brightness and on the gas-phase metallicity\footnote{According
to~\citet{narayanan12}, the CO conversion factor can be expressed as
min(6.3,10.7$\times$W$_\mathrm{CO}^{-0.32}$)/({\tt Z}/{\tt
Z$_{\odot}$})$^{0.65}$ where W$_\mathrm{CO}$ is the CO surface brightness
for a uniformly distributed molecular gas luminosity in units of (K\,\kms),
and {\tt Z}/{\tt Z$_{\odot}$ is the stellar metallicity in solar
units.}}.  Since we do not have measurements of CO surface brightness
(W$_\mathrm{CO}$) nor of the gas-phase metallicity, we approximate the
former as \LpCOone/area.  We consider a compact and an extended area given
by circular regions with radii of R$_\mathrm{gas}$\,=\,5, and 25\,kpc,
respectively, and we assume solar metallicity, even though galaxies
at $z\sim$2, and the members of one PHz protocluster for which
stellar metallicities have been measured exhibit sub-solar
metallicities~\citep{erb06b,polletta21}.  The predicted $\alpha_\mathrm{CO}$
values for our sources obtained assuming compact and extended molecular gas
distributions as a function of CO surface brightness are shown in
Fig.~\ref{fig:alpco_wco}.  We also show the value of
$\alpha_\mathrm{CO}$\,=\,3.5\,\msun\,pc$^{-2}$\,(K\,\kms)$^{-1}$ adopted in
this work. The predicted $\alpha_\mathrm{CO}$ values in case of extended
distribution range between 1.9 and 4.9, with a mean value of
3.3$\pm$0.1, close to the assumed value.  Assuming a different
$\alpha_\mathrm{CO}$, within such a range, would result in a gas mass a
factor of 1.8 smaller or 1.4 larger, at the most. Note that the
$\alpha_\mathrm{CO}$ predicted by the relation that depends only on the
metallicity~\citep[e.g.  ][]{genzel12,amorin16} would be 4.8, assuming solar
metallicity.  This choice would yield gas masses, and depletion times
systematically larger by a factor of 1.4.

Lower metallicities would entail higher $\alpha_\mathrm{CO}$ values, and
unreasonably high gas masses.  On the other hand, the inferred gas masses,
under the assumption of a $\alpha_\mathrm{CO}$\,=\,3.5, agree with the
scaling relations, suggesting that the metallicities of our sources might be
already solar.  This might be due to an observational bias since it is
challenging to detect CO emission in galaxies with low metallicities. 
We conclude that the gas mass estimates might differ, at the most, by
$+$0.15\,dex, or $-$0.27\,dex compared to those we derive assuming
$\alpha_\mathrm{CO}$\,=\,3.5.

\end{document}